\documentclass[twocolumn]{aastex61}
\usepackage{amsmath}
\usepackage{amssymb}
\usepackage{graphicx}

\usepackage{morefloats}

\usepackage{natbib}
\newcommand{\nsweeps}{339,193}
\newcommand{\nbts}{400,424}
\newcommand{\nmatched}{47,537}
\newcommand{\nkept}{9,700}

\newcommand{\alphaUse}{6.44}
\newcommand{\betaUse}{1.10}

\newcommand{\mindex}{\ensuremath{\rm{[m]}}}
\newcommand{\tindex}{\ensuremath{\rm{[t]}}}

\newcommand{\wik}{\ensuremath{w_{ik}}}
\newcommand{\probThresh}{\ensuremath{0.8}}

\newcommand{\rv}{\ensuremath{R_V}}

\newcommand{\nFeatures}{\ensuremath{q}}
\newcommand{\nComponents}{\ensuremath{K}}
\newcommand{\nParsMeans}{\ensuremath{\nFeatures \times \nComponents}}

\newcommand{\aic}{\ensuremath{AIC}}
\newcommand{\bic}{\ensuremath{BIC}}

\newcommand{\feh}{\ensuremath{\rm{\left[Fe/H\right]} }}
\newcommand{\mh}{\ensuremath{\rm{\left[M/H\right]} }}
\newcommand{\teff}{\ensuremath{T_{\rm{eff}}}}
\newcommand{\logg}{\ensuremath{\log(g)}}

\newcommand{\Si}{\ensuremath{{\mathbf{S}_i}}}
\newcommand{\thetak}{\ensuremath{\mathbf{\theta}_k}}
\newcommand{\mtindices}{\tindex, \mindex}

\newcommand{\muL}{\ensuremath{\mu_l}}
\newcommand{\muB}{\ensuremath{\mu_b}}
\newcommand{\muLcent}{\ensuremath{\langle \muL \rangle}}
\newcommand{\vL}{\ensuremath{v_l}}
\newcommand{\vLcent}{\ensuremath{\langle \vL \rangle}}

\newcommand{\muLcentMR}{\ensuremath{\muLcent_{\rm MR}}}
\newcommand{\muLcentMP}{\ensuremath{\muLcent_{\rm MP}}}

\newcommand{\vB}{\ensuremath{v_b}}

\newcommand{\vvec}{\ensuremath{\vec{v}}}
\newcommand{\muvec}{\ensuremath{\vec{\mu}}}

\newcommand{\muvecref}{\ensuremath{\vec{\mu}_{f,0}}}

\newcommand{\muvecrefshift}{\ensuremath{\Delta \muvecref}}
\newcommand{\muvecrefshiftval}{\ensuremath{(0.3, 0.1)}}

\newcommand{\kmpersecperkpc}{\ensuremath{{\rm km}~{\rm s}^{-1}~{\rm kpc}^{-1}}}
\newcommand{\trendvecmag}{\ensuremath{0.8~\kmpersecperkpc}}

\newcommand{\muMean}{\ensuremath{\vec{\mu}_0}}
\newcommand{\muCovar}{\ensuremath{\mathbf{V}_{\mu}}}

\newcommand{\EBmV}{\ensuremath{E(B-V)}}
\newcommand{\fbin}{\ensuremath{f_{bin}}}
\newcommand{\qmin}{\ensuremath{q_{min}}}
\newcommand{\basti}{BaSTI}
\newcommand{\dmbin}{\ensuremath{\Delta m_{bin}}}
\newcommand{\fdmbin}{\ensuremath{f(\dmbin)}}

\newcommand{\edGMM}{{\tt extreme-deconvolution}}
\newcommand{\xdGMM}{{\tt scikit-learn XDGMM}}

\newcommand{\MR}{``metal-rich''}
\newcommand{\MP}{``metal-poor''}

\newcommand{\rdmLong}{relative photometric parallax}
\newcommand{\rdmLongs}{relative photometric parallaxes}
\newcommand{\rdm}{\ensuremath{\pi'}}

\newcommand{\medrdm}{\ensuremath{\langle \rdm \rangle}}
\newcommand{\medrdmMR}{\ensuremath{\medrdm_{\rm EB,MR}}}
\newcommand{\medrdmMP}{\ensuremath{\medrdm_{\rm EB,MP}}}

\newcommand{\smag}{\ensuremath{s_{\rm mag}}}

\newcommand{\dKpc}{\ensuremath{D}}

\newcommand{\dzer}{\ensuremath{\dKpc_0}}
\newcommand{\dKpcCen}{\ensuremath{\dzer = 7.76~\rm{kpc}}}

\newcommand{\deltapar}{\ensuremath{\Delta \rdm}}
\newcommand{\deltaparBri}{\ensuremath{\deltapar_-}}
\newcommand{\deltaparFai}{\ensuremath{\deltapar_+}}

\newcommand{\muCenMRl}{+0.019}
\newcommand{\muCenMRb}{+0.19}
\newcommand{\muCenMRn}{381}
\newcommand{\muCenMPl}{-0.12}
\newcommand{\muCenMPb}{+0.32}
\newcommand{\muCenMPn}{290}

\newcommand{\muCenDeltal}{+0.14}
\newcommand{\muCenDeltab}{-0.13}

\newcommand{\muCenLabel}{\ensuremath{\left(\mu_{l}, \mu_{b}\right)^{0}}}
\newcommand{\muCenMRlabel}{\ensuremath{\muCenLabel_{MR}}}
\newcommand{\muCenMPlabel}{\ensuremath{\muCenLabel_{MP}}}

\newcommand{\masperyear}{mas yr$^{-1}$}

\newcommand{\trendVvec}{\ensuremath{d\vec{v} / d\dKpc}}
\newcommand{\trendVvecShift}{\ensuremath{\Delta \left(\trendVvec\right)}}

\newcommand{\symbolgrad}{\ensuremath{B}}
\newcommand{\symbolampl}{\ensuremath{A}}
\newcommand{\ratiograd}{\ensuremath{\symbolgrad_{\rm MR}/\symbolgrad_{\rm MP}}}
\newcommand{\ratioampl}{\ensuremath{\symbolampl_{\rm MR}/\symbolampl_{\rm MP}}}
\newcommand{\ratioamplL}{\ensuremath{(\ratioampl)_l}}

\newcommand{\ratiogradL}{\ensuremath{(\ratiograd)_l}}
\newcommand{\ratiogradB}{\ensuremath{(\ratiograd)_b}}

\newcommand{\slopeRatio}{\ensuremath{3.70 \pm 0.68}}
\newcommand{\slopeRatioB}{\ensuremath{0.90 \pm 0.87}}

\newcommand{\amplRatio}{\ensuremath{2.29 \pm 0.35}}

\newcommand{\slopeRatioSignif}{\ensuremath{5.4\sigma}}
\newcommand{\amplRatioSignif}{\ensuremath{6.5\sigma}}

\newcommand{\BTS}{BTS}
\newcommand{\BTSone}{\BTS v1}
\newcommand{\BTStwo}{\BTS v2}
\newcommand{\SWEEPS}{SWEEPS}
\newcommand{\HST}{HST}
\newcommand{\ACSWFC}{ACS/WFC}
\newcommand{\ACSHRC}{ACS/HRC}
\newcommand{\WFPC}{WFPC2}
\newcommand{\WFC}{WFC3}
\newcommand{\MSTO}{Main Sequence Turn-off}
\newcommand{\GaiaOne}{Gaia DR1}

\newcommand{\filtC}{F390W}
\newcommand{\filtV}{F555W}
\newcommand{\filtVbroad}{F606W}
\newcommand{\filtI}{F814W}
\newcommand{\filtJ}{F110W}
\newcommand{\filtH}{F160W}

\defcitealias{clarkson08}{Cl08}
\defcitealias{calamida14}{Ca14}
\defcitealias{calamida15}{Ca15}
\defcitealias{sahu06}{Sa06}

\begin{document}

\shorttitle{Bulge rotation curves from main-sequence proper motions}
\shortauthors{Clarkson, Calamida, Sahu, Brown, Gennaro, Avila, Valenti, et al.}

\title{Chemically-dissected rotation curves of the Galactic Bulge from Main Sequence proper motions\footnote{Based on observations made with the NASA/ESA Hubble Space Telescope and obtained from the data archive at the Space Telescope Science Institute. STScI is operated by the Association of Universities for
Research in Astronomy, Inc. under NASA contract NAS 5-26555.}}

\correspondingauthor{Will Clarkson}
\email{wiclarks@umich.edu}

\author{William I. Clarkson}
\affiliation{Department of Natural Sciences, University of Michigan-Dearborn, 4901 Evergreen Rd. Dearborn, MI, 48128, USA}

\author{Annalisa Calamida}
\affiliation{Space Telescope Science Institute, 3700 San Martin Drive, Baltimore, MD 21218, USA}

\author{Kailash C. Sahu}
\affiliation{Space Telescope Science Institute, 3700 San Martin Drive, Baltimore, MD 21218, USA}

\author{Thomas M. Brown}
\affiliation{Space Telescope Science Institute, 3700 San Martin Drive, Baltimore, MD 21218, USA}

\author{Mario Gennaro}
\affiliation{Space Telescope Science Institute, 3700 San Martin Drive, Baltimore, MD 21218, USA}

\author{Roberto Avila}
\affiliation{Space Telescope Science Institute, 3700 San Martin Drive, Baltimore, MD 21218, USA}

\author{Jeff Valenti}
\affiliation{Space Telescope Science Institute, 3700 San Martin Drive, Baltimore, MD 21218, USA}

\author{Victor P. Debattista}
\affiliation{Jeremiah Horrocks Institute, University of Central Lancashire, Preston PR1 2HE, UK}

\author{R. Michael Rich}
\affiliation{Division of Astronomy \& Astrophysics, University of California, Los Angeles, 430 Portola Plaza, Box 951547, Los Angeles, CA 90095-1547, USA}

\author{Dante Minniti}
\affiliation{Departamento de Ciencias F\`isicas, Facultad de Ciencias Exactas, Universidad Andr\'es Bello, Av. Fern\'andez Concha 700, Las Condes, Santiago, Chile}
\affiliation{Vatican Observatory, V00120 Vatican City State, Italy}
\affiliation{Millennium Institute of Astrophysics, Av. Vicu\~{n}a Mackenna 4860, 782-0436 Macul, Santiago, Chile}

\author{Manuela Zoccali}
\affiliation{Millennium Institute of Astrophysics, Av. Vicu\~{n}a Mackenna 4860, 782-0436 Macul, Santiago, Chile}
\affiliation{Instituto de Astrof\`isica, Pontificia Universidad Cat\`olica de Chile, Av. Vicu\~na Mackenna 4860, Santiago, Chile}

\author{Emily R. Aufdemberge}
\affiliation{Department of Natural Sciences, University of Michigan-Dearborn, 4901 Evergreen Rd. Dearborn, MI, 48128, USA}

\begin{abstract}

We report results from an exploratory study implementing a new probe of
Galactic evolution using archival Hubble Space Telescope imaging observations.
Precise proper motions are combined with photometric relative
metallicity and temperature indices, to produce the proper motion
rotation curves of the Galactic bulge separately for metal-poor and
metal-rich Main Sequence samples. This provides a ``pencil-beam''
complement to large-scale wide-field surveys, which to-date have
focused on the more traditional bright Giant Branch tracers. 

We find strong evidence that the Galactic bulge rotation curves drawn
from \MR~and \MP~samples are indeed discrepant. The \MR~sample shows
greater rotation amplitude and a steeper gradient against line of
sight distance, as possibly a stronger central concentration along the
line of sight. This may represent a new detection of differing orbital
anisotropy between metal-rich and metal-poor bulge objects.  We also
investigate selection effects that would be implied for the
longitudinal proper motion cut often used to isolate a ``pure-bulge''
sample. Extensive investigation of synthetic stellar populations
suggest that instrumental and observational artefacts are unlikely to
account for the observed rotation curve differences.

Thus, proper motion-based rotation curves can be used to probe
chemo-dynamical correlations for {\it Main Sequence} tracer stars,
which are orders of magnitude more numerous in the Galactic Bulge than
the bright Giant Branch tracers. We discuss briefly the prospect of
using this new tool to constrain detailed models of Galactic formation
and evolution.

\end{abstract}


\section{Introduction}

The diversity of observed properties of the Galactic bulge has
challenged attempts to provide a coherent explanation for its
formation and subsequent development. For example, while
color-magnitude diagrams suggest the majority of bulge stars are
likely older than $\sim 8$~Gy~(e.g. \citealt{zoccali03},
\citealt{kr02}, \citealt{clarkson08}, \citealt{calamida14}, although
see, e.g. \citealt{nataf12}, \citealt{haywood16} and
\citealt{bensby17} for alternative interpretation), minority
populations of younger objects have been detected
(e.g. \citealt{sevenster97}; \citealt{vanloon03}). That measurements of
even bulk parameters like bar orientation and axis ratio have not
converged with time \citep[e.g.][]{vanhollebeke09} is consistent with
a dependence of these properties on the ages of the tracers used. For
example, \citet{catchpole16} find distinct bar/bulge spatial
structures coexisting in the same volume, traced by Mira populations
of different estimated ages. As shown by \citet{ness13_part2}, the various
apparent observational contradictions may be resolved by a scenario in
which most bulge stars did indeed form early but later were rearranged
into their present-day spatial and kinematic distributions by
disk-driven evolution. Recent reviews of Galactic bulge observations
and formation scenarios include \citet{rich15}, \citet{babusiaux16},
\citet{zoccali16} and \citet{nataf17}.

Observations have long suggested a co-dependence between chemical
abundance and kinematics in the bulge, particularly as traced by
velocity dispersion, providing an observational test of formation and
evolution scenarios (e.g. \citealt{rich1990, minniti1996}). Metal-rich
samples show a steeper increase in radial velocity dispersion with
Galactic latitude than do the metal-poor objects (whose
dispersion-latitude profile at latitude $|b| \gtrsim 4^{\circ}$ is
only gently sloped and may be flat). While differences exist in the
literature as to the \feh~cuts used to define the two samples, for
latitudes $|b| \lesssim 3^{\circ}$~the metal-poor and metal-rich
samples have consistent radial velicity dispersions (Figure 4 of
\citealt{babusiaux16} presents a recent compilation for fields along
the Bulge minor axis). For the very inner-most fields in the Bulge
($|b| \lesssim 1.0^{\circ}$~and $|l| \lesssim 2^{\circ}$), a radial
velocity dispersion ``inversion'' may even be present (an expression
of a steeper dispersion gradient with longitude for metal-rich
objects), with the metal-rich stars showing {\it greater} velocity
dispersion than the metal-poor objects in bins closest to the Galactic
center \citep[e.g.][]{babusiaux14,zoccali17}.

Turning to proper motions, \citet{spaenhauer92} traced the proper
motion dispersion for a sample of 57 Bulge giants towards Baade's
window, allowing the first test of Bulge chemical and kinematic
co-dependence using proper motions. No statistically significant
discrepancy in proper motion dispersion was found between metal-poor
(defined as $\feh < 0.0$) and metal-rich ($\feh > 0.0$) objects (with
Galactic latitudinal proper motion dispersion difference $\Delta
\sigma_{\mu,l} \approx 0.5 \pm 0.6$~mas yr$^{-1}$~between the
samples), although the sample size was not large. \citet{zhao94}
combined the \citet{spaenhauer92} ground-based proper motions with
published radial velocities and metallicities to demonstrate a break
in vertex deviation near $\feh \sim -0.5$.  \citet{soto07, soto12}
demonstrated consistent variation of vertex deviation using
\HST~proper motions for bright giants (for which spectroscopic
abundances and radial velocities completed the set of observational
parameters; \citealt{babusiaux16} shows a more recent compilation of
vertex deviation as a function of metallicity).

The implications of observational chemical-dynamical correlations for
formation models of the inner Milky Way are the subject of vigorous
ongoing observational and theoretical research. For example,
\citet{debattista17} showed that samples drawn from a continuous
metallicity distribution in a pure-disk galaxy model can be
``kinematically fractionated'' by bar formation into metal-rich and
metal-poor populations with quite different morphology and dynamics,
depending on their initial (Galactocentric) radial velocity
dispersions. (In this scenario, radial velocity dispersion and
metallicity each correlate with the time at which the population
formed; thus, they correlate with each other.) This is consistent with
the tendency of the ``X''-shape to be preferentially populated by
metal-rich stars (e.g. \citealt{vasquez13}, although the magnitude of
this preference is somewhat uncertain, e.g. \citealt{nataf14}). Bias
in the ``X'' shape towards metal-rich stars has now also been
  observed in NGC 4710, a nearby disk-dominated galaxy viewed almost
  edge-on \citep{gonzalez16, gonzalez17}.

\citet{shen10} argue that the radial velocities and morphology
of Bulge stellar populations show no need for a substantial spheroidal
``Classical'' bulge component (at the level of $\lesssim 8\%$~of the
disk mass), arguing that the Milky Way can be characterized as a
pure-disk galaxy. Nonetheless, a small spheroidal component probably
{\it has} been detected, although its likely contribution to the total
Bulge mass is likely well under 10\% \citep{kunder16}. Interpretation
of this component in the context of Galactic formation is not clear;
it might, for example, represent part of the Halo population that has
also probably been detected in the inner Milky Way \citep{koch16}.

\subsection{Does bulge rotation depend on metallicity?}
\label{ss:introBifurcation}

In addition to velocity dispersion trends, the trend in bulge mean
radial velocity (against Galactic longitude or Galactocentric radius)
might also be expected to vary with metallicity, but here the
magnitude (or even existence) of such a dependence is less clear.
Earlier spectroscopic surveys suggest a clear difference between
metal-poor and metal-rich samples. For example, \citet{harding93} and
\citet{minniti1996} demonstrated that ``metal-rich'' stars show a
gradient in circular speed with Galactocentric radius, consistent with
the ``solid body''-type rotation traced by planetary nebulae
\citep{kinman88}, Miras \citep{menzies90} and SiO masers
\citep{nakada93}. In contrast, metal-poor objects (using $\feh
\lesssim -1.0$, and thus likely including a large contribution from
the inner halo) showed no strong evidence for a rotational trend. More
recently, \citet{kunder16} found that their metal-poor RR Lyrae sample
with mostly sub-solar metallicities ($-2.4 < \feh \lesssim +0.3$,
peaking at $\feh \sim -1.0$) shows no strong signature of rotation
from radial velocities in any Galactic latitude range. This is in
contrast to the majority-bulge population, which shows bulk rotation
with amplitude $v_{GC} \pm \approx 80$~km s$^{-1}$~progressing from
the first to fourth Galactic quadrant \citep[e.g.][]{howard09,
  kunder16}. This rotation-free component is estimated to be a rather
small part of the overall bulge stellar population \citep{kunder16}.

Restricting attention to $\feh \gtrsim -1.0$~(to sample mainly bulge
and disk stars), the body of more recent spectroscopic studies does
not show strong evidence for metallicity dependence of radial velocity
rotation curve (usually plotted against Galactic longitude). For
example, the {\it ARGOS} survey \citep{ness13_part3} and the {\it
  Gaia-ESO} survey \citep{williams16} each show no strong difference
between metal-rich and metal-poor bulge objects (the studies use
slightly different cuts for metal-rich and metal-poor
objects). However, the {\it Giraffe Inner-Bulge Survey} ({\it GIBS},
which is unusual among the spectroscopic studies in reaching as close
as $b=-2^{\circ}$~to the Galactic mid-plane) shows a possible
difference in rotation curve slope between objects at $\feh <
-0.3$~and $\feh > +0.2$, however at about 1.5$\sigma$~significance,
the difference is not yet compelling \citep{zoccali17}. Thus, the
radial velocity surveys focusing on the majority bulge population
(with $\feh \gtrsim -1.0$) show no strong metallicity dependence in
the trends of mean radial velocity against Galactic longitude.

Proper motions offer an independent method to kinematically chart the
bulge rotation curves, and, if information on chemical composition is
available, explore whether multiple abundance-samples really do show
distinct {\it mean} motions as well as the well-established velocity
dispersion differences. 

To-date, proper motion investigations in the context of multiple
populations (or a continuum) have mostly been performed using bright
giants. For example, in addition to the vertex deviation
investigations reported in the previous section, proper motions of
bulge giants using OGLE \citep{poleski13}~and with the Wide Field
Imager on the La Silla 2.2m telescope \citep{vasquez13} have been used
to uncover azimuthal streaming in the bulge X-shaped
structure. However, \citet{qin15} caution via N-body models that the
underlying bar pattern speed cannot directly be constrained just from
the nearside/farside longitudinal proper motion difference.

The above radial velocity and proper motion studies all use bright
giants as tracers, often Red Clump Giants (RCGs), which are much less
spatially crowded from the ground than are Main Sequence (MS)
objects. This causes them to be limited by the small intrinsic
population size per field of view. For example, {\it ARGOS} typically
observed about 600 stars at $\feh > -1.0$~per $2^{\circ}$-diameter
field of view; \citep{ness13_part3}. Thus, mean velocities interpreted
for rotation trends represent averages both over quite large angular
regions on the sky, and, more importantly, over the entire distance
range along the line of sight.

To make further progress, an independent measure of bulge rotation is
needed, using a tracer sample sufficiently populous that the sample
can be dissected by line-of-sight distance to mitigate the statistical
limitations of giant-branch tracers. MS tracers are orders of
magnitude more common on the sky, affording the opportunity to dissect
a single sight line along the line of sight, thus offering a
``pencil-beam'' complement to the wide-field surveys that use the
bright end of the color magnitude diagram.\footnote{Indeed, bulge
  giants are so bright that they can be challenging to precisely and
  efficiently measure from space.}

It is the charting of the chemically-dissected Bulge rotation curve
from MS proper motions that we report here. Because this is a
relatively new technique, we briefly review the short literature in MS
proper motion bulge rotation curve determination before proceeding
further.

\subsection{Proper motions of Main Sequence bulge populations}
\label{ss:pmMainSequence}

Proper motion-based rotation curves\footnote{Throughout, the {\it
    rotation curve} is defined as the run of the mean proper motion
  (or transverse velocity) against \rdmLong~(or distance). The run of
  proper motion dispersion (or velocity disperson) is referred to as
  the {\it dispersion curve}. The rotation curve is distinct from the
  {\it circular speed curve}~(the run of circular speed about the
  Galactic center against distance from the Galactic center), which
  requires projection to Cylindrical Galactic co-ordinates and an
  assumption of the orbit shape.}  from {\it Main Sequence} bulge
stars are relatively rare in the literature. \footnote{For
    clarity of presentation, here and throughout we define the
    ``nearside'' of the bulge to be the sample closer to the observer
    than the bulge midpoint along the line of sight, and the
    ``farside'' to be its counterpart farther than this midpoint. Main
    sequence stars on the bulge nearside can thus generally be
    distinguished from their counterparts on the farside by
    photometric parallax.} \citet{kr02} were the first to demonstrate
the approach for MS populations, for both the Baade and Sagittarius
Windows, presenting the HST/WFPC2-derived rotation and dispersion
curves against photometric parallax (with photometric parallax
determined as a linear combination of color and magnitude in order to
remove the color-magnitude slope of the MS tracer population of
interest). This demonstrated a clear sense of rotation, with the
nearside of the bulge showing positive mean longitudinal proper motion
relative to the farside \citep[a determination made {\it before} the
  much brighter RCGs were used to show Bulge rotation from proper
  motions;][]{sumi04}. The proper motion dispersion showed a slight
increase in the most populous middle bins of photometric parallax
(most strongly pronounced in the latitudinal proper motion dispersion
$\sigma_b)$~for their Sagittarius-Window field. \citet{kuijken04}
presented an extension of this work to multiple fields across the
bulge, including the use of three minor-axis fields to estimate the
vertical gravitational acceleration along the Galactic minor axis.

\citet{koz06}~were able to demonstrate similar behavior to the
\citet{kr02}~rotation curves in their analysis of proper motions in
Baade's Window. This was the only field for which a sufficiently
large sample of sufficiently precisely-measured MS stars could be
measured from their large 35-field study (which used \WFPC~for
early-epoch and \ACSHRC~for late-epoch observations). While their
dispersion curve is consistent with a flat distribution, the rotation
trend in galactic longitude was clearly observed. \citet{koz06} may
also have been the first to detect the weak trend in latitudinal
proper motion $\mu_b$~due to Solar reflex motion (see
\citealt{vieira07} for discussion of this effect, including its
detection using sets of ground-based observations of bulge giants over
a 21-year time-baseline). In any case, \citet{koz06} were the first to
detect the proper motion correlation $C_{l,b}$~at statistical
significance from any population (using the RCGs that formed their
main target population), using it to constrain the tilt-angle of the
Bulge velocity ellipsoid. As they point out, detection of
$C_{l,b}$~(or equivalently the orientation angle $\phi_{lb}$~of the
proper motion ellipsoid) allows constraints to be placed on the orbit
families for bulge populations, although the conversion from
observation to physical constraint is not simple
\citep[e.g.][]{zhao94, haefner00, rattenbury07}.

\citet[][hereafter \citetalias{clarkson08}]{clarkson08} extended the
rotation curve approach, using a much deeper dataset with
\ACSWFC~towards the Sagittarius Window, estimating photometric
parallax directly with reference to a fiducial isochrone describing
the average population in the color-magnitude diagram. Consistent with
\citet{kr02}~and \citet{koz06}, this showed a clear sense of rotation
in Galactic longitude, a clear detection of the latitudinal proper
motion trend from nearside to farside, and a pronounced peak in the
velocity dispersion of both coordinates ($\sigma_l$~and $\sigma_b$)
coincident with the most densely-populated section of the photometric
distance-range of the sample. \citetalias{clarkson08} converted proper
motions to velocities, charting the run of the mean velocity (i.e.,
the rotation curves), the semiminor and semimajor axis lengths
(i.e. the velocity dispersions) and the variation of the orientation
$\phi_{lb}$~of the projected velocty ellipse with line of sight
distance, and verified through simulation and comparison with the
behavior of RCGs that indeed distance effects {\it are}
observable in MS photometric parallax (though unlike RCG tracers,
unresolved binaries blur somewhat the inferred distances for a given
main-sequence population).

More recently, in a careful study of three off-axis Bulge fields using
\WFPC~for early-epoch observations and \ACSWFC~for the late epoch,
\citet{soto14} were able to extract the rotation curve (and associated
proper motion dispersion curves) for a field farther from the
mid-plane, at $(l,b) = (+3.58^{\circ}, -7.17^{\circ})$.\footnote{This
  was the only field of the three analyzed by \citet{soto14} with a
  sufficient number of well-measured stars to produce the rotation
  curve from proper motions.}  \citet{soto14} also computed the run of
velocity ellipse orientation $\phi_{lb}$~with photometric distance,
finding trends consistent with \citetalias{clarkson08}. The
kinematics of main-sequence objects at some distance from the plane,
were thus established to be broadly similar to those at the more
central Baade and Sagittarius Window fields.

Ground-based surveys are now starting to measure proper motions for
main sequence bulge objects. For example, proper motions from the VVV
survey have already been used to draw proper motion rotation curves
for {\it both} giant-branch and upper main-sequence populations
(although the upper main sequence population shows much higher proper
motion scatter and substantially different selection effects compared
to the giants; \citealt{smith18}).

The lack of metallicity information for MS populations has limited
both the measurement accuracy and scientific applicability of MS
proper motion rotation curves. The \feh~spread for bulge populations
contributes a scatter of up to $\sim 1$~magnitude on the main sequence
(e.g. \citealt{haywood16}), competing with the photometric parallax
signal due to the intrinsic distance distribution along the line of
sight. While comparison with the behavior of RCGs suggests that indeed
the rotation curve can be recovered, a lack of \feh~information for
the MS tracers contributes to substantial mixing in photometric
parallax that can dilute the signature of underlying rotation
(\citealt{clarkson08}).  Conversely, charting bulge proper motion
rotation curves from samples partitioned by relative metallicity
allows an independent probe of the chemical and dynamical correlations
resulting from the complex formation and evolutionary processes at
work in the inner Milky Way.

\subsection{Main-sequence proper motions for multiple populations}
\label{ss:spreadMS}

Until recently, no observational dataset existed that would allow the
proper motion-based rotation curves to be charted for {\it multiple}
spatially-overlapping main-sequence metallicity samples in the Bulge,
as the relevant tracer samples (a few magnitudes beneath the \MSTO,
and well clear of the subgiant and giant branches in the CMD) are far
too faint and spatially crowded for objects to be chemically
distinguished using current spectroscopic technology.

The situation changed with the {\it WFC3 Bulge Treasury Survey}
(hereafter \BTS; \citealt{brown09}), which used three-filter flux
ratios to construct a ``temperature'' index \tindex, (a function of
\filtV, \filtJ, \filtH~magnitudes, similar to $V,J,H$), and a
``metallicity'' index \mindex~(using \filtC, \filtV,
\filtI~magnitudes, similar to Washington-$C$, $V$,$I$), with scale
factors chosen so that \tindex~and \mindex~are relatively insensitive
to reddening. This allows stars to be chemically tagged in a relative
sense by their location in \mindex, \tindex~space, down to much
fainter limits and in regions of higher spatial density than currently
allowed by spectroscopy. \citet{brown10} showed that indeed the wide
bulge metallicity range can be traced photometrically by this method,
setting \tindex~and \mindex~indices for tens of thousands of MS
objects in each of the four observed bulge fields. Inverting the
photometric indices then produced relative \feh~distributions broadly
similar to the spectroscopic indications from much brighter objects
\citep[e.g.][]{hill11, johnson13}. Computing these indices
appropriately for objects near the bulge MS turn off, \citet{brown10}
found that the candidate exoplanet hosts of the SWEEPS field
(\citealt{sahu06}) tend to pile up at the metal-rich end of the
\mindex~distribution as expected, suggesting that \mindex~is indeed
tracking metallicity. Exploitation of this unique dataset to directly
constrain the star-formation history of the bulge is ongoing (see
\citealt{gennaro15} for an example of the techniques involved).

Here we combine the relative metallicity estimates from WFC3
BTS~photometry with ultra-deep proper motions using \ACSWFC, to
construct the proper motion-based rotation curves of candidate \MP~and
``metal-rich'' MS samples, and examine whether and how the kinematics
of the two samples differ from each other. Our work represents the
first extension of chemo-dynamical studies of the bulge down to the
Main Sequence.

This paper is organized as follows. The observational datasets are
introduced in Section \ref{s:observations}, with the techniques
  used to classify samples as \MP~or \MR~and to draw rotation curves
  described in Section \ref{s:analysis}. The rotation curves
  themselves are presented in Section \ref{s:results}. Section
\ref{s:discussion} discusses the implications of our results both for
the distribution of populations within the Bulge and proper motion
sample selection, and discusses the impact of various systematic
effects, with conclusions outlined in Section
\ref{s:conclusions}. Appendices
\ref{a:unctyPM}-\ref{s:rotCurveTables} provide supporting information,
including the full set of results in tabular form.

\section{Observations}
\label{s:observations}

By the standards of modern proper motion measurements with
\HST~\citep[e.g.][]{sahu17}, the relative streaming motions of the
near- and farside bulge populations are not small; the mean motion of
the bulge nearside being typically $\Delta \mu_l \sim
2$~\masperyear~relative to the farside, while the foreground disk is
separated from the bulge by relative proper motion $\Delta \mu_l \sim
4$~\masperyear, although the intrinsic proper motion dispersion is of
roughly similar magnitude; \citep{calamida14}. Thus, extraction of
proper motion-based rotation curves should in general be reasonably
straighforward for many bulge fields for which multiple epochs are
available.

For this exploratory study, however, we choose the deepest and most
precisely-measured sample of \HST~proper motions available towards the
Bulge, to minimize complications due to completeness effects and
varying measurement uncertainty. This is the \SWEEPS~dataset, which,
with many epochs over a 9-year time-baseline, represents the current
state-of-the-art in space-based proper motion measurement towards the
bulge with \HST~(e.g. \citealt{calamida15}, \citealt{kains17}). We
attached \SWEEPS~proper motions (\autoref{ss:obsSWEEPS}) to the
\BTS~photometry (\autoref{ss:obsBTS}), to afford the maximum
sensitivity to proper motions for populations that we can label
chemically in a relative sense. \autoref{tab:obsSummary} summarizes
the observations.

\autoref{fig:orientation} presents a finding chart. The observations
cover a single \ACSWFC~field of view ($\sim 3.4' \times 3.4'$) in the
Sagittarius Window, a low-reddening region ($\EBmV \approx$~0.5-0.7,
depending on the reddening prescription; e.g. \citetalias{calamida15})
that is close in projection to the Galactic center ($l,b =
1.26^{\circ}, -2.65^{\circ}$).

% Table: observations summary
\begin{deluxetable*}{clccccc}
\tabletypesize{\scriptsize}
\tablecaption{Provenance of the observational datasets used in this work. $N_{\rm all}$~represents the number of objects in each catalog (with measurements in all filters for \SWEEPS~and \BTS). The median Modified Julian Dates are indicated for the 2004 and the 2011-2012-2013 SWEEPS epochs. The \SWEEPS~field lies at $(\alpha, \delta)_{\rm J2000.0} \approx$~(17:59:00.7, -29:11:59.1), or $(l, b)_{\rm J2000.0} \approx (+1.26^{\circ}, -2.65^{\circ})$. \label{tab:obsSummary}}
%\tablecolumns{6}
%\label{tab:obsSummary}
\tablehead{
\colhead{Dataset} & \colhead{Program (PI)} & \colhead{Observation dates} & \colhead{Instrument} & \colhead{Filters or wavelength range} & \colhead{$N_{\rm all}$} & \colhead{Section}
}
\startdata
\SWEEPS & HST GO-9750 (Sahu)  & 2004 Feb (MJD 53060) & HST-ACS/WFC   & F606W, F814W          & 339,193 &  \autoref{ss:obsSWEEPS} \\
        & HST GO-12586 (Sahu)  & 2011 Oct - 2013 Oct &   &          &  &  \\
        & HST GO-13057 (Sahu)  & (MJD 56333)         &   &          &  &  \\
\hline
\BTS    & HST GO-11664~(Brown)      & 2010 May                      & HST-WFC3/UVIS & F390W, F555W, F814W   & 52,596 & \autoref{ss:obsBTS} \\
        &                           &                               & HST-WFC/IR    & F110W, F160W          &      &  \\
\hline
VLT & ESO 073.C-0410(A) & 2004 June                 & VLT-UT2/UVES  & $4812-5750$\AA & 123 & {\bf Appendix} \ref{ss:obsVLT}  \\
    & (Minniti) &                  &   & $5887-6759$\AA & &  \\
\hline
\enddata
\end{deluxetable*}

\subsection{\SWEEPS~photometry and proper motions}
\label{ss:obsSWEEPS}

The \SWEEPS~dataset used here consists of an extremely deep imaging
campaign with a 9-year time baseline using \ACSWFC~in \filtVbroad,
\filtI~(programs GO-9750, GO-12586~and GO-13057, PI K. C. Sahu). The
observations, analysis techniques used to produce the proper motions
and photometry used herein, are described in some detail in previous
papers (\citealt{sahu06}, hereafter \citetalias{sahu06};
\citetalias{clarkson08}; \citealt{calamida14}, hereafter
\citetalias{calamida14}; \citealt{calamida15}, hereafter
\citetalias{calamida15}, and \citealt{kains17}). Here we briefly
describe the relevant characteristics for the present study.

Stellar positions in individual images were estimated using the
distortion solution and effective-PSF methods developed by J. Anderson
for HST and implemented for \ACSWFC~in the {\tt img2xym.F}~routine
\citep{ak06} and associated utilities. This yields highly precise
position measurements in a reference frame that is nearly free of
distortion. With these techniques, per-measurement random
uncertainties are as small as $\epsilon_{X,Y} \approx 0.002$~pixels
per co-ordinate (e.g. Figure 3 of \citetalias{clarkson08}) and
residual distortion is as low as $\sim 0.01$~pixels (\citealt{ak06},
see also Appendix \ref{a:unctyPM}). Detailed discussion of the methods
can be found in \citet{ak06} as well as \citet{anderson08a,
  anderson08b}.

The 2011-2012-2013 epoch consists of 60 (61) images in
\filtVbroad~(\filtI) taken with an approximately two-week cadence,
while the 2004 epoch consists of 254~(265) exposures in \filtVbroad
(\filtI) taken over a 1-week interval in 2004 \citepalias[][all exposures
  in both programs being $\approx 5.5$~minutes each, which
  well-samples the Bulge MS and minimizes down-time for
  buffer-dumps]{sahu06}.

Because the disk and bulge stars move relative to each other, the
2011-2012-2013 images were reduced separately from those in the 2004
epoch. Proper motions were derived from the best-fit positional
differences between the 2004 and 2011-2012-2013 datasets; they thus
represent two-epoch proper motions but with positions in each
individual epoch measured to very high accuracy. The positional
differences (in \ACSWFC~pixels) were rotated into a frame aligned with
Galactic co-ordinates and converted from a displacement in pixels into
rate of positional change in \masperyear~using the \ACSWFC~plate scale
(50 mas pix$^{-1}$~in the distortion-free frame of \citealt{ak06})~and
the time-baseline between the two epochs (8.96 years;
\autoref{tab:obsSummary}). This yields transverse relative motions in
\masperyear~in a frame closely aligned to the Galactic coordinate
system.

Without absolute reference frame tracers in this crowded field
(e.g. \citealt{yelda10}, \citealt{sohn12}), we work exclusively with
relative proper motions. Zero proper motion \muvecref~is defined as
the median observed rate of positional change for bulge objects across
the entire field of view, without any selection for metallicity. The
sample defining this proper motion reference consists of stars that
are not saturated in the deep exposures
%at $\filtI \approx 18$,
(these objects are at the bulge main-sequence turn-off and
  fainter).\footnote{Some notational clarification is in order:
    although all the proper motions are reported relative the average
    motion of a sample defined by the astrometric signal to noise, we
    follow standard practice in this sub-field \citep[e.g.][]{kr02}
    and refer to these relative proper motions simply as ``proper
    motions'' \muvec=(\muL, \muB), rather than $\Delta \muvec$,~to
    avoid cluttering the notation.}

\citetalias{calamida15} conducted extensive artificial star-tests to
estimate measurement uncertainty in the proper motions, with
artificial objects injected with proper motions into {\it individual}
measurement frames to characterize the random proper motion
uncertainty as a function of apparent magnitude. Including random
measurement uncertainty, random intrinsic uncertainty due to tracer
star motion, and estimated residual distortion, the proper motion
uncertainty per co-ordinate is approximately $\lesssim 0.12$~mas
yr$^{-1}$~over the apparent magnitude range of interest (see Appendix
\ref{a:unctyPM} for details), easily sufficient to measure relative
stellar motions in this field.

The result is a set of 339,193 objects with ACS/WFC positions,
apparent magnitudes, and proper motion estimates, all with
uncertainties characterized as a function of apparent
magnitude. Exploitation of these data are presented in
\citet{calamida14, calamida15} and \citet{kains17}.

\subsection{\WFC~photometry from the WFC3 Bulge Treasury Project (\BTS)}
\label{ss:obsBTS}

The WFC3 Bulge Treasury Project (\BTS; program GO-11664, PI
T. M. Brown) visited four fields in the Bulge, with \WFC, including
the \SWEEPS~field. The observations are described in detail in
\citet{brown10}, here we briefly summarize the characteristics
relevant for the present paper.

\begin{figure}
\begin{center}
\includegraphics[width=3.5in]{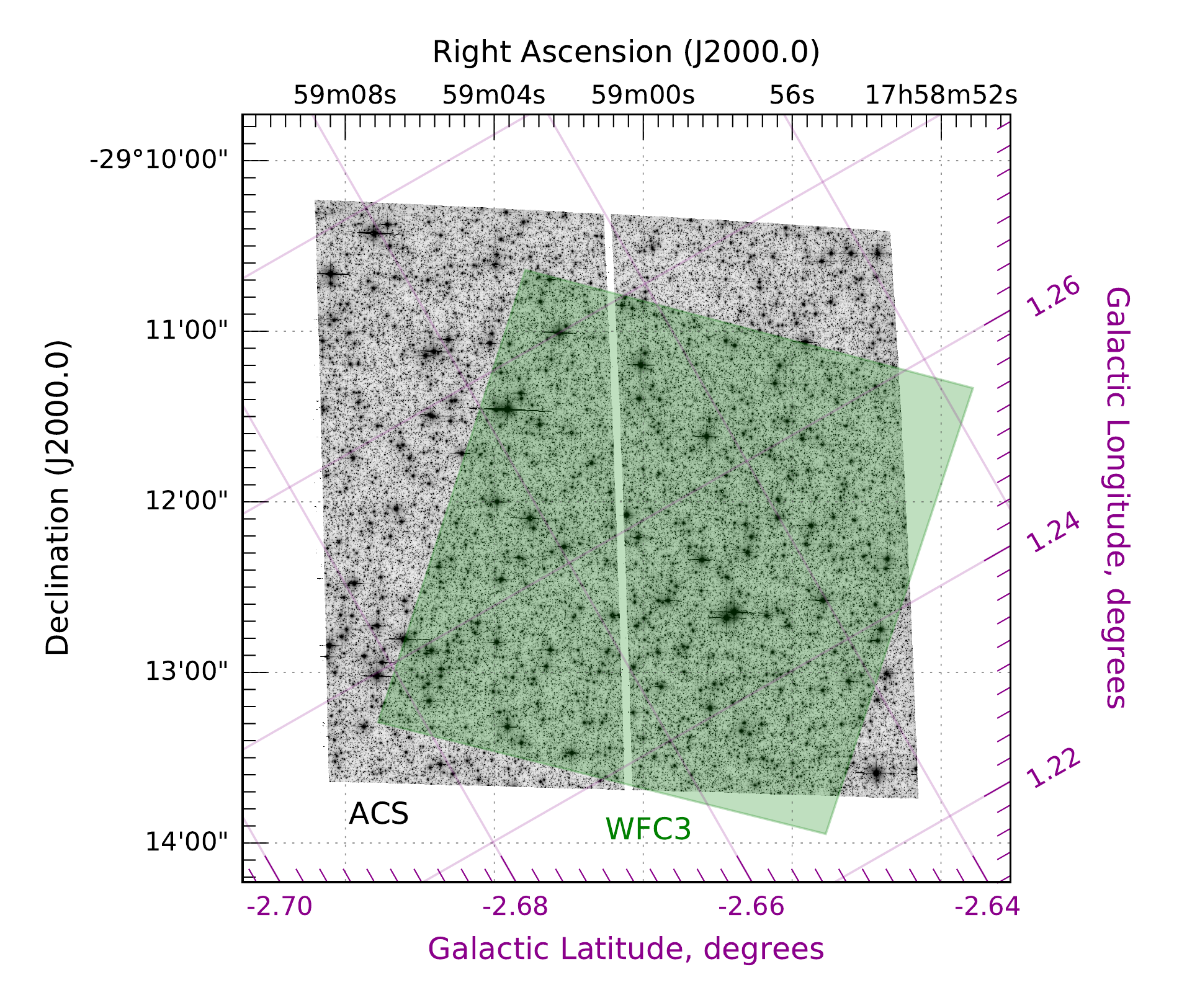}
\end{center}
\caption{Finding chart for the \SWEEPS~\ACSWFC~and \BTS
  ~\WFC~datasets used in this work. The tilted solid dark magenta grid
  shows Galactic co-ordinates, spaced at 0.02$^{\circ}$~intervals. The
  dotted gray grid shows equatorial co-ordinates, spaced at
  $1'$~intervals. The green polygon shows the \BTS~\WFC~coverage; our
  sample is drawn from the region of overlap between the two surveys.
  North is up, East left, and the \ACSWFC~field of view is
  approximately $3.4'\times 3.4'$, centered approximately at $(\alpha,
  \delta)_{\rm J2000.0}$ = (17:59:00.7, -29:11:59.1), or $(l, b)_{\rm
    J2000.0} \approx (+1.26^{\circ}, -2.65^{\circ})$. See
  \autoref{s:observations}.}
\label{fig:orientation}
\end{figure}

In each field, observations were taken in UVIS/F390W (11,180s),
UVIS/F555W (2,283s), UVIS/F814W (2,143s), IR/F110W(1,255s) and
IR/F160W (1,638s), with IR images (field of view $123'' \times 136''$)
dithered in order to fully cover the UVIS observations (field of view
$162'' \times 162''$). Good overlap was achieved with the
\SWEEPS~\ACSWFC~observations; nearly all the \BTS~objects in this
field also fall within the \SWEEPS~\ACSWFC~field of view
(\autoref{fig:orientation}).

Version 1~of the \BTS~catalog,\footnote{A second version of the
    \BTS~catalog was released when the present work was at an advanced
    stage. This second catalog version is discussed in Appendix
    \ref{app:v1v2}: while the measurement techniques are improved
    over the first version, the differences do not impact the results
    presented in this work.}  which we use here, employed
photometry and positions measured with {\tt daophotII}
(\citealt{stetson87}, \citealt{brown10}). The resulting \BTS~v1
catalog lists 400,424 objects in the Sagittarius window with reported
apparent magnitude in any of the \BTS~filters. Of these, 52,596 have
measurements in all five of the \BTS~filters that are required to
construct \mtindices~estimates.

\section{Analysis}
\label{s:analysis}

To construct the \MR~and \MP~rotation curves, we used the
\BTS~photometry to draw \MR~and \MP~samples by use of \mtindices~and
used the \SWEEPS~data to estimate the \rdmLongs~and proper
motions. Within each sample, the \rdmLong~(\rdm) for a given star is
defined as the apparent magnitude offset from the fiducial ridgeline
in the \SWEEPS~color-magnitude diagram for the sample. The
\SWEEPS~deep (\filtVbroad, \filtI) color-magnitude diagram was used to
estimate \rdm~because this choice of filters is relatively insensitive
to metallicity variations when compared to, for example, the
($C,V$-$I$)~color-magnitude diagram presented in \citet{brown10}.

This Section is organized as follows: \autoref{ss:mergeCat} describes
the merging of the \SWEEPS~and \BTS~catalogs, with the sample
selection for proper motion study discussed in \autoref{ss:cmdSel} and
the calculation of the photometric indices \mtindices~shown in
\autoref{ss:indices}. These indices require a prescription for
extinction, discussed in \autoref{ss:prescrip:EBmV}. The
classification into \MR~and \MP~samples is discussed in
\autoref{ss:mtClassify}. The kinematic behavior of the two samples was
then measured in two ways; a simple one-dimensional characterization
of the longitudinal proper motion \muL~is indicated in
\autoref{ss:rotnCurves}, while a more sophisticated dissection of the
velocity ellipse with \rdmLong~\rdm~is shown in
\autoref{ss:pmEllipse}.

\subsection{Merging the \ACSWFC~and \BTS~catalogs}
\label{ss:mergeCat}

The \BTS~and \SWEEPS~catalogs were first cross-matched by equatorial
co-ordinates. Although the absolute pointing of \HST~is accurate only
to $\sim 0.1''$~\citep{drizzlepacHandbook}, with \filtI~observations
in both datasets,\footnote{Small differences in effective bandpass of
  the F814W filter between \ACSWFC~and \WFC~do not significantly
  impact the cross-matching.} matching of similar objects in both
catalogs is straightforward (using \filtV~and \filtVbroad~measurements
in \WFC~and \ACSWFC~respectively to refine the matches). For the first
round of matching, a kd-tree approach was used to cross-match on the
sphere, with a 5-pixel radius used for initial matching. In the second
round, pixel-positions in the two catalogs were cross-matched and fit
using a general linear transformation for objects in the $18 \le
\filtI \le 26$ range. While the population of good matches transitions
to a background of mismatched objects at a radius of $\sim$2-pixels
and larger, the vast majority of cross-matches were somewhat better,
falling within a 1-pixel matching distance. The matching process
resulted in a list of \nmatched~objects with proper motions and
seven-filter apparent magnitudes, with uncertainty estimates for all
quantities.

\subsection{Sample selection for proper motion study}
\label{ss:cmdSel}

The successive selection steps isolating the sample for further study,
are detailed in \autoref{tab:sampleSel}. Of an initial sample of
\nsweeps~\SWEEPS~objects and \nbts~\BTS~objects, \nkept~($\sim 2.9\%$)
were retained for further analysis.

% Table: Selection Polygon vertices
\begin{deluxetable}{cc}
\tabletypesize{\footnotesize}
\tablewidth{700pt}
\tablecaption{Vertices of the selection polygon in the \SWEEPS~CMD that was used to select objects for further proper motion study. See \autoref{ss:cmdSel} for discussion. \label{tab:cmdPoly}}
\tablehead{\colhead{(F606W - F814W)} & \colhead{F606W}\\ \colhead{$\mathrm{mag}$} & \colhead{$\mathrm{mag}$}}
\startdata
1.40 & 24.80 \\
1.54 & 21.30 \\
1.34 & 20.50 \\
1.17 & 23.80 \\
\enddata
\end{deluxetable}

\begin{figure}
\begin{center}
\includegraphics[width=3.5in]{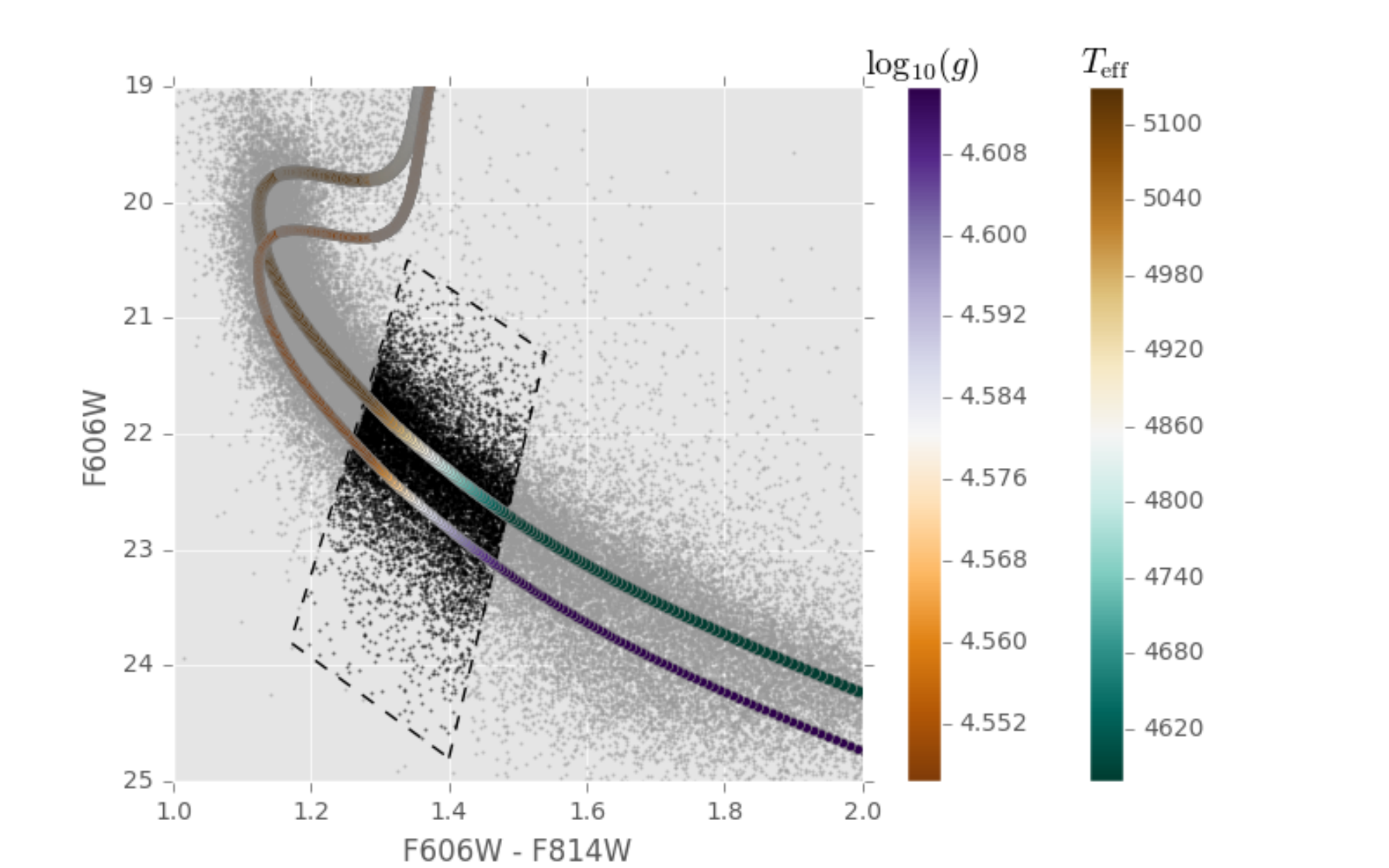}
\end{center}
\caption{Region-selection in the \SWEEPS~color-magnitude diagram. The dashed polygon shows the selection region for objects selected for proper motion study (see \autoref{ss:cmdSel} and \autoref{tab:cmdPoly}). To illustrate typical stellar parameter ranges for this sample, also overplotted is a 10 Gy isochrone at $\feh=-0.09$~from the ``canonical'' $\alpha$-enhanced set within the \basti~library (\citealt{pietrinferni04}, using the ``F05'' opacities of \citealt{ferguson05}). The isochrone is plotted twice, color-coded to show \logg~(left colorbar) and \teff~(right colorbar) and offset for clarity, with color minima and maxima set to the range of parameters across the sample of interest. See subsections \ref{ss:cmdSel} and \ref{ss:indices}. }
\label{f:CMDsel}
\end{figure}

Two aspects of the sample selection are worth highlighting. Firstly,
the selection region in the (\filtVbroad, \filtI) color-magnitude
diagram was chosen to be well clear of the \MSTO, subgiant and giant
branches, to encompass as many stars as possible with good proper
motion measurements, and finally to capture a region over which the MS
for a given population is reasonably free of curvature in the
CMD. This selection region is shown in \autoref{tab:cmdPoly} and
\autoref{f:CMDsel}. Secondly, the photometric metallicity and
temperature indices include coefficients that amplify measurement
uncertainty (particularly \filtJ~and \filtH, which appear in the
temperature index \tindex). For this reason, objects were only
selected for further study for which all apparent magnitude
uncertainties in the photometric catalog are smaller than 0.1 mag.

% Table: selection steps
\begin{deluxetable}{ccc}
\tabletypesize{\footnotesize}
\tablewidth{700pt}
\tablecaption{Selection steps used to isolate the proper motion sample for further study. The cuts are cumulative, reading from top to bottom. The {\bf third} step includes selecting out any rows for which {\it any} of the seven phometry and two proper motion measurements are listed as a "bad" value in either the \SWEEPS~or \BTS~catalogs. In practice this limits the sample to ($18.5 \le F606W \le 27.5$). The \SWEEPS~CMD selection region is shown in \autoref{f:CMDsel}. For the three instrumental configurations listed, objects must show photometric uncertainty $< 0.1$~mag in all relevant filters: (\filtVbroad, \filtI)~for ACS/WFC, (\filtC, \filtV, \filtI)~for WFC3/UVIS and (\filtJ, \filtH)~for WFC3/IR. Objects passing \mtindices~clipping satisfy $(-3.50 \le \tindex \le -1.00)$ and ($-0.60 \le \mindex \le 0.40)$. See \autoref{ss:cmdSel} for discussion. \label{tab:sampleSel}}
\tablehead{\colhead{Selection} & \colhead{N(remaining)} & \colhead{N(removed)}}
\startdata
\SWEEPS~sample \citep{calamida14} & 339,193 & - \\
Cross-matched with \BTS & 55,666 & 283,527 \\
\BTS~measurements in all filters & 47,537 & 8,129 \\
Within \SWEEPS~CMD selection region & 10,225 & 37,312 \\
$\sigma_{mag}(\rm{ACS/WFC}) < 0.1$~mag & 10,222 & 3 \\
$\sigma_{mag}(\rm{WFC3/UVIS}) < 0.1$~mag & 10,209 & 13 \\
$\sigma_{mag}(\rm{WFC3/IR}) < 0.1$~mag & 10,145 & 64 \\
Clipping far outliers in \mtindices & 9,700 & 445 \\
\enddata
\end{deluxetable}

\subsection{Production of \mtindices~for the proper motion sample}
\label{ss:indices}

The photometric indices \mtindices~take the following form \citep{brown09}:
\begin{eqnarray}
  \tindex & \equiv & (V - J) - \alpha (J-H) \nonumber \\
  \mindex & \equiv & (C - V) - \beta (V-I)  
\label{eq:indices}
\end{eqnarray}
\noindent with $\alpha \equiv E(\filtV - \filtJ)/E(\filtJ-\filtH)$~and
$\beta \equiv E(\filtC - \filtV)/E(\filtV - \filtI)$, all of which
have a dependence on stellar parameters. The median values of these
stellar parameters for the proper motion sample ($\teff \approx
4800$~K~and $\log(g) \approx 4.6$)~were estimated from an isochrone
chosen to overlap the observed sample (see \autoref{f:CMDsel}; several
combinations of metallicity, age and extinction were tried, indicating
that the parameter range for this sample is roughly $4200$~K$\lesssim
\teff \lesssim 5200$~K and $4.5 \lesssim \logg \lesssim 4.7$).

\begin{figure}[h]
\begin{center}
  \includegraphics[width=8cm]{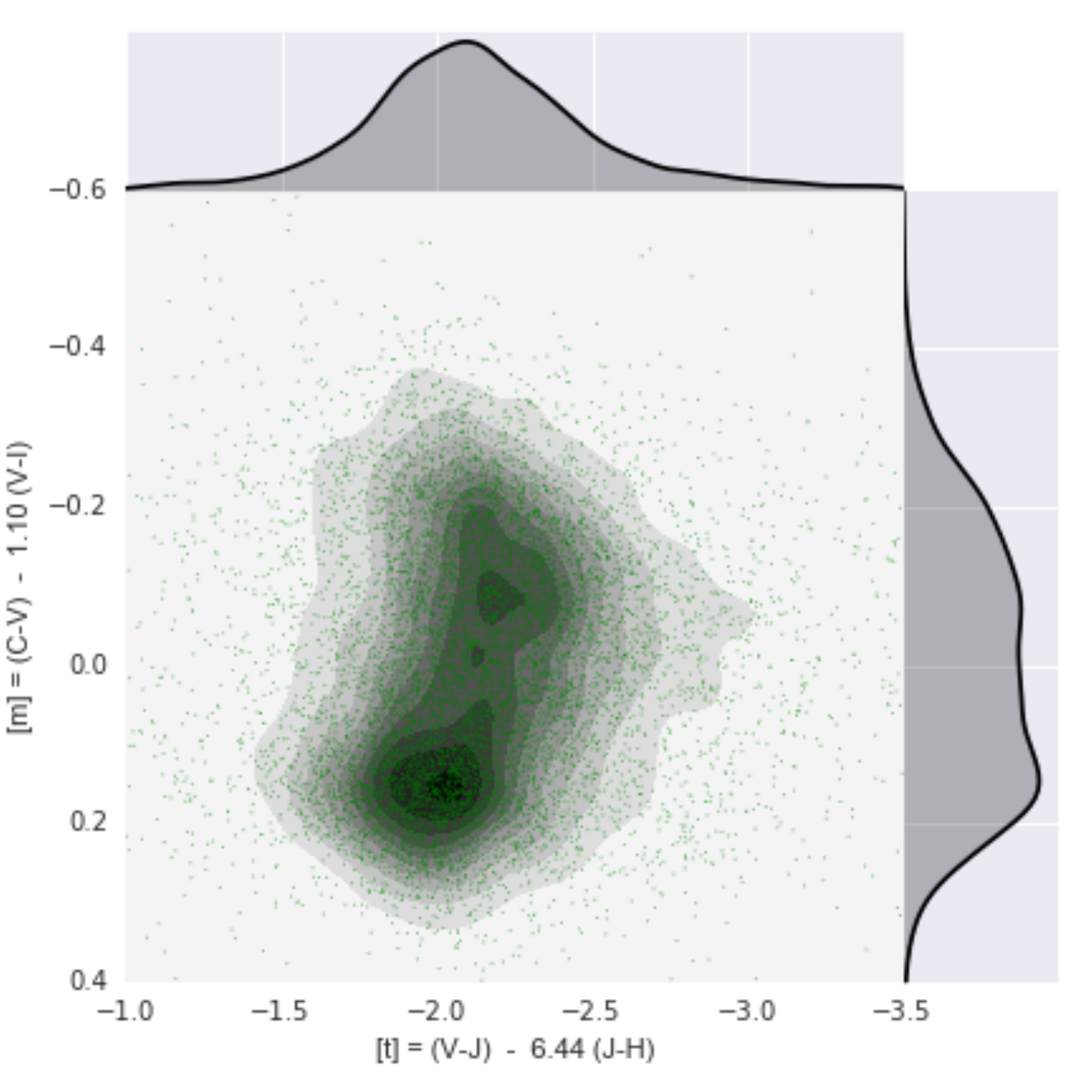}
\end{center}
\caption{\mtindices~distribution of the population selected for proper motion study. In the {\it main panel,}~green points show individual objects, black contours show the smoothed representation as a two-dimensional Kernel Densite Estimate (KDE) with ten levels plotted. Marginal distributions in \tindex~and \mindex~are shown in the {\it top} and {\it right} panels, respectively. Typical estimates for measurement uncertainty in this space are presented in \autoref{f:mtCovar}. See \autoref{ss:indices}.}
\label{f:mtKDE}
\end{figure}

\subsection{Extinction estimates for reddening-free indices}
\label{ss:prescrip:EBmV}

The factors $\alpha$, $\beta$~are three-filter extinction ratios
\citep{brown09}. Synthetic photometry was used to estimate the
relationship between reddening and extinction~for the objects of
interest, and to generate reddening vectors in the various filter
combinations of interest. For a range of \EBmV~values, {\tt pysynphot}
was used to generate synthetic stellar spectra and the run of
$A_X$~against \EBmV~was fit as $A_X = k_X \EBmV$~separately for all
seven filters used in this study, over the range 0.0 $\le$ \EBmV $\le$
1.5. The calculation was performed for \teff, \logg~appropriate to the
SWEEPS CMD region chosen for proper motion study
(\autoref{f:CMDsel}). The process was repeated for low-
  and high-metallicity objects to estimate sensitivity of the
  extinction prescription to metallicity variation within the sample
  selected for further study, and for (\teff, \logg) for objects at
  the median, minimum and maximum \teff~within this sample to estimate
  spread of $\alpha, \beta$~along the sample. 

This procedure requires a prescription for the extinction law towards
the bulge. This extinction law appears to be somewhat non-standard and
strongly spatially variable, with some doubt in the literature about
whether a single-parameter model can accurately reproduce observed
behavior from the visible to the near-infrared \citep[e.g.][and
  references therein]{nataf16}. As the \mtindices~indices use
photometry over a very broad wavelength range ($CVIJH$, or $\lambda
\approx 350$-1700 nm), systematic uncertainties in the extinction
prescription will in turn impact any inferences about the underlying
metallicity distribution (this is one reason why we use
\mtindices~only to classify objects by {\it relative} \feh~estimates).

To make progress, we adopted a single-parameter reddening
  law, but with ratio of selective to total extinction $R_V = 2.5$, as
  suggested by the investigations of \citet{nataf13}.\footnote{As a check, the entire kinematic analysis of Sections \ref{s:analysis} \& \ref{s:results} was also performed using $R_V = 3.1$. Although the mean position of objects in the \mtindices~diagram shifts slightly when $R_V=3.1$~is adopted, the kinematic trends for the \MR~and \MP~samples are similar to the trends when $R_V = 2.5$~is used.} As this value is
  not among the standard parameterizations available in {\tt
    pysynphot}, the coefficients $A_X/\EBmV$~for the seven filters were
  estimated for $R_V=2.1$~and $R_V=3.1$~and linearly interpolated to
  $R_V = 2.5$.

\autoref{tab:extinction} shows the $k_X$~estimates for each filter,
along with the coefficients $\alpha, \beta$~in the
\mtindices~indices. These are quite different from the MS coefficients
reported in \citet{brown09}, as expected since here we are targeting a
specific population some way beneath the Main Sequence turn-off, and
have used a different prescription for extinction.

For a given choice for $R_V$, the variation of all extinction-relevant
quantities appears to be small within the sample of interest; $\alpha,
\beta$~each vary by $< 0.1$~between the two abundance-sets tested,
and, for a given abundance, by $\lesssim 0.02$~across the \teff~range
of this sample. We adopt $(\alpha, \beta) = (\alphaUse, \betaUse)$~for
the rest of this work.

% Extinction prescription
\begin{table*}
\begin{center}
\scriptsize{
\caption{Estimates of $k_X \equiv A_X/E(B-V)$~and derived parameters. Here $\rm{T}_{\rm{eff}} = 4800.0$~and $\log(g) = 4.59.$~For convenience, the scale factor for the \SWEEPS~color index is also shown. The quantities $\alpha, \beta$~give the extinction ratios relevant for \mtindices. Specifically, $\alpha \equiv E(F555W - F110W)/E(F110W-F160W)$~and $\beta \equiv E(F390W - F555W)/E(F555W - F814W)$. See \autoref{ss:indices} and \autoref{ss:prescrip:EBmV}. \label{tab:extinction}}
\begin{tabular}{c|p{1.5cm}p{1.5cm}|p{1.5cm}p{1.5cm}|p{1.5cm}p{1.5cm}}
Config & CCM89, $R_V$ = 2.1: log(Z)= -3.3 & CCM89, $R_V$ = 2.1: log(Z)= -1.6 & CCM89, $R_V$ = 3.1: log(Z)= -3.3 & CCM89, $R_V$ = 3.1: log(Z)= -1.6 & CCM89, $R_V$ = 2.5: log(Z)= -3.3 & CCM89, $R_V$ = 2.5: log(Z)= -1.6 \\
\hline
ACS/WFC1/F606W & 1.847 & 1.849 & 2.786 & 2.788 & 2.222 & 2.224 \\
ACS/WFC1/F814W & 1.064 & 1.064 & 1.821 & 1.822 & 1.366 & 1.367 \\
WFC3/UVIS1/F390W & 3.507 & 3.492 & 4.489 & 4.475 & 3.899 & 3.885 \\
WFC3/UVIS1/F555W & 2.183 & 2.186 & 3.167 & 3.171 & 2.576 & 2.58 \\
WFC3/UVIS1/F814W & 1.074 & 1.075 & 1.833 & 1.834 & 1.377 & 1.378 \\
WFC3/IR/F110W & 0.560 & 0.558 & 1.025 & 1.021 & 0.746 & 0.743 \\
WFC3/IR/F160W & 0.345 & 0.345 & 0.635 & 0.634 & 0.461 & 0.460 \\
\hline
(F606W-F814W)$_{\rm ACS/WFC1}$ & 0.784 & 0.785 & 0.965 & 0.966 & 0.856 & 0.857 \\
\hline
$\alpha$ & 7.55 & 7.64 & 5.49 & 5.56 & 6.42 & 6.49 \\
$\beta$ & 1.19 & 1.18 & 0.99 & 0.98 & 1.10 & 1.09 \\
\hline
\end{tabular}
}
\end{center}
\end{table*}

% 2017-12-13 to bring the RCG table in, uncomment this line:
% \input{./tables/TEST_table_RCG_interp_4paper.tex}

\subsection{Classifying samples by relative metallicity}
\label{ss:mtPops}
\label{ss:mtClassify}
% (Doubly-defined label since we rearranged the subsections.

The resulting (\mtindices) distribution of objects is shown in
\autoref{f:mtKDE}. Two concentrations are apparent; one near (\tindex,
\mindex) = (-2.0, 0.15), with a second, more elongated concentration
with major axis angled at about $-45^{\circ}$~in \autoref{f:mtKDE},
centered near (\tindex, \mindex) $\approx (-2.2, -0.1)$.

To classify objects by relative metallicity, and thus draw \MR~and
\MP~samples for further study, the population highlighted in
\autoref{f:CMDsel} was characterized as a Gaussian Mixture Model (GMM)
in (\tindex, \mindex) space, and members of the \MP~and \MR~samples
identified by their formal membership probability \wik~(see
\autoref{a:aboutGMM}). The number $K$~of mixture components to use,
was determined by increasing $K$~until the characterization stopped
improving (see Appendix \ref{aa:nComponents} for details). At least
two components seem to be required, but a four-component mixture model
appears to provide the best representation of the
\mtindices~distribution.

\begin{figure}
  \includegraphics[width=9cm]{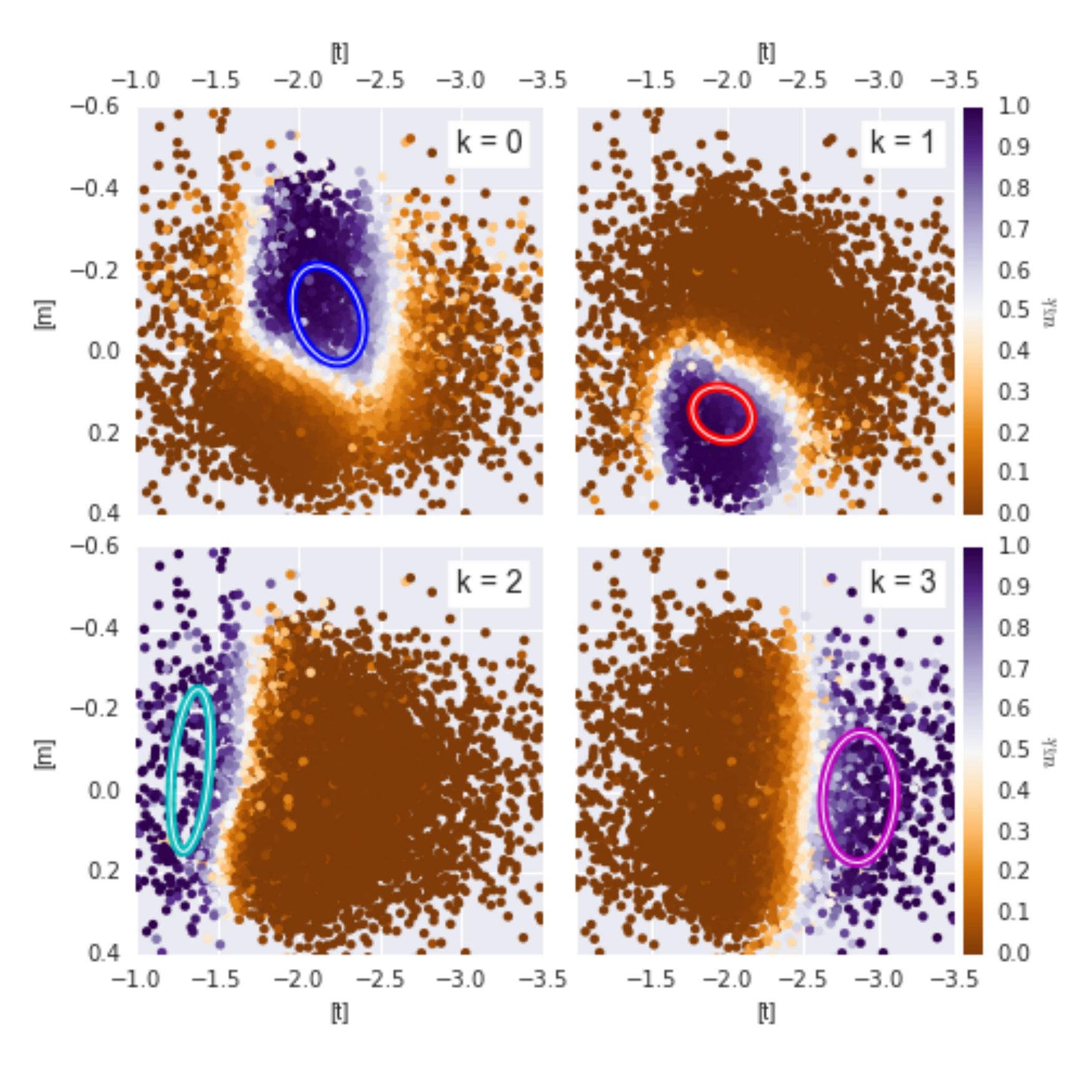}
\caption{The \mtindices~sample color-coded by membership probabilities \wik~(\autoref{eq:memProb}) for the $k$'th model component in the GMM characterization of the observed distribution. The $1\sigma$~ellipse for the $k$'th model component is overplotted in each case as a colored ellipse. Reading clockwise from top-left, panels show the \MR the \MP, and the two background components. See the discussion in \autoref{ss:mtClassify}.}
\label{f:popProbs}
\end{figure}

We therefore adopt a four-component Gaussian Mixture Model to
characterize the observed distribution in \mtindices~space for the
rest of this work. \autoref{tab:gmm} shows the GMM parameters, while
\autoref{f:popProbs} presents the model components visually. The two
most significant components correspond roughly to visually apparent
concentrations in \autoref{f:mtKDE}, together accounting for 91\%~of
the mixture; these form our \MR~and \MP~samples. The remaining two
components, making up about $6\%$~and $3\%$, do not correspond to any
physically obvious population. These two components might represent
populations of outlier objects, or structure in the background in
(\mtindices). We retain these low-level components in the GMM for all
subsequent work using the \BTS~catalog, but do not interpret them as
representing any intrinsic population component.

This four-component GMM provides the basis for our classification of
objects by relative metallicity, with \MR~and \MP~objects
corresponding to the two most significant components of the GMM
(\autoref{tab:gmm}).\footnote{We are {\it not} at this stage
  suggesting that the bulge sample of \BTS~is intrinsically bimodal in
  metallicity (as opposed to a continuum of populations,
  e.g. \citealt{gennaro15, debattista17}). Instead, we are using the
  photometric indices \mtindices~to draw samples near the extremes of
  relative abundance.}

A rough estimate for the centroid \feh~values of the two samples may
be drawn by charting \feh~contours in the \mtindices~diagram for
synthetic stellar populations and interpolating to estimate \feh~at
the \mtindices~locations of the corresponding GMM component centroids
(see Appendix \ref{ss:testBasti:median} for more details on the
synthetic stellar populations used). The GMM component centroids
presented in \autoref{tab:gmm} correspond to $\feh_0 \approx
+0.18$~for the \MR~sample (using scaled-to-solar isochrones) and
$\feh_0 \approx -0.24$~for the \MP~sample (using $\alpha$-enhanced
isochrones for this model component). These centroids are roughly
consistent with values suggested from spectroscopic surveys
\citep[e.g.][]{zoccali17, hill11}.

% Table: GMM parameters
\begin{deluxetable}{cccccccc}
\tabletypesize{\footnotesize}
\tablewidth{700pt}
\tablecaption{Parameters of the Gaussian Mixture Model in \mtindices~space for stars beneath the main sequence selected for further study. Reading left-right, columns indicate the component index $k$, its label (if any), its (rounded) mixture fraction $\alpha_k$, the two components of its centroid, and the three unique components of the covariance matrix $\mathbf{V}_{k}$. See \autoref{ss:mtPops}. \label{tab:gmm}}
\tablehead{\colhead{$k$} & \colhead{Name} & \colhead{$\alpha_k$} & \colhead{$\tindex_0$} & \colhead{$\mindex_0$} & \colhead{$\sigma^2_{\tindex \tindex}$} & \colhead{$\sigma^2_{\mindex \mindex}$} & \colhead{$\sigma^2_{\tindex \mindex}$}\\ \colhead{$\mathrm{}$} & \colhead{ } & \colhead{ } & \colhead{$\mathrm{mag}$} & \colhead{$\mathrm{mag}$} & \colhead{(mag$^2$)} & \colhead{(mag$^2$)} & \colhead{(mag$^2$)}}
\startdata
0 & \MP & 0.557 & -2.18 & -0.09 & 0.0479 & 0.0143 & -0.00742 \\
1 & \MR & 0.358 & -1.97 & 0.15 & 0.0384 & 0.0043 & -0.00187 \\
2 & - & 0.026 & -1.34 & -0.05 & 0.0153 & 0.0397 & 0.00886 \\
3 & - & 0.059 & -2.87 & 0.01 & 0.0573 & 0.0255 & 0.00240 \\
\enddata
\end{deluxetable}

For an object to be classified with the \MR~or \MP~sample, it must
show formal membership probability $\wik \ge \probThresh$~(see
\autoref{eq:memProb}; note that an object need not be classified with
either sample when there are four model components). The shading in
\autoref{f:popProbs} visualizes the membership probabilities
\wik~associated with each mixture component. The threshold $\wik \ge
\probThresh$~was chosen as a tradeoff between sample purity (typical
objects should not fall into more than one model component at the
chosen threshold) and the need to have a sufficient sample size (at
least a few thousand) to permit the dissection of the proper motions
by \rdmLong~with sufficient resolution to chart the rotation curves.

\begin{figure}
  \includegraphics[width=8cm]{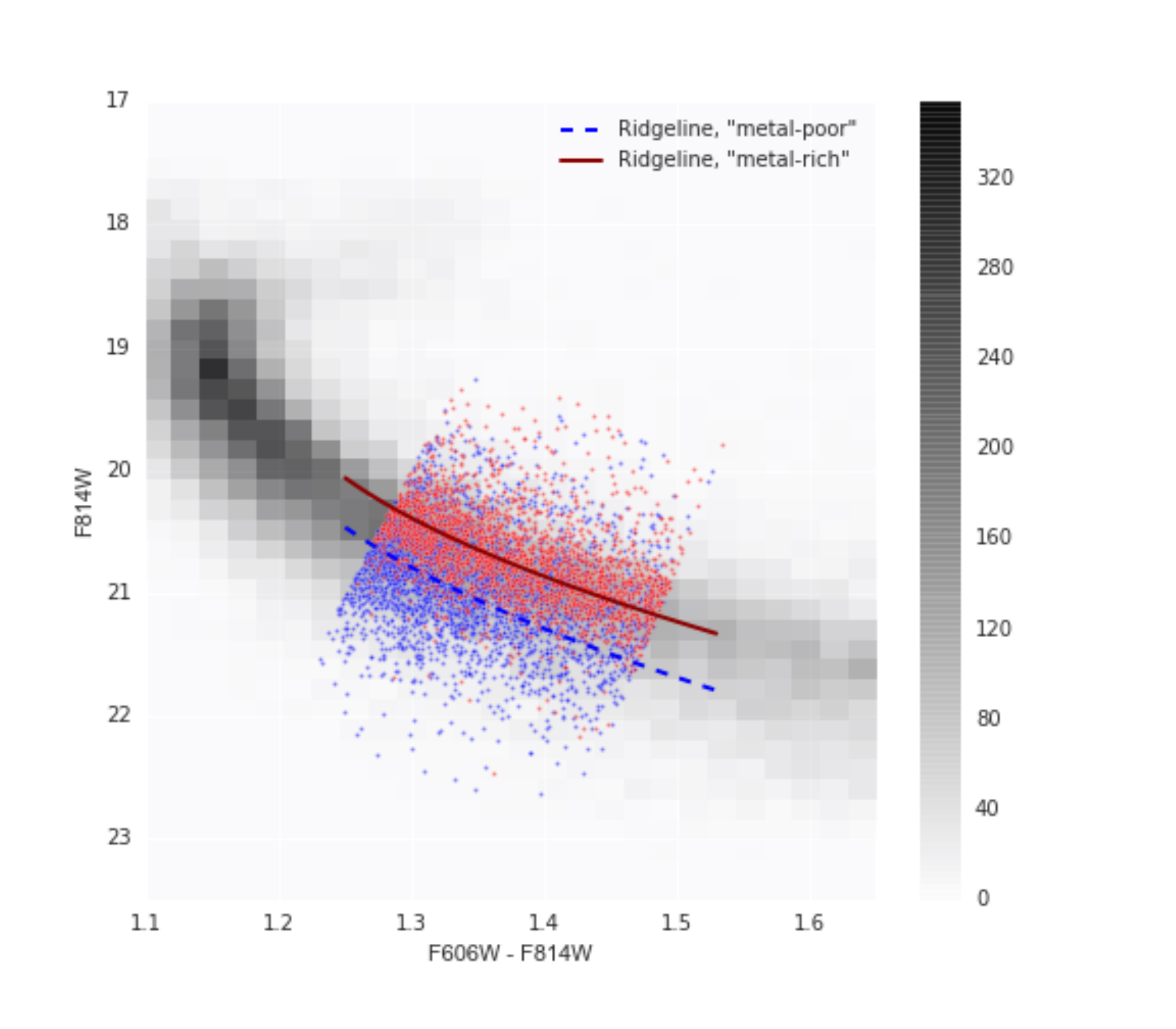}
\caption{Ridgelines for the \MR~and \MP~samples. The grayscale shows the \ACSWFC (\filtVbroad, \filtI) Hess diagram for the larger \SWEEPS~sample. Objects falling within the region of interest for our kinematic study are presented as points, color-coded by \MR~(red) or \MP~(blue). The empirical median-sample ridgelines for the \MR~(dark red solid line) and \MP~(blue dashed line) are overlaid. See \autoref{ss:mtClassify} and \autoref{tab:ridgelines}.}
\label{f:mtRidgelines}
\end{figure}

Assigning \rdmLong~(\rdm) is the final step required before
  proper motion rotation curves can be charted, with reference to
  fiducial ridgelines for the \MR~and \MP~samples. The fiducial
ridgelines themselves were determined by a simple empirical fit
to the density of each sample in the \SWEEPS~CMD. A second-order
polynomial adequately represents the median samples, and allows
very rapid evaluation of relative photometric
parallax. \autoref{f:mtRidgelines} shows the adopted fiducial
  ridgelines for the \MR~and \MP~samples~in the \SWEEPS~CMD,
  while their parameters are given in \autoref{tab:ridgelines}.

% Table: Ridgeline parameters in the CMD for MP and MR
\begin{deluxetable}{ccccc}
\tabletypesize{\footnotesize}
\tablewidth{700pt}
\tablecaption{Ridgeline parameters in the \SWEEPS~color-magnitude diagram, for the \MP~and \MR~samples. These purely empirical ridgelines are used to rapidly evaluate photometric parallax for objects in each sample, and take the form $\filtI = \Sigma_{j} a_j x^j$~with $x$~the (\filtVbroad - \filtI) color. See \autoref{ss:mtClassify} for discussion. \label{tab:ridgelines}}
\tablehead{\colhead{$k$} & \colhead{Name} & \colhead{$a_{0}$} & \colhead{$a_{1}$} & \colhead{$a_{2}$}\\ \colhead{$\mathrm{}$} & \colhead{ } & \colhead{$\mathrm{mag}$} & \colhead{$\mathrm{}$} & \colhead{(mag$^{-1}$)}}
\startdata
0 & \MP & -16.906 & 48.557 & -14.954 \\
1 & \MR & -5.720 & 33.458 & -10.293 \\
\enddata
\end{deluxetable}

\subsection{Proper motion Rotation curves}
\label{ss:rotnCurves}

\begin{figure}
  % 2018-02-19 updated with pdf figures at native width
  \begin{center}
  \includegraphics[width=3.3in]{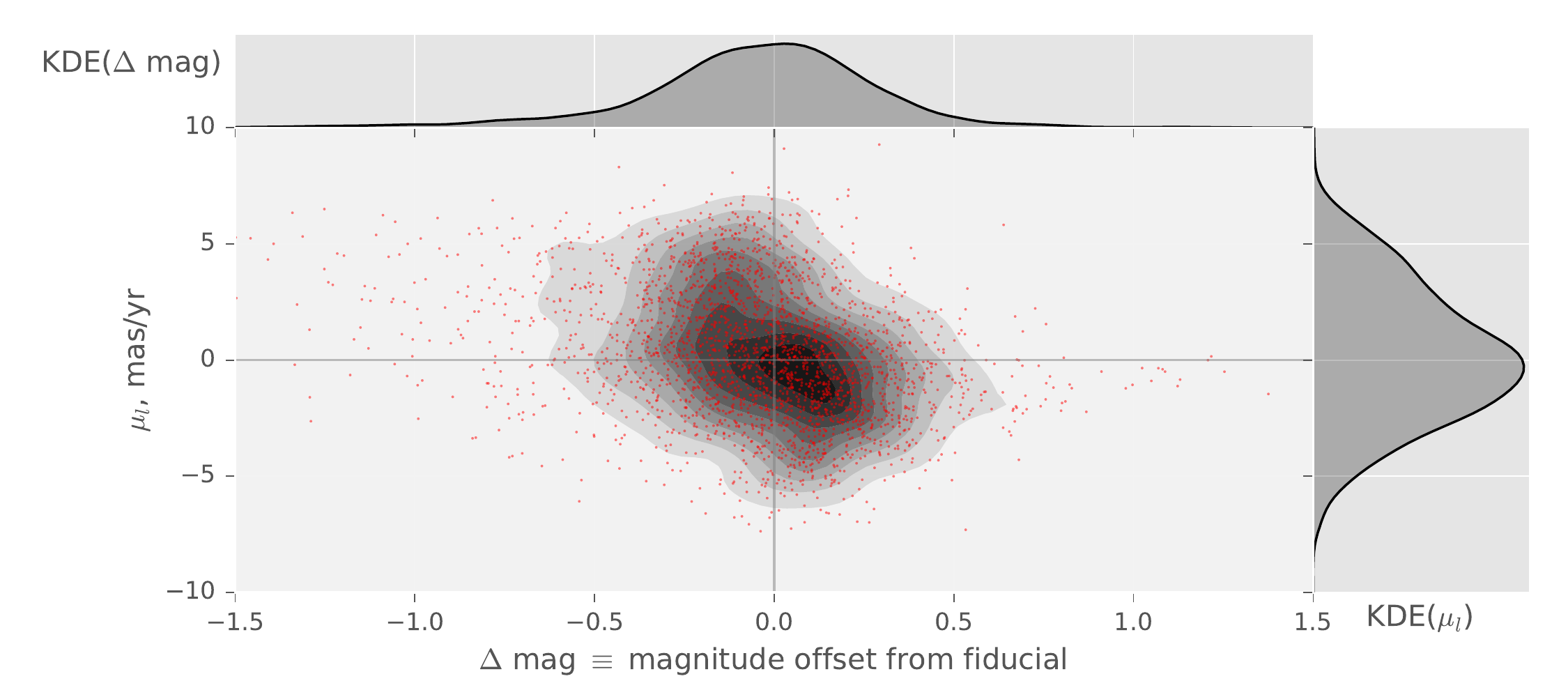}
  \includegraphics[width=3.3in]{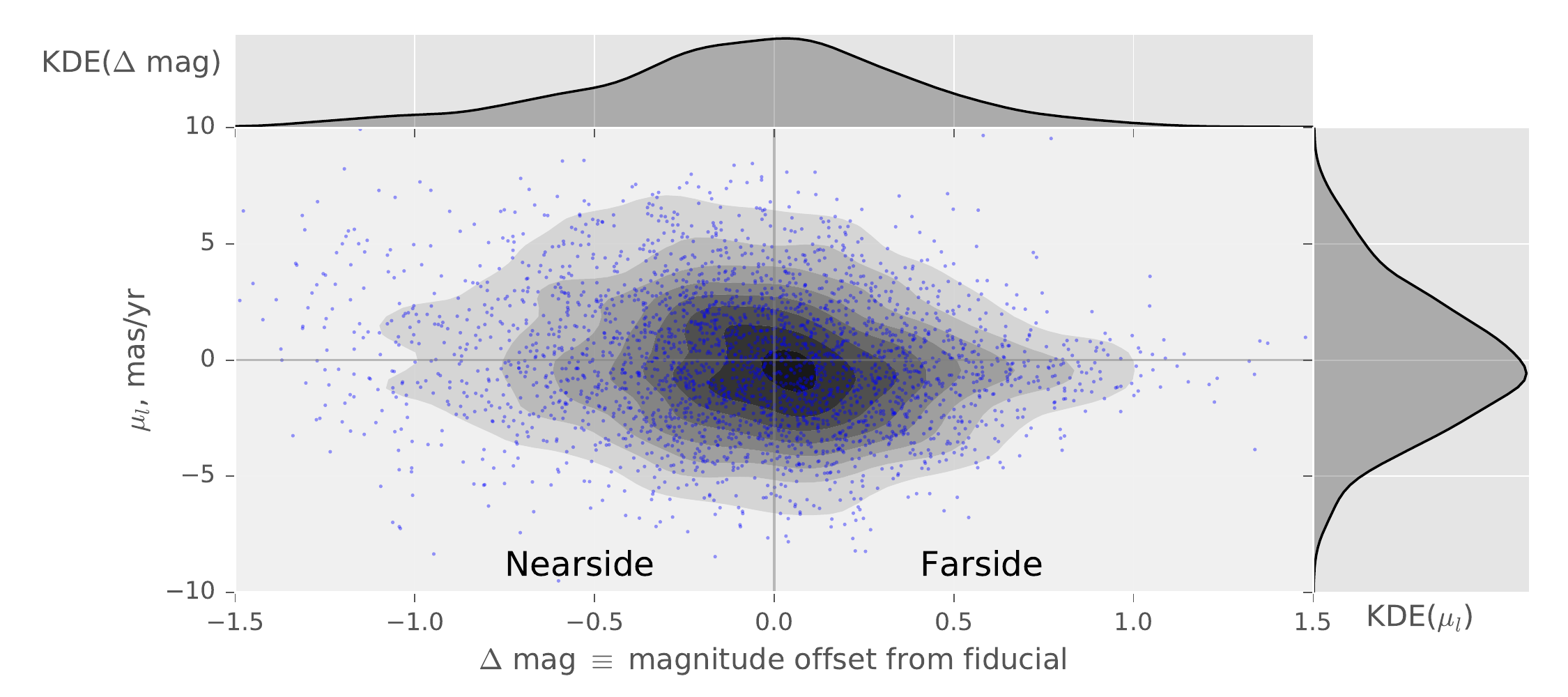}
  \end{center}

\caption{Raw distribution of $\mu_l$~against \rdmLong~(\rdm), for the
  \MR~(red) and \MP~(blue) populations. The nearside of the population
  is to the left in both panels. The points themselves are illustrated
  by colored scatterplots in the main panels, with density contours
  indicated in grayscale. The top- and right-panels show the marginal
  distributions of \rdm (top panels) and $\mu_l$~(right panels). See
  \autoref{ss:rotnCurves}.}
\label{f:simpleContour}
\end{figure}

\begin{figure}
\begin{center}
  % plotting routines changing the aspect ratio between iterations (hmm)
  \includegraphics[width=8cm]{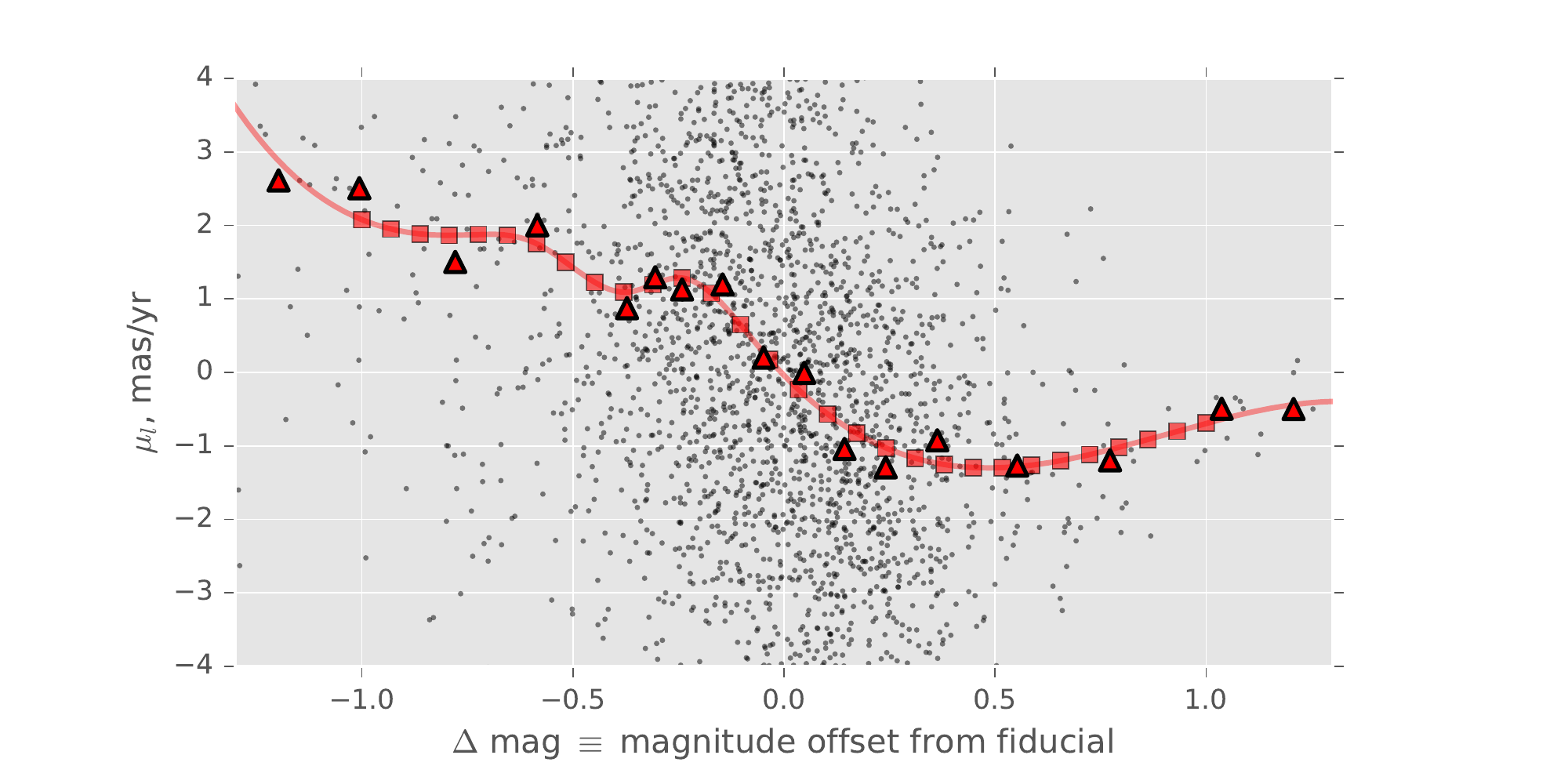}
  \includegraphics[width=8cm]{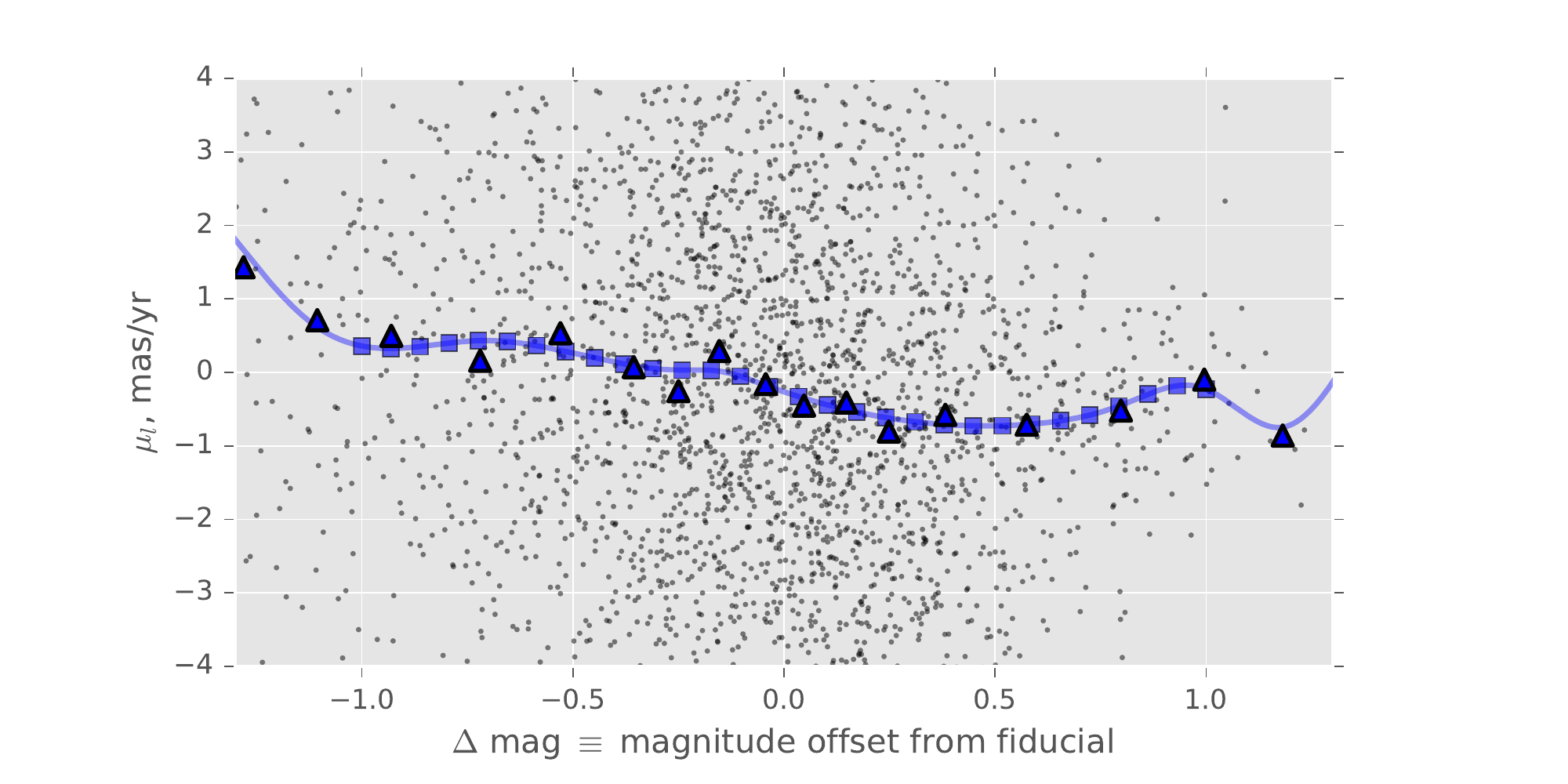}
\end{center}
\caption{Proper motion rotation curves. The \MR~sample is denoted in red in the top-panel, \MP~in blue in the bottom panel. The population is broken into bins in relative distance-modulus and the median value $\overline{\mu_l}$~determined for each bin (triangles). Faint continuous lines show a third-order smoothed spline approximation fit to the binned proper motions $\overline{\mu_l}$, while squares indicate equally-spaced evaluations of the spline approximation over the range of relative moduli ($-1.0 \le (m - m_0) < + 1.0$). See \autoref{ss:rotnCurves}.}
\label{f:simpleTrends}
\end{figure}

Having drawn \MR~and \MP~samples from the \BTS~photometry, along with
fiducial sequences in the \SWEEPS~color-magnitude diagram for the two
samples, the \MR~and \MP~rotation curves can be charted.
\autoref{f:simpleContour} shows the raw distribution of longitudinal
proper motion $\mu_l$~and \rdmLong~for the \MR~and \MP~samples, with
trends presented in \autoref{f:simpleTrends}.

All proper motions in this work were measured relative to the same
proper motion zeropoint, defined without reference to any selection by
metallicity (\autoref{ss:obsSWEEPS}). To the extent that the \MR~and
\MP~samples trace bulge objects with different spatial distribution
and/or kinematic behavior, however, the average proper motions of
bulge objects in the two samples might differ.

We therefore estimated the proper motion corresponding to the fiducial
sequence for each sample. For this ``central'' proper motion, we used
the median proper motion over those sample members with
\rdmLong~between \deltaparBri~and \deltaparFai~magnitudes nearer to
and farther than the fiducial, respectively.\footnote{The near
    limit \deltaparBri~~was set from the far limit \deltaparFai~using
    the relation $\deltaparBri = 5\log_{10} ( 2 - 10^{\deltaparFai /
      5})$, corresponding to a symmetric selection by distance.}  We
adopted \deltaparFai=0.05, corresponding roughly to $\Delta \dKpc$
$\approx$0.18~kpc~at the distance of the bulge
(\autoref{ss:resultsCurves}). The central proper motion for the
\MR~sample is then $\muCenMRlabel = (\muCenMRl,
\muCenMRb)$~\masperyear~from \muCenMRn~surviving objects, while for
the \MP~sample we found $\muCenMPlabel = (\muCenMPl,
\muCenMPb)$~\masperyear~from \muCenMPn~surviving objects. The proper
motion corresponding to the \MR~fiducial was thus found to be offset
from that of the \MP~fiducial by about $\muCenLabel_{MR - MP}$
$\approx$ $(\muCenDeltal, \muCenDeltab)$~\masperyear.

\subsection{Proper motion ellipse dissected by \rdmLong}
\label{ss:pmEllipse}

With a difference in rotation curves suggested from the behavior of
$\mu_l$~against \rdmLong, the next step is to chart the
distance-variation of the $(l,b)$~proper motion ellipse. The approach
shares several similarities to that reported in
\citetalias{clarkson08}; \rdmLongs~were assigned to each star with
reference to the fiducial sequence (appropriate for the
metallicity-sample with which the star was identified) and
the sample partitioned into bins of \rdmLong~\rdm, with bin-widths
adjusted so that each bin has the same number of objects.

The proper motion distribution within each bin was fit as a
two-dimensional Gaussian, with centroid proper motion \muMean~and
covariance matrix \muCovar. Uncertainties in fitted quantities were
estimated by parametric bootstrapping: synthetic samples for each bin
were drawn from the best-fit model, perturbed by the estimated proper
motion uncertainty, and the distribution of recovered parameters over
the bootstrap trials adopted as the estimated parameter
uncertainties. Because this process can be sensitive to outliers, a
single pass of sigma-clipping was applied to the proper motion sample
within each distance bin using a $\pm 3\sigma$~threshold; this
typically removed roughly 1-2\%~of the points per bin, with the
exeption of the most distant \rdm~bin (see Tables
\ref{tab:stats:Metal-rich} \& \ref{tab:stats:Metal-poor} in
  Appendix \ref{s:rotCurveTables}).

Several improvements were made over the analysis reported in
\citetalias{clarkson08}. For example, rather than subtracting the
estimated proper motion uncertainty in quadrature from the model
covariances after fitting, the ``extreme deconvolution''~formulation
of \citet{bovy11}~was used, which incorporates estimated measurement
uncertainty as part of the fitting process (see
\autoref{a:aboutGMM}). We experimented with a multi-component GMM
within each \rdm~bin for each sample, but found a single component
adequate (see also \autoref{ss:multiComponents}). The estimates of
proper motion uncertainty themselves have also been improved compared
to \citetalias{clarkson08}, in both the characterization of random
uncertainty through the artificial star tests of
\citetalias{calamida15} and through improved characterization of
residual relative distortion \citep{kains17}. Details of the adopted
uncertainty estimates are presented in Appendix \ref{a:unctyPM}; for
the apparent magnitude range of interest, the total proper motion
uncertainty estimates ($\epsilon_i \lesssim 0.12$~mas
yr$^{-1}$)~are much smaller than the intrinsic proper motion
dispersion of the bulge ($\sim 3$~mas yr$^{-1}$).

\section{Results}
\label{s:results}

The trends in observed motions are shown graphically in Figures
\ref{f:meanMotionMu} - \ref{f:anglMotionMu}, while
\autoref{f:meanMotionVel} shows the trends after conversion from
\rdmLong~\rdm~and proper motion $\mu$~to distance \dKpc~and velocity
$v$. This information is presented in tabular form in
\autoref{s:rotCurveTables}. Section \ref{ss:resultsCurves} presents
the rotation curves, both observed (i.e., $\rdm, \mu$)~and after
conversion (to $\dKpc, v$), and shows a simple characterization of the
trends. Section \ref{ss:resultsDispersions} presents the evolution of
the velocity ellipse with distance along the line of sight.

\subsection{Distance conversion and rotation curves for the \MR~and \MP~samples}
\label{ss:resultsCurves}

%@arxiver{fig8_rev.pdf, fig11b_rev.pdf}

%%% The results -- kinematics vs photometric parallax
% reworked to fit in the 2-column format and use the space more efficiently.
\begin{figure}
\includegraphics[width=8cm]{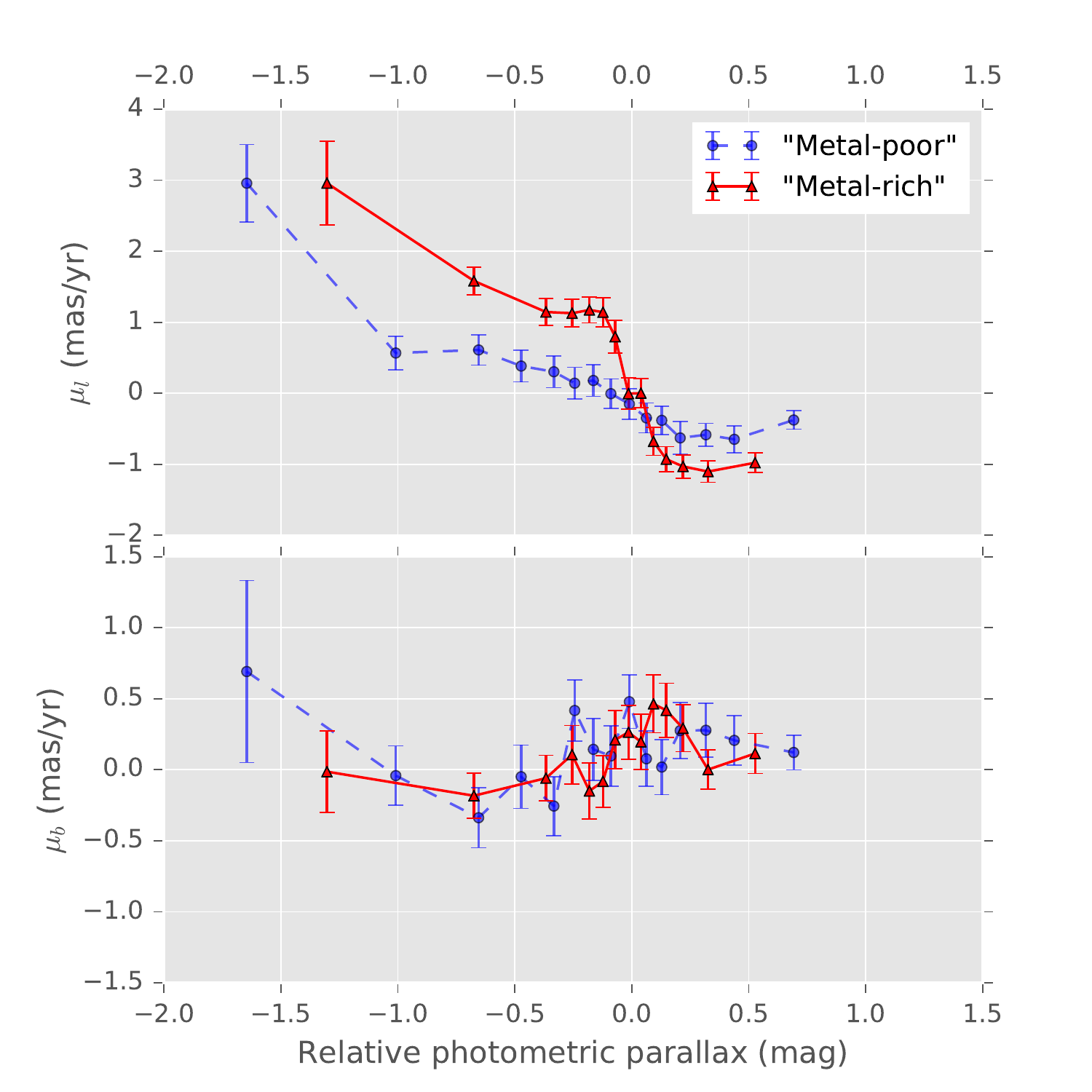}
\caption{Variation of proper motion centroid with \rdmLong, for \MR~(red triangles) and \MP~(blue circles) samples, using a binning scheme with 200 objects per bin. The top row shows the proper motion centroid in Galactic longitude, the bottom row shows the proper motion centroid in Galactic latitude. Errorbars show $1\sigma$~uncertainties from parametric bootstrapping, using the best-fit parameters and measurement uncertainties to generate 1000 trial datasets for each distance bin. See \autoref{ss:resultsCurves}.}
\label{f:meanMotionMu}
\end{figure}

\begin{figure}
  \includegraphics[width=8cm]{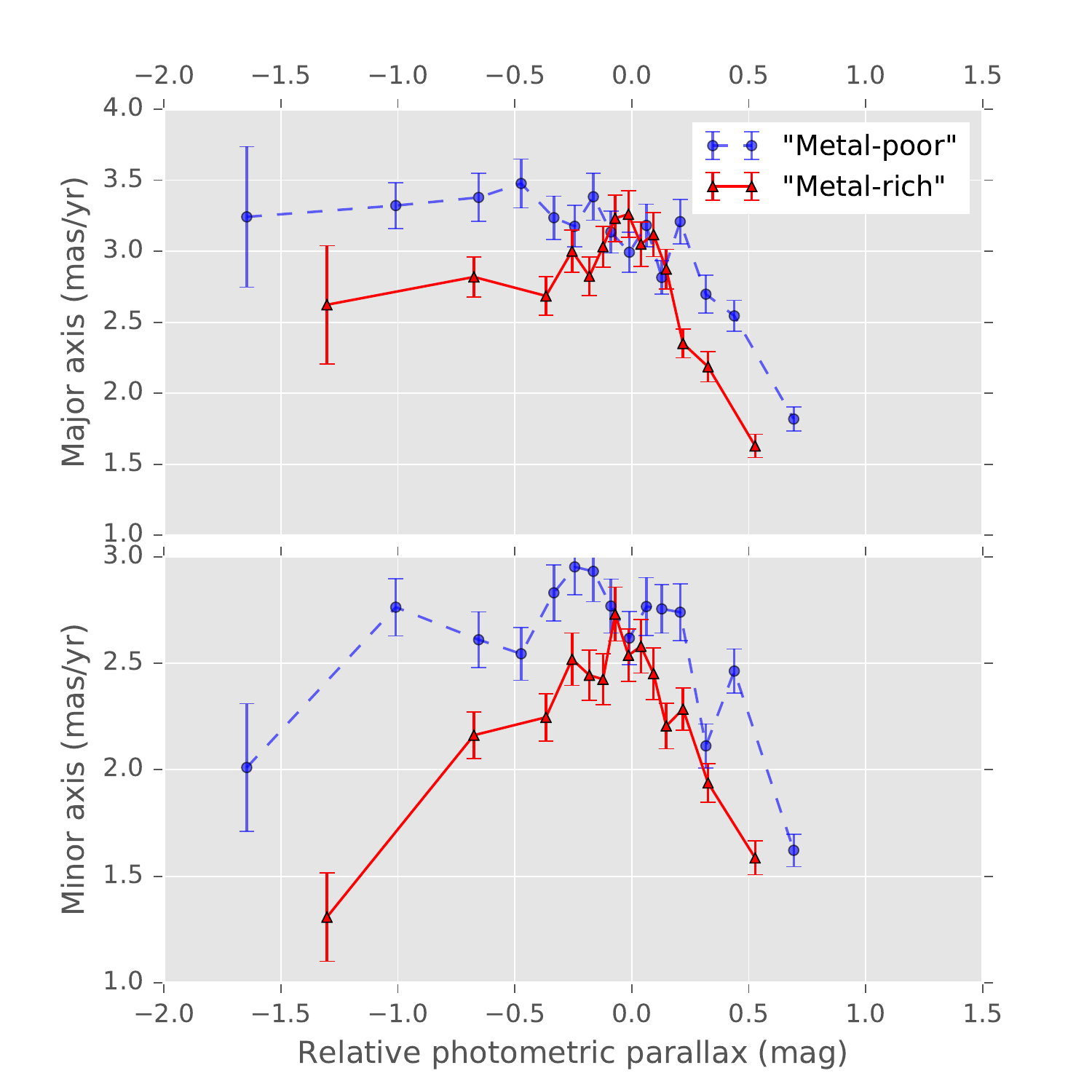}
\caption{Semimajor (top) and semiminor (bottom) axis-lengths for the proper motion ellipse. Symbols, colors and errorbars as for \autoref{f:meanMotionMu}. See \autoref{ss:resultsCurves}.}
\label{f:varMotionMu}
\end{figure}

\begin{figure}
  \includegraphics[width=8cm]{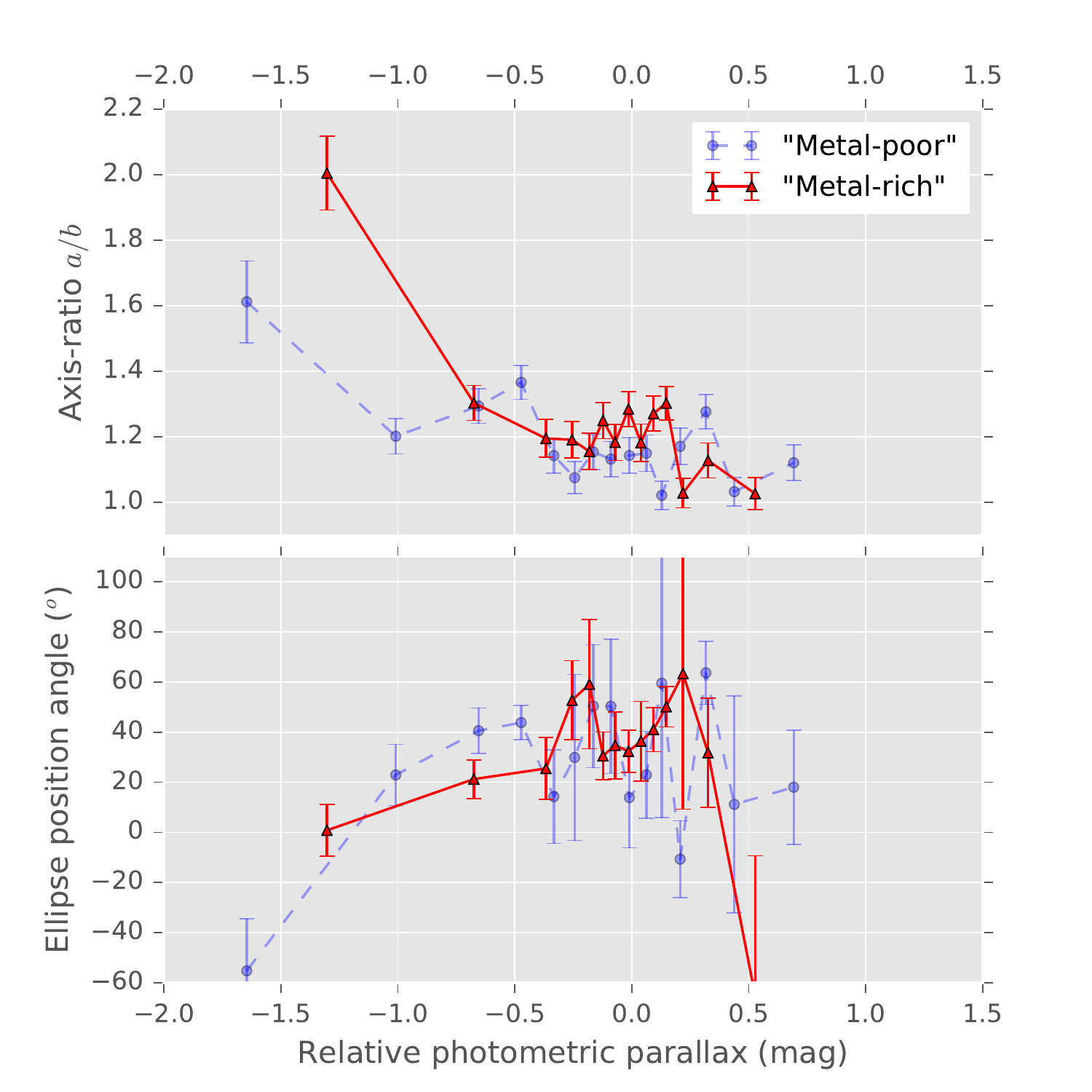}
\caption{Variation of the proper motion ellipse axis ratio (top) and the position angle of its major axis (bottom) as a function of \rdmLong. Position angle $\theta = 0^{\circ}$~would mean the proper motion ellipse major axis aligns with the Galactic longitude axis. Symbols as \autoref{f:meanMotionMu}, with the \MP~sample shown more faintly to avoid cluttering the plots. See \autoref{ss:resultsCurves}.}
\label{f:anglMotionMu}
\end{figure}

\begin{figure*}
\centerline{
  \includegraphics[width=8cm]{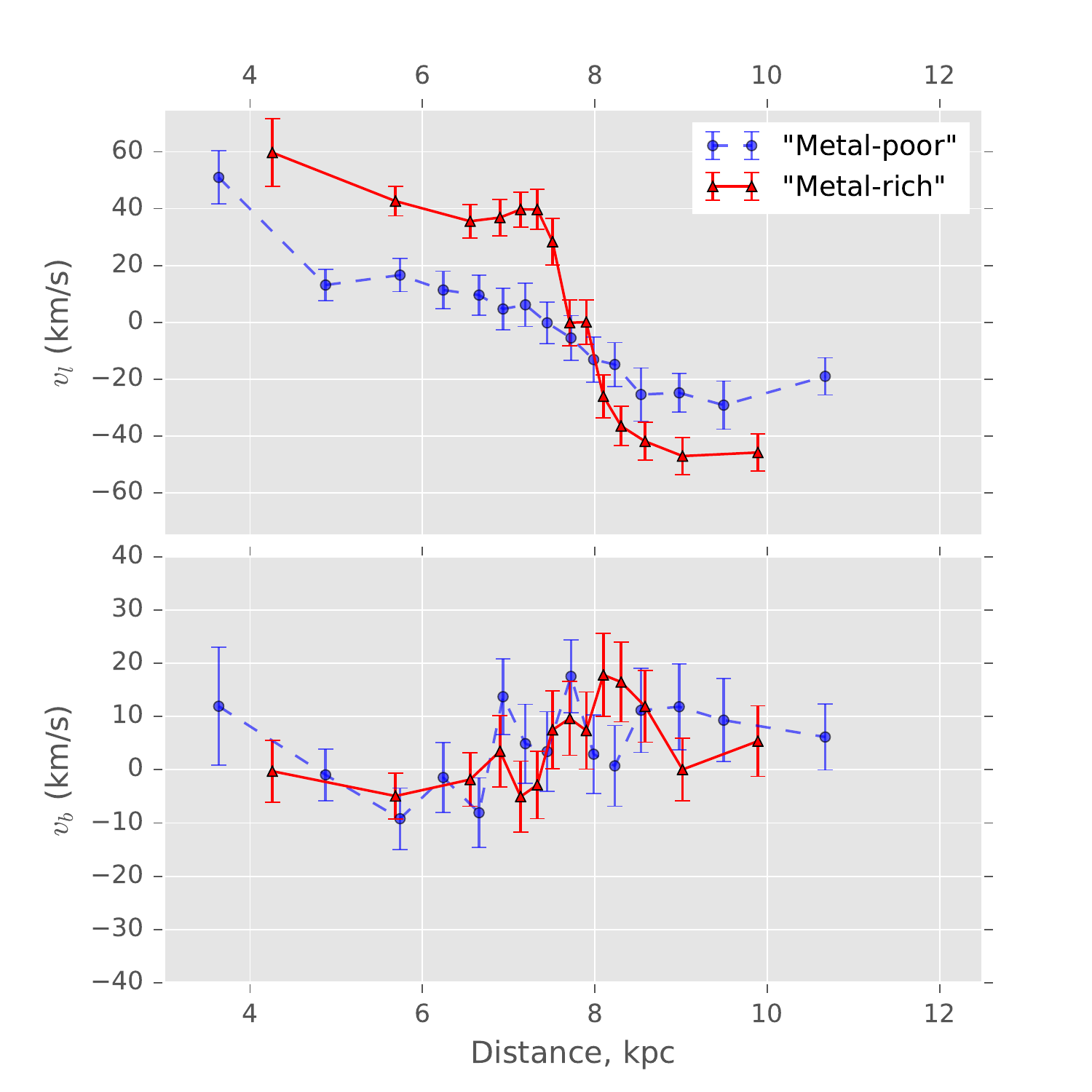}
  \includegraphics[width=8cm]{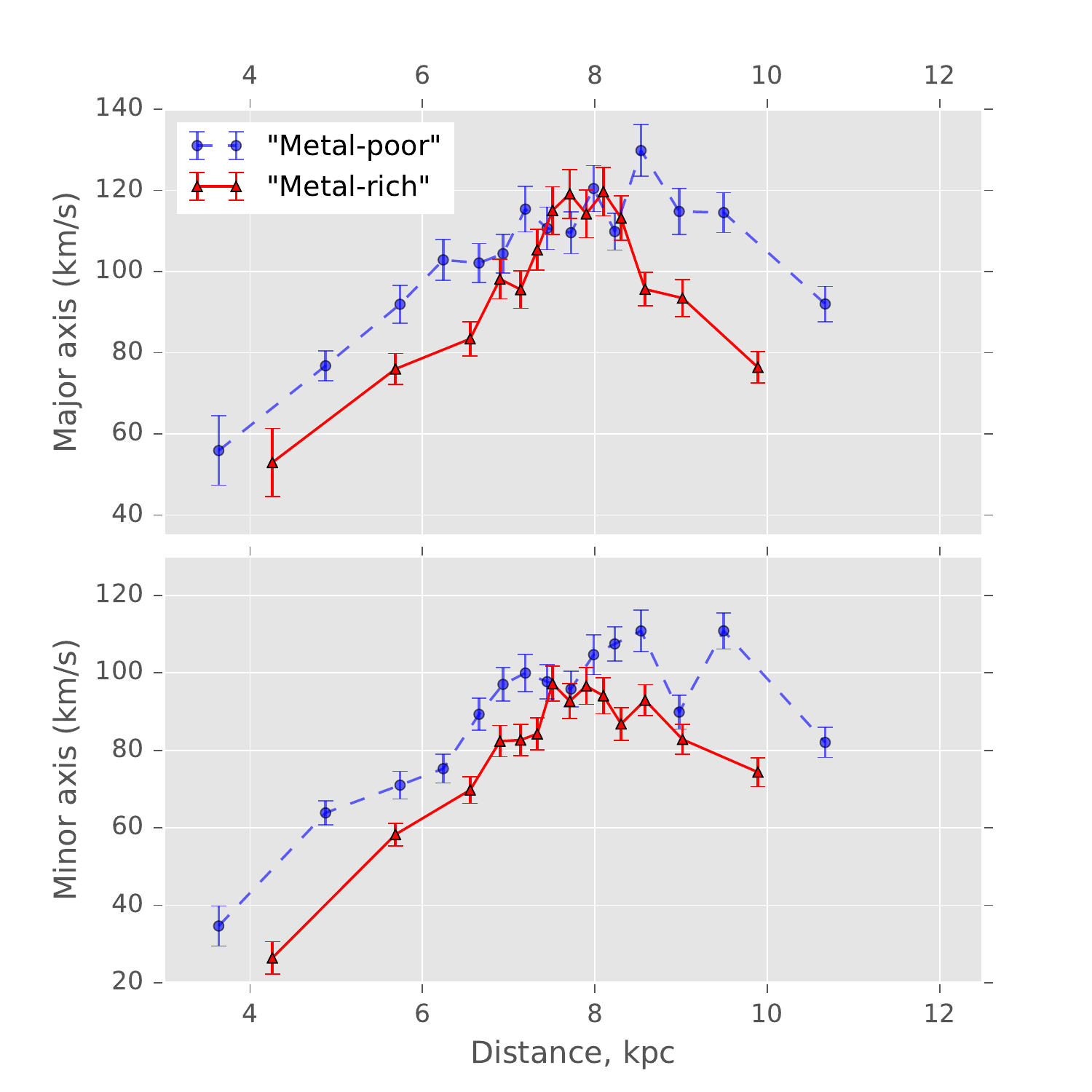}
}
\caption{Transverse velocity ellipse centroids (left column) and axis lengths (right column) as a function of estimated line of sight distances. Symbols as Figures \ref{f:meanMotionMu} \& \ref{f:varMotionMu}, except distance moduli have been converted to line of sight distances, and proper motions converted to velocities in km s$^{-1}$. See \autoref{ss:resultsCurves}.}
\label{f:meanMotionVel}
\label{f:varMotionVel}
\end{figure*}

\autoref{f:meanMotionVel} presents the rotation and dispersion curves
of the \MR~and \MP~samples expressed in terms of $(\dKpc, v)$. The
conversion of these quantities from the measured ($\rdm, \mu$)
requires the reference distance \dzer~corresponding to the fiducial
sequences for the two samples. The reference distance was set by
taking literally the distance modulus $(m-M)_0=14.45$~suggested by
studies of the SWEEPS CMD~\citepalias{calamida14}, which in turn
suggests reference distance (\dKpcCen). We assigned this reference
distance to both the \MR~and \MP~samples (a choice we examine
critically in \autoref{ss:refFrames}).

Consistent with the simple treatment in \autoref{f:simpleTrends} and
\autoref{ss:rotnCurves}, the \MR~sample shows a higher-amplitude
rotation curve than does the \MP~sample, both with a steeper slope and
about a factor $\sim 2$~greater difference in mean transverse velocity
\vLcent~between nearside and farside of the bulge than for the
\MP~sample.

%%%%%% Figures against velocity

%%%%%% end of figures against velocity

\begin{figure}
\begin{center}
    \includegraphics[width=3.5in]{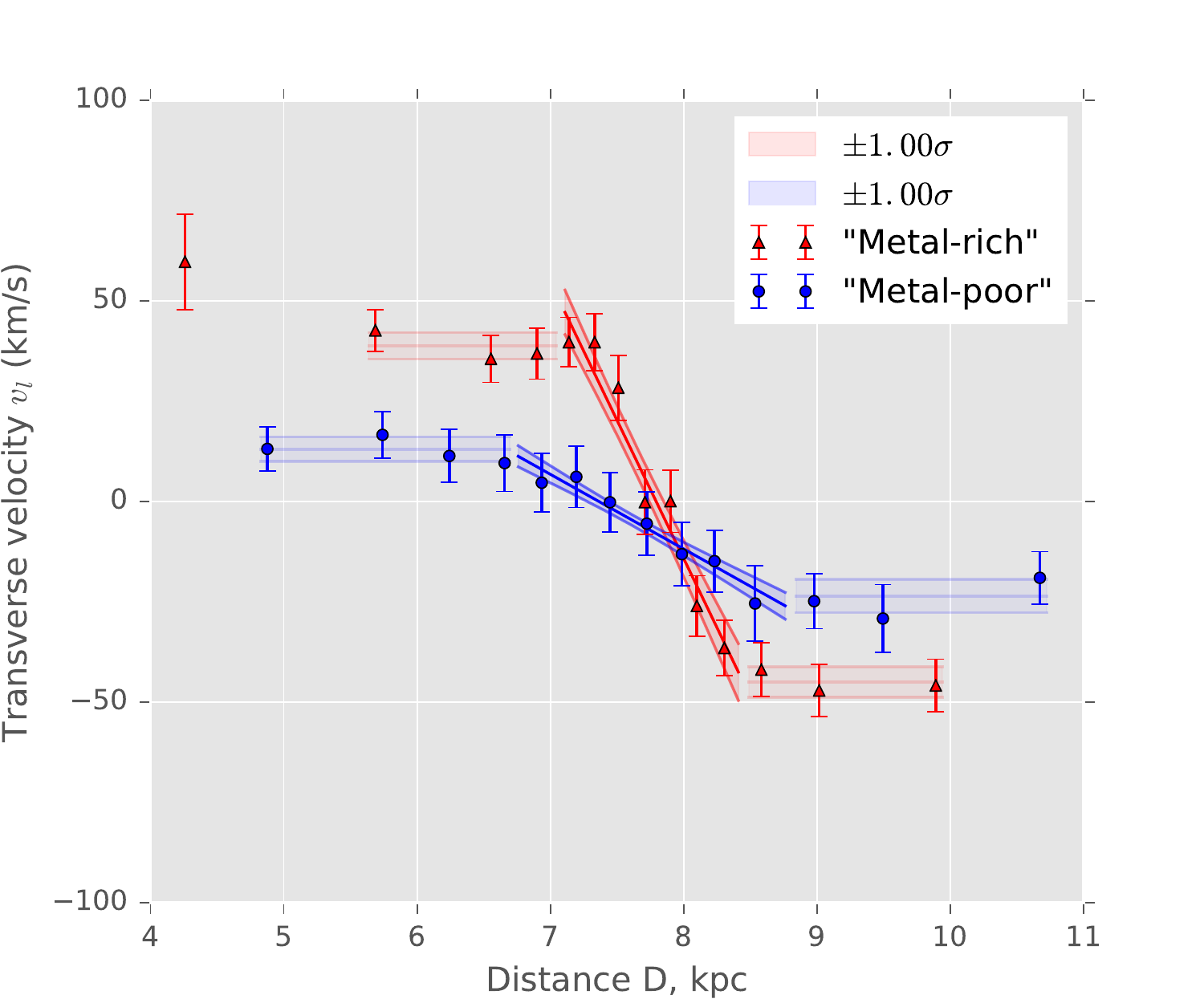}
\end{center}
\caption{Straight-line fits to the inner region of the longitudinal rotation curve along the line of sight, after conversion to velocities and physical distances. The comparison-sequences for both samples are assumed to lie at distance \dKpcCen. The filled regions indicate $\pm 1.0\sigma$~regions for each sample. The horizontal shaded regions show the intervals assumed to be ``flat'' to estimate the rotation amplitude for each sample. See \autoref{ss:resultsCurves} and \autoref{tab:trends}.}
\label{f:rotTrends}
\end{figure}

To quantify the rotation curve discrepancies between the samples, a
simple straight-line model was fit to their rotation curves for
distances close to the fiducial for each sequence. This interval was
estimated separately for the two samples since their rotation curves
appear to level off at different distances from the fiducial
(\autoref{f:meanMotionVel}). For the \MR~sample the gradient was
estimated over the interval $\dzer \pm 0.80$~kpc in the (\dKpc, \vL)
curve (corresponding to $-0.24 \la \rdm \la +0.21$~magnitudes). The
rotation curve of the \MP~sample remains sloped over a broader range,
so the fitting interval $\dzer \pm 1.4$~kpc was used (so $-0.44 \la
\rdm \la +0.36$~magnitudes). For both samples the rotation curve
amplitude was estimated from the intervals where the rotation curves
level off, covering 2-3 bins each outside the sloped region
(\autoref{f:rotTrends} indicates the regions used to estimate the
rotation curve slopes and amplitudes).
The $1\sigma$~ranges of \muLcent~and \vLcent~from the parametric
bootstrap trials were used as estimates of measurement uncertainty in
each distance-bin, and the trends were fitted to each of the (\rdm,
\muLcent) and (\dKpc, \vLcent) rotation curves separately (rather than
transforming the proper motion trends into velocity trends after
fitting). We did not attempt to deproject velocities to circular
speeds (as discussed in \citetalias{clarkson08}) but merely attempted
to characterize observed trends.

\autoref{f:rotTrends} and \autoref{tab:trends} show the results. The
ratio of the gradients \symbolgrad~was found to be \ratiogradL =
\slopeRatio, while the ratio of amplitudes \symbolampl~is \ratioamplL
= \amplRatio. Thus, a ratio in rotation curve slopes was detected at
approximately $\slopeRatioSignif$~while for the velocity amplitude the
ratio was detected at roughly \amplRatioSignif.

% Table: Trend parameters for bulge rotation curves. 
\begin{table*}
\caption{Trend parameters for the inner Bulge region. See \autoref{ss:resultsCurves}. \label{tab:trends}}
\begin{center}
\begin{tabular}{c|cc|cc}
Sample & Gradient$~(\mu_l)$ & Amplitude$~(\mu_l)$ & Gradient$~(v_l)$ & Amplitude$~(v_l)$ \\
 & (mas yr$^{-1}$ mag$^{-1}$) & (mas yr$^{-1}$) & (km s$^{-1}$ kpc$^{-1}$) & (km s$^{-1}$) \\
\hline
"Metal-poor" (MP) & $-1.78 \pm 0.23$ & $0.48 \pm 0.07$ & $-18.6 \pm 2.66$ & $18.3 \pm 2.58$ \\
"Metal-rich" (MR) & $-6.85 \pm 0.73$ & $1.16 \pm 0.07$ & $-68.9 \pm 8.04$ & $41.9 \pm 2.52$ \\
\hline
MR - MP & $-5.07 \pm 0.77$ & $0.68 \pm 0.10$ & $-50.3 \pm 8.5$ & $23.6 \pm 3.6$ \\
MR/MP & $3.85 \pm 0.64$ & $2.42 \pm 0.38$ & $3.70 \pm 0.68$ & $2.29 \pm 0.35$ \\
\hline
\end{tabular}
\end{center}
\end{table*}

\begin{table}
\caption{{\bf Gradient of straight-line fits in Galactic Latitude, for stars within $\pm 2.0$~kpc of the fiducial distance \dKpcCen~adopted in this work.} See \autoref{ss:resultsCurves}. \label{tab:trendsB}}
\begin{center}
\begin{tabular}{c|c|c}
Sample & Gradient$~(\mu_l)$ & Gradient$~(v_l)$ \\
 & (mas yr$^{-1}$ mag$^{-1}$) & (km s$^{-1}$ kpc$^{-1}$)  \\
\hline
"Metal-poor" (MP) & $0.26 \pm 0.25$ & $4.2 \pm 2.76$ \\
"Metal-rich" (MR) & $0.24 \pm 0.28$ & $3.8 \pm 2.65$ \\
\hline
MR - MP & $-0.02 \pm 0.38$ & $-0.4 \pm 3.8$ \\
MR/MP & $0.92 \pm 1.40$ & $0.90 \pm 0.87$ \\
\hline
\end{tabular}
\end{center}
\end{table}

The rotation curves in Galactic latitude (\autoref{f:meanMotionMu},
bottom panel) visually suggest gentle trends from nearside to farside,
consistent with previous measurements (e.g. \citetalias{clarkson08},
\citealt{soto14}). \autoref{tab:trendsB} reports the straight-line
characterization of the Galactic latitude rotation curves, where here
we characterized the trend as a straight line fit within $\pm 2.0$~kpc
from the fiducial distance \dzer. The behaviors of the two samples in
$\mu_b$~are statistically similar, with gradient ratio \ratiogradB =
\slopeRatioB. We do not consider this to represent a secure detection
of differing rotation curves in the direction of Galactic longitude.

\subsection{Proper motion ellipse morphology and amplitudes}
\label{ss:resultsDispersions}

The velocity dispersion profiles (measured as major and minor axis
lengths of the proper motion and velocity ellipsoids; Figures
\ref{f:varMotionMu} \& \ref{f:varMotionVel}) also show differences
between the samples. Both samples show a broadly centrally-peaked
velocity dispersion pattern against line-of-sight distance
(\autoref{f:varMotionVel}), with the \MR~sample showing a narrower
peak, particularly in the major-axis dispersion. The \MR~sample also
shows generally lower velocity dispersion by $\sim 10\%$, particularly
in terms of the velocity ellipse minor axis. 

Consistent with previous studies \citep[e.g.][]{soto14}, the proper
motion ellipse appears to be weakly elongated, with the \MR~population
possibly the more elongated of the two samples (with axis-ratio $b/a
\approx 1.29 \pm 0.05$~at $\rdm = 0$~compared to $b/a \approx 1.13 \pm
0.05$; see the top panel of \autoref{f:anglMotionMu}). However, the
two axis-ratio trends show considerable bin-to-bin scatter.

The proper motion ellipse major axis position angle also shows trends
with \rdmLong, although possibly at lower statistical significance
than the trends reported in \citetalias{clarkson08} despite a much
longer time-baseline for proper motions \citepalias{calamida14}. This
reduced significance may be due to the reduced sample size admitted by
the cuts in \mtindices~employed in this work. It may be that only the
\MR~sample substantially shows the proper motion ellipse tilt with
distance, with position angle rising to the $20^{\circ} -
40^{\circ}$~range (this tilt is strongly influenced by projection
effects; see Section 5.1 and particularly equation (2) of
\citetalias{clarkson08}). Because the \MP~population tends to be less
elongated, its position angle trends are also detected at lower
significance.

The very nearest \rdmLong~bins show behavior consistent with a
foreground population dominated by Galactic rotation. This seems
particularly clear for the \MR~sample, which shows a much more
strongly elongated proper motion ellipse for the nearest bin ($a/b
\approx 2.0 \pm 0.11$) and position angle consistent with zero
(consistent with differential rotation in Galactic latitude).

\section{Discussion}
\label{s:discussion}

The trends indicated by the union of the \BTS~and \SWEEPS~datasets,
particularly the rotation curves (presented in Figures
\ref{f:simpleTrends}, \ref{f:meanMotionMu} \& \ref{f:meanMotionVel}),
are quite striking. The \MR~rotation curve appears to show
systematically greater rotation amplitude than the ``metal-poor''
sample, shows a greater degree of central concentration along the line
of sight (see Figure \ref{f:varMotionVel} as well as the raw
distributions in \autoref{f:simpleContour}), and, with the exception
of the middle distance bins, may show systematically lower velocity
dispersion (\autoref{f:varMotionVel}).

Before attempting to interpret the trends, however, we examine the
magnitude and impact of several potential systematics that might bias
the samples, whether by amplifying or even artificially generating the
apparent differences in rotation curve (\autoref{ss:distanceMixing}),
or by reducing them due to mixing in the (\mtindices) space used to
draw the \MR~and \MP~samples (\autoref{ss:mtMixingDiscussion}). The
impact of extinction variations along the line of sight, including
additional extinction on or past the far side of the bulge, are
discussed in \autoref{ss:extinctionFarside}. Systematic uncertainties
in the final velocity rotation curves due to the proper motion
zeropoint and fiducial distance are discussed in
\autoref{ss:refFrames}.

In \autoref{ss:twoPopns} we address the question of whether the bulge
rotation curve from proper motions indeed depends on relative
abundance, and briefly assess trends in proper motion dispersion in
\autoref{ss:resDisp}. Implications of the \rdmLong~distributions for
the spatial distributions of the \MR~and \MP~samples are discussed in
\autoref{ss:meanDists}. Because a metal-poor kinematically-hot
``Classical bulge'' and/or ``halo'' bulge component may be present in
the inner Milky Way (perhaps more likely among \MP~objects), we
attempt in \autoref{ss:multiComponents}~to dissect each of the \MR~and
\MP~populations into two proper motion components per sample. Finally,
\autoref{ss:pmSelection} discusses the implications of our results for
the traditional selection of a ``clean-bulge'' sample using cuts on
longitudinal proper motion \muL.

\subsection{Difference amplification by photometric parallax mixing}
\label{ss:distanceMixing}

Differences in apparent magnitude distribution other than due to
distance spread would contribute to differences in the inferred
\rdm~distributions for the \MR~and \MP~samples. If sufficiently
severe, this differential blurring in \rdm~might cause two
intrinsically identical rotation curves to be erroneously measured as
discrepant. In the sense of our findings, the \MP~sample might be
artificially blurred in \rdm~compared to the \MR~sample, which would
produce an {\it apparent} rotation curve discrepancy where none were
present.

Several phenomena might lead the \MP~sample to exhibit greater
apparent magnitude scatter than the \MR~sample. Firstly, since the
\MP~ridgeline in the SWEEPS CMD is slightly fainter than the
\MR~ridgeline, the \MP~objects may be subject to increased photometric
uncertainty. Secondly, at least in principle, if the extinction
distribution experienced by the two samples were in some way
different, this could lead to a broader apparent magnitude
distribution for the \MP~sample. Thirdly, differences in binary
fraction between the samples might cause the relative photometric
parallax distribution of the two samples to differ, although the
nature, magnitude and direction of such effects may be complex and
indeed depend on the class of binaries probed (e.g. \citealt{gao14}).

Finally, differences in the {\it intrinsic} photometric scatter
  between the \MR~and \MP~samples might amplify differences between
  the rotation curves. Our own VLT spectroscopy, as well as
  spectroscopic campaigns from the literature
  \citep[e.g.][]{zoccali17, hill11} suggest that the \feh~spread for
  the \MP~population is greater than for the \MR~population, which
  would in turn contribute greater \rdm~scatter in the
  \MP~population.

We have performed simple Monte Carlo tests to determine whether
perturbations in the inferred distance distribution can be responsible
for the differences in rotation curves between \MR~and
\MP~samples. Appendices \ref{s:appDist} and \ref{s:appMetBlur} provide
details.

In the course of investigating the impact of the differential
  \feh~distribution on the \rdm~distribution, it became apparent that
  the \basti~set of artificial stellar population methods used to
  generate synthetic \feh~distributions, were (at the time of this
  work) imposing an apparently artificial population
  truncation. \autoref{s:testBasti} provides details, with the
  method we adopted to mitigate this selection effect discussed in
  \autoref{s:appMetBlur}.

Perturbations were tested due to additional photometric uncertainty or
differential extinction variations (Appendix
\ref{ss:distBlur:photom}), differences in the fraction of unresolved
binaries (Appendix \ref{ss:distBlur:binaries}) and in the photometric
parallax spread caused by differing intrinsic spreads in metallicity
(\autoref{s:appMetBlur}). In all scenarios, the effect is either too
small to bring the rotation curves into agreement (for binaries), or
the required perturbation is too large to have gone un-noticed in
previous studies (for extinction), possibly by an order of magnitude
(for photometric uncertainty). The strongest single contributor
of \rdmLong~mixing is intrinsic difference in metallicity spread
between the samples; this likely contributes differential
distance-mixing up to a third the amount required to artificially
reproduce the observed discrepancy in rotation curves. Since
independent sources of additional photometric scatter would presumably
add in quadrature, their combination is very unlikely to be sufficient
to bring about the observed discrepancies in trends.

We therefore conclude that differential distance scatter is not
responsible for the difference in rotation curves or
  \rdm~distributions, due to additional photometric uncertainty,
differential extinction, differences in the unresolved binary
populations, or in the differences in metallicity spread between
  samples.

\subsection{Difference reduction by sample cross-contamination}
\label{ss:mtMixingDiscussion}

While blurring in \rdmLong~would tend to artificially increase
the difference between trends in the \MR~and \MP~samples,
cross-contamination of the samples in (\mtindices) would tend to
artificially reduce these differences. While we have used reasonably
conservative thresholds in drawing our \MR~and \MP~samples, genuinely
metal-rich objects might be moved into the \MP~sample by measurement
uncertainty, and vice versa.

Because of the complexities involved in rigorous reconstruction of the
observed distributions \citep[e.g.][]{gennaro15}, full exploration of
this cross-contamination is deferred to future work. We have performed
a simple Monte Carlo contamination test for the formal membership
probability threshold $\wik > \probThresh$~used in this work
(\autoref{s:appContam}). Under the assumptions of that test, we find
that the \MR~sample is contaminated at the $\sim 5\%$~level (mostly
from the \MP~sample), while the \MP~sample is contaminated at the
$\lesssim 1\%$~level (mostly due to the \MR~sample, but with some
  contribution from background component $k=3$~in
  \autoref{tab:gmm}). This is not severe enough for the observed
low-amplitude \MP~rotation curve to be due to sample contamination
from a small population of objects following the kinematics of the
\MR~sample.

\subsection{Trend modification by line-of-sight extinction variations}
\label{ss:extinctionFarside}

Our treatment of the impact of extinction on the photometry (and
therefore the \rdmLong) assumes the extinction is constant over the
line of sight distances of interest. Violations of this assumption
might in principle influence the trends we observe, by artificially
broadening the line-of-sight distribution (with stars more affected by
extinction appearing farther from the observer).\footnote{Indeed,
    superposition of extinction in two separate spiral arms along the
    line of sight might be partially responsible for the difficulties
    characterizing extinction law towards the bulge with simple models
    \citep[e.g.]{nataf16}.} Here we examine the likely impact on our
main results of extinction variations along the line of sight.

A few studies have mapped the three dimensional extinction
distribution out to the far side of the bulge
(e.g. \citealt{schultheis14, marshall06}). Particularly for
sight-lines close to the Sagittarius Window we study here, most of the
extinction at these distances takes place at distances $\dKpc \lesssim
5$~kpc from the Sun, possibly broken into two foreground
concentrations (at $\dKpc \approx 3, 5$~kpc;
e.g. \citealt{marshall06}). Thus, along our sight-line, extinction
variations within the bulge are likely to be small compared to variations in
the foreground disk, an interpretation consistent with the photometry
of Red Clump stars in this field \citepalias{clarkson08}. So, the
trends we find for line of sight distances $5 \lesssim \dKpc \lesssim
11$~kpc - the main sample of interest - are likely unaffected.

By symmetry we might expect additional extinction from at least one
dust screen at distances $\dKpc \gtrsim 11$~kpc due to spiral
structure on the far side of the bulge (see, e.g. Figure 10 of
\citealt{schultheis14}). This would be mitigated somewhat by the
slightly tilted path of our line of sight compared to the Galactic
plane; at Galactic latitude $b_{\rm J2000.0} = -2.65^{\circ}$, our
sight-line is already $\approx 480$~pc below the Galactic midplane
when it reaches $\dKpc=11$~kpc, roughly where it might intersect the
first dust concentration on the far side of the bulge (compared to
$\approx 210$~pc at 5 kpc), suggesting far-side extinction may likely
be somewhat weaker than experienced in the foreground.

We therefore conclude that indeed the photometric parallaxes for
objects closer than $\dKpc \approx 5$~kpc and farther than $\dKpc
\approx 11$~kpc may have been assigned photometric parallaxes that are
artificially close and far, respectively. However, as those distances
are outside our main regions of interest, this does not impact any of
the trends that we report.

\subsection{The influence of the proper motion and distance zeropoints}
\label{ss:refFrames}

All proper motions in this work are reported relative to the same
proper motion zeropoint \muvecref, which is defined as the average
proper motion of astrometrically well-measured stars, whatever their
metallicity (\autoref{ss:obsSWEEPS}). We have chosen not to apply
separate proper motion zeropoints to the two samples, for example by
forcing the two samples to each show \muL = 0.0~\masperyear~at \rdm =
0.0~magnitudes, but opt to keep the proper motions in the same
reference frame for each sample to allow direct comparison between
(\rdm, $\mu$)~rotation curves.

We then find that the central proper motions for the \MR~and
\MP~samples (i.e., the median proper motions for stars near their
fiducial sequences) differ by $\muCenLabel_{MR - MP}$~$\approx$
$(\muCenDeltal, \muCenDeltab)$~\masperyear~(\autoref{ss:rotnCurves});
equivalently, the rotation curves do not meet at $(\rdm, \muL)$=(0
mag,0 \masperyear).

These discrepancies could be due to differences in the mean intrinsic
velocities of the fiducial stars between the samples, or differences
in the line of sight distance at which the fiducual stars are found,
or a combination of the two. Since the fiducial sequences for the two
samples are determined from their CMD population densities see
\autoref{f:mtRidgelines}), their central proper motions could well
differ if their densest observed regions occurred at different
distances.
This could occur naturally if the two
  samples are oriented differently in the Galactic plane (in which
  case the relationship between the mean velocity \vvec~and its
  transverse velocity components \vL~\&~\vB~would also differ between
  the samples).

\begin{figure}
  \begin{center}
    \includegraphics[width=3.5in]{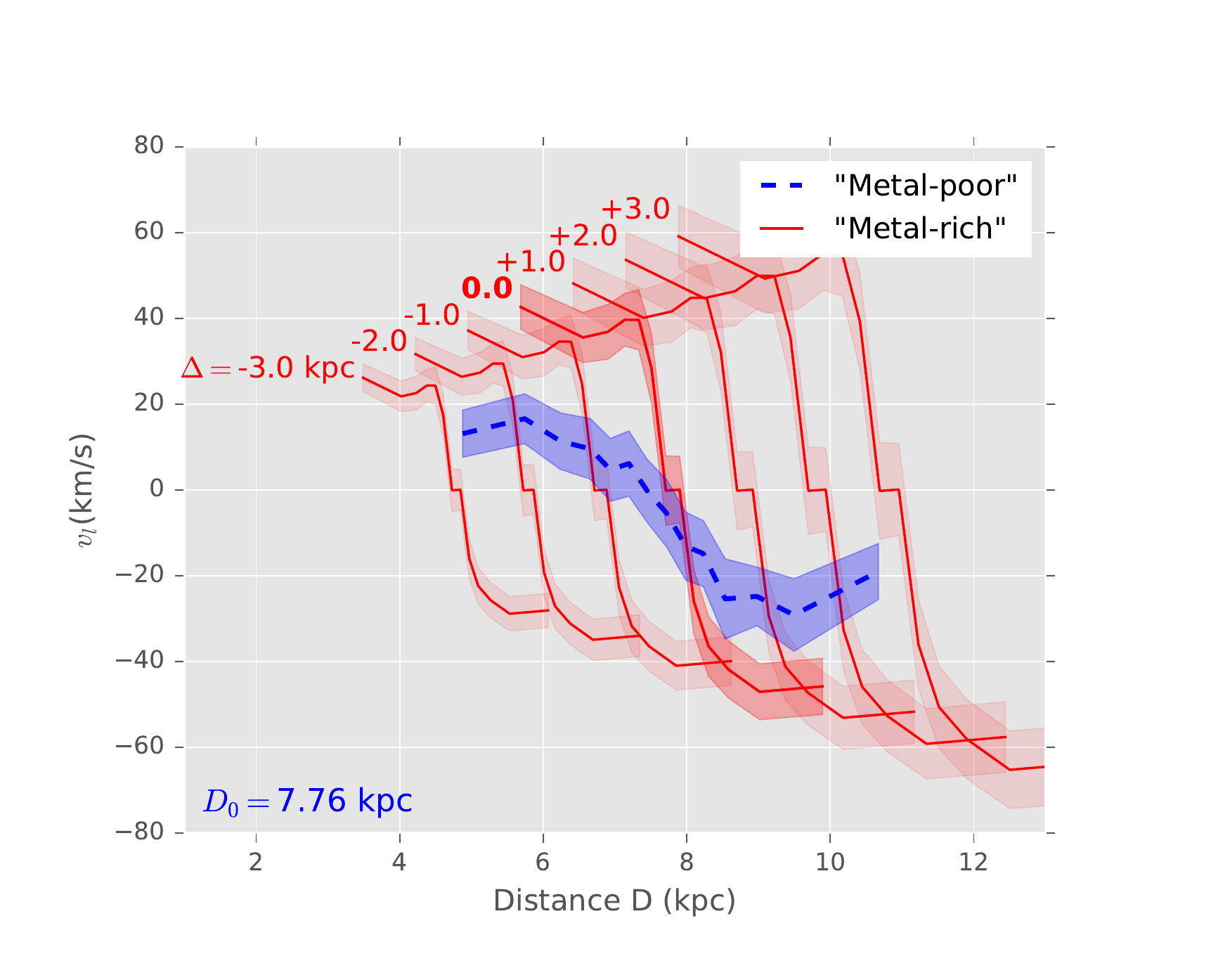}
    \end{center}
\caption{Estimating the impact on the velocity rotation curves of
    allowing the fiducial distance to differ between the samples. The
    blue dashed line shows the velocity rotation curve of the
    \MP~sample, converting from (\rdm, \muL) to (\dKpc, \vL) using
    fiducial distance \dKpcCen~(the shading encompasses $\pm
    1\sigma$~uncertainties at each distance bin). The red solid lines
    show velocity rotation curves for the \MR~sample, using fiducial
    distance $\dzer + \Delta$. Reading left-right, the distance offset
    varies over the range -3.0 kpc $\le \Delta \le$ 3.0 kpc~in 1~kpc
    increments. The case $\Delta = 0$~is highlighted for reference. See
    \autoref{ss:refFrames}.}
\label{f:tweakDist}
\end{figure}

If the fiducial stars for the two samples do indeed lie at different
distances $\dKpc_{0,k} = \dzer + \Delta_k$~(with $\Delta_k$~giving the
distance offset for a particular sample) then the appropriate
conversion from (\rdm, \muL)~to (\dKpc, \vL) will also differ, in turn
impacting the difference in rotation curve velocity amplitude for the
two samples. \autoref{f:tweakDist} shows how the velocity rotation
curve is impacted by shifting the \MR~fiducial by distance offset
$\Delta$~kpc from the \MP~fiducial. Bringing the \MR~fiducial closer
than the \MP~does reduce the velocity amplitude discrepancy between
the two samples. However, to bring the velocity amplitudes of the two
samples into agreement, the \MR~fiducial would need to be brought
closer by $|\Delta| \gtrsim 2$~kpc, which seems unlikely for samples
so close to the Galactic rotation axis (at $l \approx
+1.26^{\circ}$). Unless the spatial distributions of the two samples
really are radically different, then, we consider it unlikely that a
difference in fiducial distance between the samples can by itself
produce the observed difference in velocity rotation curve amplitude
we are measuring.

While an offset $\Delta$~in fiducial distance scales the velocity
amplitude by a corresponding amount, an offset \muvecrefshift~in the
proper motion zeropoint produces a systematic shift $\trendVvecShift =
-4.74\muvecrefshift$~in the velocity gradient. If \muvecref~(the
average proper motion of well-measured bulge stars of all
metallicities) and \dzer~(the average distance to bulge stars of all
metallicities) are both determined from the same set of stars, then we
would have $\muvecrefshift=0$~\masperyear~and thus $\trendVvecShift =
0$~\kmpersecperkpc.

However, in reality the sets of stars used to estimate \muvecref~and
\dzer~will in general differ. The fiducial distance \dzer~is estimated
from the distribution of ``extreme-bulge''~(EB) stars showing $\mu_l <
-2.0$~\masperyear~(e.g. \citetalias{calamida15}) while \muvecref~is
estimated from stars below the \MSTO~without any proper motion
selection.
Thus, although stars well-measured astrometrically tend also to be
well-measured photometrically, differing selection effects in the
determination of \dzer~and \muvecref~will still lead to a global
systematic offset in \muvecrefshift.

The true value of \muvecrefshift~is unknown, however we can form a
rough estimate as the median proper motion of the population traced by
the EB objects used to estimate \dzer~\citepalias{calamida14}. The EB
objects have highly negative proper motions by construction, but we
can estimate the median proper motion of the underlying population
that they trace by estimating the median photometric parallax
\medrdm~of the EB tracers and applying the proper motion rotation
curve characterization $\muL(\rdm)$~of
\autoref{ss:resultsCurves}. Since we have performed this
characterization separately for \MR~and \MP~samples, we can estimate
\muvecrefshift~separately from the two samples. Applying the kinematic
cut $\muL \le -2.0$~\masperyear~to extract EB~tracers for the \MR~and
\MP~populations, we find median photometric parallax $\medrdmMR
\approx +0.025$, $\medrdmMP \approx -0.002$~magnitudes for EB objects
in the \MR ~\& \MP~samples, respectively (so that objects with $\muL =
0$~\masperyear~lie slightly in front of the EB population, as
expected; see \autoref{ss:pmSelection}). This suggests that the
underlying population traced by the EB objects - corresponding to the
fiducial distance \dzer - has median longitudinal proper motion
$\muLcentMR \approx -0.17, \muLcentMP \approx 0.00$~\masperyear. These
figures are likely also sensitive to differing intrinsic proper motion
and distance distributions between the samples, but this estimate
suggests that the proper motion difference between the sample from
which the proper motion reference frame was set, and the sample from
which \dzer~was estimated, is not larger than \muvecrefshift~$\lesssim
0.17$~\masperyear. Thus, the systematic velocity gradient uncertainty
\trendVvecShift~may be on the order of $\sim \trendvecmag$.

Systematic uncertainty in the proper motion zeropoint \muvecref~may
therefore impact the ratio of longitudinal velocity gradients reported
in \autoref{ss:resultsCurves} by $\lesssim
10\%$~(\autoref{tab:trends}), which is too small to materially affect
the main results or conclusions we report.

\subsection{Does the proper motion rotation curve vary with \feh?}
\label{ss:twoPopns}

We are finally in a position to answer the question posed by
\autoref{ss:introBifurcation}. Our \MR~and \MP~rotation curves are
inconsistent with each other at $\sim \slopeRatioSignif$~for the
rotation curve slope and $\approx \amplRatioSignif$~for the
nearside-farside rotation amplitude (\autoref{ss:resultsCurves}).

These proper motion-based results stand in strong contrast to
determinations of the radial velocity rotation trends, which either
find weak if any discrepancy between \MR~and \MP~samples
\citep[e.g.][]{ness13_part3, williams16, zoccali17}, or require a
large contribution from samples with $\feh < -1.0$~to produce a
discrepancy in rotation curves (e.g. \citealt{minniti1996, kunder16};
note that the fraction of stars in our sight-line with $\feh <
-1.0$~is likely low; \citealt{zoccali17}). It seems unlikely that this
discrepancy between our proper motion- and these radial velocity-based
studies can be due purely to any differences between the uses of giant
and dwarf stars as tracers, since microlensed {\it dwarf} stars also
show no strong differences in mean radial velocity between metal-poor
and metal-rich stars (or, for that matter, between stars younger and
older than $\sim$7 Gy; \citealt{bensby17}).\footnote{See
  \citet{cohen10}~for discussion of possible differences in
  metallicity distribution between dwarfs and giants due to
  stellar-evolutionary effects.}

The most likely explanation for the difference between proper motion-
and radial velocity results is the strong difference in the way the
studies sample the inner Milky Way. Detailed comparison of our new
observational indications with model prediction is deferred to future
work; however, a likely scenario to explain the differences can be
outlined as follows. In the simulations of \citet{debattista17}, the
final orbital configuration of bulge stars depends on their (radial)
velocity dispersion before bar formation. To the extent that \feh~and
radial velocity dispersion correlate with each other (or,
equivalently, each correlate with time of formation), the ``kinematic
fractionation'' resulting from bar formation might well leave
metal-rich stars with a higher fraction of elongated orbits than for
metal-poor objects. That the ``X''-structure is {\it observed} to
preferentially contain metal-rich objects (e.g. \citealt{vasquez13})
supports the notion that stars within our \MR~and \MP~samples which on
average move along differently-shaped orbits, while the spatial
structures traced by Miras of different pulsation period ranges
suggest that samples with differing spatial configuration can co-exist
in the same volume at the present day; \citep{catchpole16}. The
\MR~and \MP~samples may then show quite different transverse velocity
distributions as a function of line of sight distance, even if the
distributions produce similar mean velocities when averaged along the
line of sight due to radial velocity survey selection effects. In this
scenario, only by dissecting the population by line of sight distance
(or its proxy, \rdm), can the differing velocity distributions of the
two co-existing samples be distinguished.

Since stars in the \MR~and \MP~samples all move through the same
present-day potential, by detecting differences in the proper
motion-generated rotation curve, we may well be detecting differences
in orbital anisotropies between metal-rich and metal-poor bulge
objects. While detailed prediction is a topic of ongoing work, the
differences we detect seem qualitatively reasonable at present.

Having shown that the proper motion-based rotation curve does show
discrepancy between \MR~and \MP~populations, the necessary next step
is to extend our approach to more sight-lines within the inner
bulge. By comparing metallicity-dissected proper motion-based rotation
curves between fields, the trends with location in the bulge can be
charted empirically, allowing a sharper test of the true variation of
bulge rotation with the metallicity of the sample probed. This work is
deferred to a future communication.

\subsection{Proper motion dispersion trends with photometric parallax}
\label{ss:resDisp}

Both the \MR~and \MP~samples show a clear central peak in velocity
ellipse major axis length near the distance interval where the samples
are the most densely populated (\autoref{f:varMotionVel}). The peak
persists in the velocity ellipse minor axis~for the \MR~sample, but is
rather less clear in the \MP~sample. This is broadly similar to the
trends found from the combined population in previous studies
(e.g. \citetalias{clarkson08}, \citealt{soto14}). We note a rough,
qualitative similarity with the curve of radial velocity against
Galactic longitude for Galactic latitude $b = -2^{\circ}$~(see the
middle-left panel of \citealt{zoccali17} Figure 12), however we remind
the reader that the radial velocity and proper motion trends cannot
directly be compared because they suffer from differing selection
effects. That the proper motion dispersion of the \MP~component is
generally slightly larger than that of the \MR~(particularly along the
minor axis), is qualitatively consistent with expectations that a
metal-poor, less rotationally-supported population should show higher
velocity dispersion \citep[e.g.][]{debattista17, ness13_part3}.

We may also be detecting the velocity-dispersion ``inversion''
detected at the inner-most fields in radial velocity studies
\citep{babusiaux14, zoccali17}. Consistent with the low-latitude
radial velocity dispersion trends, the proper motion-based velocity
dispersion might also be greater for the \MR~sample than for the
\MP~sample at the distance-bins closest to the center of the Bulge
(see \autoref{f:varMotionVel}). For the inner-most bulge regions, the
proper motion-based \MR~velocity dispersions also show steeper gradient
than the \MP, but with the gradient against line-of-sight distance
rather than Galactic longitude, with the inner-most distance bin
possibly showing slightly greater velocity dispersion for
the \MR~sample. 

\subsection{The line of sight distance distributions of the two samples}
\label{ss:meanDists}

The tendency of the \MP~sample to show greater dispersion in
\rdmLong~(or, correspondingly, in distance \dKpc), is qualitatively
consistent with the ``kinematic fractionation'' of
\citet{debattista17}. Under that mechanism, more \MP~populations also
initially had greater radial velocity dispersion, leading to a
distinct (and broader) present-day spatial distribution when compared
to the most \MR~objects.

The difference we find in line-of-sight distribution between the
\MR~and \MP~samples might also be consistent with the observations of
\citet{catchpole16}, who find differing bar angles and degrees of
central concentration for Mira variables of different ages. However,
the interplay between age and metallicity of bulge stars is likely not
simple. For example, while a gentle relationship may exist between
\feh~and the fraction of stars younger than about 8 Gy (e.g. Figure 14
of \citealt{bensby17} or Figure 10 of \citealt{bernard18}), the
microlensing spectroscopic surveys suggest that stars can take any
metallicity value (for $\feh \gtrsim -1.0$) for any age. How the
predictions of \citet{catchpole16} translate into predictions for the
two samples here is deferred to future work.

\subsection{Are the metallicity-samples themselves composite?}
\label{ss:multiComponents}

\begin{figure*}
\centerline{
  \includegraphics[width=8cm]{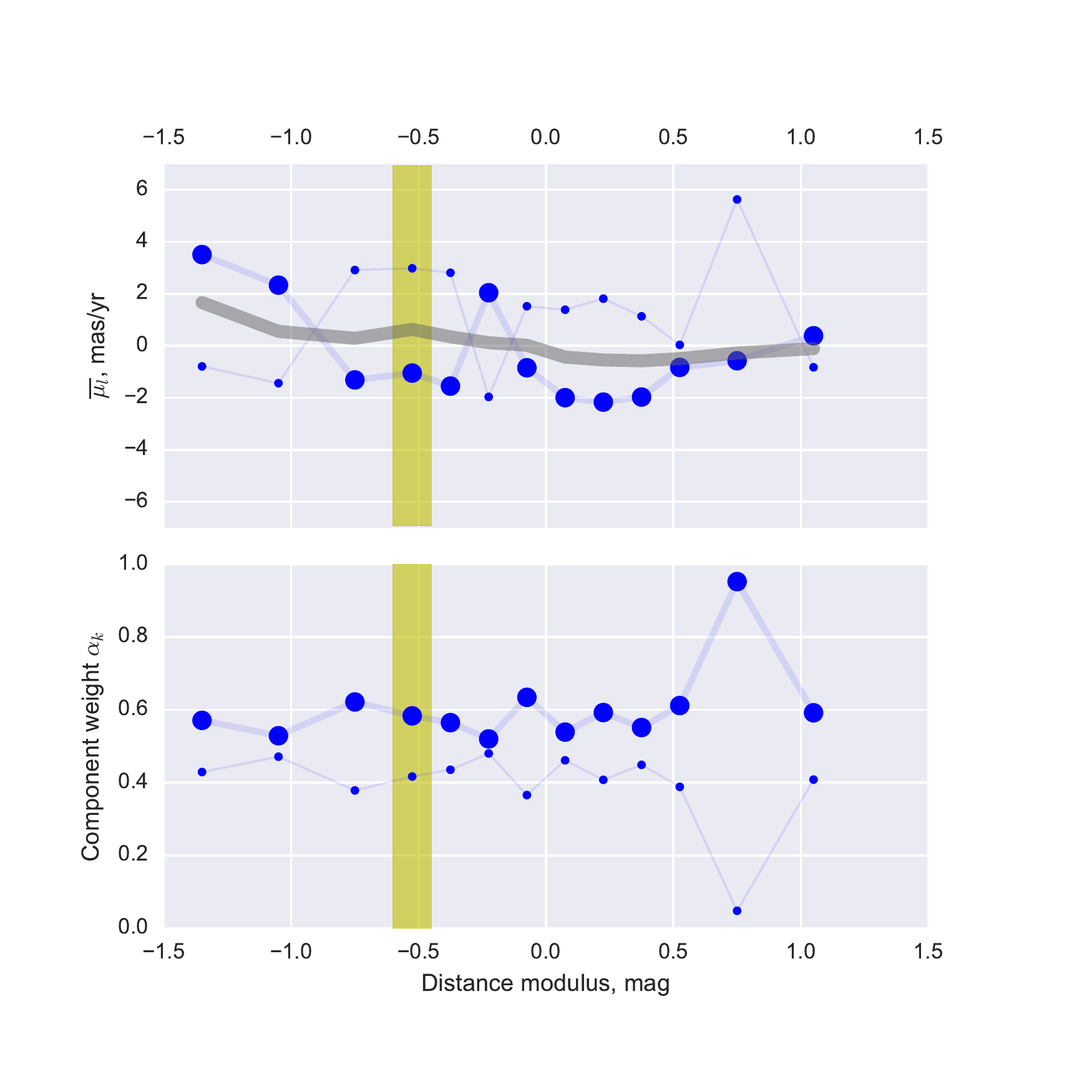}
  \includegraphics[width=8cm]{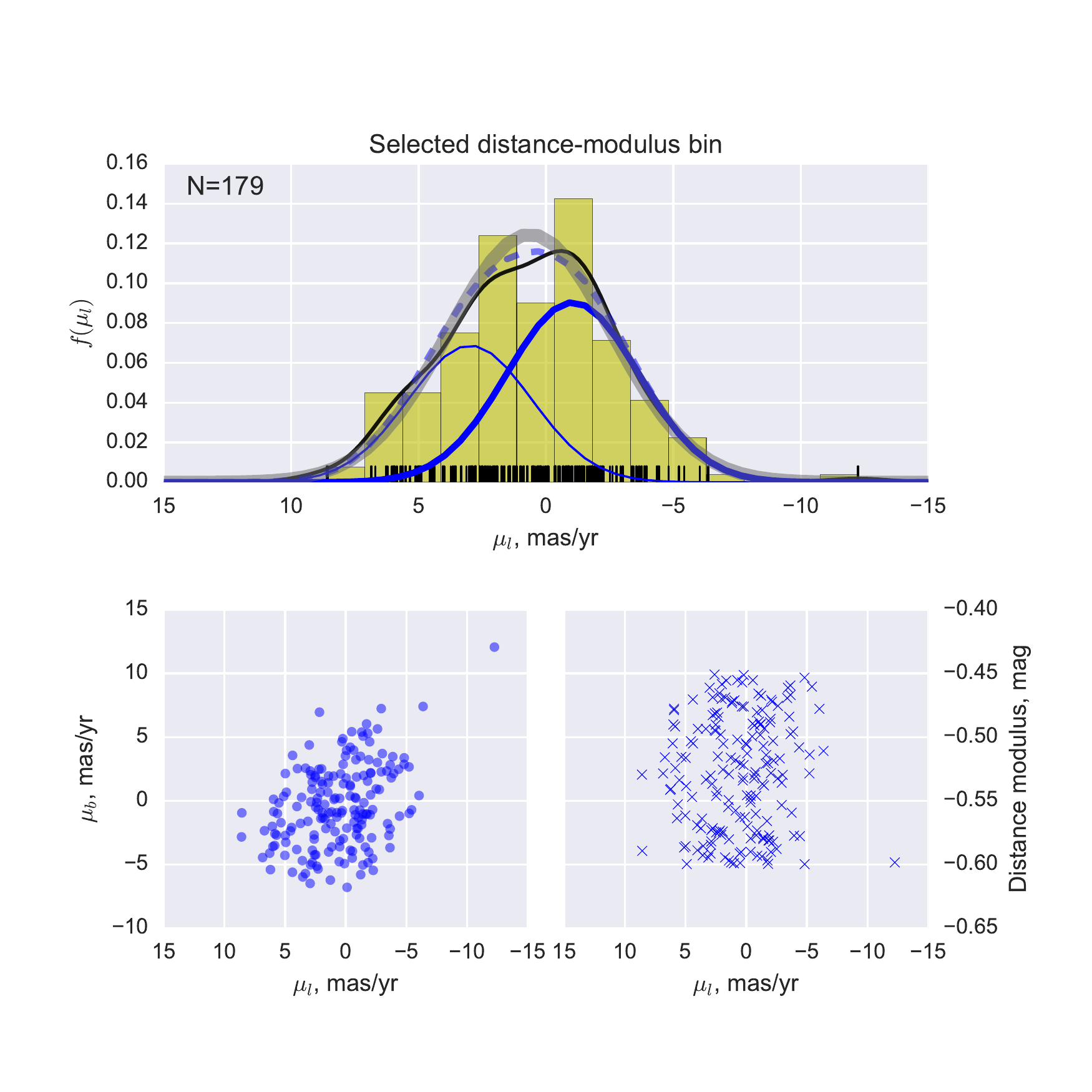}
}
\caption{Representation of the \MP~rotation curve as a two-component
  GMM in \muL, to test the hypothesis that the \MP~sample might itself be composite. {\it Left column:} rotation curve in \muL~as tracked by
  the two model components ({\it left top}), and the relative weights of the
  two model components ({\it left bottom}). Large dots indicate the more
  populous of the two model components in each bin ($\alpha_k = 0.75$~would mean three quarters of the sample came from model component $k$). The gray line in the left-top panel shows the rotation curve inferred using a single model component at each distance-bin. {\it Right column:} \muL~distribution for the bin indicated for the shaded distance-bin in the left-column. {\it Right top:} \muL~distribution (shaded histogram and thin black line), with the prediction of the two-component GMM (the thick and thin blue lines indicate the more- and less-populous model component, respectively, while the blue dashed line indicates the sum of the two). The gray thick line shows the prediction of the single-component model. The {\it bottom left} and {\it bottom right} panels show the vector point diagram and distance modulus distribution, respectively. See \autoref{ss:multiComponents}.}
\label{f:multiComponentsMP}
\end{figure*}

In addition to any continuous metallicity-velocity correlation, the
samples may include populations from distinct entities within the
Bulge region, whether interlopers from the Halo
\citep[e.g.][]{koch16}, any Thick-disk component
\citep[e.g.][]{ness13_part2} or a small ``Classical'' bulge component
\citep{kunder16}.

\begin{figure*}
\centerline{
  \includegraphics[width=8cm]{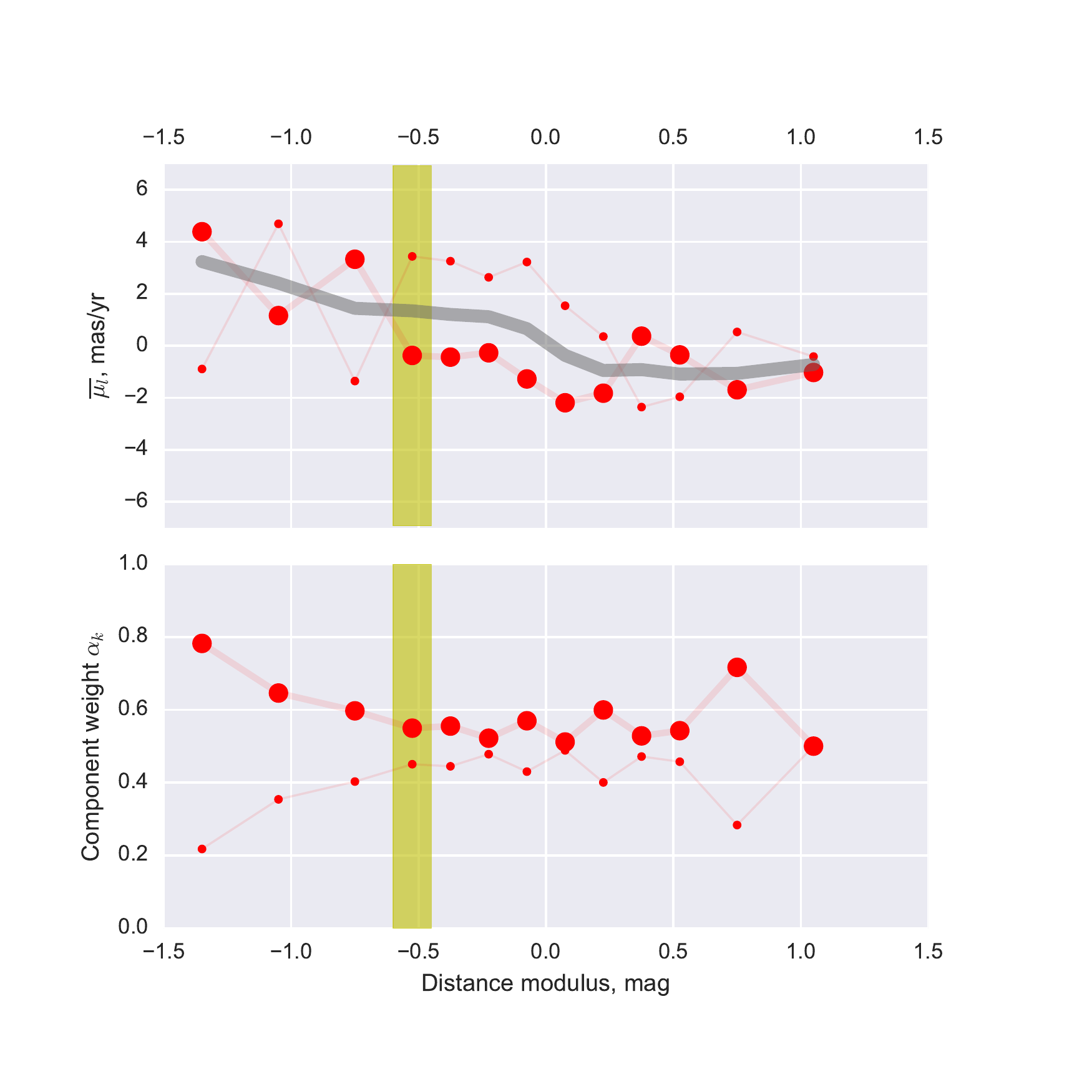}
  \includegraphics[width=8cm]{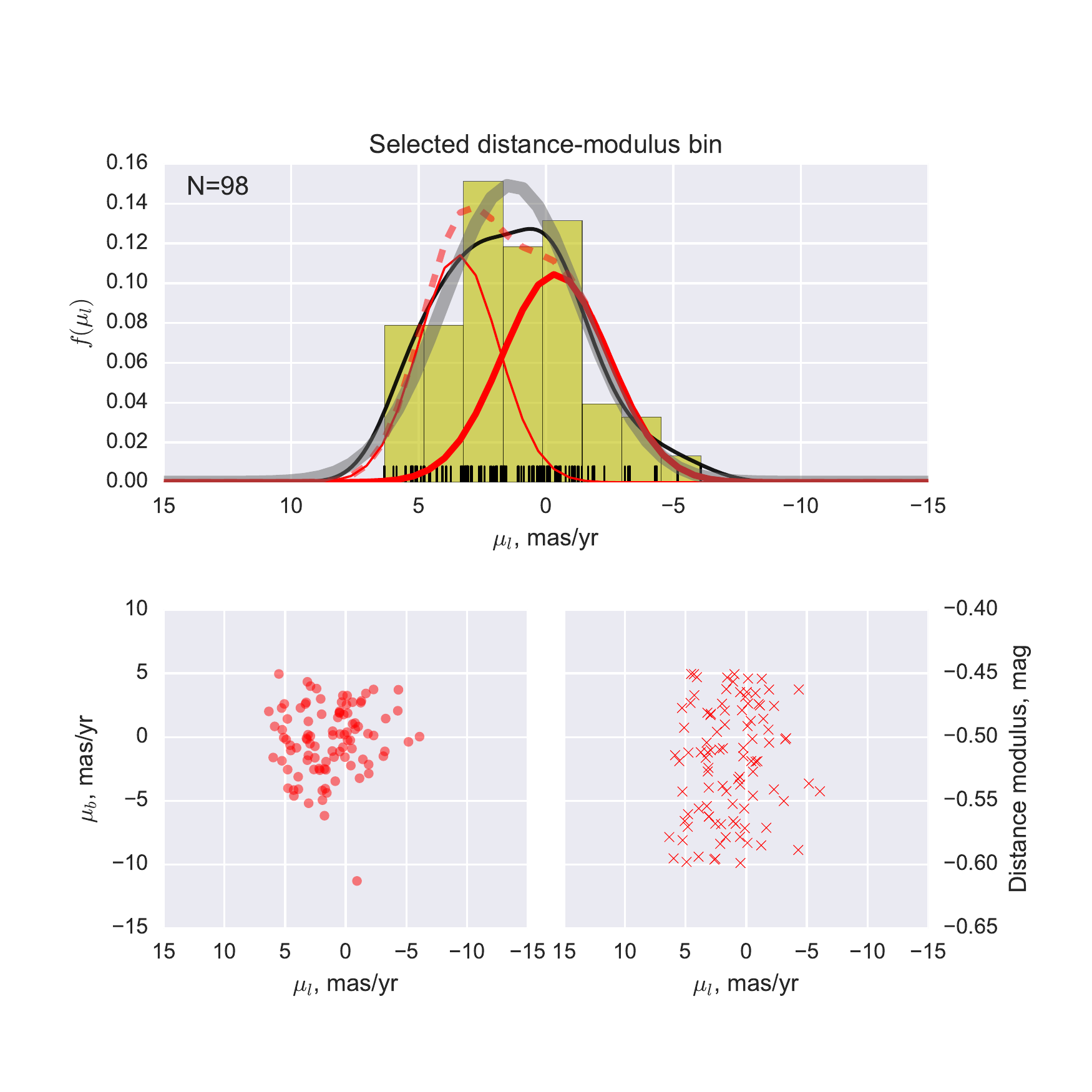}
}
\caption{Representation of the \MR~rotation curve as a two-component
  GMM in \muL, to test the hypothesis that this sample might be a composite of two sub-populations. Symbols similar to \autoref{f:multiComponentsMP},
  except red symbols and lines are substituted for blue. See
  \autoref{ss:multiComponents}.}
\label{f:multiComponentsMR}
\end{figure*}

We have therefore performed the exercise of decomposing each of the
\MR~and \MP~samples into two-component GMM's (in \muL~only), to
determine if any minority component is distinguishable within the
rotation curves formed from the two samples (figures
\ref{f:multiComponentsMP} \& \ref{f:multiComponentsMR}). No minority
population is detected in either sample; indeed, when a two-component
GMM is used, the two centroids track the mean rotation curve within
each sample roughly symmetrically about the mean rotation curve, while
each sub-component has roughly equal weight in the mixture.

We therefore conclude that a minority component with discrepant
rotation curve is not required in either the \MR~or \MP~sample, but
due to the small sample size ($\approx 2,000$~stars in total per
sample), we cannot at this stage rule out its presence. Direct
comparison with population models may allow upper limits to be set on
the presence of any minority component within each sample, but this is
deferred to future work.

\subsection{Implications for proper motion selection}
\label{ss:pmSelection}

Photometric studies of the Bulge typically impose a condition $\mu_l <
-2.0$~mas yr$^{-1}$~to isolate a clean bulge sample for further study
\citep[e.g.][]{kr02, calamida14}, although there are exceptions
(e.g. \citealt{bernard18}).\footnote{(Here the symbol \muL~takes
  exactly the same meaning as elsewhere in the present report,
  referring to proper motion relative to mean bulge objects rather
  than relative to the Sun. Thus, the proper motion cut $\muL <
  -2.0$~mas yr$^{-1}$~selects objects on the far side of the
  longitudinal proper motion distribution from foreground Disk
  objects.)} This procedure is appropriate because in the sight-lines
  typically studied near the Galactic center, the foreground disk
  population typically shows proper motion relative to the mean-bulge
  population of $\Delta \mu_l \approx +4$~mas yr$^{-1}$, as suggested
  by direct comparison of the proper motions of bulge giant branch
  stars with those of the upper main-sequence population of (mostly)
  disk foreground stars (e.g. \citetalias{clarkson08},
  \citealt{soto14}).

\begin{figure*}
\centerline{
  \includegraphics[width=8cm]{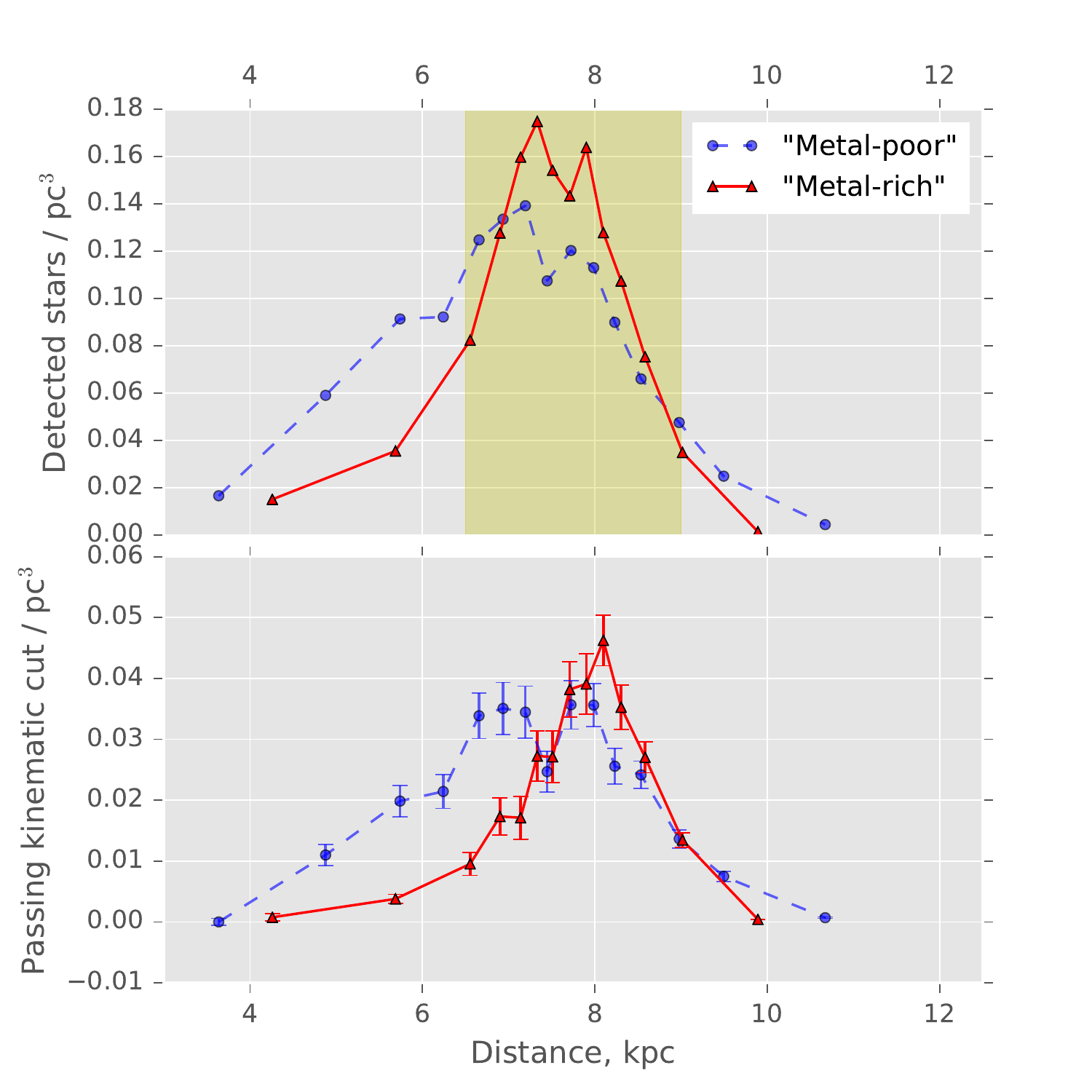}
  \includegraphics[width=8cm]{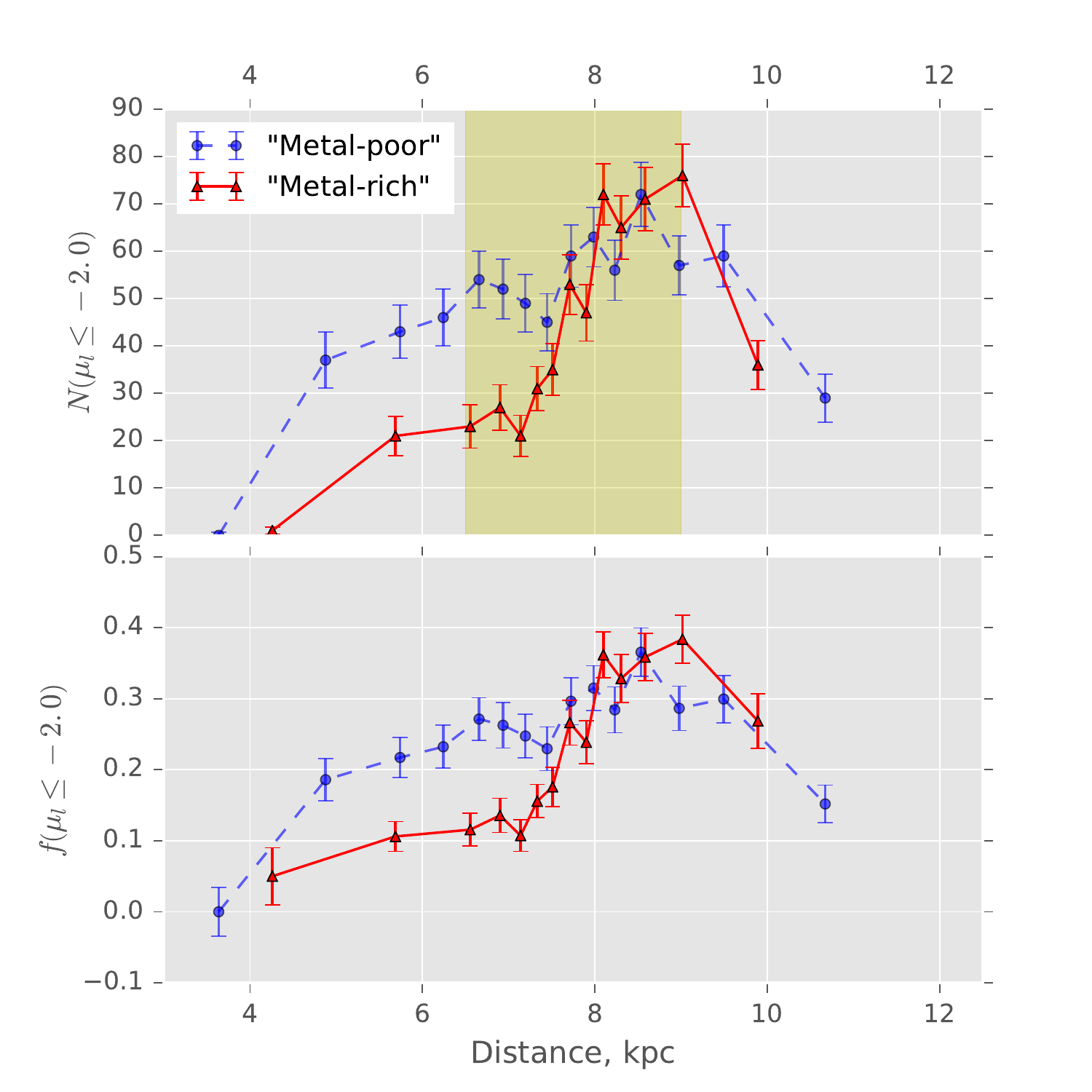}
}
\caption{Selection functions imposed by the traditional kinematic cut
  $\mu_l \le -2.0$~mas yr$^{-1}$, for the \MR~(red triangles, solid
  line) and \MP~(blue circles, dashed line) samples. {\it Left
    column:} volume density of all objects assigned to each sample
  (left-top) and which would pass the kinematic cut (bottom
  panel). {\it Right column:} number of objects per sample that would
  pass the kinematic cut (right-top) and the fraction (right-bottom)
  that would pass the cut. Each distance bin has the same number of
  detected objects by construction (\autoref{ss:pmEllipse}), so
  bootstrap uncertainty estimates are not applicable to the top-left
  panel. Because of this binning scheme, bins with the same counts
  generally do not indicate the same volume density of detected
  objects (between bins or between samples). The shaded yellow region
  represents inferred distances $(6.5 \leq d \leq 9.0)$~kpc. For
  discussion, see \autoref{ss:pmSelection}.}
\label{f:curveNdetectMu}
\end{figure*}

To investigate whether and how a simple cut on \muL~imposes selection
effects on the two samples, we computed the sample counts, fractions
and volume densities for objects that would pass the longitudinal
proper motion cut ($\mu_l < -2.0$~mas yr$^{-1}$). The results are
plotted in \autoref{f:curveNdetectMu} and presented in tabular form in
Tables \ref{tab:stats:Metal-rich} and \ref{tab:stats:Metal-poor} in
\autoref{s:rotCurveTables}.

We find that the cut ($\mu < -2.0$~mas yr$^{-1}$) admits very few
foreground objects from either sample; both samples show fewer than
two objects passing this cut for the closest distances ($\dKpc
\lesssim 4.53$~kpc and $\lesssim 4.10$~kpc for \MR~and \MP~samples,
respectively; see \autoref{tab:stats:Metal-rich} and
\autoref{tab:stats:Metal-poor} in \autoref{s:rotCurveTables}), while
the mean proper motion \muLcent~of the foreground population climbs
strongly for the closest distance bins (\autoref{f:meanMotionMu}). We
therefore confirm that the traditional proper motion cut ($\mu <
-2.0$~mas yr$^{-1}$) does indeed remove nearby objects cleanly for the
\SWEEPS~field.\footnote{Strictly speaking, the classification of the
  nearest objects into \MR~and \MP~samples may suffer different
  selection effects to the rest of the samples because either or both
  of the stellar parameters and extinction might be different for the
  very nearest objects compared to the majority sample at more
  bulge-like distances. We are making the assumption that the relative
  classification of objects in the nearest distance bins, is still
  valid.}

Beyond this, however, the dissection by relative abundance has
revealed several interesting selection effects among the
kinematically-cleaned sample (\autoref{f:curveNdetectMu}).

Firstly, as expected, there is a bias towards the far side of the
bulge, but this bias is much stronger in the \MR~sample than for the
\MP; indeed the fraction of \MP objects passing the kinematic cut is
almost flat with inferred distance between $d \lesssim 5$~kpc
$\lesssim 9$~kpc.

Secondly, the raw counts of sources thus isolated in the \MR~and
\MP~samples are of similar orders of magnitude. Considering sample
sizes that pass the kinematic cut at inferred distances between 6.4
and 9.1 kpc (chosen to encompass the bulge populations; see Tables
\ref{tab:stats:Metal-rich} and \ref{tab:stats:Metal-poor}), the total
  counts in each sample are $521 \pm 19$~and $507 \pm 19$~for the
  \MR~and \MP samples, respectively (the uncertainties, estimated from
  the quadrature sum of parametric bootstrap uncertainty estimates in
  these counts for each bin, are almost certainly
  underestimates). With total sample sizes within this distance range
  of 2181 (1783)~for the \MR~(\MP)~samples, this translates into
  fractions $24\% \pm 1\%$~($28\% \pm 1\%$)~of the \MR~(\MP) samples
  that pass the kinematic cut. Thus, {\it of objects in this distance
    range}, the kinematic cut appears to slightly favor the \MP
  sample, although the difference is small.

In principle, a population of compact objects among the foreground
population, might fall into the farther distance-bins for the
\MP~sample,\footnote{At colors typical of the \MP~sample, the quiescent dwarf
  novae found by \citetalias{calamida14} at the distance of the bulge
  show \filtVbroad~$\sim 28$. Similar objects in a very nearby
  foreground disk population ($\lesssim 3$~kpc) might fall within the
  faintest bins of our chosen sample.} polluting a sample with
bulge-like motions with a small population showing disk-like
motions. However, with the foreground disk population at $\sim
10\%$~of the total \citepalias{calamida14} and with a substantial WD population perhaps
unlikely for a typical ``young'' foreground population, we do not
consider this a significant contaminant, and leave exploration of the
impact of foreground WDs to future work.

\section{Conclusions}
\label{s:conclusions}

We have performed an exploratory study to determine the utility of
\HST~proper motions in charting the kinematic behaviors of \MR~and
\MP~samples within the Galactic bulge from their proper motions,
extending the rotation-curve technique first pioneered by
\citet{kr02}. The ultra-deep \SWEEPS~photometric and astrometric
dataset communicated in \citet{calamida14}~was merged with the WFC3
Galactic Bulge Treasury Survey \citep{brown10}, \MR~and \MP~samples
were drawn using the \mtindices~indices of \citet{brown09}, recomputed
for the stellar parameters appropriate to the proper motion sample of
interest and assuming $\rv=2.5$. The proper motion-based rotation
curves were determined from the \MR~and \MP~samples separately, using
relative distance modulus as the depth co-ordinate. While detailed
comparison to population models is deferred to future work, we draw
the following conclusions at present:

\begin{itemize}
  \item{The union of \SWEEPS~and \BTS~datasets has revealed that
    indeed the \MR~and \MP~rotation curves are clearly discrepant from
    each other.} 
\item{Characterizing the rotation curves for the inner bulge regions
  with straight-line fits, the \MR~population shows a steeper rotation
  curve in Galactic longitude, with gradient ratio \ratiogradL =
  \slopeRatio~(a $\approx$ \slopeRatioSignif~detection).}
\item{The nearside-farside velocity amplitude is also determined to be
  discrepant; the rotation curve amplitude \symbolampl~of the
  \MR~sample is greater than that of the \MP~sample by a factor
  $\ratioamplL = \amplRatio$~(a $\approx
  \amplRatioSignif$~detection).}

\item{While selection effects are likely complex, it does not appear
  to be possible to force the rotation curve of the \MR~sample into
  consistency with that of the \MP~sample by any reasonable
  observational perturbation of the \MR~sample. Therefore, the
  differences in rotational behavior likely represent intrinsic
  behavior, {\it not}~instrumental or observational artefacts.} 

\item{The velocity dispersion curve of both samples shows a clear peak
  at the line of sight distance where the samples are most dense. At
  the innermost distance-bins, the velocity dispersion of the
  \MR~sample shows a steeper gradient than does the \MP~sample,
  consistent with recent radial velocity studies.}

\item{These results may indicate differences in orbital anisotropy
  between metal-rich and metal-poor objects within the bulge, in turn
  providing a new observational criterion for testing models of bulge
  formation and evolution.}

\item{The traditional proper motion cut used to isolate a clean-bulge
  sample, $\mu_l < -2.0$~mas yr$^{-1}$, slightly over-selects
  \MP~objects compared to \MR, at the level of $28\%$~compared to
  $24\%$.}

\item{However, this selection effect is a function of \rdmLong; with
  this cut, the fraction of \MP~objects selected is roughly constant
  while for the \MR~population, the selection strongly prefers objects
  on the far side of the Bulge.}

\end{itemize}

In addition, while exploring population systematics, we have found that:
\begin{itemize}
  \item{The current version (v5.0.1) of the widely-used \basti~set of synthetic stellar population methods and isochrones appears to be imposing a truncation on populations near the edges of the \feh~distribution found in the bulge; this includes a large part of the metallicity range traced by stellar halo models (e.g. \citealt{an13}). Studies using \basti~version 5.0.1. or earlier may be vulnerable to this truncation.}
\end{itemize}

The Galactic bulge thus joins the list of stellar populations
suspected to show distinct rotation curves depending on the chemistry
of the tracer stars used, including at least one Globular cluster
(M13; \citealt{cordero17}) and the Sculptor dwarf spheroidal galaxy
(e.g. \citealt{zhu16} and references therein).

While the \SWEEPS~dataset represents the deepest (by far) set of
images ever taken by \HST~towards the inner bulge, the typical
apparent magnitude range probed by this study is shallow enough that
we expect the techniques presented herein to be applicable to other
fields for which (\mtindices) are available. The extension of this
work to the other fields in the \BTS~dataset is deferred to a future
communication. This will provide a relatively assumption-free set of
observational constraints against which the trends from the most
recent set of models can be compared, subjecting them to direct
test. This will finally enable the Galactic bulge to be used as a
quantitative test-case for the formation and development of galactic
structure.

\acknowledgements
\section{Acknowledgements}

This work made use of the Hubble Legacy Archive, which is a
collaboration between the Space Telescope Science Institute
(STScI/NASA), the Space Telescope European Coordinating Facility
(ST-ECF/ESA) and the Canadian Astronomy Data Centre (CADC / NRC /
CSA). Support for programs 9750, 11664, 12020, 12586 and 13057 were
provided by NASA through grants from STScI, which is operated by AURA,
Inc., under NASA contract NAS 5-26555. This work is partly based on
observations collected at the European Organisation for Astronomical
Research in the Southern Hemisphere under ESO programme 073.C-0410(A).

This work has made use of data from the European Space Agency (ESA)
mission {\it Gaia} (\url{https://www.cosmos.esa.int/gaia}), processed
by the {\it Gaia} Data Processing and Analysis Consortium (DPAC,
\url{https://www.cosmos.esa.int/web/gaia/dpac/consortium}). Funding
for the DPAC has been provided by national institutions, in particular
the institutions participating in the {\it Gaia} Multilateral
Agreement.

WIC acknowledges support from the University of Michigan-Dearborn
through departmental startup funds (project U039878), and from the
Office of Research and Sponsored Programs (project U042549, {\it The
  Milky Way Bulge at UM-Dearborn}), and acknowledges partial support
from HST program GO-12020 (PI Clarkson). WIC acknowledges equipment
funding from a Theodore Dunham, Jr. Grant from the Foundation
Center. VPD is supported by STFC Consolidated grant ST/M000877/1. DM
and MZ acknowledge support by the Ministry of Economy, Development,
and Tourism's Millennium Science Initiative through grant IC120009,
awarded to The Millennium Institute of Astrophysics (MAS), by Fondecyt
Regular grants 1170121 and 1150345, and by the BASAL-CATA Center for
Astrophysics and Associated Technologies PFB-06.

All the external software packages and methods used in this work are
freely available to the community. This research made use of {\tt
  Astropy}, a community-developed core Python package for Astronomy
(Astropy collaboration,
2013).
This work
made use of the {\tt astroML} suite of tools for machine learning in
Astronomy. 
This work made use of {\tt
  scikit-learn}.
This work has made use of the {\tt pysynphot} synthetic photometry
utilities.
This work has made use of BaSTI web tools.

WIC thanks Jay Anderson, Jo Bovy, Dana Casetti-Dinescu, Oscar
Gonzalez, No\'{e} Kains, Andreas Koch, Vera Kozhurina-Platais and
Laura Watkins for enlightening interaction at various stages of this
analysis. This work was only possible thanks to the distortion
solution and astrometric measurement methods developed by Jay
Anderson. We thank Santi Cassisi for assistance with the
\basti~synthetic stellar population tools, and for kindly providing
custom synthetic populations at high metallicity.

Finally, we thank the anonymous referee, whose thorough reading and
insightful comments led to substantial improvement of the manuscript.

\facilities{HST(ACS), HST(WFC3), VLT(UVES)}

\software{Astropy \citep{astropy18}\footnote{\url{http://www.astropy.org/index.html}}, scikit-learn \citep{scikit-learn}\footnote{\url{http://scikit-learn.org/stable/}}, astroML \citep{astroMLText}\footnote{\url{http://www.astroml.org/}}, BaSTI \citep{pietrinferni04}\footnote{\url{http://basti.oa-teramo.inaf.it/index.html}}, pysynphot \citep{pysynphot}\footnote{\url{http://pysynphot.readthedocs.io/en/latest/}}}

\bibliography{rotcurves}

\appendix

\section{\SWEEPS~proper motion measurement uncertainty}
\label{a:unctyPM}

Proper motion uncertainties from the 2004-2013 \SWEEPS~data are
impacted by random uncertainties, by intrinsic velocity dispersion of
the objects used to fit frame transformations when estimating proper
motions, and by residual relative distortion between epochs. Here we
discuss these sources of uncertainty in turn.

As part of the
investigation of the faintest detectable objects in the \SWEEPS~field,
\citetalias{calamida15} performed extensive artificial star-tests including
the injection of proper motions across the entire set of 2004-2013
epochs, yielding the run of random proper motion uncertainty in each
co-ordinate with apparent magnitude, which we denote here as
$\xi(\filtI)$. While \citetalias{calamida15} thus produced separate
estimates for uncertainties in the detector-X and detector-Y
directions, for the apparent magnitude range of interest to this work
the characterizations in the two directions are similar; in practice
we use the two runs in detector-X and detector-Y as separate samples
of a symmetric underlying uncertainty distribution, characterizing
$\log_{10}(\xi)$~as a fifth-order polynomial in \filtI~for rapid
evaluation.

Improved characterization of residual distortion has also become
available, as the datasets used to characterize ACS/WFC distortion
have grown. In the SWEEPS filters, residual distortion is on the order
of $\approx 0.01-0.02$~ACS/WFC pixels (0.5-1.0 mas at $\approx 50$~mas
pix$^{-1}$), with a complex pattern of variation with spatial scale
roughly 150 ACS/WFC pixels \citep{kozPlatais15, ak06}. This is
consistent with a recent high-precision astrometric characterization
of the full set of SWEEPS epochs for astrometric microlensing
\citep{kains17}, which indicated residual distortion corrections of
$\approx \pm 0.02$~ACS/WFC pixels for the candidate astrometric
microlensing sources (evaluated within 200 ACS/WFC pixels of each
candidate; see \citealt{kains17} for details), with the residual
changing sign seasonally due to the mid-year $180^{\circ}$~flip in
HST's orientation angle for observations of this field. The
observation dates of the 2011-2012-2013 epoch sample both HST
orientations roughly equally, so the residual distortions in this
epoch were to some extent averaged through when mean positions were
computed per star, while central pointings in this epoch are typically
within $\sim 50$~ACS/WFC pixels of the central pointing of the 2004
epoch. We therefore adopt $\Delta \approx 0.015$~pix (0.75 mas) as a
reasonable estimate for the differential residual distortion suffered
when proper motions are estimated across the two epochs.

For each object, then, the per-coordinate proper motion uncertainty
$\epsilon_i$~can be estimated from the relation
\begin{equation}
  \epsilon^2_i \approx \xi(\filtI_i)^2 + \frac{\sigma^2_{\rm pm}}{N_{\rm tr}-2} + \left( \frac{\Delta_i}{\tau} \right)^2
\label{eq:pmUncty}
\end{equation}

\noindent where $\xi(\filtI_i)$~is the artificial star-test random
proper motion uncertainty estimate evaluated at the apparent magnitude
of the object. $N_{\rm tr}$~is the number of tracer stars used to map
the reference frames between epochs, and $\sigma_{\rm pm}$~the proper
motion dispersion (in mas yr$^{-1}$) of the tracer stars (assumed to
be estimated from the observed data, although if $N_{\rm tr}$~is large
this assumption has little effect).\footnote{Because the artificial
  star tests inject few enough stars per trial to avoid altering the
  image crowding, they do not significantly alter the sample of moving
  tracer stars used to map reference frames between epochs when
  recovering injected proper motion; thus, artificial star tests are
  only minimally sensitive to $\sigma_{\rm pm}$.} $\tau$~is the
time-baseline for the two-epoch proper motions, and $\Delta_i$~is the
positional offset (in mas) incurred at the detector due to
differential residual distortion between the epochs, discussed
above. (The third term $\Delta_i/\tau$~in \autoref{eq:pmUncty} does
not appear in equation (1) of \citetalias{clarkson08} because
local-transformations were used for that work to mitigate residual
distortion.)

The random uncertainties $\xi(\filtI_i)$~are small for most of the
sample. Most of the objects selected for rotation curve analysis are
in the range $19.5 \le \filtI \le 23.3$~(e.g. \autoref{f:CMDsel}), for
which the artificial star-tests of \citetalias{calamida15} suggest
proper motion random uncertainty $0.008 \lesssim \xi_i \lesssim
0.07$~mas yr$^{-1}$~per co-ordinate. For the second term in
\autoref{eq:pmUncty}, the number of tracers $N_{\rm tr}$~is large (on
the order of $N_{\rm tr} \approx 4 \times 10^4$~since the full field
of view was used to relate the reference frames of the 2004 and
2011-2012-2013 epochs), so the second term in \autoref{eq:pmUncty}
evaluates to $\approx$ (0.015 mas yr$^{-1}$)$^2$. Finally, as
discussed above, the typical magnitude and spatial scale of variation
of residual distortion suggests $\Delta \approx 0.75$~mas, while the
time baseline $\tau \approx 8.96$~years (\autoref{tab:obsSummary})
then suggests the third term in \autoref{eq:pmUncty} can be estimated
as $(\Delta_i/\tau)^2 \approx$ (0.08 mas yr$^{-1}$)$^2$.

\begin{figure}
  \begin{center}
  \includegraphics[width=12cm]{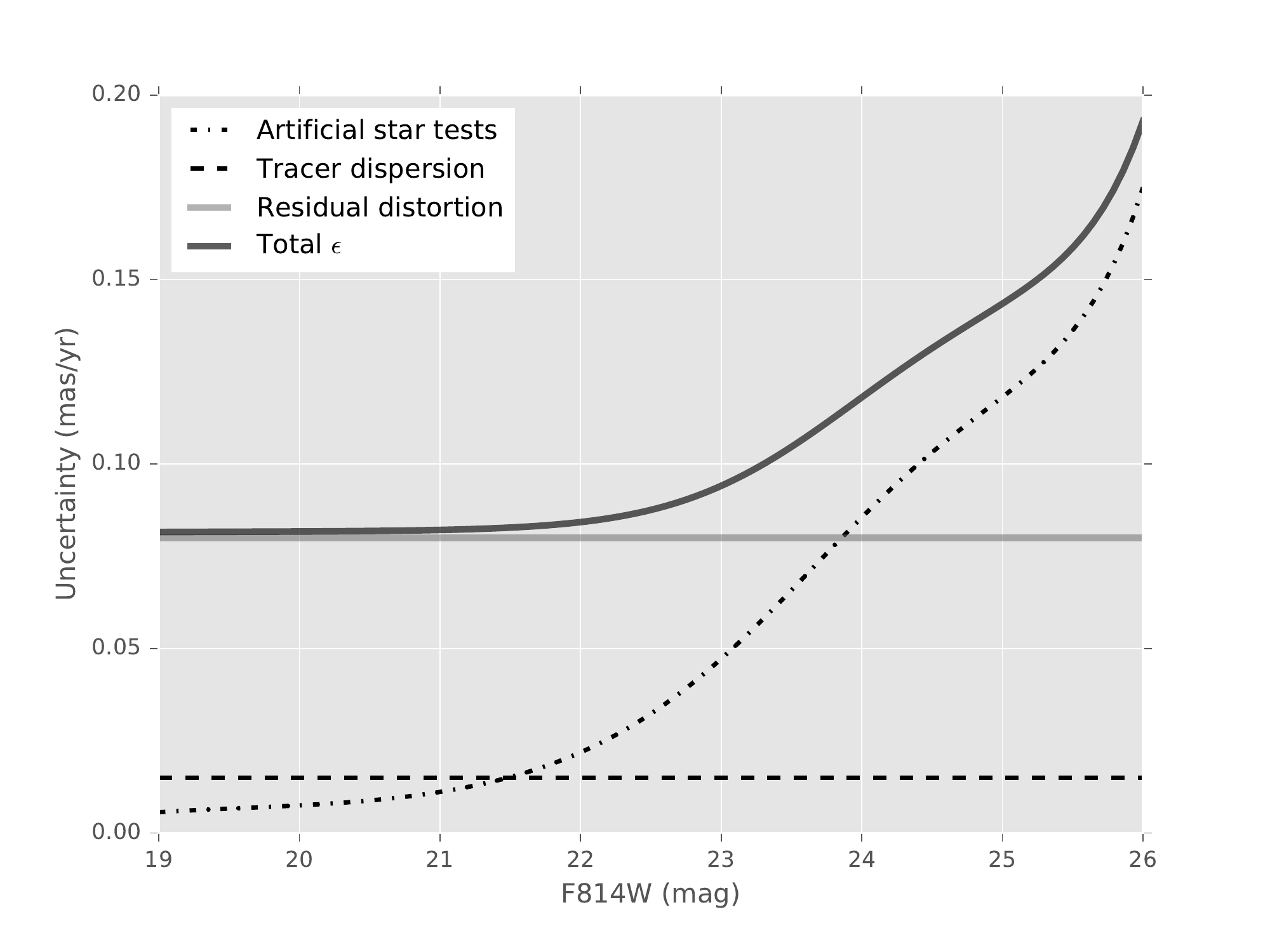}
  \caption{The run of adopted proper motion uncertainty (per
    co-ordinate) against \filtI~apparent magnitude, including random
    uncertainty suggested by artificial star tests
    (\citealt{calamida15}; dot-dashed line), the contribution due to
    intrinsic motion of the reference-frame tracer stars (dashed), and
    the estimated effect of residual differential distortion (grey
    solid line). Nearly all of our proper motion sample falls in
      the range $19.5 \le \filtI \le 23.3$~(\autoref{f:CMDsel}). See
    the discussion in Appendix \ref{a:unctyPM}.}
\end{center}
\label{f:pmUncty}
\end{figure}

\autoref{f:pmUncty} shows the adopted characterization of the proper
motion uncertainty, plotted over an apparent magnitude range that
encompasses the proper motion sample used herein. Differential
residual distortion is likely the largest contributor to the proper
motion uncertainty for most of the proper motion sample, although the
random uncertainty becomes roughly as large at the faint end of the
proper motion sample considered here.\footnote{\autoref{f:pmUncty}
  shows that random uncertainty dominates the proper motion
  uncertainty for $\filtI \gtrsim 25$; thus the artificial star tests
  of \citetalias{calamida15} do indeed capture nearly all of the
  proper motion uncertainty appropriate for the white dwarf sample of
  \citetalias{calamida14} and the sample at the low-mass end of the MS
  charted in \citetalias{calamida15}.} Since the magnitude of the
residual distortion $\Delta_i$~actually suffered by each object is
unknown, some caution is warranted when interpreting the magnitude of
the proper motion based velocity dispersion from these data. However,
the total proper motion uncertainty estimates ($\epsilon_i \lesssim
0.12$~mas yr$^{-1}$)~are still far smaller than the intrinsic proper
motion dispersion of the bulge ($\sim 3$~mas yr$^{-1}$) and so the
reported trends should be reasonably robust against proper motion
measurement uncertainty.

\section{Gaussian Mixture Modeling}
\label{a:aboutGMM}

This work makes heavy use of Gaussian Mixture Modeling (GMM) to
characterize overlapping populations in various spaces
(e.g. Sections \ref{ss:indices}, \ref{ss:mtClassify} \&
\ref{ss:pmEllipse}). GMM is a standard technique in unsupervised
machine learning \citep[e.g.][]{bishop06}, with growing use in
Astronomical data analysis (\citealt{astroMLText} and \citealt{bovy11}
provide particularly clear and authoritative presentations of GMM in
an astronomical context, including the extension of the methods to
strongly non-uniform measurement uncertainty). Briefly, the
sample is modeled as a sum
of $(k=1... K)$~Gaussian components, with the mixture weight
$\alpha_k$~of each component (where $\Sigma^{K}_k
\alpha_k=1$)~estimated by treating the unknown component
  identification of each object as a latent variable, fitting the
  mixture model components $\theta_k$~iteratively along with the mixture weights,
  usually using the Expectation Maximization algorithm or a variant
  thereof.

Under the GMM framework, we can write the formal membership
probability \wik~that a given object belongs to each model
component~(the ``responsibility'' in the language of
\citealt{bishop06}), as
\begin{equation}
      \wik = \frac{\alpha_k p(\vec{x}_i | \thetak, \Si)}
                {\sum^K_{m=1} \alpha_m p(\vec{x_i}| \mathbf{\theta}_m, \Si ) } 
\label{eq:memProb}
\end{equation}
\noindent (as has been common practice for decades in the field of
  globular cluster studies, under slightly different notation). Here
  $\vec{x_i}$~represents the measured co-ordinates of the $i$'th
  object, \thetak~the components of the $k$'th model in the mixture
  (i.e., its mean and covariance matrix), $\alpha_k$~is the relative weight of the $k$'th
  model component, \Si~the covariance matrix due to measurement
  uncertainty for the $i$'th object, and $p(\vec{x_i} | \mathbf{\theta}_k, \Si )$~the likelihood of measuring $\vec{x_i}$~given the $k$'th
  model parameters, assuming the object does belong to that
  component.

\subsection{Measurement uncertainties in \mtindices}
\label{aa:unctyMT}

\begin{figure}
\begin{center}
  \includegraphics[width=13cm]{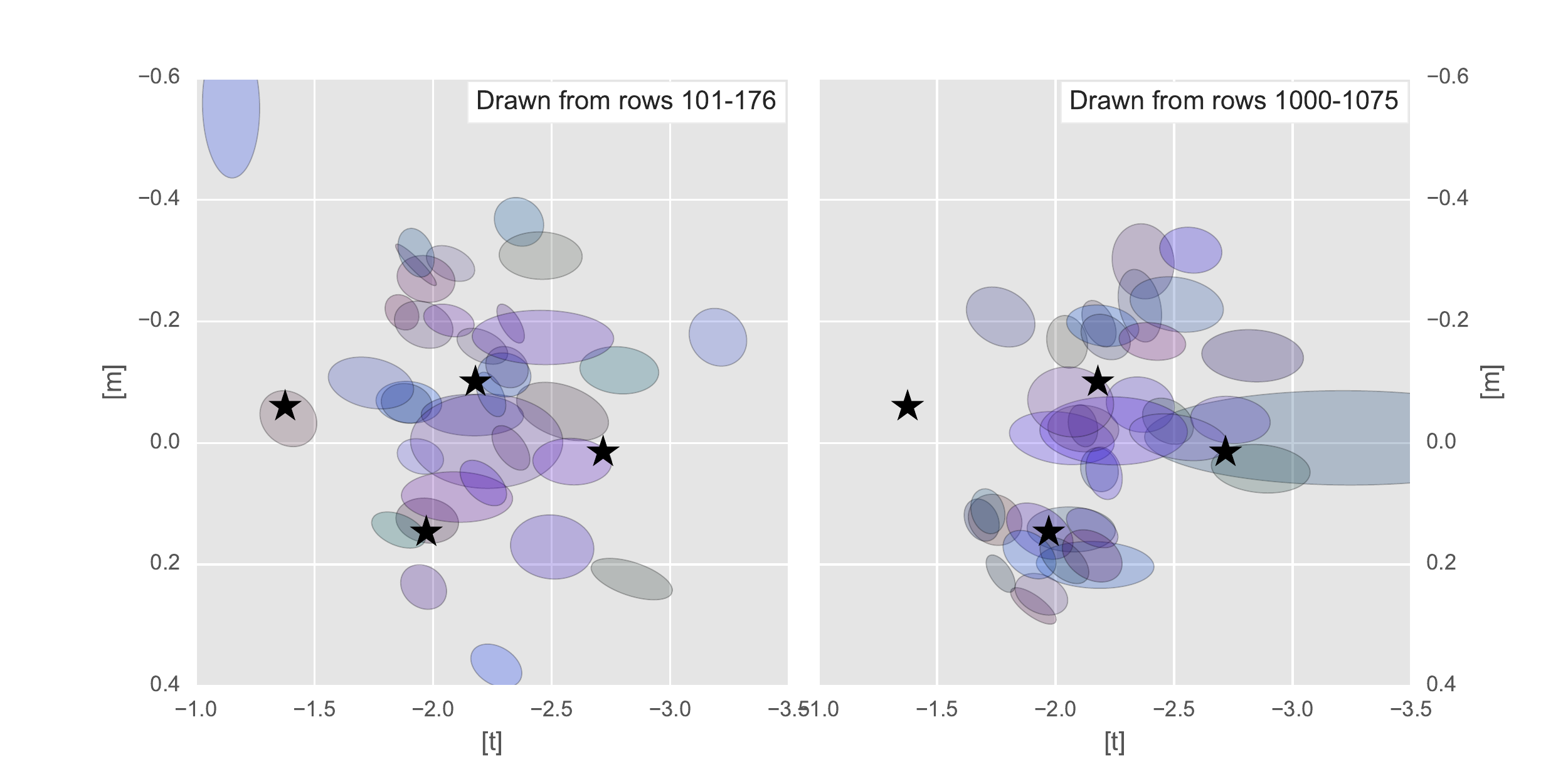}
\caption{Covariance matrices \Si~due to measurement uncertainty (following \autoref{eq:mtCovar}), for a randomly-chosen selection of \BTS~measurements within the population selected for rotation-curve study (\autoref{f:CMDsel}). Black stars show the central locations of the mixture-model components. Because both \mindex~and \tindex~contain \filtV~measurements, an appreciable tilt in the covariance matrices is often present. In many cases, the near-infrared measurements dominate the uncertainty, as expected given the large value of the scale factor $\alpha$~in the definition of \tindex~(\autoref{eq:indices}). See the discussion in \autoref{a:aboutGMM}.}
\end{center}
\label{f:mtCovar}
\end{figure}

From the definition of the \mtindices~indices (\autoref{eq:indices}), uncertainty
propagation produces an approximation for the appropriate measurement
uncertainty covariance \Si~for each datapoint, which we reproduce here
for convenience. We adopt
\begin{equation}
  \Si = \left(
\begin{array}{cc}
\sigma^2_{t} & \sigma^2_{tm} \\
\sigma^2_{mt} & \sigma^2_{m} \\
\end{array}
\right)_i
 = 
\left(
\begin{array}{cc}
\sigma^2_V + \left(1+\alpha\right)^2 \sigma^2_J + \alpha^2 \sigma^2_H & -\left( 1 + \beta \right)\sigma^2_{V} \\
-\left( 1 + \beta \right)\sigma^2_V & \left(1 + \beta \right)^2 \sigma^2_V + \sigma^2_C + \beta^2 \sigma^2_I \\
\end{array}
\right)_i 
\label{eq:mtCovar}
\end{equation}
\noindent where $\left(\sigma^2_C, \sigma^2_{V}, \sigma^2_{I},
  \sigma^2_J, \sigma^2_H\right)$~are the individual photometric
  uncertainty estimates in the \BTS~filters, and ($\alpha, \beta$)~the appropriate scale factors for the indices (\autoref{eq:indices}). Since $\alpha^2 >> (1+\beta)$~for these indices \citep[for all populations of interest;][]{brown09}, we expect the covariance matrices for most of the stars to generally align with the \tindex~direction, with only weak uncertainty covariance. Indeed, this is usually the case, though there are exceptions (\autoref{f:mtCovar}). 

We are also assuming the apparent magnitudes and their relevant linear
combinations are Normally distributed, working in apparent magnitude
space rather than flux space because the photometric uncertainties are
already reported in magnitudes in the \BTS~catalog. We impose a
photometric uncertainty cut of $\sigma < 0.1$~mag
(\autoref{tab:sampleSel}) to reduce the number of objects that
strongly violate this assumption. Nevertheless, long tails in the
observed \mtindices~distribution for objects with relatively high
photometric uncertainty may be expected.

\subsection{How many mixture components?}
\label{aa:nComponents}

To estimate the number of components required to best represent the
\mtindices~distribution, we employ two commonly-used measures, the
Akaike Information Criterion (AIC) and Bayesian Information Criterion
(BIC). These measures quantify the badness-of-fit while penalizing
more complex models, with the BIC penalizing overly complex models
more severely. More information can be found in \citet{astroMLText};
these measures take the forms

\begin{eqnarray}
  {\rm AIC} & = & 2p - 2 \ln{L} \\
  {\rm BIC} & = & p \ln{N} - 2\ln{L}
  \label{eq:AIC}
\end{eqnarray}

\noindent where lower values indicate a formally better fit. Here
$p$~is the number of parameters in the model, $N$~the number of
datapoints and $L$~the likelihood (data given model) returned by the
mixture modeling procedure. For a GMM consisting of a mixture of
$K$~model components representing \nFeatures-dimensional datapoints,
the number of parameters $p$~is given by
\begin{equation}
  p = \left(\nParsMeans\right)  + \left(\frac{\nParsMeans \times \left(\nFeatures + 1\right)}{2}\right) + \left(\nComponents - 1 \right)
  \label{eq:nParams}
\end{equation}
\noindent so that mixtures with $K=1,2,3,4...$~model components
consist of $p = 5,11,17,23...$~parameters when fitting the
2-dimensional \mtindices~distribution. When characterizing the
\mindex~or (\mtindices)~distribution with a GMM, we allow $K$~to vary
up to large values (usually $K=9$) and look for models in which the
AIC and BIC stop improving as $K$~is increased.

\begin{figure*}
\centerline{
  \includegraphics[width=5cm]{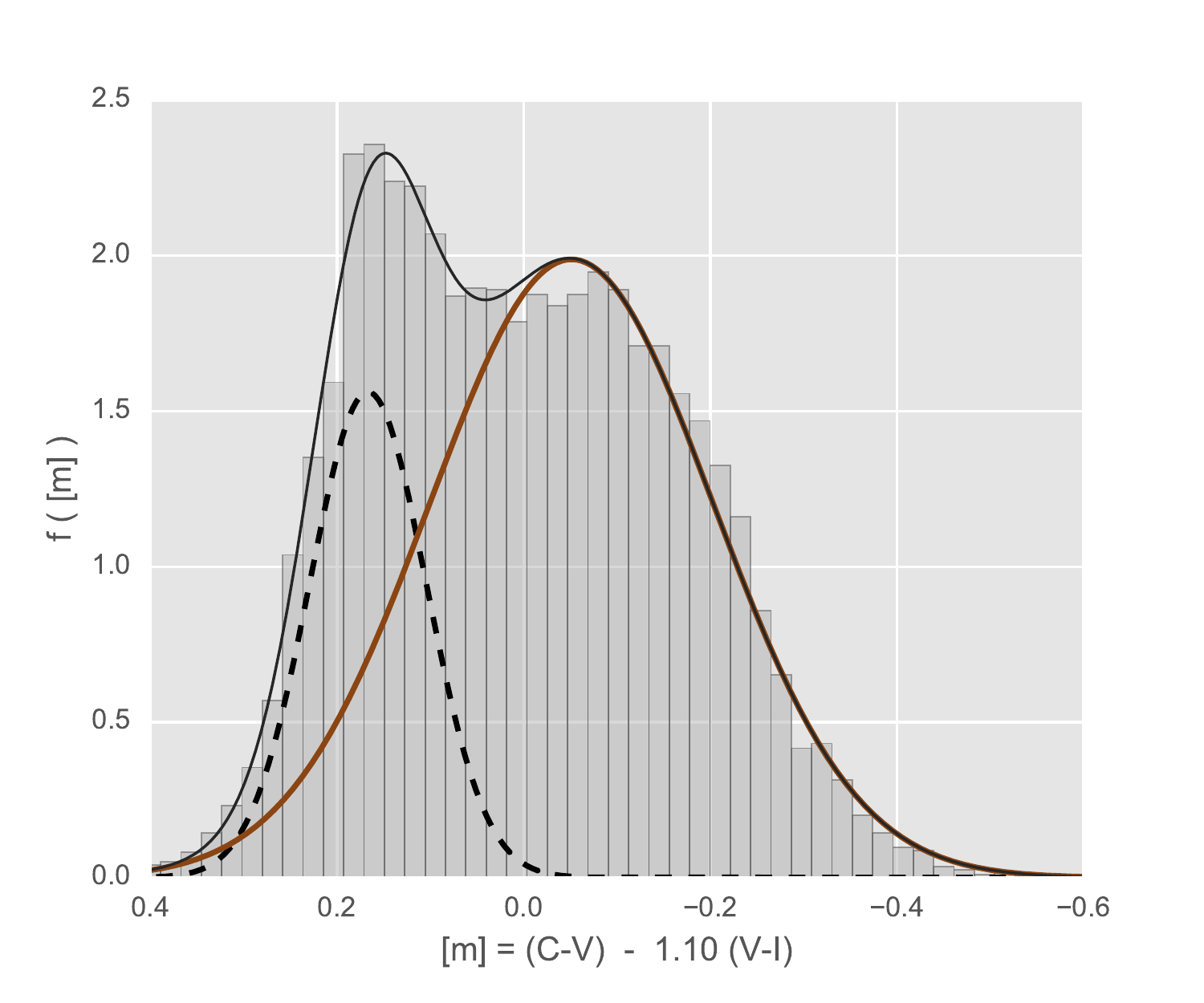}
  \includegraphics[width=5cm]{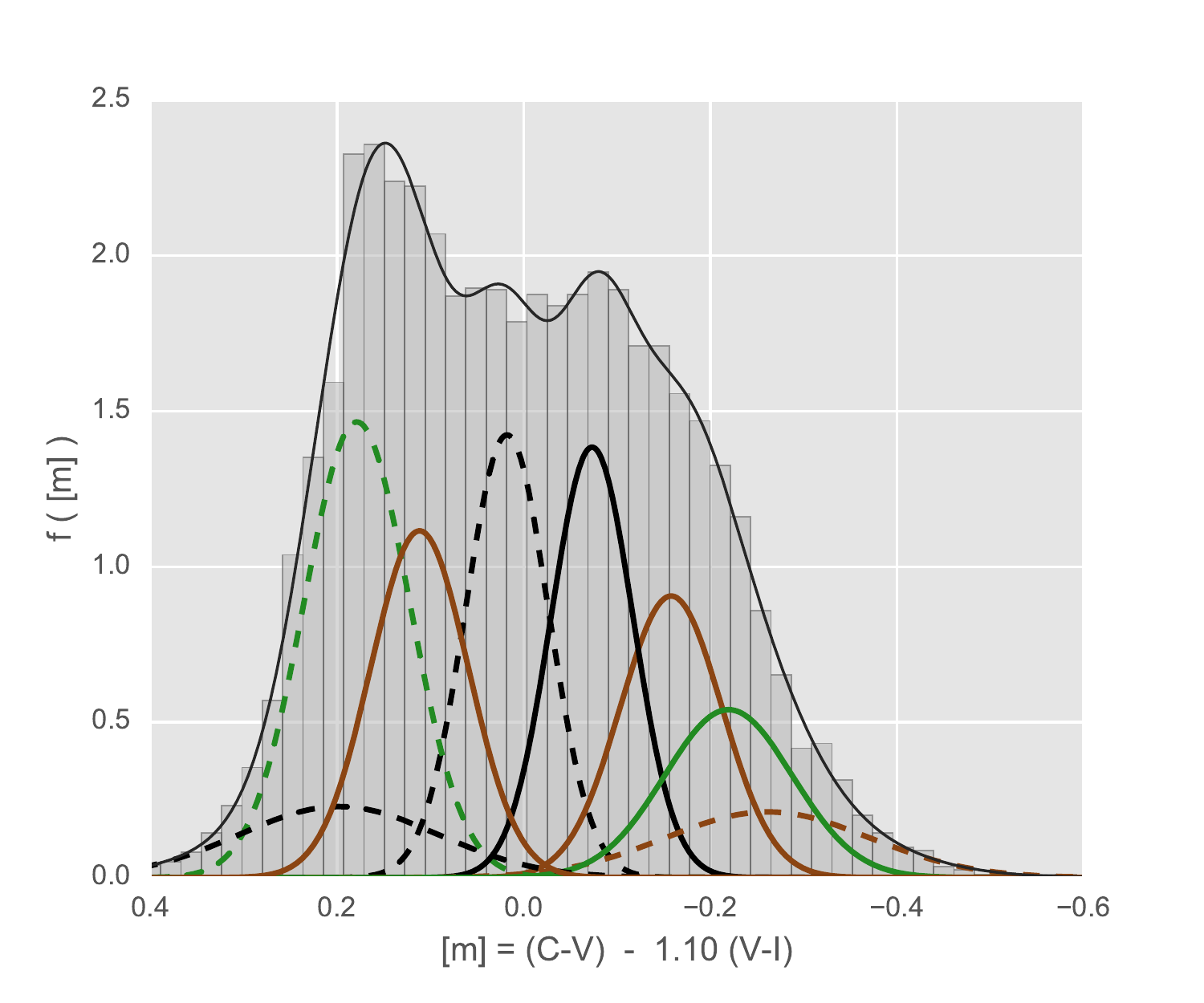}
  \includegraphics[width=5cm]{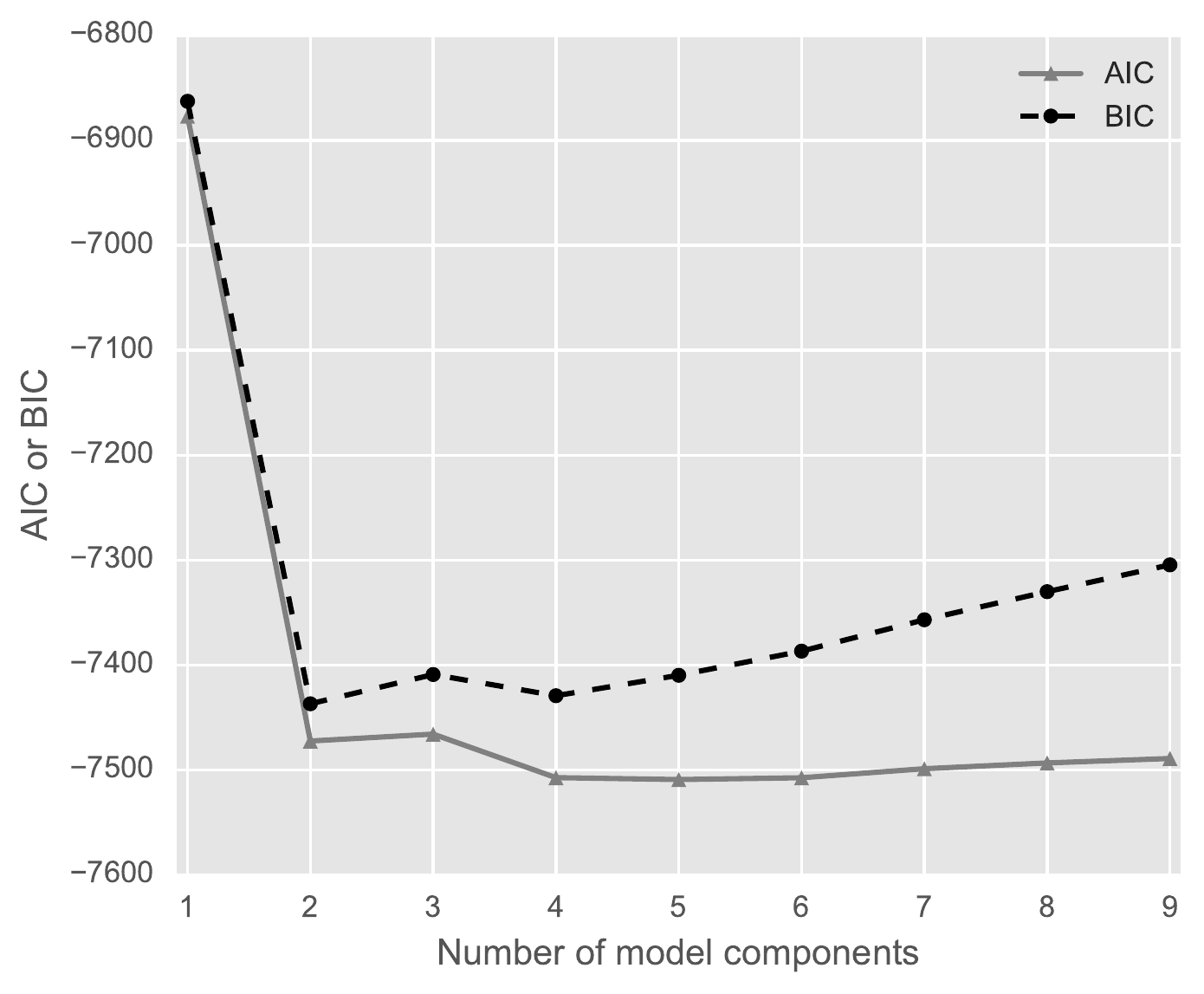}
}
\caption{{\it Left panel:} distribution of \mindex, for objects
  satisfying $-2.8 \le \tindex~\le -1.4$, representing roughly the
  population within the outer contour in \autoref{f:mtKDE}. The gray
  shaded region shows the observed \mindex~distribution. The upper
  gray solid line shows a Gaussian Mixture Model trained on the
  \mindex~distribution. The colored solid and dashed curves show
  realizations of the individual model components. {\it Middle panel:}
  as in the left panel, but with an eight-component Gaussian mixture
  model (GMM) specified as an ansatz for a continuum of
  populations. {\it Right panel:} Formal assessment of the number of
  parameters required to reproduce the observed
  \mindex~distribution. Standard figures of merit, the Bayesian
  Information Criterion (BIC, black dashed line) and the Akaike
  Information Criterion (AIC, gray solid line; see
  e.g. \citealt{astroMLText}) are plotted as a function of the number
  of model components. A GMM representation of the
  \mindex~distribution seems to require at least two components, with
  little improvement for more complex models. See Appendix
  \ref{aa:nComponents}.}
\label{f:mOnlyMix}
\end{figure*}

\autoref{f:mOnlyMix} shows an attempt to reproduce the distribution of
\mindex~only as a Gaussian Mixture Model (GMM; see
\autoref{a:aboutGMM}~for discussion of the technique). At least two
components seem to be required, although the data do not discriminate
between the simplest model that fits the data (two components) and a
continuum (e.g. 8 components). In early trials using data selected
only on photometric measurement uncertainty, a mixture model with more
than three components would usually include an extremely broad,
low-significance Gaussian component. On plotting the \mindex~counts on
a log-scale, this component was seen to be fitting handfuls of far
outliers in the \mindex~distribution (with $|\mindex| > 0.5$; compare
with the range in \autoref{f:mOnlyMix}). This may be expected if the
outliers are not well-represented by the model form; nevertheless, the
GMM implementation would attempt to assign a model component to the
outliers once the model grew sufficiently complex, which in turn would
distort model components much nearer to the location of the main
population of objects. Circumventing this outlier problem was the main
motivator for outlier removal in \mtindices~when selecting objects for
further analysis (\autoref{tab:sampleSel}).

\begin{figure*}
\begin{center}
\includegraphics[width=18cm]{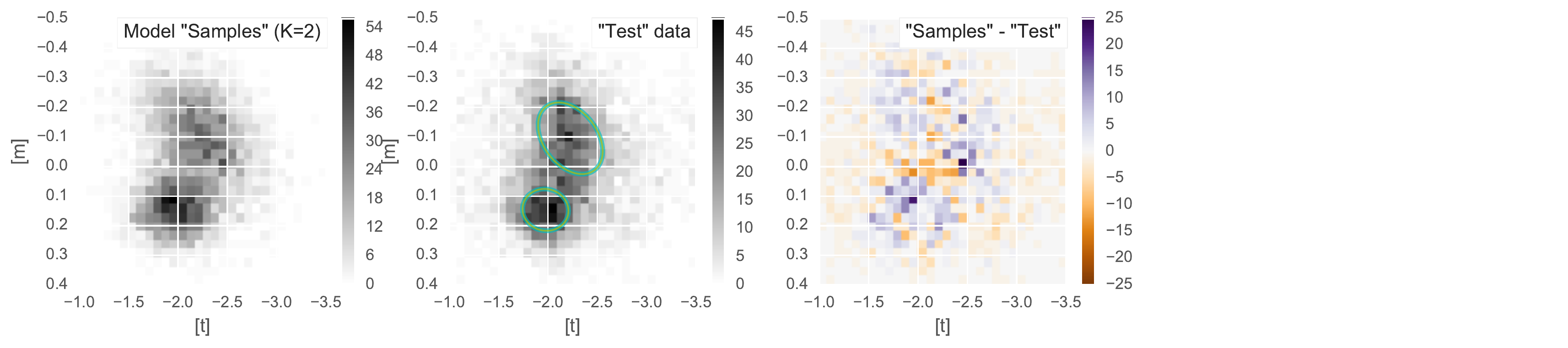}
\includegraphics[width=18cm]{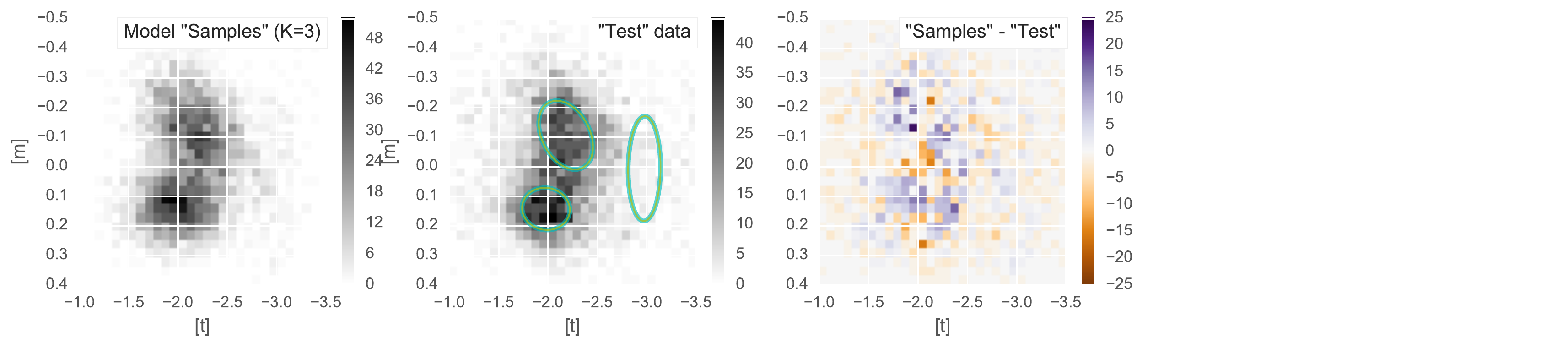}
\includegraphics[width=18cm]{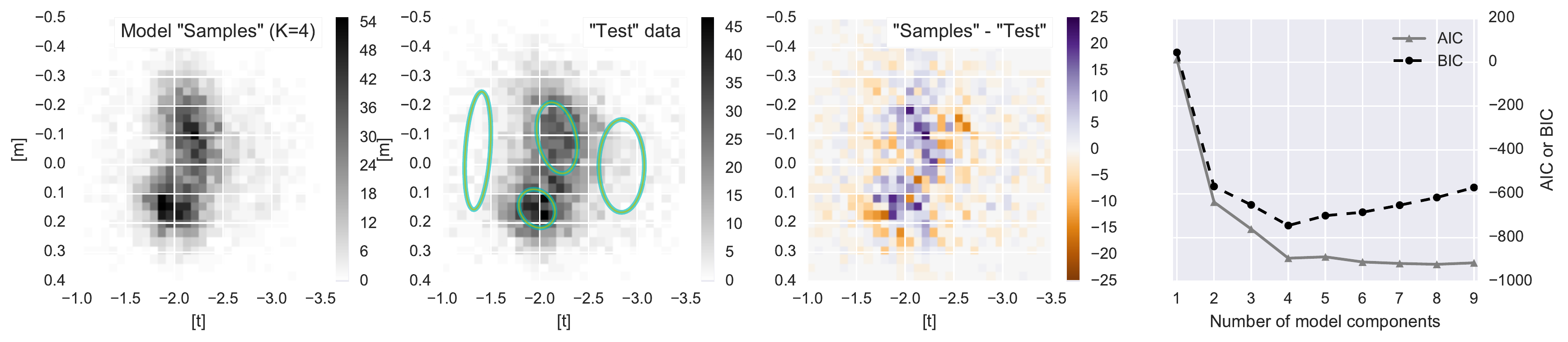}
\end{center}
\caption{Gaussian Mixture Model (GMM) of the population selected for rotation curve study. Reading top-bottom, panels show the GMM characterization for $K=2,3,4$~mixture components. {\it Left panels} show the histogram of samples drawn from a GMM fit to a randomly selected sample of half the data (the ``training set''). The {\it middle-left panels} show the other half of the data (the ``test set''), with the $1\sigma$~contours of the model components overplotted as thick cyan ellipses. The {\it middle-right panels} show the residuals (samples from the model minus the observed counts in the ``test set''). The lower-right plot shows formal fit statistics as a function of the number of model components. See Appendix \ref{aa:nComponents}.}
\label{f:mtMaps}
\end{figure*}

\autoref{f:mtMaps} shows the characterization of the (\mtindices)
distribution with a 2-dimensional GMM as the number of model
components is increased. To examine the impact of changing the number
of model components $K$, the \mtindices~data were split into two
equal-size samples (the ``training'' and ``test'' sets), and the GMM
fit using the ``training'' set. Samples (of \mtindices) were then
drawn from the model and perturbed by measurement covariances \Si~from
the ``test'' set, and the (\mtindices) distribution of this predicted
set compared with the ``test'' set. While models with $K=2,3,4$
components each provide a reasonable visual match to the observed
\mtindices~distribution, the \aic~and \bic~both indicate
$K=4$~provides the best representation of the data, while increasing
the number of components beyond $K$=4~does not improve the fit further
(indeed the \bic~suggests models with $K > 4$~fit the data more
poorly).

\section{Spectroscopic estimate of the \feh~spread in SWEEPS-field bulge stars}
\label{s:appFeH}

An estimate of the spectroscopic metallicity distribution in this
field is useful to calibrate synthetic stellar populations when
investigating possible systematic effects. To perform this estimate,
we use a deep set of VLT spectroscopic observations originally
performed to provide radial-velocity follow-up to the
\SWEEPS~transiting planet candidates; details can be found in
\citetalias{sahu06}, here we outline the relevant features for the present
paper.

\subsection{Spectroscopic observations of the SWEEPS field}
\label{ss:obsVLT}

Fiber-fed echelle spectroscopy were taken using {\it UVES} between
2004 June 22-25 (ESO program 073.C-0410(A), PI Dante
Minniti). \mh~estimates were produced in a similar manner to the
analysis in \citet{fischerValenti05}~and \citet{valentiFischer05};
typically $\sim 50$~absorption features from a Solar spectrum
(numerically degraded to the spectral resolution of the observations)
are scaled and shifted to find the best match to the observed
spectra. In addition to radial velocities, this process also yielded
estimates for \mh~(as well as \logg~and \teff). The \mh~determination
used mainly metal lines, with very few C and O lines in the templates
used, which reduces sensitivity in the \mh~estimates to systematic
differences between giants and main sequence objects
\citep{valentiFischer05}.

The 123 objects in the resulting catalog were trimmed by longitudinal
proper motion ($\mu_l < -2.0$~mas yr$^{-1}$) to produce a sample of 93
likely-bulge objects with spectroscopic \mh~estimates.

\subsection{GMM characterization of the VLT spectroscopic sample}
\label{ss:fehMixture}

\begin{figure*}
  \centerline{\hbox{
  \includegraphics[height=10cm]{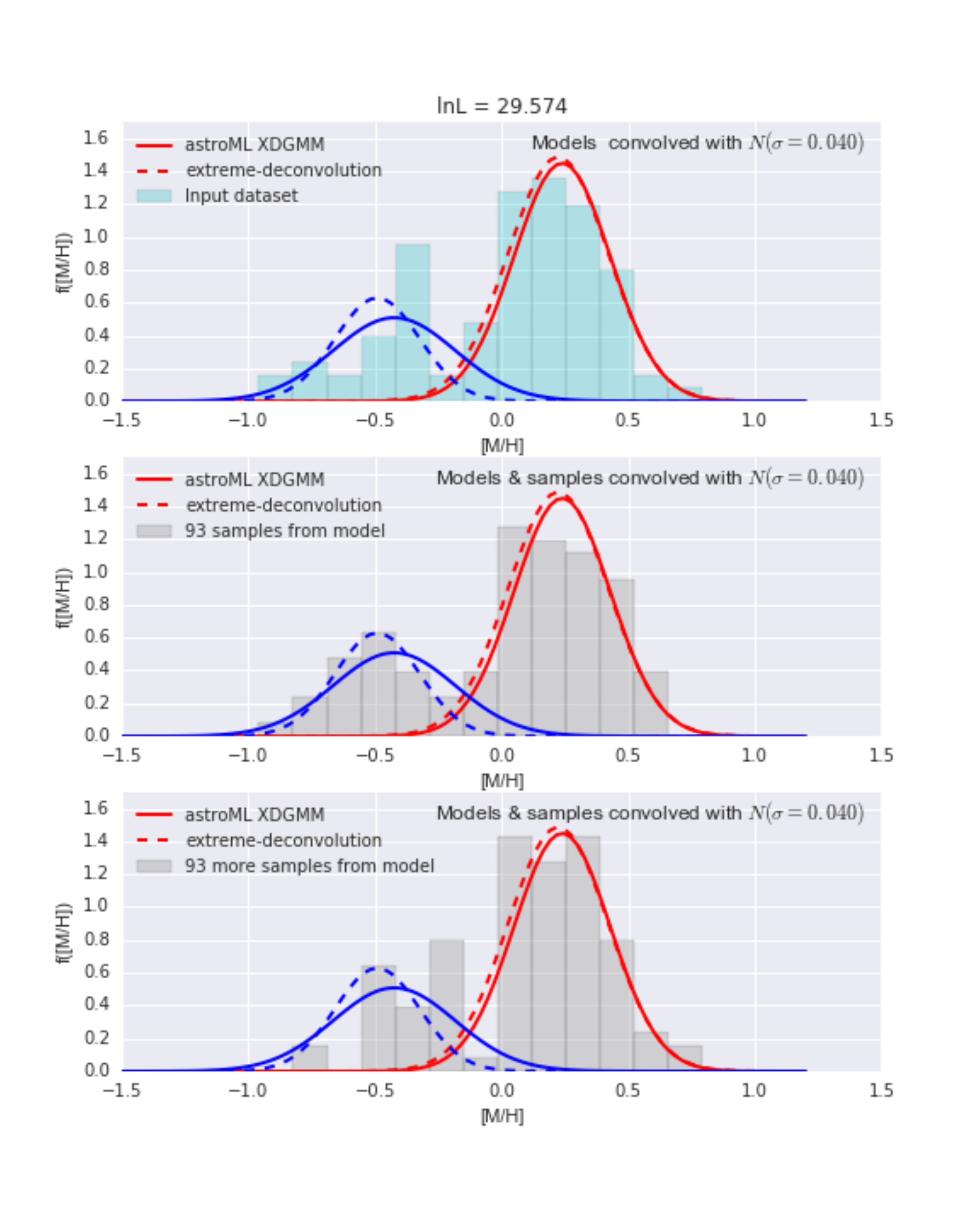}
  \includegraphics[height=10cm]{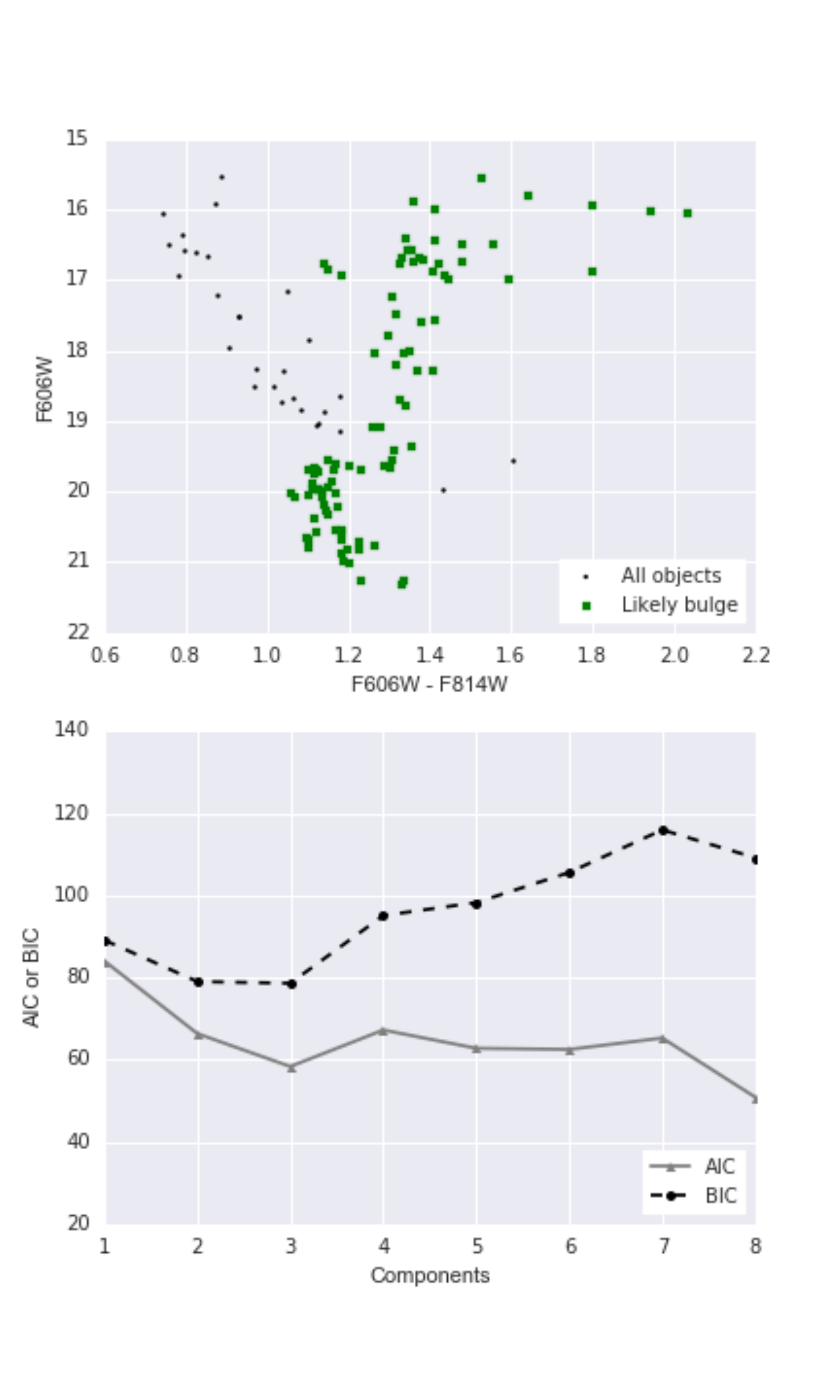}
  }}
\caption{{\it Left column:} GMM decomposition of the 93 kinematically-identified bulge objects with spectroscopic \feh~estimates (Appendix \ref{ss:obsVLT}). {\it Left:} visualization of a two-component GMM fit to the individual datapoints, over the histogram of the samples (left-top) and two realizations of the GMM model (left-middle and left-bottom). Solid lines refer to the model fit with the \xdGMM~implementation \citep{scikit-learn}, while dashed lines show the parameteres fit using the \edGMM~implementation of \citet{bovy11}. The algorithm fits the underlying model distribution after correction for measurement uncertainty; the models and model samples have therefore been convolved with a Gaussian with the median measurement uncertainty for visualization. See \autoref{f:fehMixSel} and \autoref{s:appFeH}. {\it Right column:} Sample selection and mixture fit-criteria for the characterization of the {\it VLT} spectroscopic abundance estimates (Appendix \ref{ss:obsVLT}). {\it Top panel:} \SWEEPS~color-magnitude diagram showing all 123 spectroscopically-sampled objects (black points) and the subset of 93 objects kinematically identified with the Bulge (green squares). The bulge main sequence turn-off, giant branch, and disk main sequence are each apparent. {\it Bottom:} the variation of formal figures of merit as a function of the number of model components. See \autoref{f:fehMixture} and \autoref{s:appFeH}.}
\label{f:fehMixSel}
\label{f:fehMixture}
\end{figure*}

Following previous works, which use multi-component Gaussian mixtures
to model the \feh~distributions \citep[e.g.][]{zoccali17,
  schultheis17, hill11}, we also characterize the
abundance~distribution of the 93 spectroscopically-measured
likely-bulge objects as a Gaussian mixture (Figure
\ref{f:fehMixture}). Two implementations of GMM with uncertainties are
used; the \edGMM~method of \citet{bovy11}, and
\xdGMM~\citep{scikit-learn}. The parameters fitted by the two
implementations are generally consistent with each other, and are
shown in \autoref{f:fehMixture} and \autoref{tab:fehMixture}.

% Table: GMM decomposition of the spectroscopic sample
\begin{deluxetable*}{c|rrr|rrr}
\tabletypesize{\footnotesize}
\tablewidth{700pt}
\tablecaption{GMM fits to the SWEEPS spectroscopic sample of 93 likely-bulge objects.Two GMM implementations are reported: "XD" refers to the \xdGMM~implementation while ``ED'' refers to the \edGMM~method of \citet{bovy11}. Reported ranges denote the standard deviation over 500 non-parametric bootstrap resampling trials. Parameter-sets are reported for 2- and 3-component mixture models.
\label{tab:fehMixture}}
\tablehead{\colhead{$k$} & \colhead{$\alpha_k$~(XD)} & \colhead{$[Fe/H]_0$~(XD)} & \colhead{$\sigma_{[Fe/H]}$~(XD)} & \colhead{$\alpha_k$~(ED)} & \colhead{$[Fe/H]_0$~(ED)} & \colhead{$\sigma_{[Fe/H]}$~(ED)}}
\startdata
1 & $0.31 \pm 0.049$ & $-0.42 \pm 0.079$ & $0.24 \pm 0.059$ & $0.26 \pm 0.057$ & $-0.49 \pm 0.056$ & $0.16 \pm 0.042$ \\
2 & $0.69 \pm 0.049$ & $0.24 \pm 0.025$ & $0.19 \pm 0.020$ & $0.74 \pm 0.057$ & $0.22 \pm 0.027$ & $0.19 \pm 0.023$ \\
\hline
1 & $0.28 \pm 0.046$ & $-0.48 \pm 0.049$ & $0.17 \pm 0.041$ & $0.28 \pm 0.059$ & $-0.48 \pm 0.061$ & $0.17 \pm 0.044$ \\
2 & $0.36 \pm 0.086$ & $0.13 \pm 0.068$ & $0.12 \pm 0.062$ & $0.27 \pm 0.202$ & $0.11 \pm 0.110$ & $0.11 \pm 0.075$ \\
3 & $0.36 \pm 0.086$ & $0.34 \pm 0.052$ & $0.17 \pm 0.039$ & $0.45 \pm 0.209$ & $0.31 \pm 0.176$ & $0.18 \pm 0.074$ \\
\enddata
\end{deluxetable*}

%%% Mixture model of the metallicities

%2017-12-13 force the single-column for the appendix
\onecolumngrid

Although the 93 objects have somewhat limited statistical power
  to distinguish models, it does appear that at least a two-component
  mixture is preferred. At four or more components, both
  implementations always include a very broad, almost insignificant
  component, which suggests over-fitting - and indeed the AIC and BIC
  do not suggest more than two components are required by these data
  (\autoref{f:fehMixSel}, right column). 

The parameters of the two-component GMM are consistent with those
reported by spectroscopic surveys of nearby fields
\citep[e.g.][]{zoccali17, schultheis17}, both of which find at least
two spectroscopic components with similar fractions $\alpha_k$,
centroids, and dispersions. The sample does not include a more
metal-poor component that might be suggestive of a Halo component
\citep[e.g.][]{schultheis17, ness13_part2}.

\section{Differential spread in photometric parallax}
\label{s:appDist}

Since the distance determination
is based on \rdmLong~(\rdm), in principle the \MP~population might be
subject to additional photometric scatter that causes it to be more
mixed in {\it apparent} distance than the \MR~population
(\autoref{ss:distanceMixing}). Might differential distance blurring be
responsible for the apparent differences in rotation curves, even if
the {\it intrinsic} kinematic trends for both samples were identical?

To address this question, we perform simple Monte Carlo tests,
communicated in this section. Differences in {\it absolute}
  magnitude distribution due to the differing stellar parameter ranges
  between the selected samples - particularly \feh~- require a more
  sophisticated analysis and are discussed in \autoref{s:appMetBlur}.

Individual objects in the \MR~sample are perturbed in apparent
magnitude and the proper motion rotation curve for the
distance-blurred \MP~sample compared to the observed rotation curve
for the \MP sample, by computing and comparing the smoothed rotation
curves between distance moduli $(-1.0 \le \rdm \le +1.0)$~for both
samples.

For each form of distance-modulus blurring, a run of 30 effect scales
are considered. A thousand realizations were run at each of the effect
scales, and the match between the distance-blurred \MR~and the
observed \MP~rotation curves evaluated. Three figures of merit are
assessed: (i) The root-mean-square difference between the two trends
is used as the primary badness-of-match statistic, where the
longitudinal proper motion offset between the two observed trends
(\muCenDeltal~\masperyear; \autoref{ss:rotnCurves})
is subtracted from the \MR~sample to ease interpretation (so that a
perfect match between the two samples would produce badness-of-match
value zero). In addition, the difference in \rdm~distribution between
the blurred-\MR~and observed \MP~samples is quantified by the
difference in (ii) the \rdm~standard deviations for each distribution,
and (iii) the skewness of the two \rdm~distributions, since the
observed \MP~distance modulus distribution does exhibit an asymmetry
towards the nearside of the median population (e.g. Figures
\ref{f:simpleContour} \& \ref{f:simpleTrends}).

To determine the ranges of these figures of merit that would be
consistent with a match, for every trial a control test is
performed. A set of \rdm~values is drawn following the observed
\MP~\rdm~distribution, and the observed \MP~rotation curve (and proper
motion dispersion curve) sampled at the generated \rdm~values. For
this generated sample, the rotation curve and comparison statistics
are obtained exactly as for the blurred-\MR~sample. In this way, the
figures of merit are also produced for a set of samples when the
\MP~distribution is compared against a statistical clone of itself,
allowing the range of badness-of-fit values to be charted that suggest
the underlying samples are drawn from the same distribution.

Two forms of potential distance-modulus blurring are considered
independently. Additional scatter in the intrinsic flux distribution
is discussed in Appendix \ref{ss:distBlur:photom}, which accounts for
additional photometric uncertainty or differences in extinction (or
indeed any perturbation that would lead to an additional flux
perturbation of the same general form). The impact of differing binary
fraction is discussed separately in Appendix \ref{ss:distBlur:binaries},
because its imprint on the flux distribution takes a different form.

\subsection{Additional photometric scatter in the \MP~population}
\label{ss:distBlur:photom}

Additional photometric scatter is simulated as a perturbation in
flux. The apparent magnitudes in the \MR~sample are perturbed by
amount $\Delta m_p$, defined as
\begin{eqnarray}
  \Delta m_{p,i} & = & -2.5 \log_{10}\left(\frac{F_{0,i} + \Delta F_{i}}{F_{0,i}}\right) \nonumber \\
  & = & -2.5\log_{10}\left(1 + s \mathcal{N}(0,1)_i \right) 
\label{eq:mPert}
\end{eqnarray}
\noindent where $\Delta F_i$~is the perturbation in flux, assumed
Normally distributed, $s$~the scale of the additional flux uncertainty
as a multiple of the original unperturbed flux $F_{0,i}$~and
$\mathcal{N}(0,1)_i$~a draw from the unit Normal distribution. For
large values of $s$, the Normally distributed flux perturbation can
cause the perturbed flux values for some simulated objects to go
negative; the simulation treats these cases as nondetections and
removes affected objects from consideration, thus penalizing
simulations with very large simulated flux uncertainty.

\begin{figure}
\centerline{\hbox{
  \includegraphics[width=6cm]{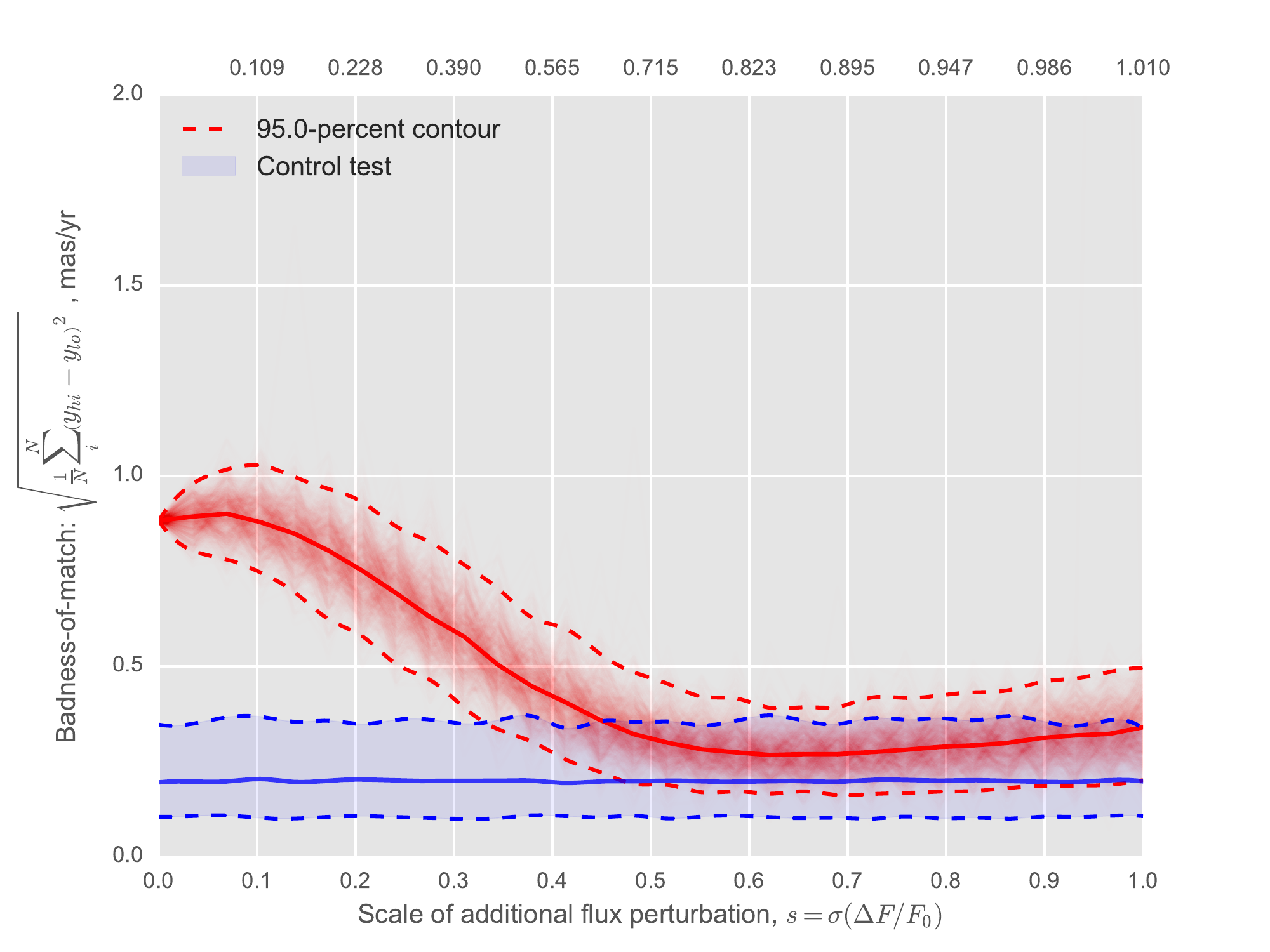}
  \includegraphics[width=6cm]{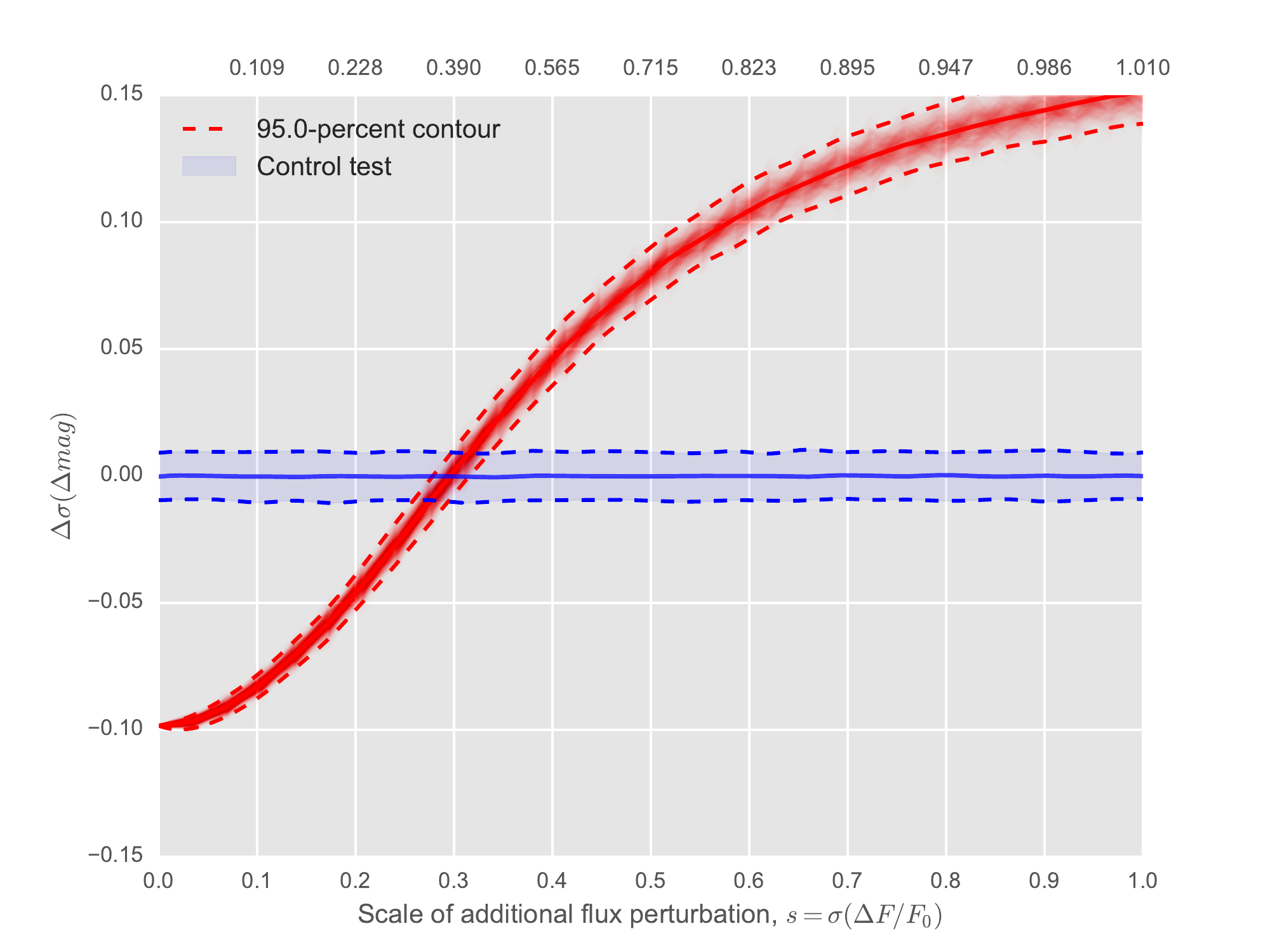}
  \includegraphics[width=6cm]{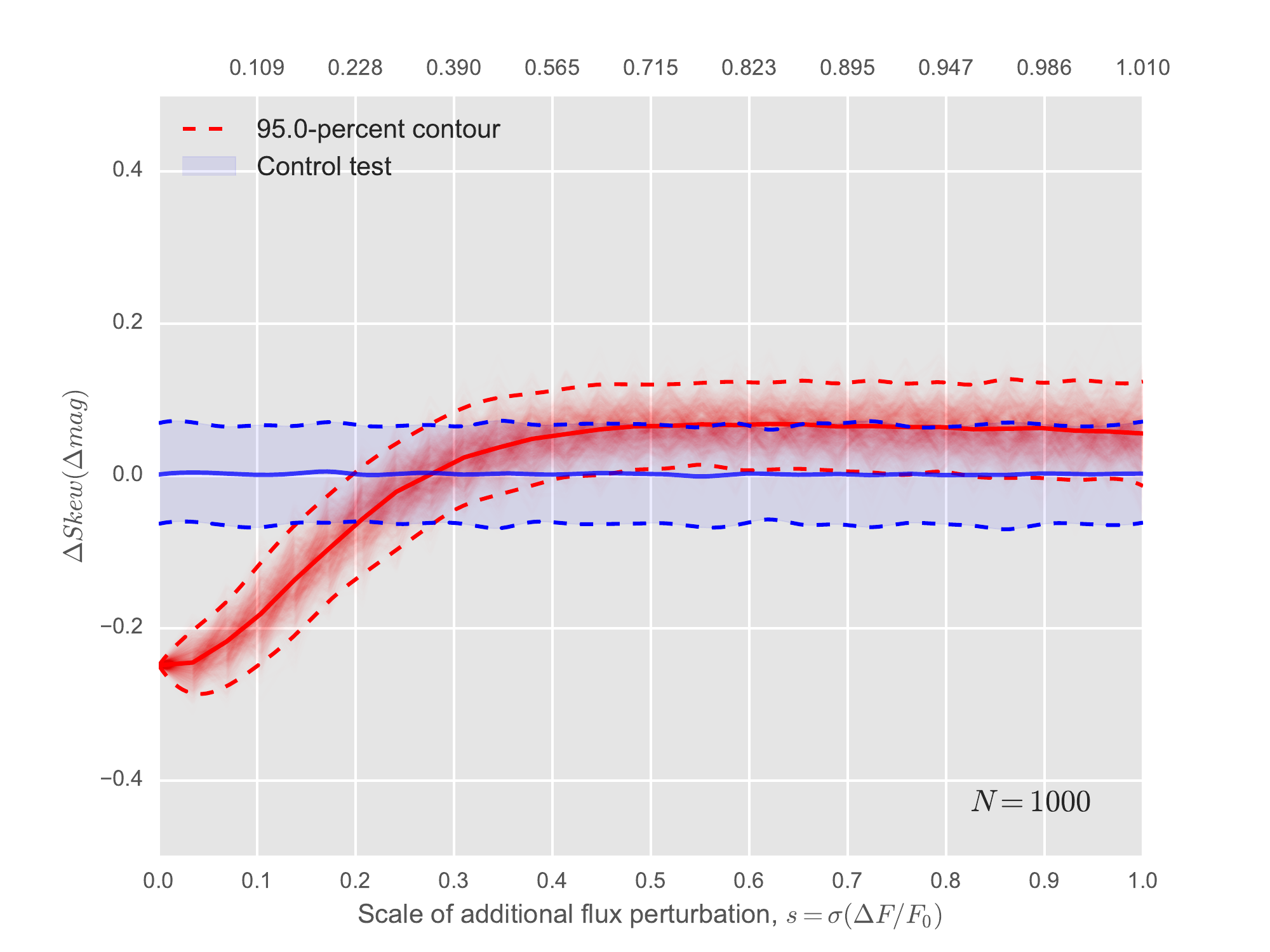}
}}
\caption{Testing the hypothesis that additional flux scatter $\Delta
  F$~can by itself cause identical rotation curves and
  \rdm~distributions to exhibit the observed discrepancies between
  \MP~and \MR~samples. In each panel, the lower horizontal axes each
  show the scale $s$~of the fractional flux perturbation, while the
  upper horizontal axes show \smag, the corresponding sample standard
  deviation in apparent magnitude. Reading left-right, panels show the
  badness-of-match statistic, the difference in distance modulus
  standard deviations, and the difference in distance modulus
  skewness, respectively. Solid red lines show the median of each
  statistic, and 95\% of the samples fall within the dashed
  contours. The blue shaded region and contours show the control
  test. See Appendix \ref{ss:distBlur:photom}.}
\label{f:distFoMsPhotom}
\end{figure}

\autoref{f:distFoMsPhotom} shows indications from this test. To aid
interpretation in terms of apparent magnitude, we also characterize
the sample standard deviation in apparent magnitude caused by the
perturbation (which we denote \smag), displaying it alongside the
input scale $s$~of flux perturbation; the quantity \smag~is plotted
along the top axes in \autoref{f:distFoMsPhotom}. The rotation curve
badness-of-match statistic suggests observed rotation curve
discrepancy can result from increased photometric scatter for scale
factor $s \gtrsim 0.35$~(in apparent magnitude, $\smag \gtrsim
0.48$)~while the \rdm~distribution of the \MP~sample is brought into
rough agreement with that observed, for scale factor range $0.25
\lesssim s \lesssim 0.35$~(or in magnitudes, $0.30 \lesssim \smag
\lesssim 0.48$).

\begin{figure}
\begin{center}
\includegraphics[width=10cm]{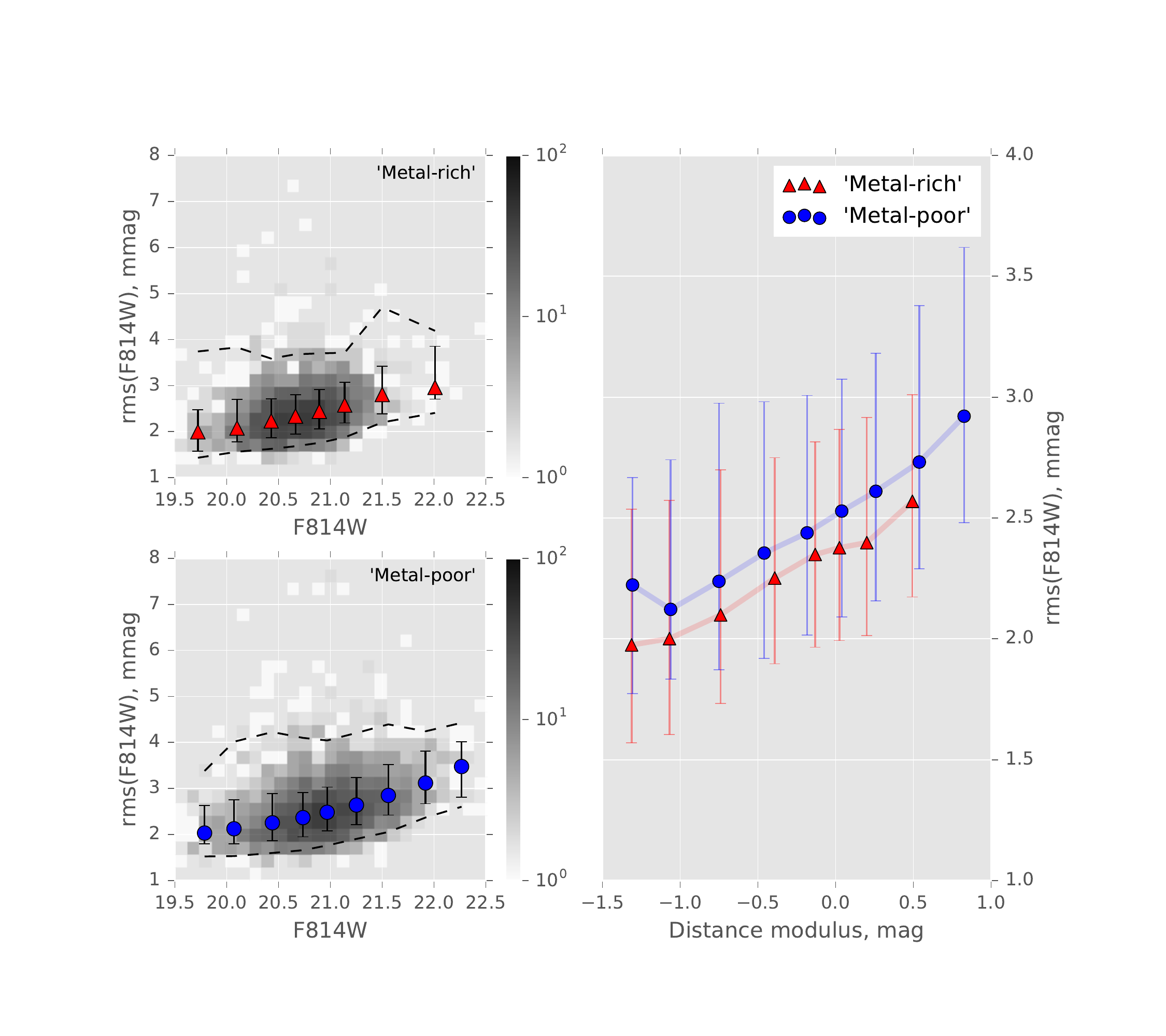}
\end{center}
\caption{Comparison of internal photometric precision for the \MR~and \MP~samples as a function of apparent magnitude (left) and distance modulus (right). Errorbars indicate the upper- and lower-bounds within which 68\%~of objects are found in each bin, the dashed contours encompass 95\% of objects per bin and the plot symbols show the medians. The grayscale shows object counts in the two samples, on a logarithmic scale. See Appendix \ref{ss:distBlur:photom}.}
\label{f:photRMS}
\end{figure}

It is difficult to see how the \MP~sample might be subject to such a
large additional photometric scatter. For example, the additional
photometric scatter is likely far larger than the difference in
photometric precision in the two samples from the
\SWEEPS~measurements.  \autoref{f:photRMS} shows the {\it internal}
  photometric precision (defined as the root-mean-square of the
  apparent magnitude measurements along the set of images) as a
  function of apparent magnitude and \rdm~for objects in the \MR~and
  \MP~samples. The \MP~population shows only a slight increase in
  internal photometric uncertainty compared to the \MR~population, and
  both are very small (on the order of a few mmag; these objects are
  well above the photometric completeness limit for the
  \SWEEPS~survey). While indeed the internal precision refers to the
         {\it random} component of photometric uncertainty and not the
         absolute photometric accuracy, a sample difference in
           photometric uncertainty of $\sim 0.3-0.5$~magnitudes seems
           highly unlikely for these data.

A difference in extinction distribution between the samples,
  characterized in any way \footnote{e.g. by change in \EBmV, in $R_V$, or by
  functional form such as introducing and varying a second parameter}, if large
  enough to bring about the $\smag \sim 0.3-0.5$-magnitude additional
  scatter required, would surely have led to additional observational
  consequences that are not seen in these data. For example, the observed \filtI~dispersion of the Red Clump Giants
(RCG) in the SWEEPS dataset is close to $\sigma(\filtI) \approx
0.17$~magnitudes \citepalias{clarkson08}. Even if {\it all} this
dispersion were due to extinction, which seems unlikely, this would
still be a factor $\gtrsim 2$~too low to bring about the observed
discrepancies between \MR~and \MP~samples.

\begin{figure}
\begin{center}
  \includegraphics[width=14cm]{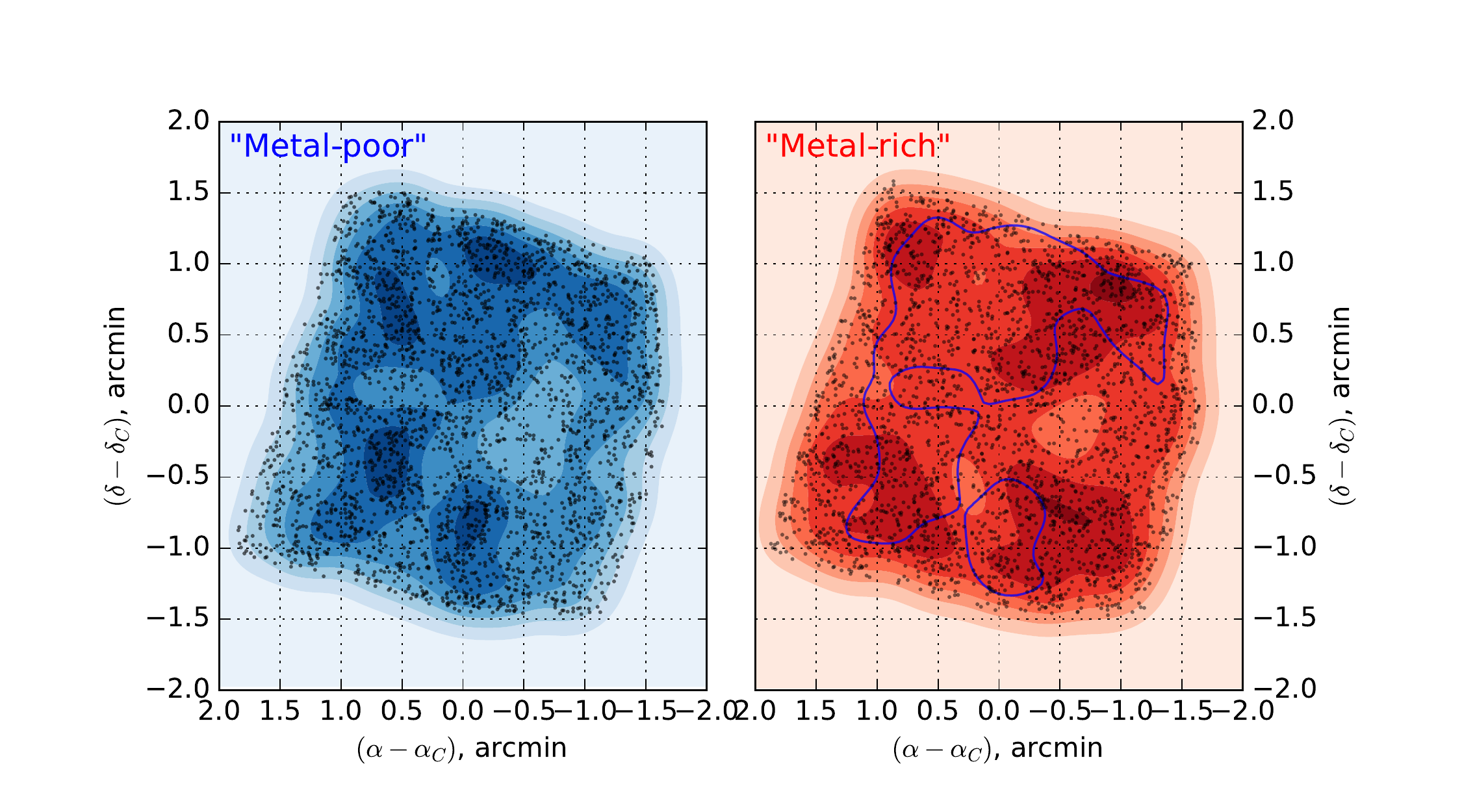}
\end{center}
\caption{Spatial distributions of the \MP~(blue, left) and \MR~(red, right) populations discussed in \autoref{ss:mtClassify}, over the $\sim (3'\times 3')$~of the \BTS-SWEEPS cross-matched field. In each panel, points represent the individual objects, while the filled contours indicate the KDE representation of the local density at each point. In both panels, contours correspond to six equally-spaced density levels. To aid visual comparison, the outer contour of the second-highest level from the metal-poor population is plotted over the metal-rich distribution in the right-hand panel. See Appendix \ref{ss:distBlur:photom}.}
\label{f:popSpatial}
\end{figure}

When the depth of the bulge along the line of sight is
  considered, the allowed contribution of differential extinction to
  \rdm~blurring becomes somewhat smaller. For example, assuming the
  bulge RCG are scattered along this line of sight by $\pm 0.50$~kpc
  allows room for only $0.1$~mag of photometric blurring due
  to extinction of any prescription. Since extinction effects would need to apply
  {\it differentially} to the \MP~sample compared to the \MR~sample to
  bring the two rotation curves into agreement, we conclude that
  differential extinction effects are likely at least a factor 3-5 too
  small to account for the observed rotation curve discrepancy.

It is also not clear why the \MP~sample would be subject to a strongly
discrepant extinction distribution (however parameterized) in the
first place. The two populations are not strongly different in their
projected distributions on the sky (\autoref{f:popSpatial}), which
would seem to argue against, say, the \MP~sample being located within
a region on the sky showing stronger, clumpier extinction than the
\MR~sample. Additionally, the RCG apparent magnitude distribution in
this field does not appear to be bimodal \citep[e.g.][]{nataf13,
  clarkson11}.

We point out that this test applies to the {\it dispersion}~of
differential extinction, not to differences in the median extinction
between the two samples. Although a difference in median $R_V$~might
affect the drawing of the \MR~and \MP~samples using \mindex,
\tindex~(because those indices are computed in terms of extinction
ratios, which are dependent on the prescription for extinction), it
would not by itself change the \rdm~dispersion for a given population
($R_V$~variations are considered in more detail in Appendix
\ref{ss:rvvar}). The \rdm~values for \MR~and \MP~populations are both
constructed by reference to fiducial ridgelines fit to the {\it
  observed} populations in the \SWEEPS~color-magnitude diagram. While
the interpretation of a given ridgeline with a particular set of
population parameters (like \feh, \EBmV, \fbin, \qmin,~and, to a
lesser extent for this population, age) does depend on the median
\EBmV, this does not impact the fiducial ridgelines of the observed
median populations on the \SWEEPS~color magnitude diagram.

We thus reject additional photometric scatter as a cause for the
\MR~and the \MP~samples to be drawn from the same kinematic
population, because, whatever the cause, its likely magnitude is much
too low to have gone unnoticed elsewhere in these data.

\subsection{Differences in binary fraction}
\label{ss:distBlur:binaries}

\begin{figure}
\begin{center}
  \includegraphics[width=14cm]{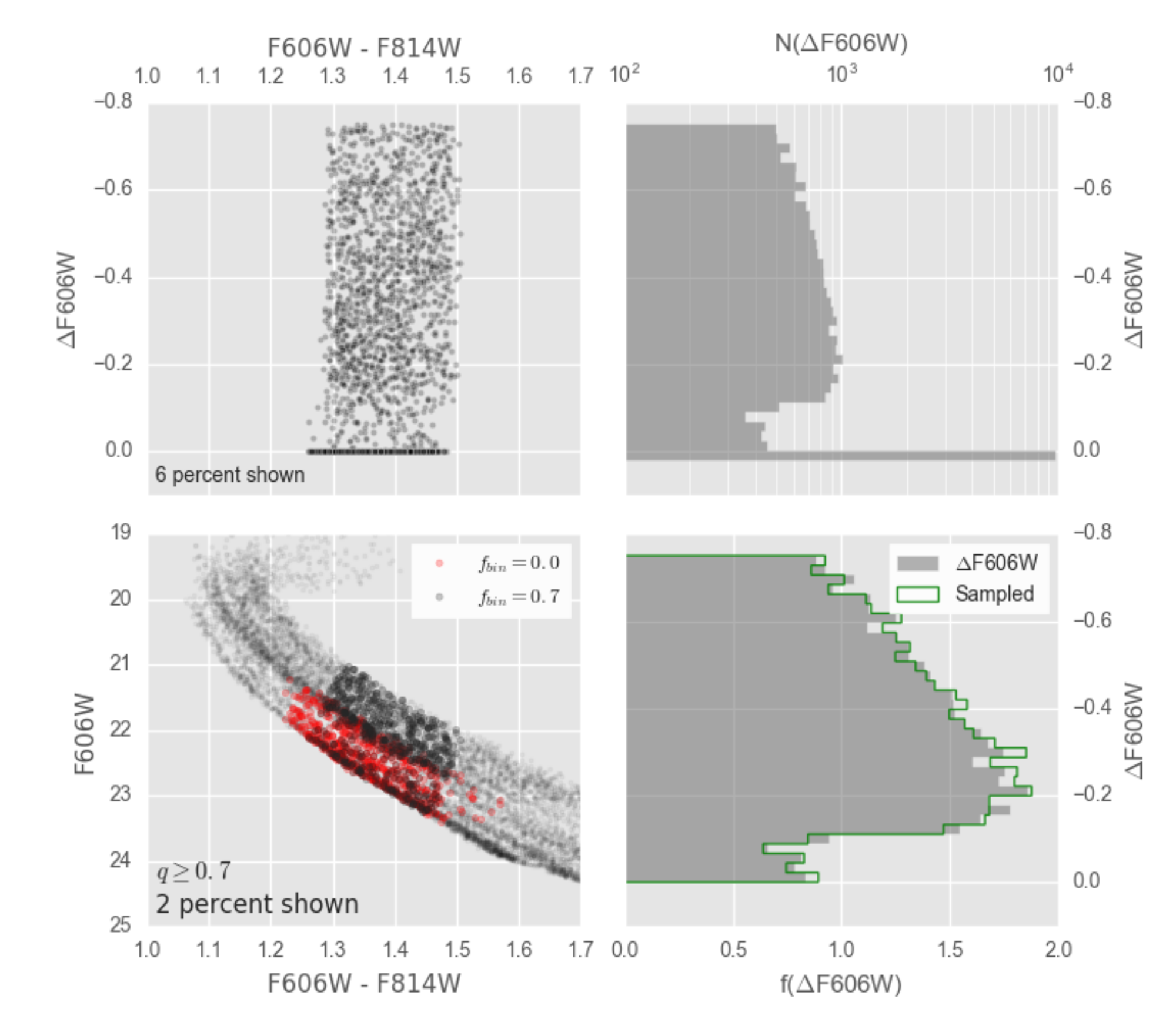}
\end{center}
\caption{Characterization of the distribution of apparent magnitude
  perturbation due to unresolved binaries, using the \basti~suite of
  models and stellar population tools. Reading clockwise from
  lower-left: {\it Lower-left:} synthetic stellar populations in the
  \SWEEPS~filter-set. Red points show the simulation
  without binaries, gray the population with binaries. Faint points
  show a representative set of the entire simulation in each case,
  dark points show the objects which fall within the CMD selection
  region in the presence of unresolved binaries. {\it Upper-left:}
  \dmbin~due to the presence of unresolved binaries, for objects only
  within the selection region. {\it Upper-right:} the distribution of
  \dmbin~(on a log scale), with histogram boundaries at the upper end
  of each bin. This panel includes objects not assigned a binary
  companion in the simulation. {\it Lower-Right:} Normalized
  distribution of \dmbin~for objects assigned a binary companion (gray
  shading). The green open histogram shows the distribution of draws
  from a non-parametric resampling of \fdmbin. See
  Appendix \ref{ss:distBlur:binaries}.}
\label{f:binariesCharacterize}
\end{figure}

If the \MP~sample has a highly discrepant binary fraction or binary
companion mass ratio distribution from the \MR~sample, then this might
produce a population with larger distance-spread, where the additional
inferred distance scatter would be biased to closer distances than the
mean-population - qualitatively similar to the trends observed
(e.g. \autoref{f:simpleTrends}). 

The binary fraction \fbin, minimum binary (initial) mass ratio
\qmin~and indeed the shape of the distribution of mass ratio $q$, are
not known for the bulge (see, e.g. \citealt{calamida15}), and are
difficult to constrain observationally for the sample selected for the
present proper motion study (e.g. \autoref{f:CMDsel}). A complete
search of $(\fbin, \qmin)$~parameter space, and indeed of the form of
the mass ratio distribution, is beyond the scope of the present
investigation. Instead, we characterize statistically the distribution
of \dmbin~due to unresolved binaries, for the CMD region of interest
to this study (\autoref{f:CMDsel}), and draw from this distribution
\fdmbin~for each realization of the Monte Carlo trial.

To maximize the impact of a difference in binary fraction between the
\MR~and \MP~samples, we assume for the purposes of this test that the
\MR~population has no binaries at all, and perturb it using an
unresolved binary fraction to approximate the \MP~population. (This
thus allows the excess binary fraction to be tested in the range $0
\le \fbin \le 1$; if we assume the \MR~sample has a binary fraction of
0.3, then only tests in the range $\fbin < 0.7$~would be
meaningful). We also assume for this test that it is {\it only} the
population of unresolved binaries that differs between the two samples
(i.e. there is no difference in metallicity distribution between the
two samples).

Version 5.0.1. of the
\basti \footnote{\url{http://basti.oa-teramo.inaf.it/}}~suite of
simulation tools and stellar population models \citep{pietrinferni04,
  pietrinferni06} is used to produce a representative set of
distributions \fdmbin, to characterize for the Monte Carlo
draws. Thanks to the capability of \basti~to accept user-defined
random number seeds, simulations that are almost identical but for
small changes in input parameters can be run. This allows us to
compare synthetic populations on a star-by-star basis, with and
without the addition of unresolved binaries.\footnote{The populations returned by \basti~are not quite
  identical for identical random number seeds; $\sim 1/1000$~of the
  objects in the binary-free simulation are missing in the
  binary-equipped simulation. Thus re-matching of rows across
  simulations is required even for identical seeds.}

To combine the sophistication of \basti~with the speed necessary for
Monte Carlo trials, the distribution \fdmbin~itself is characterized
non-parametrically, using the method outlined in \citet[][their Section
  3.7]{astroMLText} - and thus does not depend on a functional form
for \fdmbin. This resampling is $10^{5-6}$~times faster than
running a \basti~simulation for each iteration, and brings into reach
Monte Carlo exploration of the impact of binaries for our purposes
here.

\basti~simulations are run for four choices of the minimum binary
initial mass ratio: $\qmin = (0.0, 0.3, 0.5, 0.7)$. The
``bulge''~star-formation history \citep{molla00}~within \basti~is used
to populate the sample, with Scaled-to-solar heavy element abundances
and the \citet{kroupa93} initial mass function. Absolute magnitudes
are converted to apparent magnitudes using a fiducial distance and
reddening. This allows \fdmbin~to be characterized specifically for
the population we have selected for proper motion study. For
$\qmin=0.0$, the distribution \fdmbin~turns out to closely resemble
$\fdmbin = 1/\dmbin$, while for $\qmin >0$~the distribution becomes
more complicated and nonparametric resampling is preferred
(\autoref{f:binariesCharacterize}).

\begin{figure*}
\begin{center}
\centerline{
  \includegraphics[width=6cm]{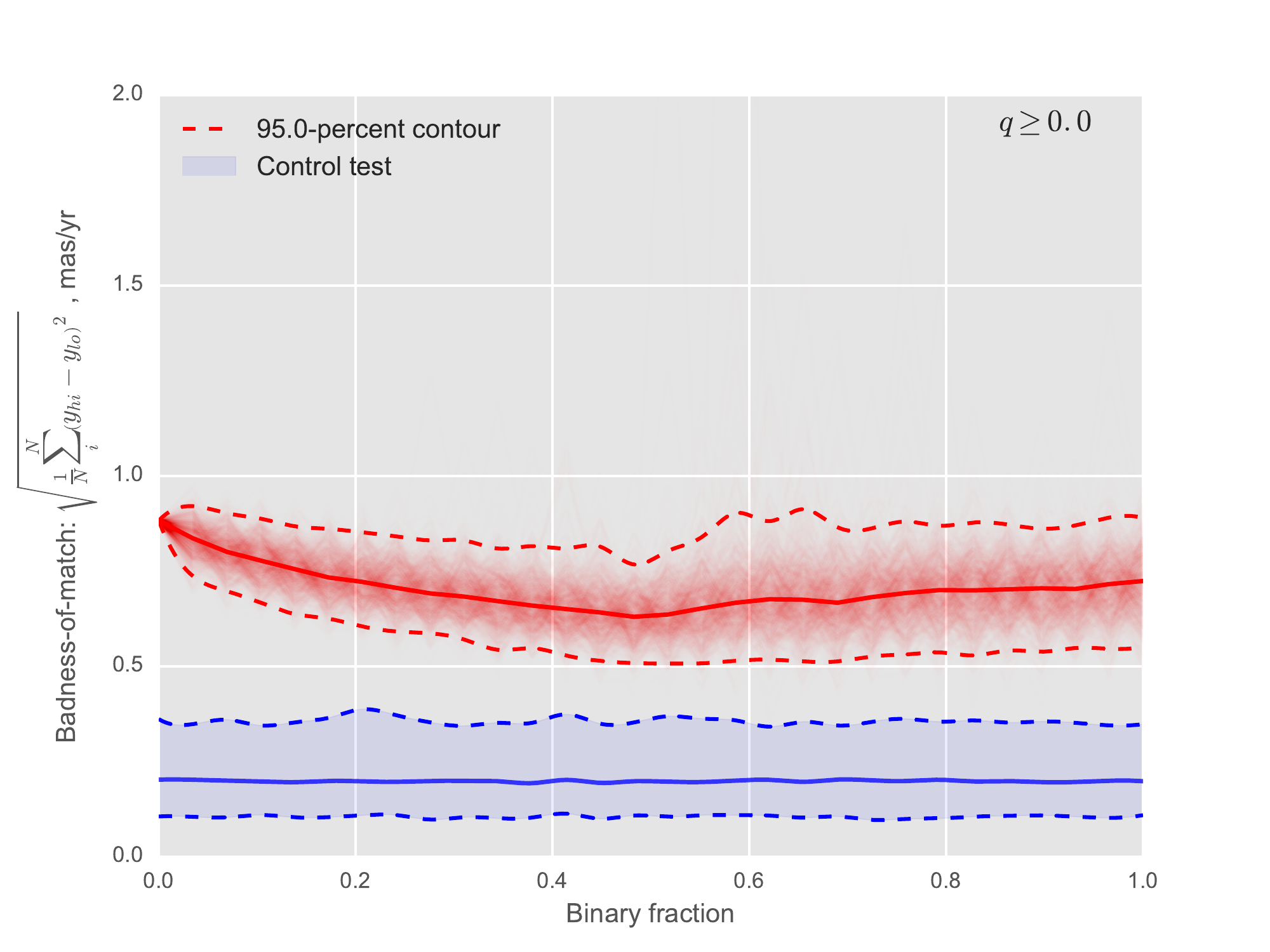}
  \includegraphics[width=6cm]{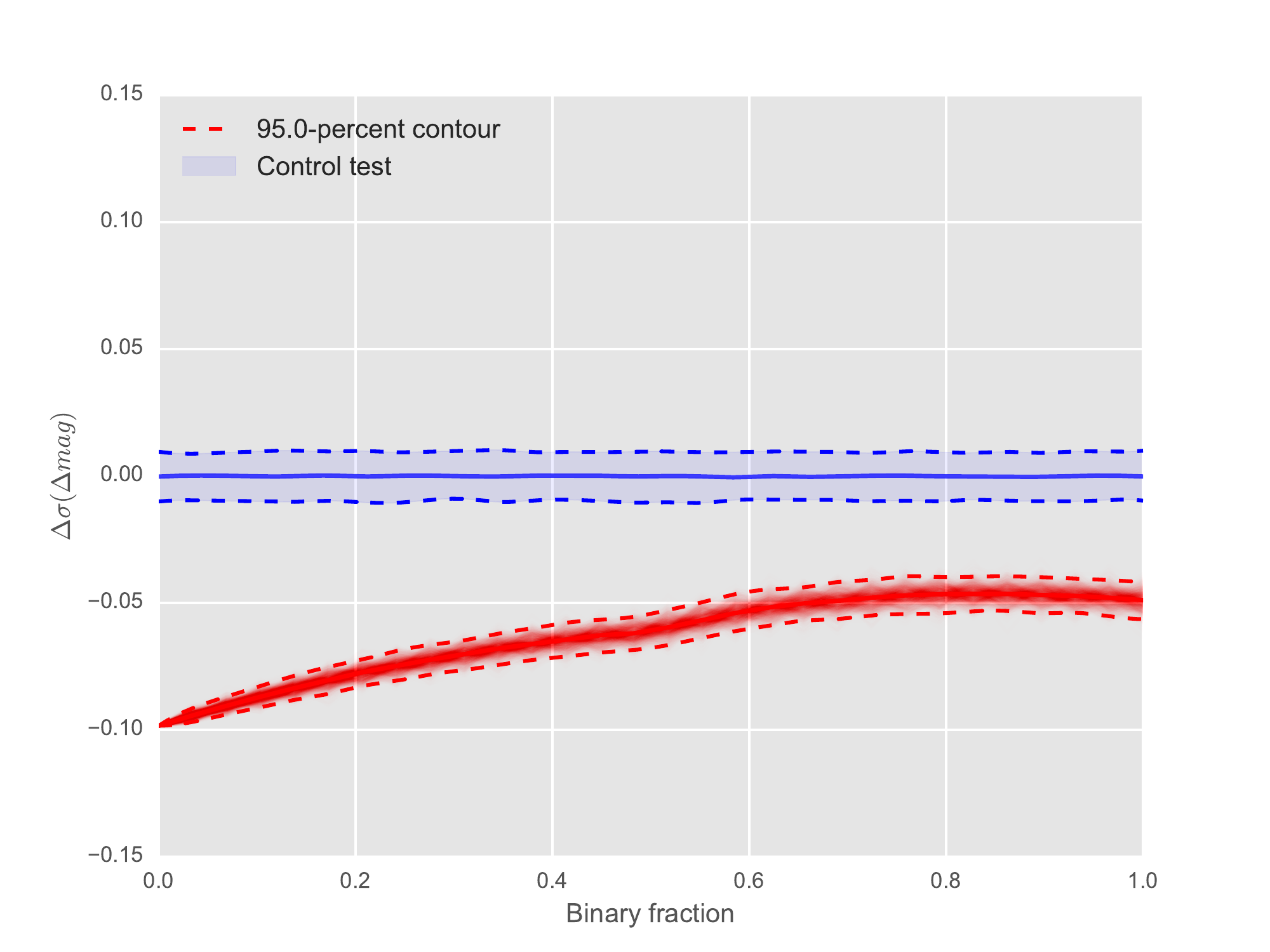}
  \includegraphics[width=6cm]{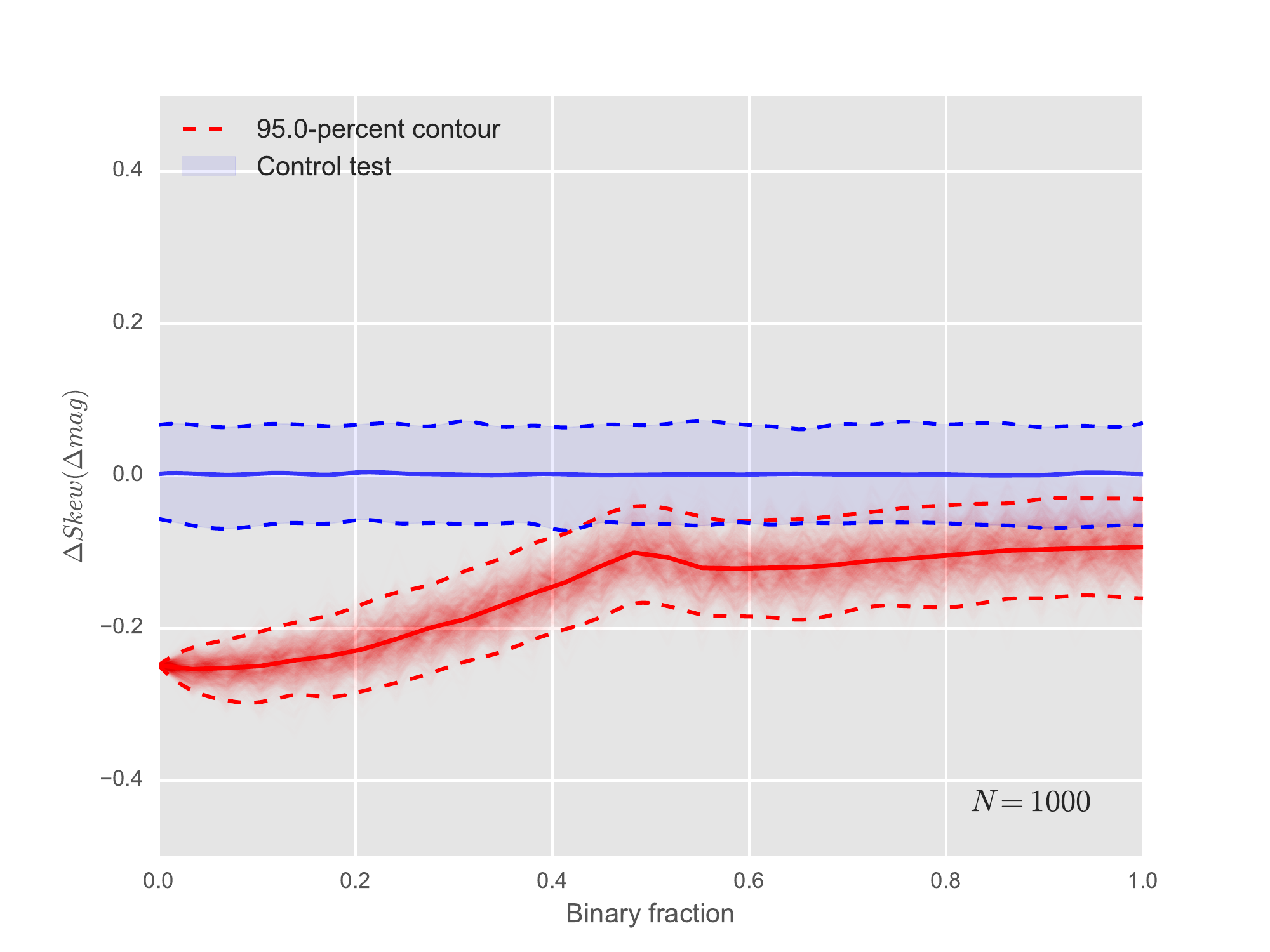}
}
\centerline{
  \includegraphics[width=6cm]{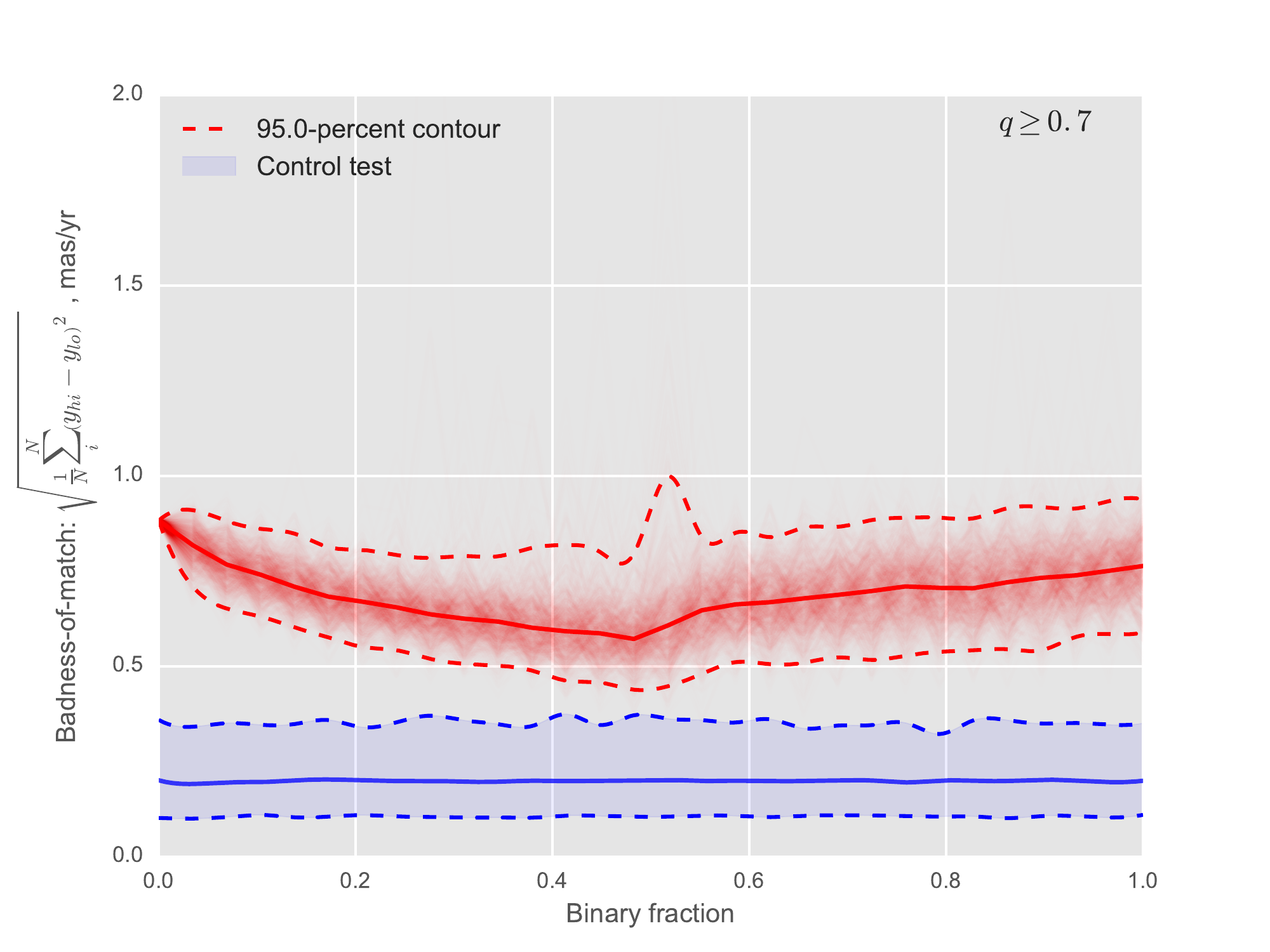}
  \includegraphics[width=6cm]{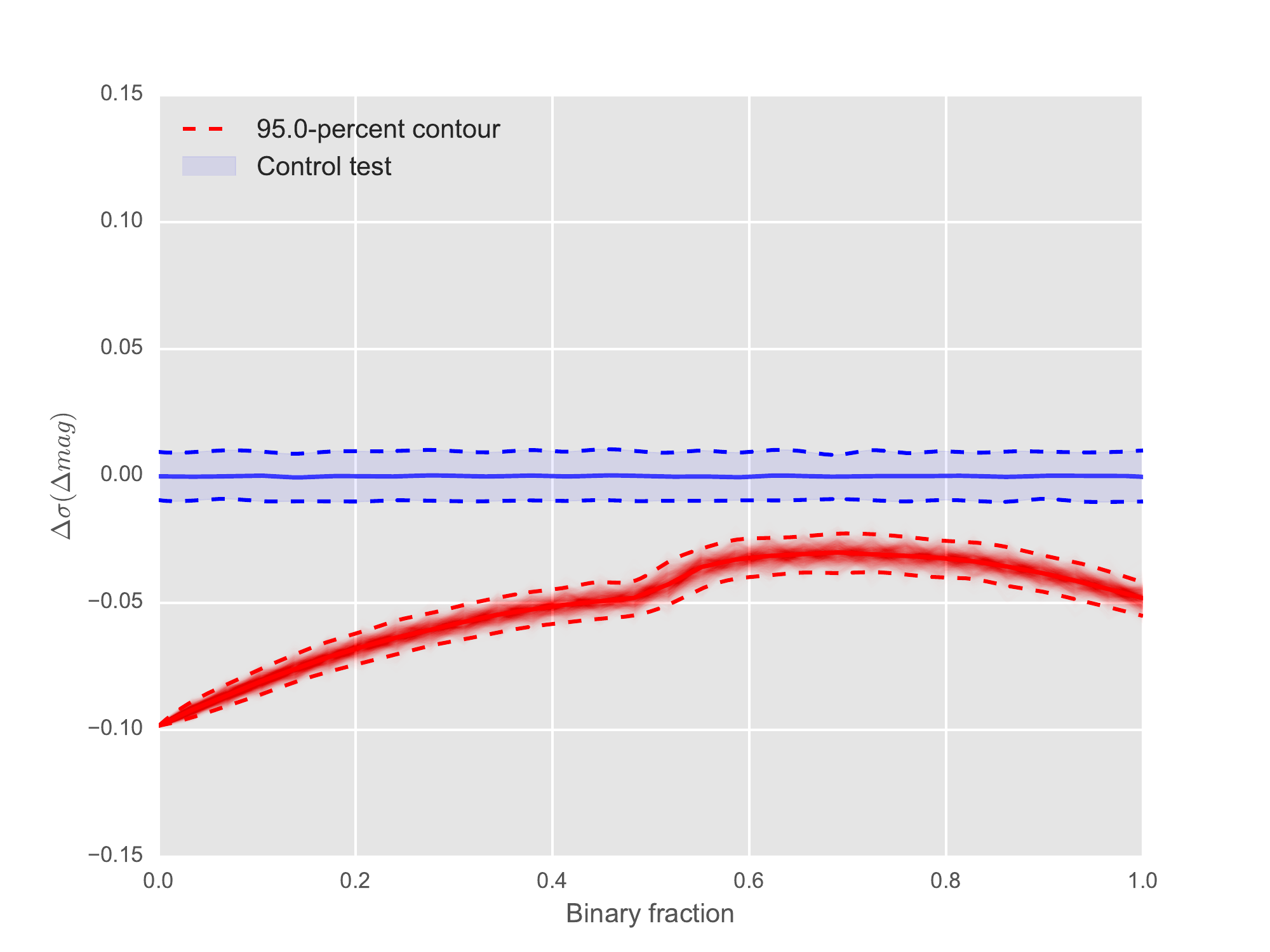}
  \includegraphics[width=6cm]{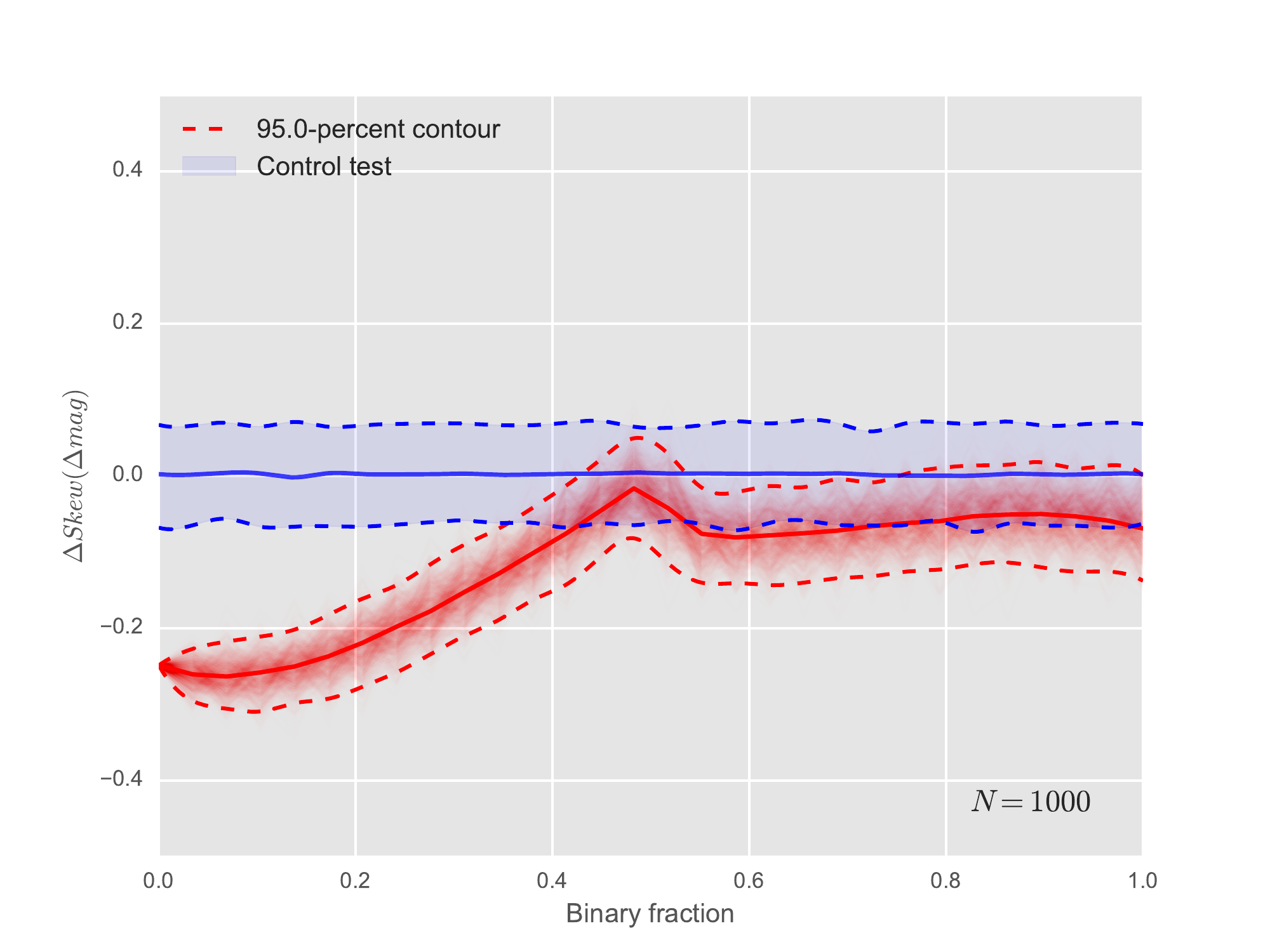}
}
\end{center}
\caption{Evaluation of the impact of a difference in binary fraction
  between \MR~and \MP~populations, in this case for $q \ge 0$~(top row) and $q \ge 0.7$~(bottom row). Panels, colors and symbols
  as with \autoref{f:distFoMsPhotom}; here, the binary fraction
  \fbin~is varied between trials. As with \autoref{f:distFoMsPhotom}, 95\% of the samples fall within the
  dashed contours. See Appendix \ref{ss:distBlur:binaries}}
\label{f:distFoMsBinaries}
\end{figure*}

In none of the cases ($\qmin = 0.0, 0.3, 0.5, 0.7$)~do we find that
the presence of an additional binary population can account for the
difference between the observed \MR~and \MP~rotation
curves. (\autoref{f:distFoMsBinaries} shows the cases $\qmin=0.0$~and
$\qmin = 0.7$). Only the skewness of the \rdm~distribution ever
approximates that of the \MP~population (at $\fbin \gtrsim 0.5$),
while the rotation curve and \rdm~spread do not overlap for any binary
fraction.

We therefore conclude that an excess of unresolved binaries in the
\MP~over the \MR~population is highly unlikely to be responsible for
the difference in rotation curves.

\section{The impact of differential \feh~dispersion on photometric parallax}
\label{s:appMetBlur}

Under a model in which the bulge contains at least two metallicity
components, with differing \feh~dispersions, the spread in inferred
photometric parallax within identified \MP~and \MR~samples will also
differ, even if there is no difference in intrinsic distance
distribution along the line of sight. Here we examine the likely
magnitude of this systematic.

The method is
  outlined in Appendix \ref{ss:appMetBlur:method}, with simulated
  population components described in
  Appendix \ref{ss:appMetBlur:popComponents}. In the course of this
  investigation, it became apparent that the widely-used
  \basti~simulation framework truncates samples at \feh~values well
  within the limits of likely values in the \SWEEPS~field; the
  technique used to characterize absolute magnitude spread in the
  presence of this truncation is described in
  Appendix \ref{ss:appMetBlur:varTruncated}. Finally, the differential
  scatter between \MR~and \MP~populations is presented in
  Appendix \ref{ss:appMetBlur:results}. (The \basti~truncation itself
  is characterized in \autoref{s:testBasti}.)

\subsection{General method}
\label{ss:appMetBlur:method}

To estimate the differential scatter in photometric parallax
  produced by differing \feh~dispersions between \MR~and
  \MP~samples, a synthetic composite stellar population is produced
  for the SWEEPS field by sampling \basti~simulations (computed for
  all three cameras and resampled in the manner of
  Appendix \ref{ss:distBlur:binaries}), which include the effects of age,
  \feh~spread, and unresolved stellar binaries. The synthetic
  populations are perturbed by photometric uncertainty (in all seven
  filters), photometric parallax, and reddening, where the width of
  the distributions in all three quantities can be specified
  separately for each population.

\begin{figure*}[h]
\centerline{
  \includegraphics[height=5.0cm]{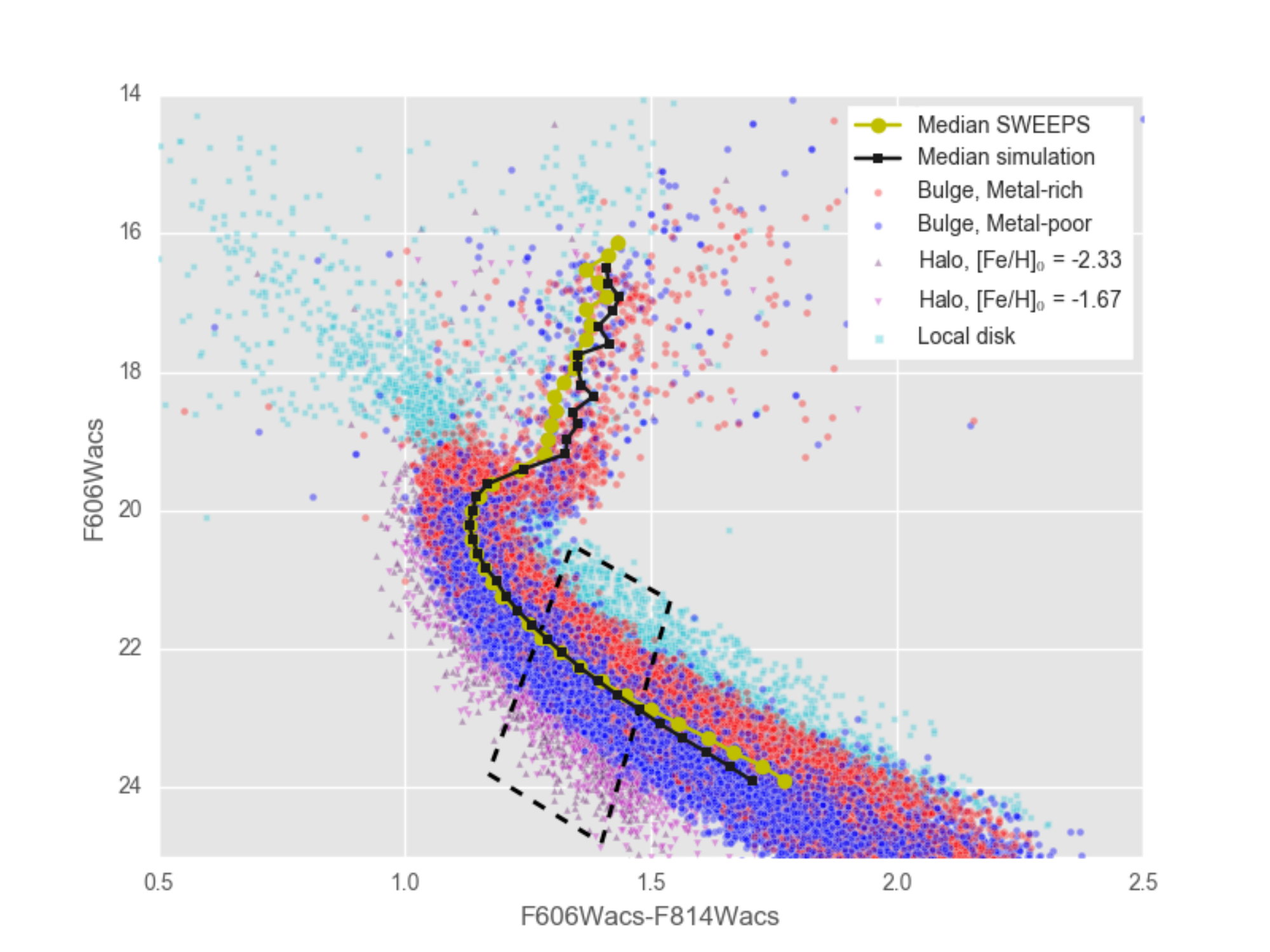}
  \includegraphics[height=5.0cm]{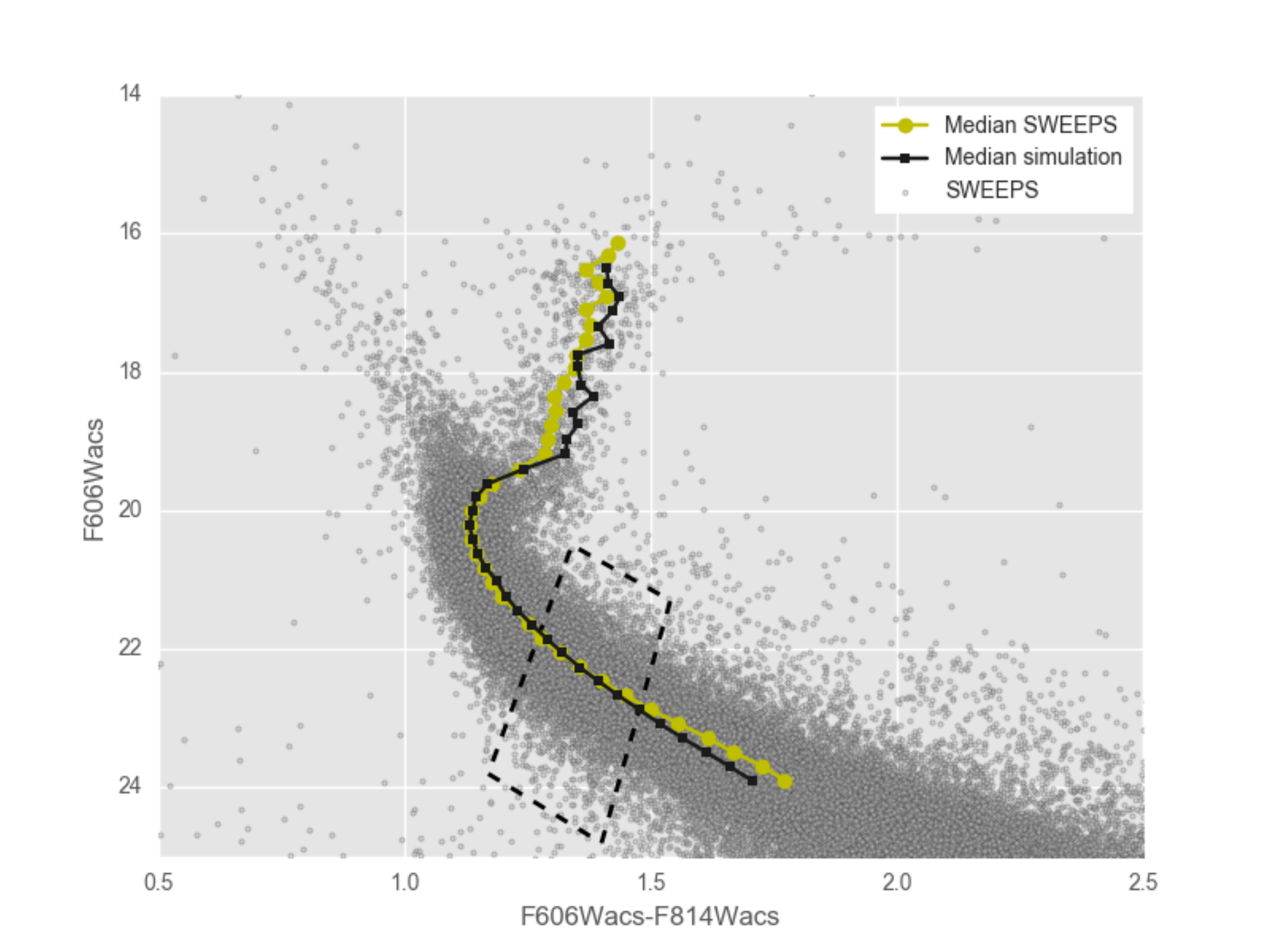}
  \includegraphics[height=5.0cm]{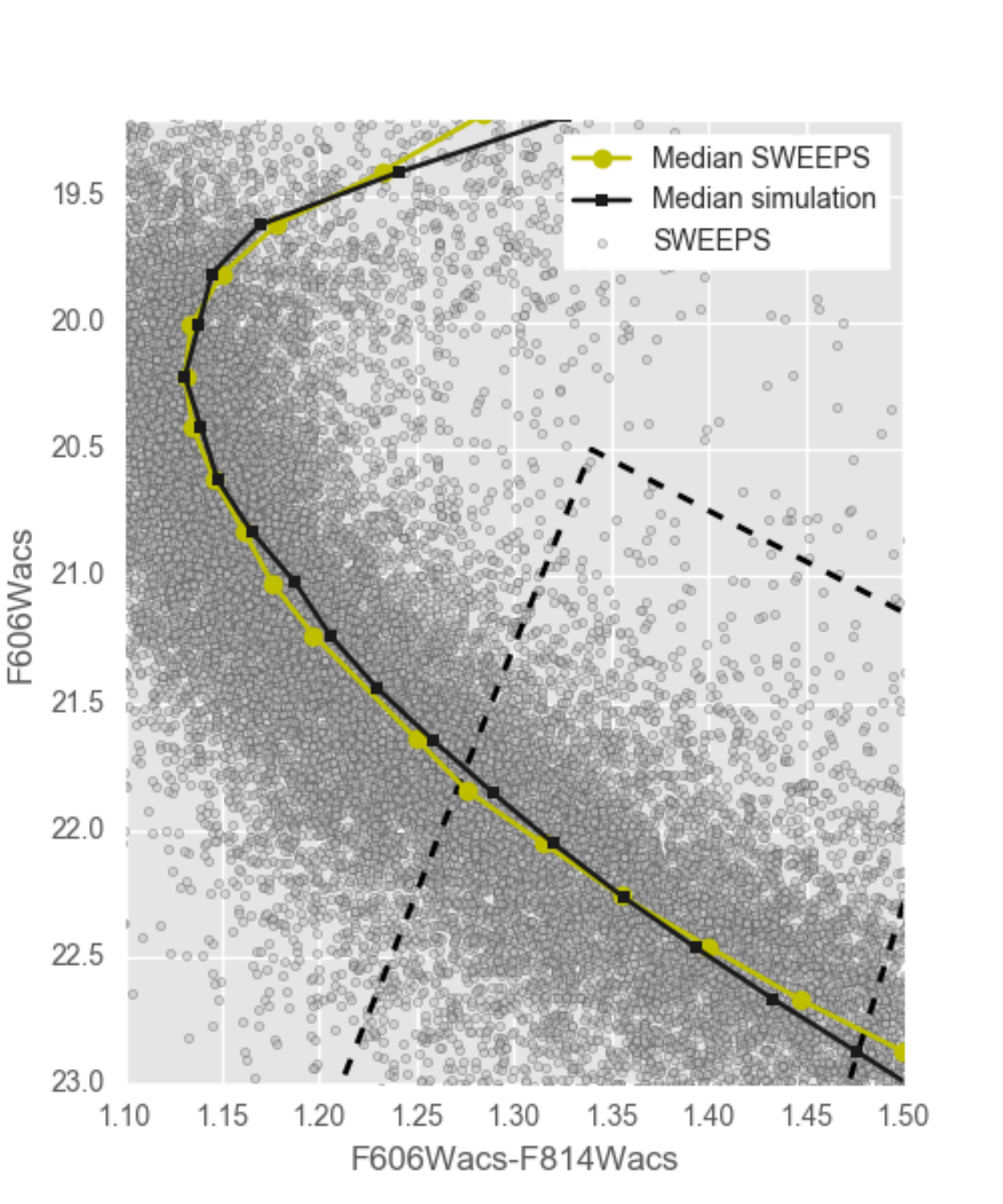}
}
\caption{Composite simulated \SWEEPS~population. {\it Left:} the synthetic populations. Red and blue circle points show metal-rich and metal-poor Bulge components, respectively, violet and gray triangle points the Halo components, cyan squares the Local Disk populations. {\it Middle \& Right panels:} the observed \SWEEPS~CMD, with the median simulated (black line and squares in all three panels) and \SWEEPS~(yellow line and circles in all three panels) populations. See Appendix \ref{ss:appMetBlur:popComponents} for details.}
\label{f:metBlur:CMDs}
\end{figure*}

This produces a \SWEEPS~CMD and \mtindices~distribution for the
  synthetic population. Synthetic objects are selected for further
  ``study'' in a similar manner as for the real data
  (e.g. \autoref{tab:sampleSel}); in particular, synthetic \SWEEPS~CMD
  objects must fall within the selection box in the \SWEEPS~filters
  (\autoref{f:CMDsel}). The surviving synthetic objects are then
  classified as likely \MP~and \MR~populations in the same manner as
  for the observed data (using the GMM components in \mtindices~that
  were fitted to the real data), isolating ``observed'' samples of
  \MR~and \MP~objects. In this manner, the synthetic samples are
  isolated in a similar fashion to those drawn from the real data.

Finally, best-fit loci are determined for the model {\it
    absolute} magnitudes of the synthetic \MR~and \MP~samples, and the
  differences $\Delta M_V$~from these loci determined for every object
  in the samples. The model absolute magnitude is used rather than the
  apparent magnitude because we wish to isolate the impact of
  metallicity spread on {\it intrinsic} magnitude scatter - i.e.,
  before distance, reddening, and photometric uncertainty have
  perturbed the measurements (which impacts the sample selection), but
  {\it including} the intrinsic effects of age, \feh, and binarity.

Modeling the selection cuts on the synthetic samples requires
  simulating the composite stellar population of the
  \SWEEPS~field. This field is somewhat complex, consisting of at
  least three distinct populations (bulge, local disk, halo), each of
  which could well consist of multiple sub-populations or a continuum.

Full population decomposition presents a formidable challenge
\citep[e.g.][]{gennaro15}, and is complicated by the difficulty in
adequately accounting for extinction across the broad wavelength range
of the \BTS~photometry in the inner Bulge region
\citep[e.g.][]{nataf16}. To produce a reasonable approximation to the
selection effects at work in the \SWEEPS~field, a multi-component
stellar population is instead simulated with parameters drawn from the
literature and the \feh~spread estimated in this work
(\autoref{s:appFeH}). A set of about a dozen synthetic populations
with various parameter settings are simulated using \basti, with
typically 5 components from this set combined appropriately to produce
a synthetic composite population for the \SWEEPS~field, with mixture
parameters tuned by hand to provide an approximate match to the
observed \SWEEPS~CMD and \mtindices~distribution.

\subsection{Synthetic population components}
\label{ss:appMetBlur:popComponents}

All population components used the same prescription for binaries,
with binary fraction 0.35 and minimum binary mass ratio 0.0. The
Initial Mass Function followed the \citet{kroupa93} prescription for
all components over the \basti~default mass range ($0.1 \le
M/M_{\odot} \le 120$). Convective core overshooting was not selected
for any model component, and mass-loss parameter $\eta = 0.4$~was used
throughout. When not using a pre-determined star formation history
supplied by \basti, the star-formation histories were specified as a
series of single bursts at given ages, with \feh~described as a
Gaussian with user-specified centroid and standard deviation. Specific
details for various population components follow below.

\begin{figure*}
  \centerline{
    \includegraphics[width=6cm]{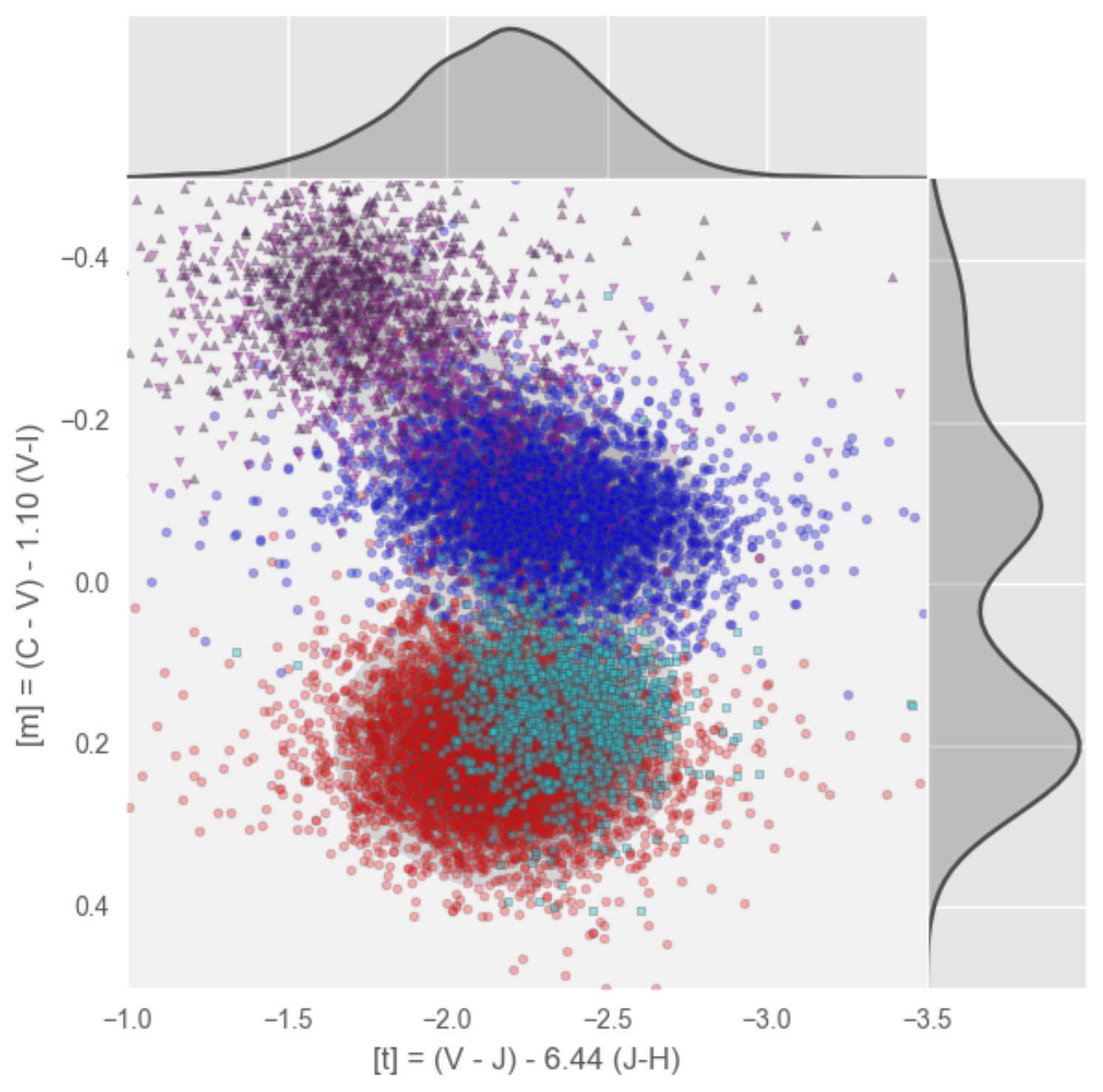}
    \includegraphics[width=6cm]{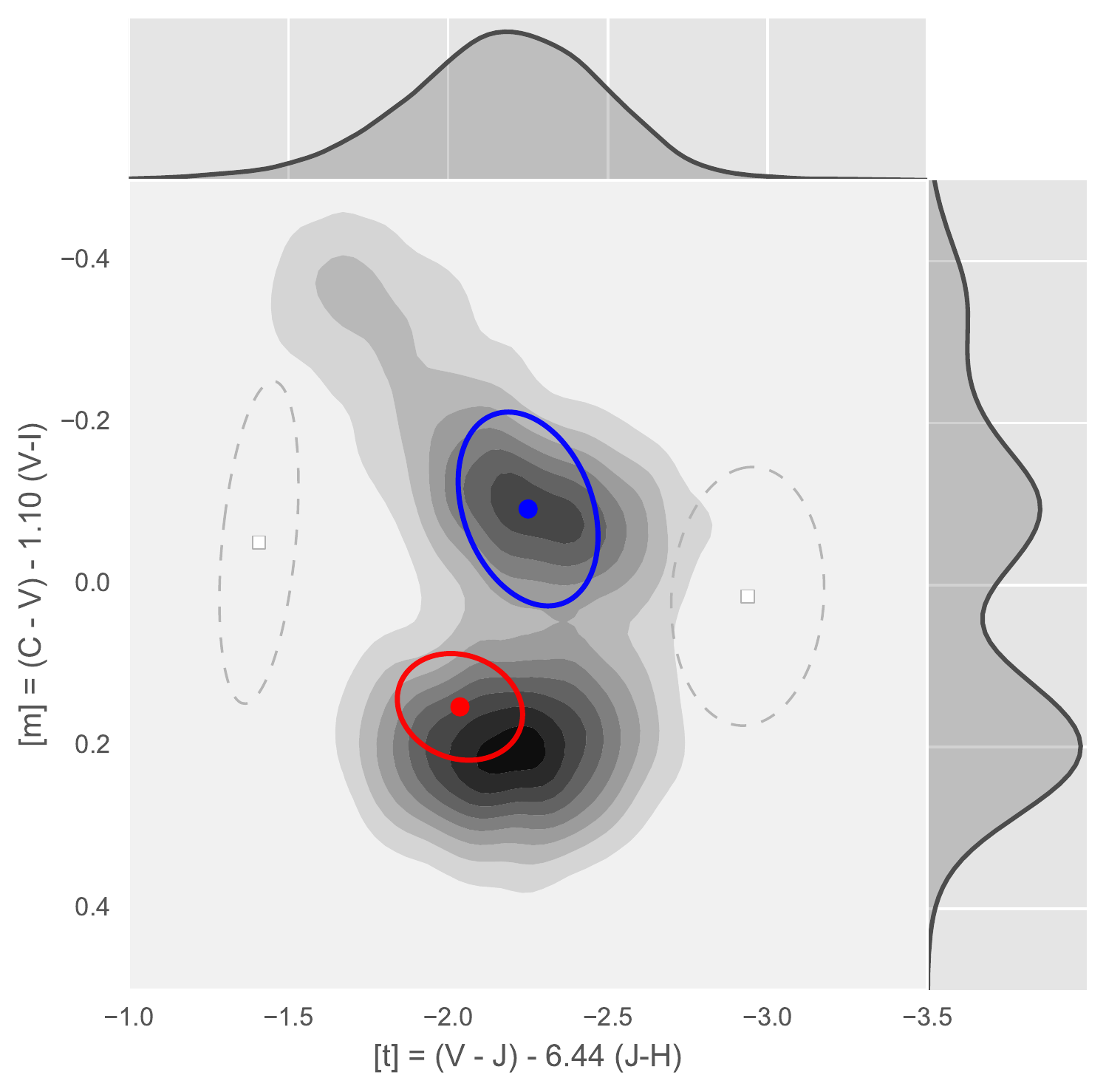}
    \includegraphics[width=6cm]{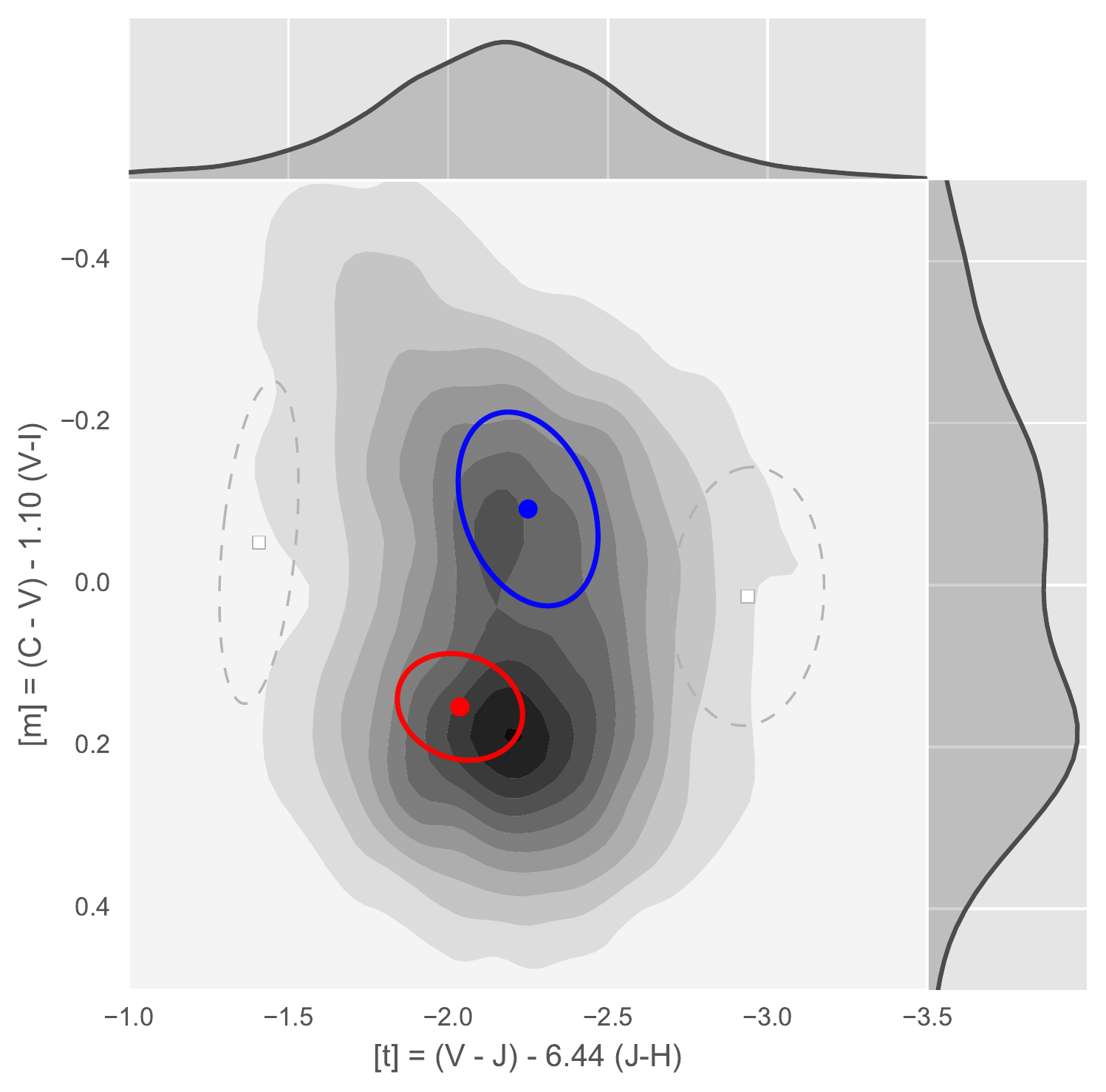}
  }
\centerline{
  \includegraphics[width=6cm]{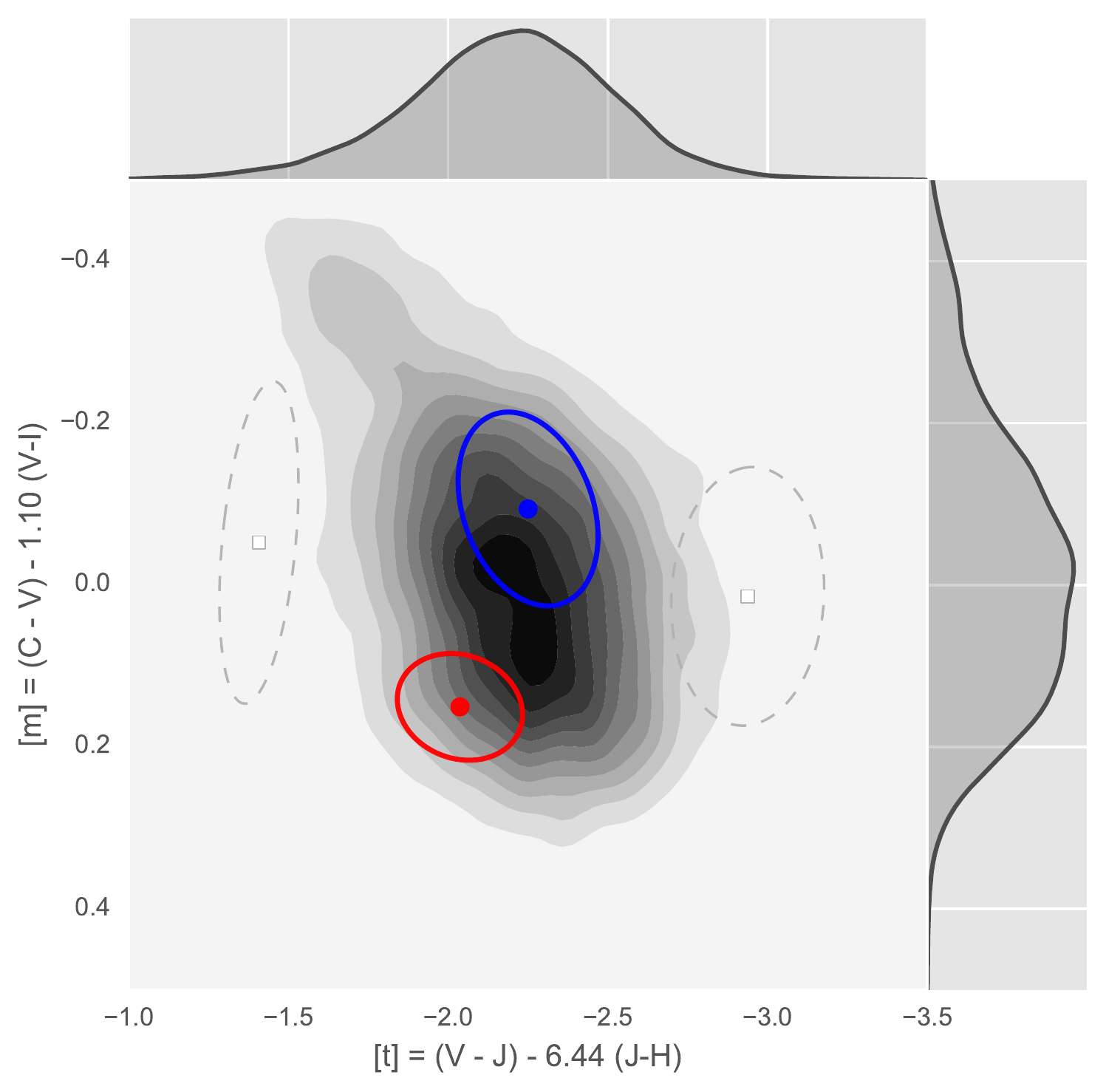}
  \includegraphics[width=6cm]{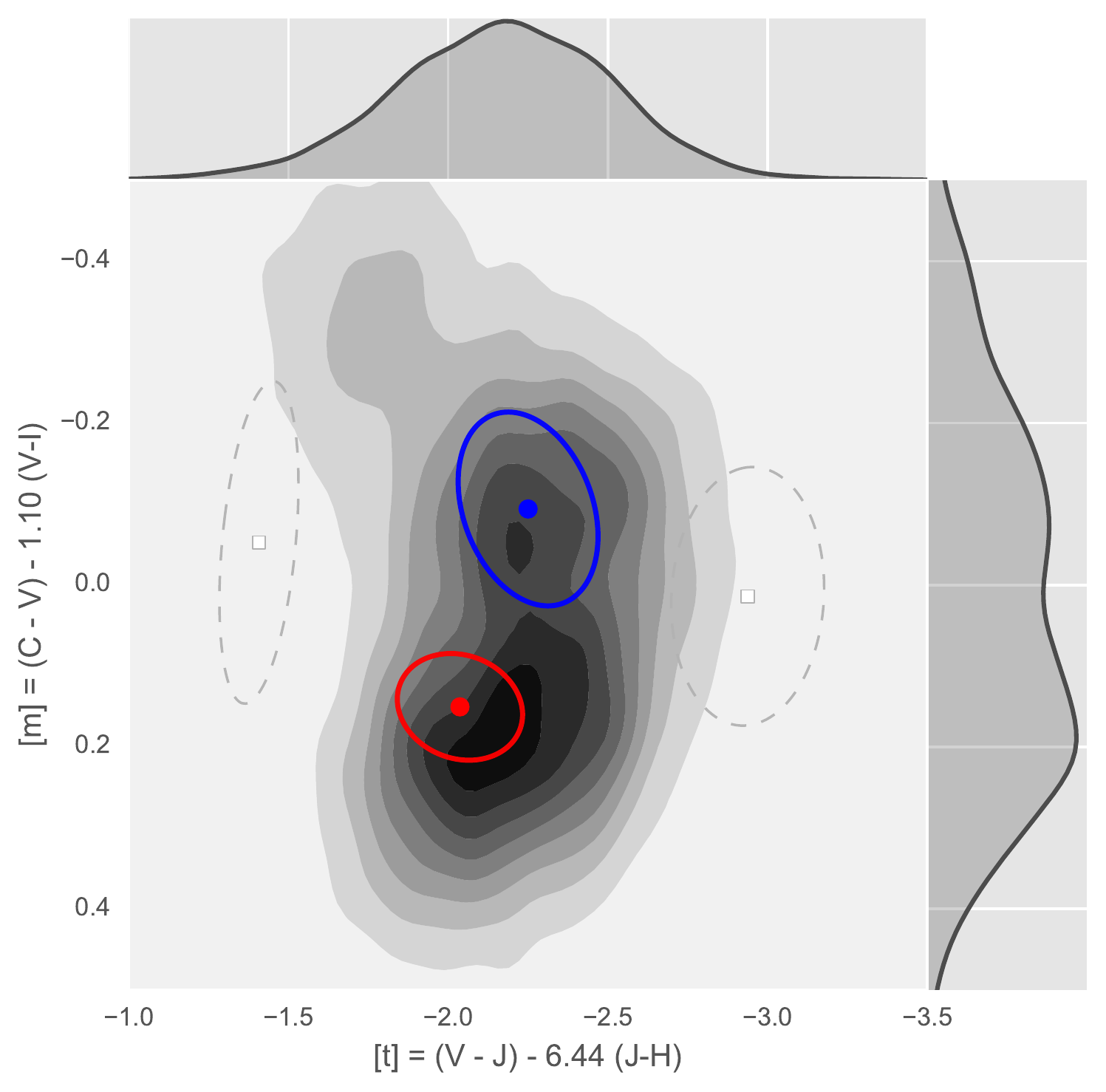}
  \includegraphics[width=6cm]{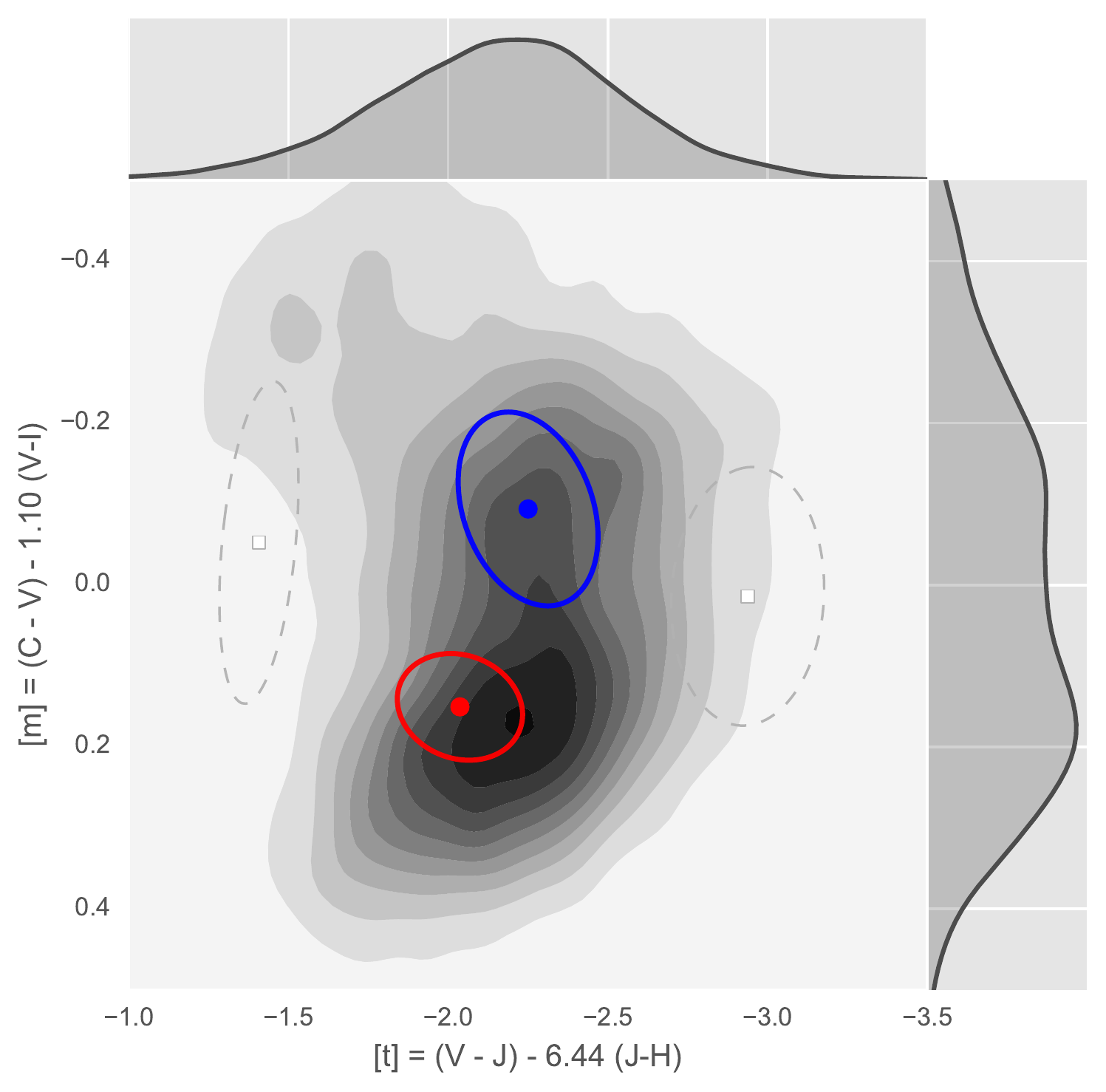}
}
\caption{Example synthetic populations used to estimate metallicity
  effects on sample selection and measured \rdmLong~distributions; all
  panels show the \mtindices~distributions of the model populations
  with marginal distributions of \tindex~and \mindex~plotted over the
  top and right axes, respectively (compare with
  \autoref{f:mtKDE}). All but the bottom-left panel show populations
  with the same intrinsic parameters, modified observationally in
  different ways. {\it Top-left:} Example \mtindices distribution;
  symbols label components of origin in the same way as in
  \autoref{f:metBlur:CMDs}. {\it Top-middle:} KDE representation of
  the simulation in the top-left panel, with the ``metal-rich'' and
  ``metal-poor'' model components that were fitted to the observed
  data overlaid as ellipses for reference. {\it Top-right:} population
  parameters as with the top-middle panel, but with photometric
  uncertainties in the \BTS~filters multiplied by a factor two to
  enhance scatter. {\it Bottom-left:} bulge components drawn from
  \basti's ``bulge'' star formation history \citep{molla00}, using one
  set each with \basti's $\alpha$-enhanced and scaled-to-Solar
  isochrone-sets. {\it Bottom-middle:} population parameters as for
  the top-middle panel, but with \mtindices~each blurred by
  independent Gaussians, with width parameter $\sigma_{\tindex} =
  0.15$~and $\sigma_{\mindex} = 0.05$. {\it Bottom-right:} population
  parameters as per the top-middle panel, but with $R_V$~Normally
  distributed at $R_V = 2.5 \pm 0.52$~in order to bring the marginal
  distributions roughly into line with those observed. See Appendix
  \ref{ss:appMetBlur:popComponents} for details.}
\label{f:metBlur:mtDiagrams}
\end{figure*}

For the foreground disk, the formation history of
  \citet{rocha-pinto00} was used (the default ``Local Disk''~scenario
  within \basti), typically forming $5\%-10\%$~of the stars in the
  simulation sets.

Stellar halo components were simulated using the bimodal
  \feh~distribution reported by \citet{an13}~from SDSS photometry;
  this model consists of a very metal-poor component centered at
  \feh~$\approx -2.33$~and another slightly less metal-poor component
  centered at \feh~$\approx -1.67$. For a bimodal bulge population
  following any of the GMM fits to our spectroscopic data, or for the
  \feh~distribution of \citet{zoccali17} near the \SWEEPS~field, this
  separate halo component is necessary to populate the regions in \mtindices~space for objects with $\feh \lesssim -2.0$.

Bulge components were constructed separately as
    Normally-distributed \feh~distributions specified through the
    \basti~web interface, using the characterization presented in
    \autoref{s:appFeH}, both for the two-
    and three-component GMM decompositions. For components less metal-rich than
    $\feh_0 < +0.3$, separate runs were simulated using the
    ``Scaled-to-Solar'' and ``$\alpha$-enhanced'' options within
    \basti~ in order to allow some exploration of $\alpha$-enhancement
    on population spread in the \mtindices~diagram. A variety of age
    prescriptions were attempted, mostly to improve the match at the
    bright end of the \SWEEPS~CMD, by ascribing either a single burst
    of star formation to each metallicity, or by assigning several
    bursts to each metallicity (e.g. bursts at 5.0, 6.0 and 7.0 Gy for
    a component with $\feh_0 = -0.42$). We have not yet explored more
    sophisticated age-metallicity prescriptions through user-defined
    star formation histories \citep[e.g.][]{bensby17, haywood16}.

The more continuous bulge star formation history of \citet[][used
    as a default in \basti]{molla00} was also tried, for ``Scaled-to-Solar,'' ``$\alpha$-enhanced'' isochrones, and for varying admixtures of the two.

We have not yet explored a separate ``thick-disk'' component in
  this context. The metal-poor wing of the bulge distribution or the
  metal-rich wing of the halo component could mimic such a population
  in the \mtindices~diagram and we do not make the distinction here.

Figures \ref{f:metBlur:CMDs} and \ref{f:metBlur:mtDiagrams} show
  examples of the synthetic populations thus produced. None of the
  population mixtures that we have produced quite reproduces both the
  observed \SWEEPS~CMD and the \mtindices~diagram, although in view of
  both the challenges of extinction characterization and apparent
  simulation truncations imposed by \basti~itself
  (Appendix \ref{ss:appMetBlur:varTruncated}), full reproduction is likely
  to be difficult. The basic two-component bulge we simulate here
  produces an \mtindices~distribution that is much more strongly
  bimodal than that observed (e.g. \autoref{f:mtKDE}), while the
  ten-component ``Bulge'' star-formation history within
  \basti~\citep{molla00} produces an \mtindices~distribution that is
  too smooth compared to that observed. 

Several methods were attempted to bring the simulated
  \mtindices~distribution into closer agreement with that of the
  observed data in \autoref{f:mtKDE}. One simple ansatz is to simply
  multiply the \BTS~estimated photometric uncertainties by a factor
  two before selection and computation of
  \mtindices~(\autoref{f:metBlur:mtDiagrams}, upper-right
  panel). Another is to apply Gaussian blurring in \tindex~and
  \mindex~separately (lower-middle panel of
  \autoref{f:metBlur:mtDiagrams}). Varying $R_V$~with a Gaussian of
  width $\sigma_{\rv} = 0.52$~does bring the marginal distribution
  reasonably close to that observed (lower-right panel of
  \autoref{f:metBlur:mtDiagrams}), although the
  \mtindices~distribution that results is distorted compared to the
  observed sample (particularly the \MR~sample), and in addition the
  required $\sigma_{\rv}$~is at least a factor $\sim 2$~larger than
  that suggested by the \SWEEPS~color-magnitude diagram
  (Appendix \ref{ss:rvvar}, which also shows the \mtindices-blurring effect
  due to $R_V$~variations that are compatible with the \SWEEPS~data).

For the purposes of estimating the impact of varying
  \feh~distribution on \rdmLong~variations, we retain the
  two-component bulge model with and without \BTS~uncertainty scaling,
  for further investigation; the former is consistent with estimated
  \feh~distributions and estimates of photometric uncertainty, while
  the latter is the ``broadened'' option among those tried that
  closely resembles the observed distribution (\autoref{f:mtKDE}).

\subsection{Characterizing excess variability in the presence of truncation}
\label{ss:appMetBlur:varTruncated}

While conducting tests on the simulated datasets, it quickly
  became apparent that samples generated with the current version of
  \basti\footnote{\basti~version 5.0.1.} show truncation at extremes of both high- and low-metallicity, leading to a hard edge in the CMD of the simulated population that has no counterpart in the reported \feh~distribution. This truncation, characterized in \autoref{s:testBasti}, impacts the metal-rich simulated bulge sample more strongly than its metal-poor simulated counterpart and thus could artificially enhance the discrepancy in absolute magnitude breadth between the metal-rich and metal-poor simulated components.

This hidden systematic complicates efforts to characterize the excess
  magnitude scatter due to differing \feh~distributions, with much of
  the most metal-rich end of the metal-rich simulated sample assigned
  apparently incorrect magnitudes (absolute and apparent). We
  therefore adopt a restricted-sample estimate of the magnitude scatter, by
  sampling only the fainter side of the magnitude distribution for
  both samples in the comparison. Specifically, we use the quantity $\sigma_{\rm hi}$~defined by\footnote{(\autoref{eq:sigmaHi} uses 1/$N$~instead of $1/(N-1)$~because the median $\overline{m}$~is determined from a fit to a larger sample than the set over which $\sigma_{hi}$~is evaluated. In practice, with $N(m \ge \overline{m})$~always larger than a few hundred objects, the distinction is unimportant.)} 

\begin{equation}
  \sigma^2_{\rm hi} \equiv \frac{1}{N(m \ge \overline{m})}\sum_{m \ge \overline{m}} \left( m_i - \overline{m}\right)^2
  \label{eq:sigmaHi}
\end{equation}

\noindent where, for the special case of a large, strictly symmetric
distribution, $\sigma_{\rm hi}$~closely approximates the sample
standard deviation. A practical challenge is to identify the median
magnitude $\overline{m}$~from a truncated asymmetric distribution. For
these simulations, $\overline{m}$~is estimated by discarding the most
negative $\Delta m$~samples (thus discarding objects near and outside
the truncation limits) and fitting a Gaussian function to the {\it
  histogram} of $\Delta m$~values. This fit is only used to estimate
$\overline{m}$, which thus allows $\sigma_{\rm hi}$~to be estimated
following \autoref{eq:sigmaHi}. This then allows the restricted-sample
scatter $\sigma_{\rm hi}$~to be estimated for the metal-poor and
metal-rich samples separately, and the excess difference characterized
as the quadrature difference between the two.

\begin{figure*}
  \begin{center}
  \includegraphics[width=14cm]{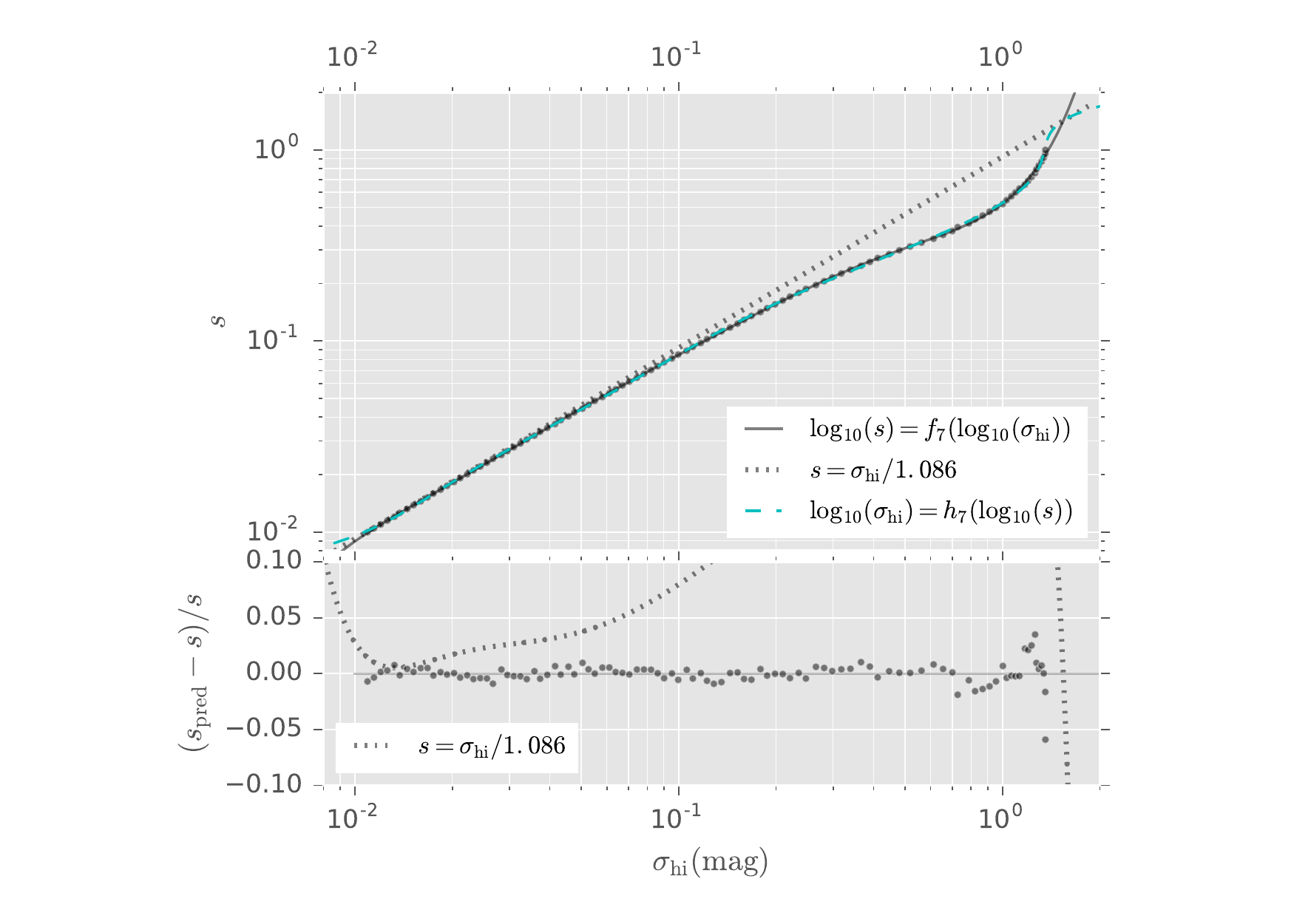}
\caption{Charting the relationship between the flux standard deviation $s$~(Appendix \ref{ss:distBlur:photom}) and the apparent magnitude scatter $\sigma_{\rm hi}$~for truncated samples (\autoref{eq:sigmaHi}). Simulated perturbed populations are generated following \autoref{eq:mPert} and the absolute magnitude distribution of the resulting sample is characterized by $\sigma_{\rm hi}$. Standard uncertainty propagation predicts $s \approx \sigma/1.086$~(with $\sigma$~the apparent magnitude standard deviation); in practice, we fit functional forms to transform between $s$~and $\sigma_{\rm hi}$. The top panel shows $s$~and $\sigma_{\rm hi}$~along with the functional forms in both directions (seventh-order polynomials in $\log_{10}$-space). The bottom panel shows fractional residuals when $\sigma_{\rm hi}$~is used to predict $s$~(residuals in the reverse direction are not shown); the polynomial approximation $f_7$~is accurate to better than 2\% over most of the range of interest. See Appendix \ref{ss:appMetBlur:varTruncated} and \autoref{tab:scatTransf}.}
\end{center}
\label{f:metBlur:transfSig}
\end{figure*}

The final step is then to convert the excess scatter $\sigma_{\rm
  hi}$~estimated from the simulated population components, to the
additional flux scatter $s$~felt by the metal-poor sample compared to
the metal-rich sample. To enable this conversion, the relationship
between restricted-sample scatter $\sigma_{\rm hi}$~and the flux
perturbation scale $s$~that generated it, was determined by
simulation. Synthetic populations with perturbation flux distribution
were produced following \autoref{eq:mPert}, subject to the same
censoring for negative flux as before
(Appendix \ref{ss:distBlur:photom}). The apparent magnitude scatter
$\sigma_{\rm hi}$~was then found for each synthetic population as
described above (and as performed for the simulated
\basti~datasets). Finally, the relationship between $\sigma_{\rm
  hi}$~and $s$~was characterized by fitting a 7th-order polynomial in
both directions. \autoref{tab:scatTransf} and
\autoref{f:metBlur:transfSig} show this characterization. This allows
us to relate the restricted-sample scatter found from
\basti~simulations, back to the flux ratio perturbation scale $s$, and
finally to compare the scale of the perturbation suggested by
differing \feh~distributions to the additional scale of flux
perturbations $s$~that our observational data would require if the
\MP~sample really were a blurred version of the \MR~sample.

% Table: polynomial coefficients for sigma_hi characterization
\begin{deluxetable*}{c|cccccccc}
\tablecaption{Polynomial coefficients relating the flux spread $s$~({\bf Appendix} \ref{ss:distBlur:photom}) to apparent magnitude scatter $\sigma_{\rm hi}$~(\autoref{eq:sigmaHi}), over the domain $(0.01 \le \sigma_{\rm hi} \le 1.0)$. The forms used are: $\log_{10}(\sigma_{\rm hi}) = \sum b_i \log_{10}(s)^i$~and $\log_{10}(s) = \sum a_i \log_{10}(\sigma_{\rm hi})^i$. See \autoref{f:metBlur:transfSig} and {\bf Appendix} \ref{ss:appMetBlur:varTruncated}. \label{tab:scatTransf}}
\tablehead{\colhead{Coeff} & \colhead{$i=7$} & \colhead{$6$} & \colhead{$5$} & \colhead{$4$} & \colhead{$3$} & \colhead{$2$} & \colhead{$1$} & \colhead{$0$}}
\startdata
$a_i$ & 0.2621 & 1.9750 & 6.0299 & 9.5353 & 8.1928 & 3.4773 & 1.2929 & -0.2817 \\
$b_i$ & 0.6190 & 4.3495 & 11.7411 & 14.7695 & 7.7084 & 0.0234 & 0.1323 & 0.1298 \\
\enddata
\end{deluxetable*}

%2017-12-13 forcing the appendix to stay in one-column format
\onecolumngrid

\subsection{Differential photometric parallax dispersion due to differential \feh~dispersion}
\label{ss:appMetBlur:results}

\begin{figure}
  \begin{center}
  \includegraphics[width=13cm]{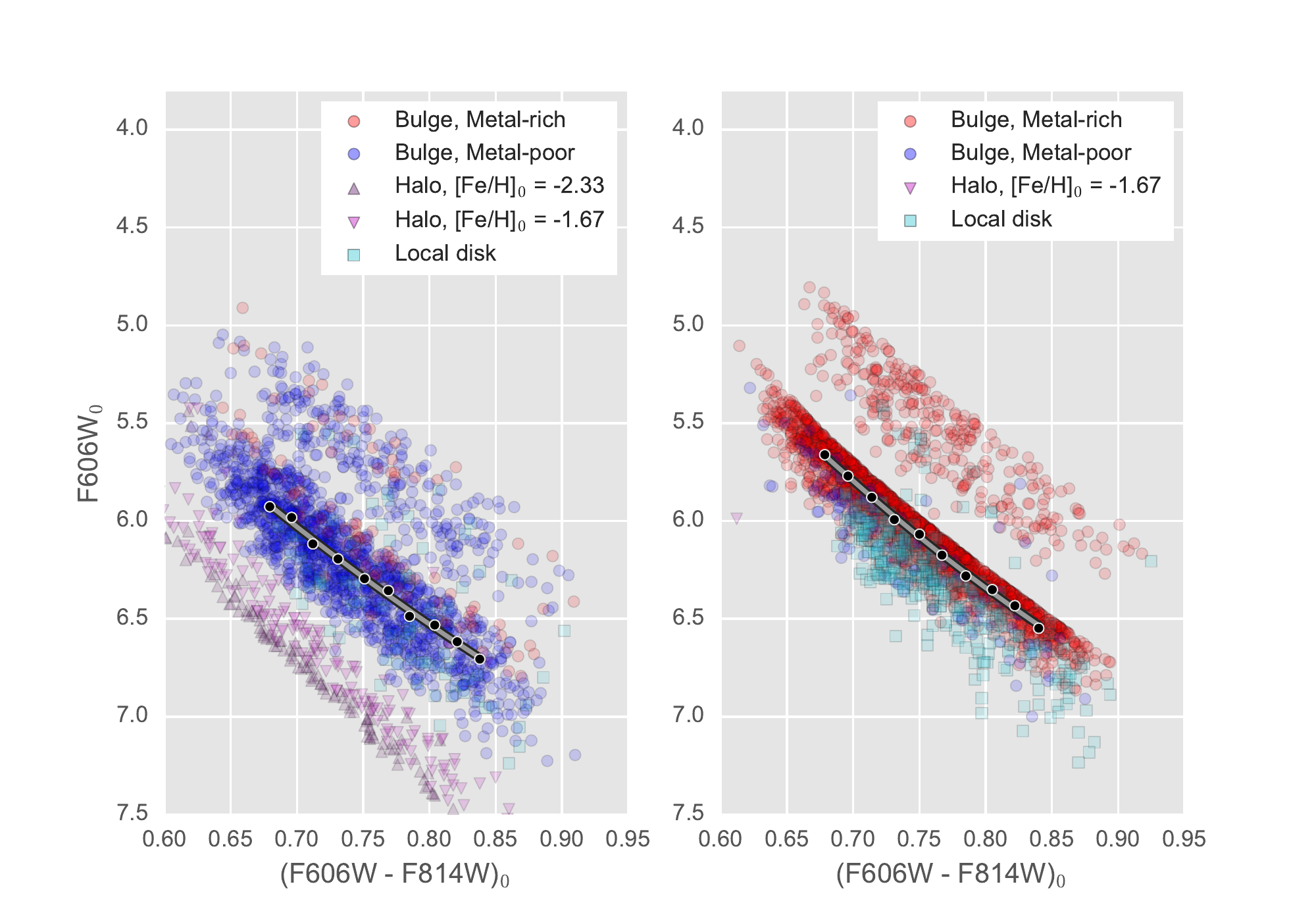}
\caption{Estimating the ridgelines for simulated objects that would be selected in the \MP~(left panel) or \MR~(right panel) samples. In each case absolute magnitudes are plotted in the \SWEEPS~filters. The component of origin for each simulated object surviving selection criteria, is indicated by color and plotting symbol. See Appendix \ref{ss:appMetBlur:results}.}
\end{center}
\label{f:metBlur:deltaUnred}
\end{figure}

We are finally in a position to estimate the additional scatter in
absolute magnitude due to differential metallicity
scatter. \autoref{f:metBlur:deltaUnred} shows the results of applying
the selection criteria to the simulation including the two-component
bulge model, a two-component halo, and local disk component.

\begin{figure*}
\centerline{
  \includegraphics[width=8cm]{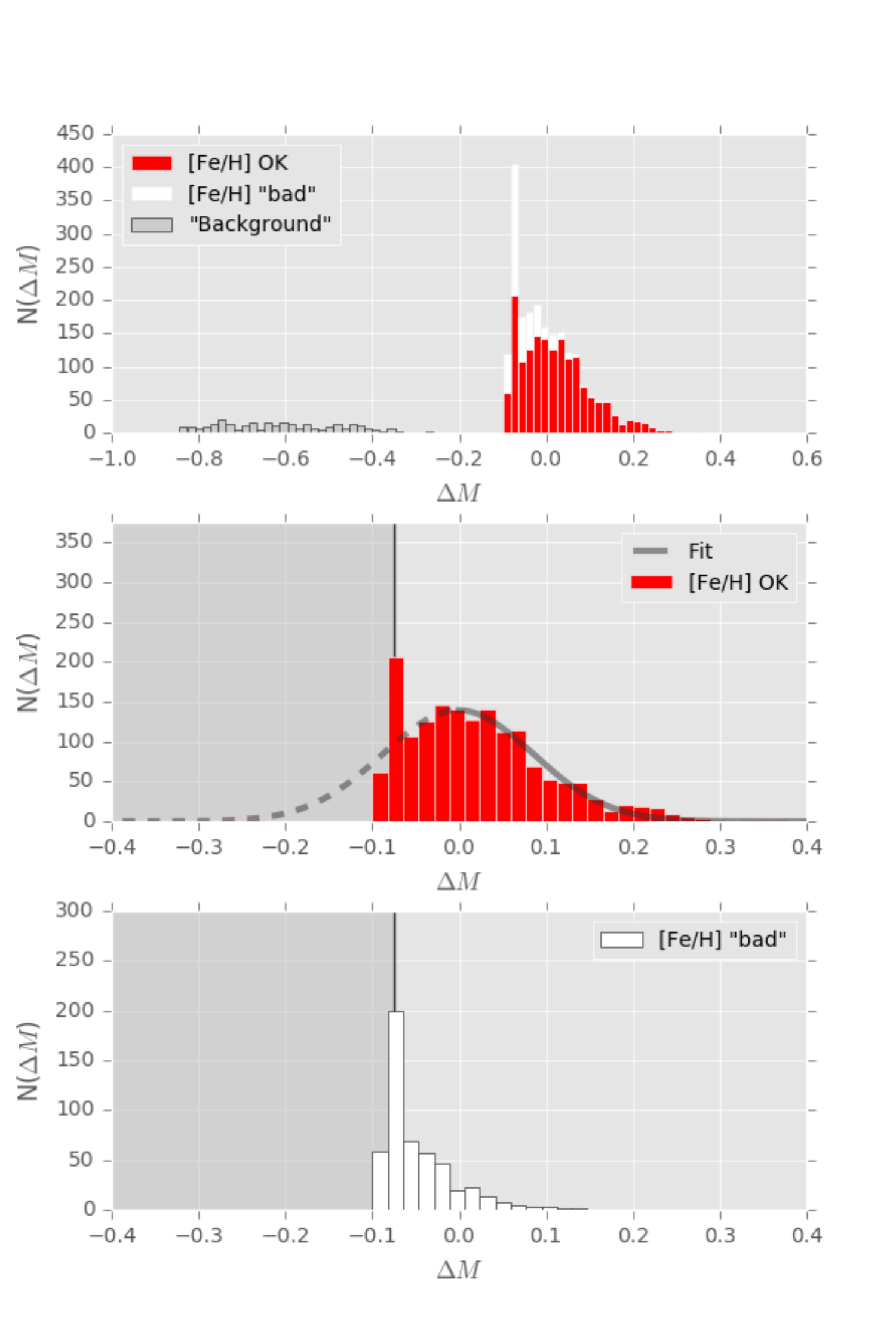}
  \includegraphics[width=8cm]{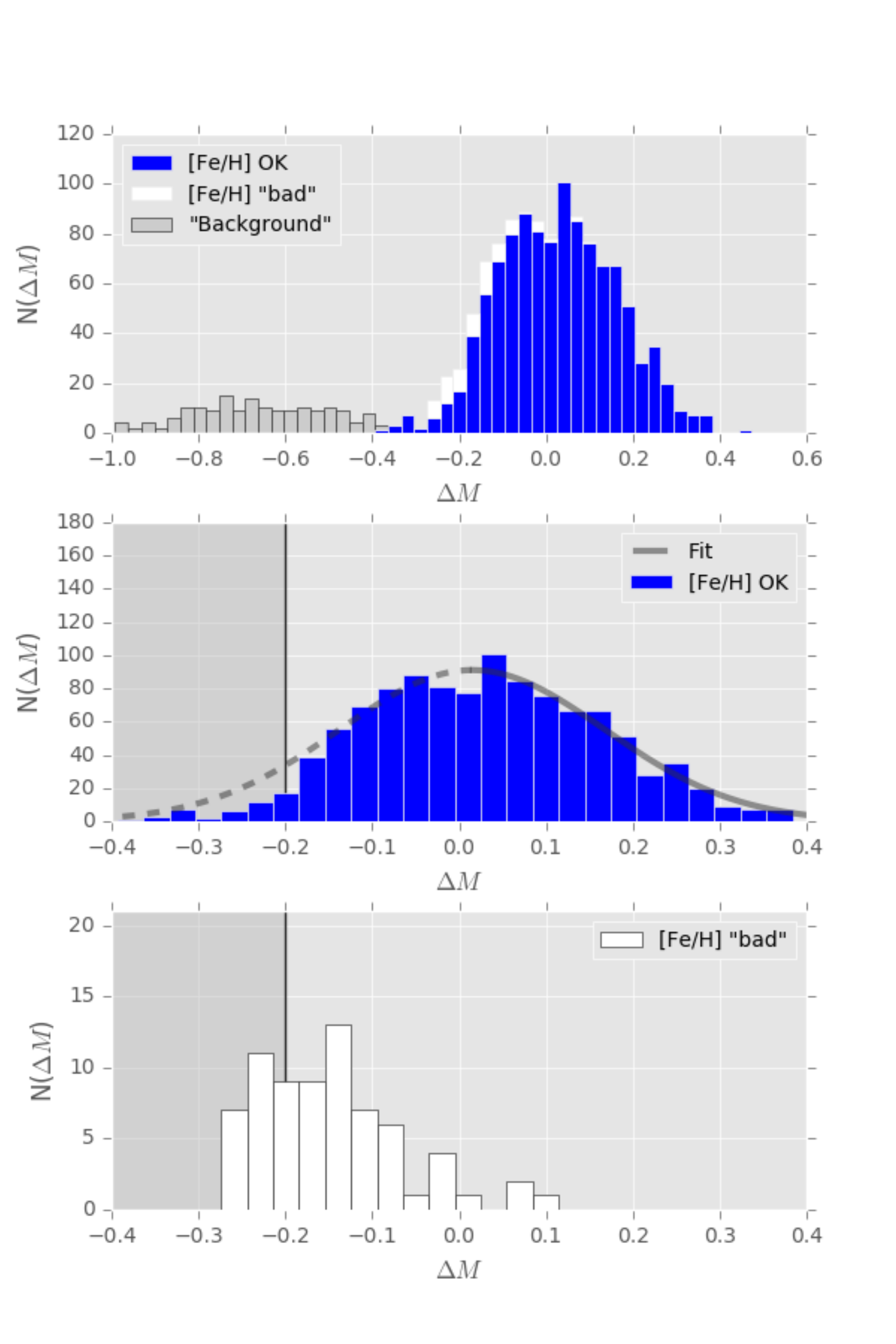}
}
\caption{Characterizing the magnitude scatter $\sigma_{\rm hi}$~(\autoref{eq:sigmaHi}) for simulated populations in the presence of truncation. The left column shows the simulated metal-rich population, the right column the simulated metal-poor. White-shaded bars in each figure show objects with \feh~outside the adopted \basti~metallicity range (using scaled-to-Solar isochrones for the metal-rich column, $\alpha$-enhanced for metal-poor). Reading top-bottom, rows show: the full distribution (top), objects with simulated \feh~within the nominal ranges (middle), and those outside the nominal ranges (bottom). The gray regions in the middle and bottom rows (delimited by the solid vertical line) show regions of $\Delta M$~excluded from the Gaussian fits to the distributions (smooth lines). The fitted median of $\Delta M$~is marked by a transition from solid to broken line in the curves. Before characterization, each simulated sample is classified into a dominant and secondary component; the secondary component, mostly made up of unresolved binaries and labeled ``Background'' in the panels here, is excluded from further consideration. Note that (i). {\it both}~the metal-rich and metal-poor samples include objects with reported \feh~above the adopted upper limit; (ii). the truncation appears to impact objects even with \feh~nominally within the adopted \feh~limits, particularly for the metal-rich simulated population, and (iii). the strong truncation in the metal-rich sample leads to a large gap between the dominant and secondary component. See Appendix \ref{ss:appMetBlur:results}.}
\label{f:appMetBlur:scatters}
\end{figure*}

\autoref{f:appMetBlur:scatters} illustrates the characterization of absolute magnitude scatter $\sigma_{\rm hi}$, while \autoref{t:appMetBlur:results} shows the evaluation of the excess flux scatter $s$~for \MP~compared to \MR~samples. Two simulated populations were evaluated in this manner; one including the two-component bulge model; the other with the \BTS~uncertainties multiplied by a factor 2 before selection to broaden the distribution in \mtindices. In both cases, the excess fractional flux scatter $s$~is less than 0.1; we find $s \approx 0.09$~for the two-component bulge model, while $s \approx 0.07$~for the enhanced-uncertainty version of this model.

% Table: Additional absolute magnitude scatter.
\begin{deluxetable}{c|cc|cc}
\tablecaption{Characterization of the additional absolute magnitude scatter due to \feh~for simulated Metal-rich and Metal-poor populations. The quadrature difference between the two samples is reported in the final line. $\sigma_{\rm hi}$~reports an estimate of the asymmetrically-sampled absolute magnitude scatter ({\bf Appendix} \ref{ss:appMetBlur:varTruncated}), while $s$~reports the scatter in the flux perturbation due to \feh~spread. The first column-pair shows results for the simulated populations and estimated uncertainties; the final column-pair shows results for \mtindices~distribution broadened to more accurately match the observed distribution. See {\bf Appendix} \ref{ss:appMetBlur:results}. \label{t:appMetBlur:results}}
\tablehead{\colhead{Component} & \colhead{$\sigma_{\rm hi}$} & \colhead{$s$} & \colhead{$\sigma_{\rm hi}$(broadened)} & \colhead{$s$(broadened)}}
\startdata
Metal-poor & 0.137 & 0.112 & 0.153 & 0.124 \\
Metal-rich & 0.103 & 0.087 & 0.119 & 0.099 \\
\hline
Excess & 0.090 & 0.071 & 0.097 & 0.075\\
\enddata
\end{deluxetable}

We contacted the authors of the \basti~web tools regarding its
internal truncation (detailed in \autoref{s:testBasti}). In response,
Santi Cassisi (2017, private communication) kindly added a
high-metallicity point to \basti's internal metallicity grid (since in
\basti~version 5.0.1, the metallicity range covered by the simulator
is more restrictive than that covered by the isochrone set), and
re-computed sets of synthetic populations using the updated version of
the simulator.\footnote{We refer to these new simulations as the
  ``Cassisi'' simulations, and the simulations ran using the current
  publicly-available \basti~suite as ``v5.0.1''} Visual inspection of
the \mtindices~distribution and the SWEEPS CMD drawn from the Cassisi
simulations indicates similar behavior to those from v5.0.1, except
without the sharp edges truncating the metal-rich end of the synthetic
population.

In this paper we retain the statistics derived using \basti~v5.0.1
since that is the version currently available to the
community. However, the comparison with the Cassisi version is
instructive. Application of the half-sample techniques of
Appendix \ref{ss:appMetBlur:varTruncated} to both the Cassisi and v5.0.1
simulations yielded highly similar results ($\sigma_{\rm
  hi}$~differing by $< 4\%$), as might be expected since this measure
uses the side of the $\Delta M$~distibution far from the truncation
limit. The Cassisi simulations also allow a direct estimate of the
accuracy of the one-sided measure adopted in
Appendix \ref{ss:appMetBlur:varTruncated}, by comparing $\sigma_{\rm
  hi}$~to the $\Delta M$~standard deviation of the objects in the
dominant component of the Cassisi simulation (see
\autoref{f:metBlur:deltaUnred} for the dominant and ``background'' components for metal-rich and metal-poor simulated populations). In
the Cassisi simulations, the $\Delta M$~standard deviation is roughly
20\% {\it smaller} than the estimate $\sigma_{\rm hi}$, suggesting our
estimates of the excess photometric scatter in
\autoref{t:appMetBlur:results} may be over-estimates.

We therefore find that the combination of differing metallicity spreads between \MP~and \MR~samples, with differing selection effects in both the \mtindices~distribution and \SWEEPS~CMD, together contribute differential flux scatter that is not larger than $\sigma_{\rm hi} \approx 0.1$~magnitudes, or additional flux standard deviation $s \approx 0.08$. This additional scatter is a factor $3$~too small to bring the observed \MP~and \MR~proper motion-based rotation curves into agreement by itself (\autoref{f:distFoMsPhotom}), and we conclude that the apparent difference in proper motion rotation curves between the two samples is {\it not} an artefact of differences in the underlying \feh~distribution.

\begin{figure}
  \begin{center}
  \includegraphics[width=12cm]{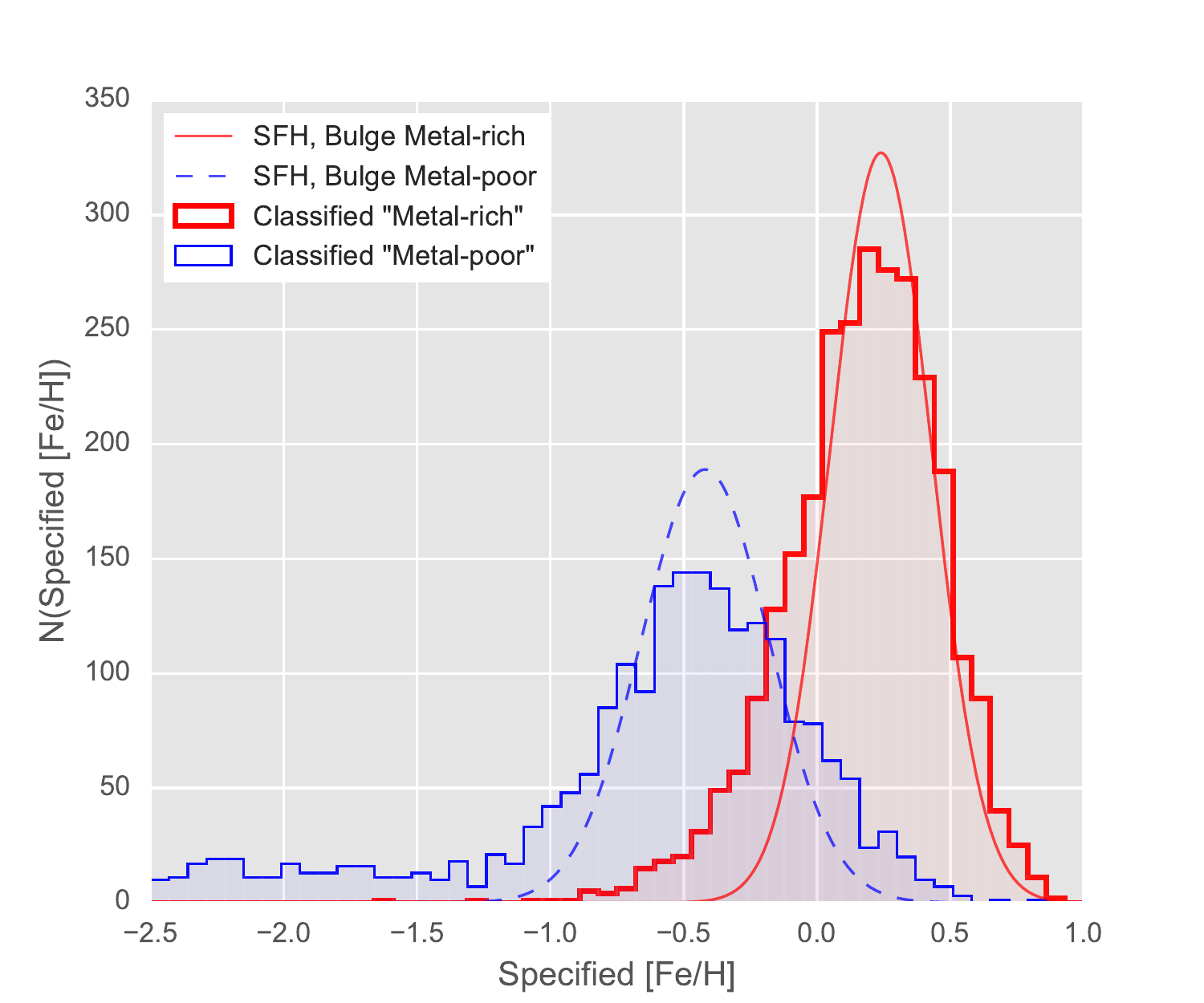}
\caption{Comparison of recovered and input samples for the five-component \basti-based synthetic composite \SWEEPS~field population (e.g. \autoref{f:metBlur:deltaUnred}). The histograms show the objects classified with the \MP~(blue, thin stepped line) and \MR~(red, thick stepped line) samples. The smooth Gaussian \feh~distributions that were specified for the two Bulge components are overlaid; the Metal-poor (blue-dashed curve) and the Metal-rich (red solid curve) components. The \feh~values are those reported in the \basti~output tables (see discussion in Appendix \ref{s:testBasti}). See Appendix \ref{s:appMetBlur}.}
\end{center}
\label{f:metBlur:recovSim}
\end{figure}

As a second check, we can compare the \feh~distribution of the
  objects classified as \MP~and \MR~with the simulated \feh~values for the
  relevant Bulge model components. We find that indeed the
  mis-classification rate in this synthetic population-based
  simulation appears to be low
  (\autoref{f:metBlur:recovSim}). Possible contaimination is explored
  further in a purely empirical manner in \autoref{s:appContam}.

\subsection{The impact of $R_V$~variations}
\label{ss:rvvar}

The framework of this Appendix also allows us to investigate the impact of $R_V$~variations on \mtindices-based determinations. The extinction-free indices \mtindices~assume a
  particular extinction prescription (\citealt{cardelli89}~using
  \rv=2.5). While \mtindices~are therefore
  insensitive to variations in \EBmV~for a particular value of $R_V$, variations in \rv~could impact the distribution of points in the \mtindices~diagram, by
  altering the relationships between apparent magnitudes in the
  \BTS~filters from those assumed when computing \mtindices.

\begin{figure}
  \begin{center}
  \includegraphics[width=8cm]{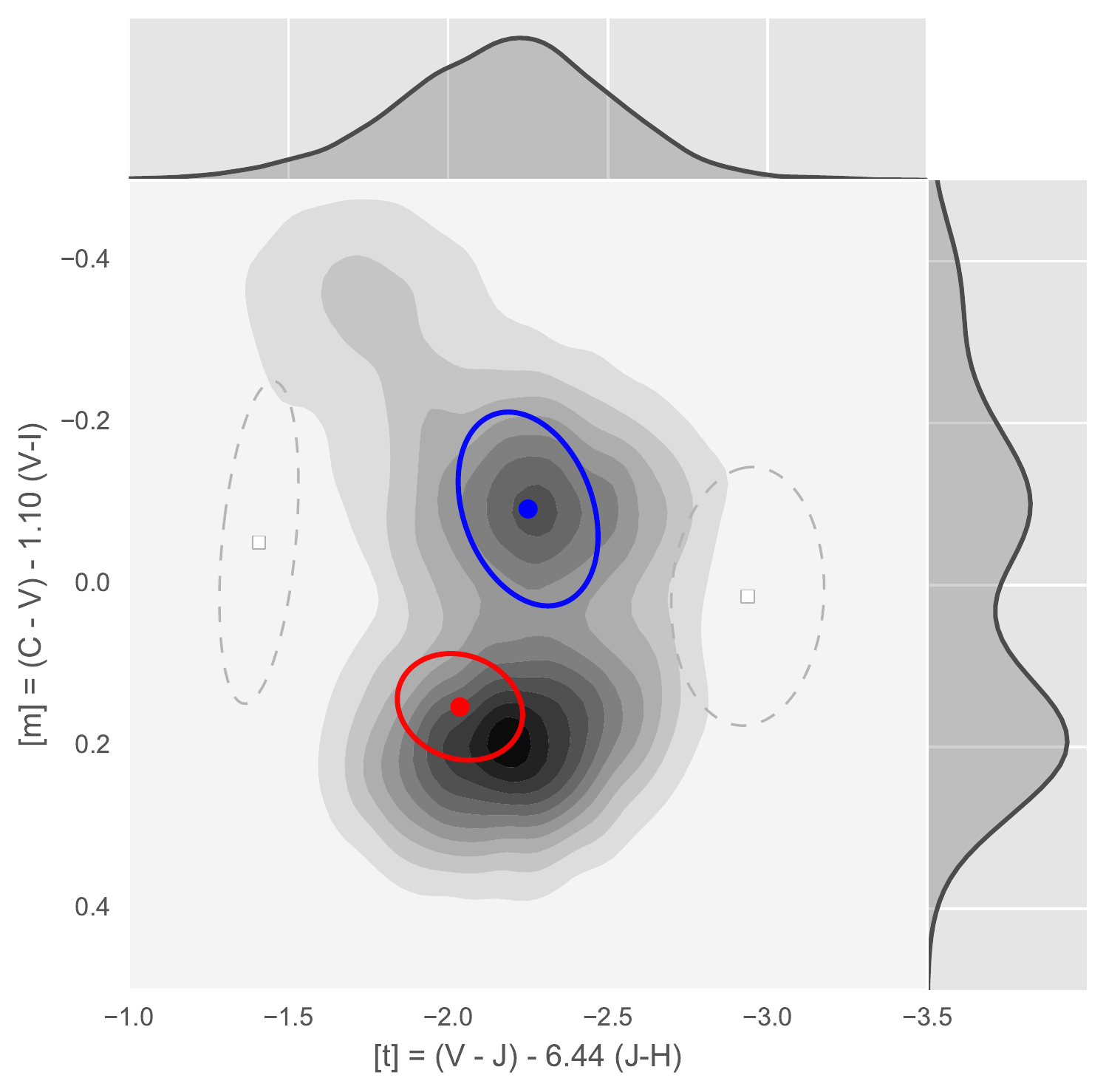}
\caption{Estimating the impact of \rv~variations in the
  \mtindices~diagram. The middle-top pattern of
  \autoref{f:metBlur:mtDiagrams} shows a simulated
  \mtindices~distribution using estimated photometric uncertainties
  and \feh~distribution, and with $R_V = 2.5$~for all objects. This
  figure shows the same simulation but this time varying \rv~by
  $\sigma_{\rv} = 0.25$. The $1\sigma$~ellipses from the GMM
  decomposition of the observed data are shown to allow rough
  comparison between this simulation and the true dataset, and the top
  and side panels show the marginal distributions of \tindex~and
  \mindex, respectively. See Appendix \ref{ss:rvvar}.}
\end{center}
\label{f:rvvar}
\end{figure}

We appeal to the \SWEEPS~color-magnitude diagram to estimate
  limits on the magnitude of $R_V$~variations in this field. Assuming
  the distance~distribution due to the physical depth of the bulge can
  in this field be characterized by a Gaussian with width parameter
  $\sigma_{d}$~kpc, the observed apparent magnitude scatter of Red
  Clump Giants (RCG) in this field then sets an upper limit on
  \rv~variations for assumed \EBmV. In the SWEEPS dataset, the
  observed \filtI~dispersion of the RCG is $\sigma(\filtI) \approx
  0.17$~magnitudes \citepalias{clarkson08}.

For this Appendix we adopt \EBmV=0.5~\citepalias{calamida14} as a
representative value (the implied $R_V$~variations would become
smaller for larger \EBmV). The extreme case of distance dispersion,
$\sigma_d = 0$, then admits \rv~variation of $\sigma_{\rv} \approx
0.45$. However, the bulge has nonzero depth along the line of sight;
picking a representative distance distribution of $\sigma_d \approx
0.5$~kpc, suggests variation closer to $\sigma_{\rv} \approx 0.25$~is
more likely. Both estimates for $\sigma_{\rv}$~are conservative {\it
  upper} limits, since they ascribe none of the observed RCG apparent
magnitude dispersion to photometric uncertainty, luminosity variations
within the RCG sample, or \EBmV~variation.

To estimate the impact of \rv~variation on the
  \mtindices~distribution (and thus sample selection and
  cross-contamination), a synthetic population was constructed using
  \basti~population components tuned to the estimated metallicity
  distribution for this field. Full details of this procedure, which
  was implemented to explore metallicity-dependent selection and
  characterization systematics (\autoref{ss:distanceMixing}), can be
  found in Appendix \ref{s:appMetBlur}.

\autoref{f:rvvar} shows the comparison of a simulated
  \mtindices~population, with and without \rv~variations at the
  $\sigma_{\rv} = 0.25$~level admitted by the \SWEEPS~dataset. For
  each relevant WFC3 filter, the scale factors $A_X/\EBmV$~were
  estimated by linear interpolation in \rv~using information shown in
  \autoref{tab:extinction}. The simulated magnitudes were thus
  perturbed into ``observed'' magnitudes using different \rv~values
  for each star, but the \mtindices~were computed using the $\alpha,
  \beta$~values appropriate for \rv=2.5. This then mimics the use of a
  single \rv~value to compute \mtindices~for a population that in
  reality shows \rv~variations.

Comparing the synthetic \mtindices~distributions with and without
  \rv~variations (\autoref{f:rvvar}), it seems unlikely that
  \rv~variations at the level admitted by the \SWEEPS~color magnitude
  diagram can contribute a strong effect on GMM fitting or sample
  selection in \mtindices; the impact of \rv~variations is simply too
  small. We therefore proceed under the assumption that indeed $\rv
  \approx 2.5$~for all objects in the \SWEEPS~field of view.

\section{Testing the behavior of the \basti~stellar evolutionary models}
\label{s:testBasti}

Because stars in the \SWEEPS~field likely span a very wide
  \feh~range, including possibly objects outside the ranges traced by
  the \basti~evolutionary models, we test the behavior of the
  \basti~synthetic population framework when objects with very low or
  very high metallicities are simulated.

We find that \basti~v5.0.1 appears to be imposing an internal
  truncation on the simulated populations, probably on \feh~or on an
  internal variable that correlates with metallicity (for clarity, we
  refer to internal limits as \feh~limits throughout this
  section). This in turn leads to a discrepancy between the requested
  and simulated population, and between the reported \feh~values in
  the simulated output and the resulting population. Since \basti~is
  used very widely in studies of resolved stellar populations (with
  over 600 refereed citations), we report here our investigation into
  this truncation.\footnote{The analysis and figures in
      Appendix \ref{s:testBasti} can be reproduced using the notebook {\tt
        2017-09-08\_quicklookBaSTi\_truncation.ipynb} in the
      repository at
      \url{https://github.com/willclarkson/bastiTest}. This repository
      includes the full set of simulations and input parameters, as
      well as relevant methods used to generate the figures in this
      section.}

A variety of synthetic populations were simulated using \basti's
  ``user-specified SFH'' option. This allows the user to build a
  population from a series of bursts of star formation, with the mean
  and standard deviation \feh~specified for each population, as well
  as the number of years elapsed since the burst took place. In
  addition to the components that might make up the scene in the
  \SWEEPS~field of view (e.g. Appendix \ref{ss:appMetBlur:popComponents}),
  we simulated a number of ``test-pattern'' populations, with
  components regularly (or nearly-regularly) spaced in \feh.

The behavior of the color-magnitude diagram in the
  \SWEEPS~filters is then examined for consistency with the specified
  \feh~distribution and also the \feh~values reported in the simulated
  population. For regions in the CMD~approximately near the selection
  region used in this communication, the absolute magnitude difference
  $\Delta M$~is computed from a fitted median sequence (in much the
  same manner as is done for the observed population), and the
  distribution of $\Delta M$~examined for hard edges that are not
  present in the requested \feh~distribution.

The \basti~documentation was used to estimate median \feh~values near the limits of its metallicity range.\footnote{See
      \url{http://basti.oa-teramo.inaf.it/main_mod.php}~and links
      therein.} Specifically, we assumed the appropriate \feh~limits
  to be $(-2.27 \le \feh \le +0.40)$ for scaled-to-Solar models,
  and $(-2.62 \le \feh \le +0.05)$~for $\alpha$-enhanced
  models. Results for a representative set of test-cases are reported below, which suggest the following effects:

\begin{itemize}
  \item{Any bursts of star formation with specified median \feh~outside internal limits, are clipped to these limits before generation of the stellar population (Appendix \ref{ss:testBasti:median});}
    \item{If the specified \feh~distribution leads to individual objects with \feh~outside the limits, the absolute magnitudes of these objects are truncated internally, but the reported \feh~values appear to be unaffected, leading to a discrepancy between reported and applied \feh~values (Appendix \ref{ss:testBasti:spread});}
      \item{The truncation behavior appears more complex than a simple clipping or substitution; discrepant objects can appear quite deep into the main body of the selected population, and the effective \feh~limits might differ from those suggested by the documentation (Appendix \ref{ss:testBasti:metalRich}).}
\end{itemize}

\subsection{\basti~selection applied to median populations}
\label{ss:testBasti:median}

\begin{figure*}
    \centerline{\hbox{
    \includegraphics[height=12cm]{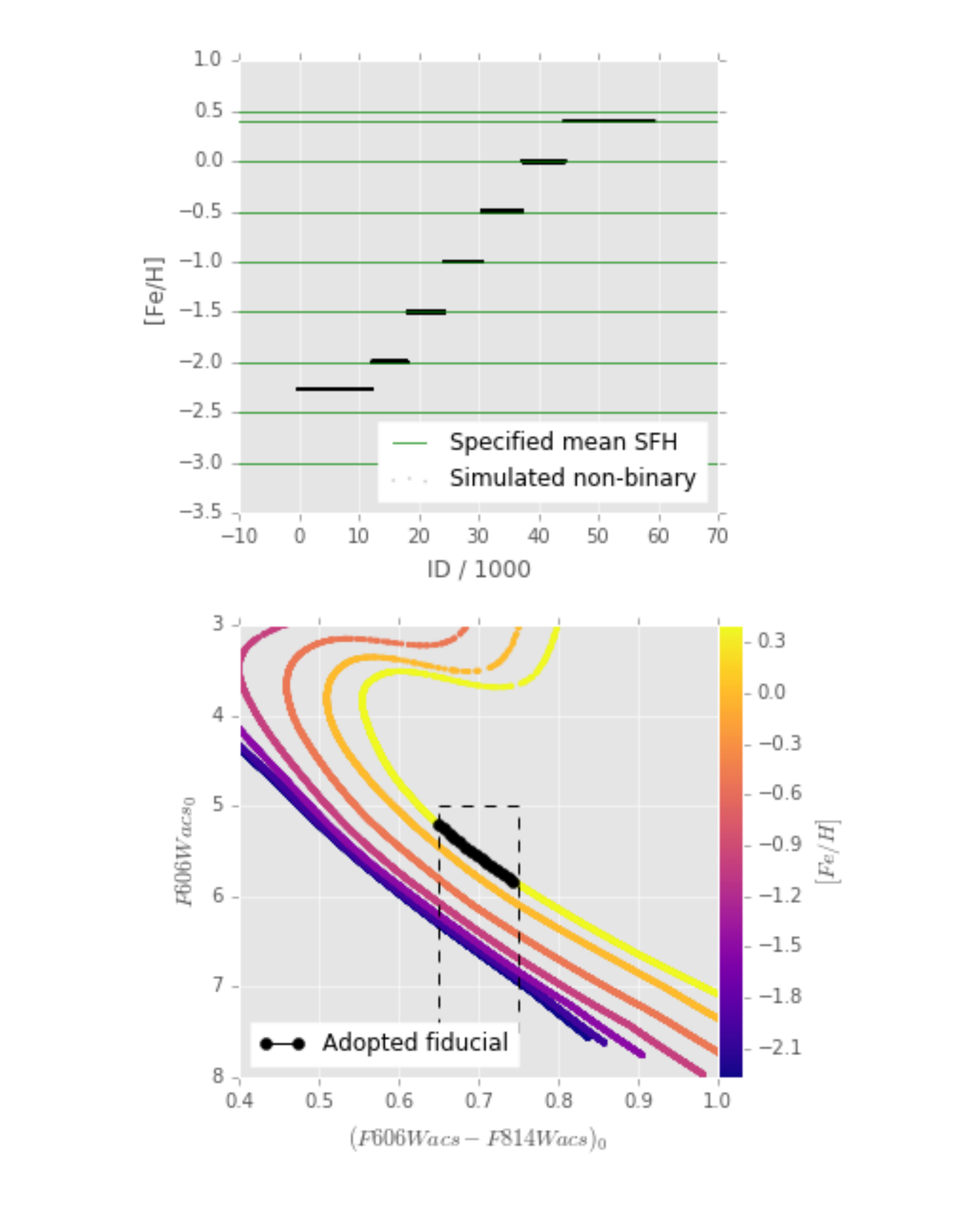}
    \includegraphics[height=12cm]{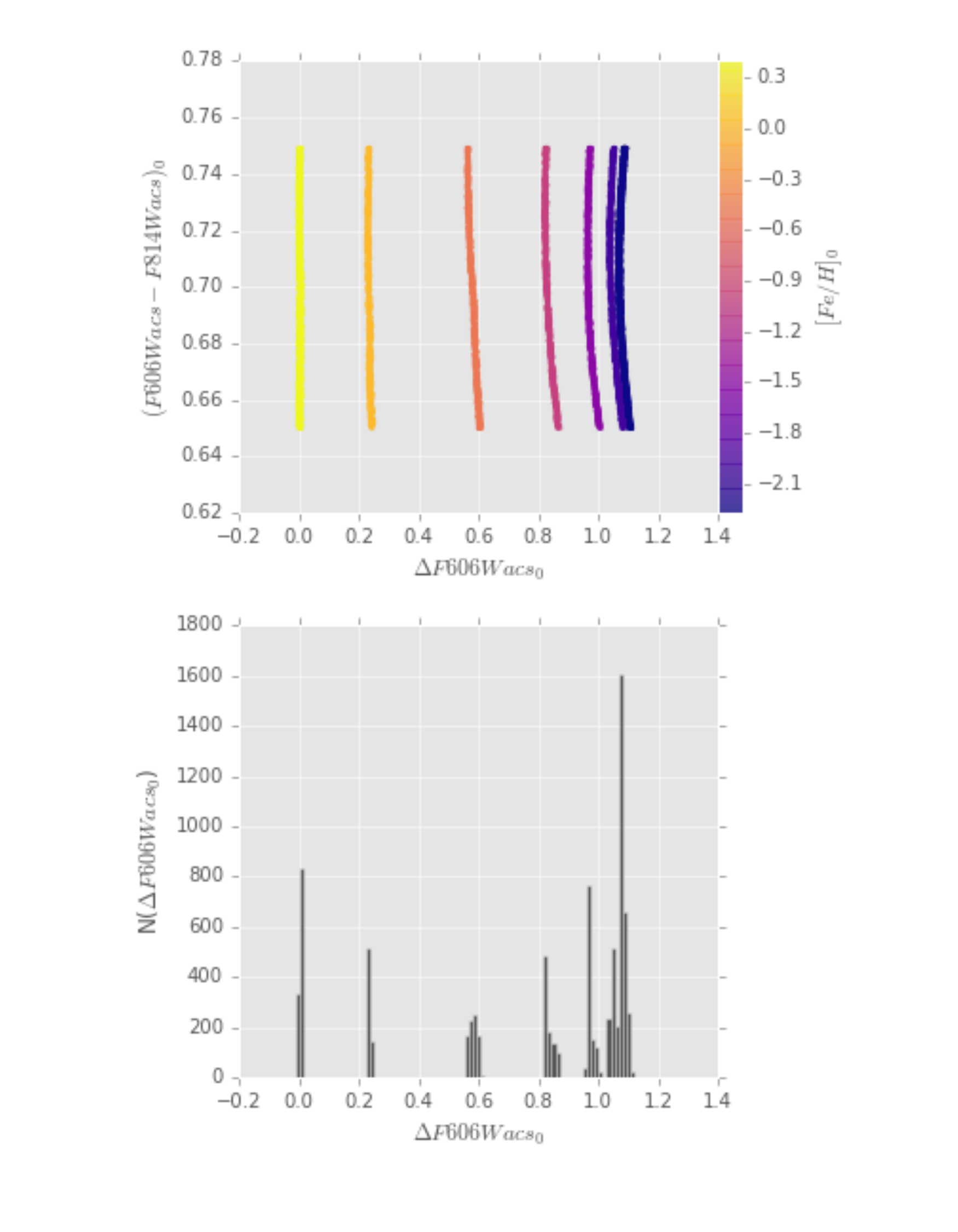}
    }}
\caption{Testing the relationship between specified and simulated metallicities when bursts of star formation with a very wide range of median \feh~values is requested with the \basti~interface. {\it Left top:} specified and simulated \feh~median values. The green horizontal lines show the median \feh~values for the bursts of star formation, with the gray points indicating \feh~values reported in the output simulation. In this example, specified bursts are ordered from bottom to top and left to right in the simulated objects. {\it Left bottom:} absolute magnitude CMD in the \SWEEPS~filters of the resulting population, color coded by reported \feh. The black dots and line refer to the fitted fiducial in the selection region and a polynomial fit to the fiducial, respectively. {\it Right top:} absolute magnitude offsets $\Delta M$~from the adopted fiducial, ordered by \SWEEPS~color, with symbols color-coded by \feh~reported in the simulated population. {\it Right bottom:} histogram of $\Delta M$. Here the specified median \feh~values were $\left\{-3.0, -2.5, -2.0, -1.5, -1.0, -0.5, +0.0, +0,4, +0.5\right\}$, all with specified spread $\sigma_{\feh} = 0.0001$~dex. Populations with $\feh \lesssim -2.3$~or $\feh \gtrsim +0.4$~seem to have been wrapped by \basti~to the metallicity limits. See Appendix \ref{ss:testBasti:median}.}
\label{f:testBasti:median}
\end{figure*}

To investigate whether \basti~is applying the truncation to the
  median population in a requested sample, test-populations were
  simulated for bursts of star formation of equal magnitude but with
  very narrow \feh~distributions. \autoref{f:testBasti:median} shows an
  example for a scaled-to-Solar set of isochrones, with $\feh =
  \left\{-3.0, -2.5, -2.0, -1.5, -1.0, -0.5, +0.0, +0,4, +0.5\right\}$, all with spread $\sigma_{\feh} = 0.0001$~dex~to isolate selection effects applied to the mean populations in each case. The two most metal-poor and the single most metal-rich populations are found to be forced away from their specified values, probably to some internal limit. Reading off the figure, the most metal-poor populations seem to be brought up $\feh \approx -2.3$~with the most metal-rich brought down to $\feh \approx +0.40$. These values are entirely consistent with the \feh~limits suggested by the \basti~documentation referenced earlier.

This suggests that \basti~enforces \feh~limits on the median populations requested in a simulation.

\subsection{\basti~truncation near the \feh~limits}
\label{ss:testBasti:spread}

\begin{figure*}
  \centerline{
    \includegraphics[height=12cm]{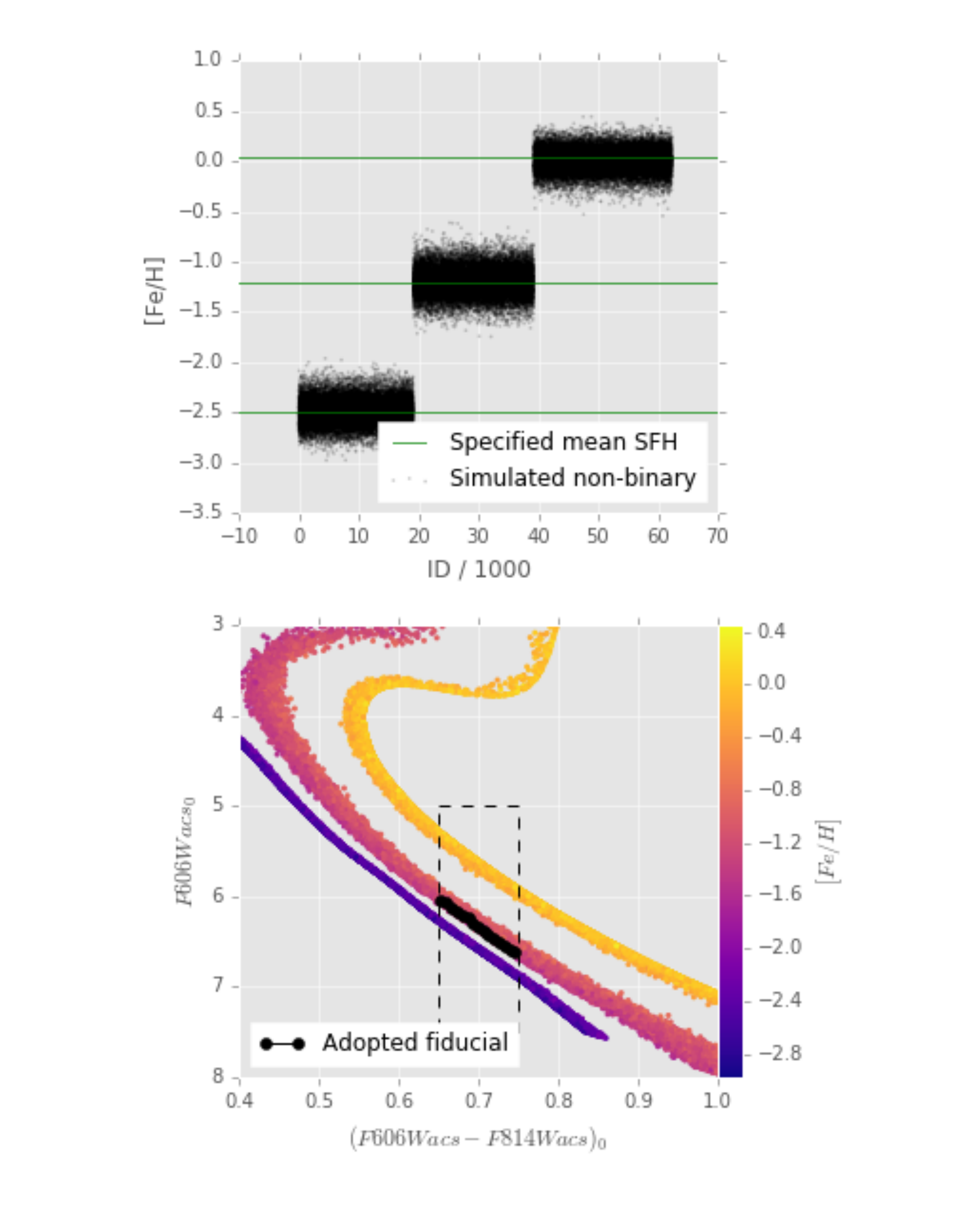}
    \includegraphics[height=12cm]{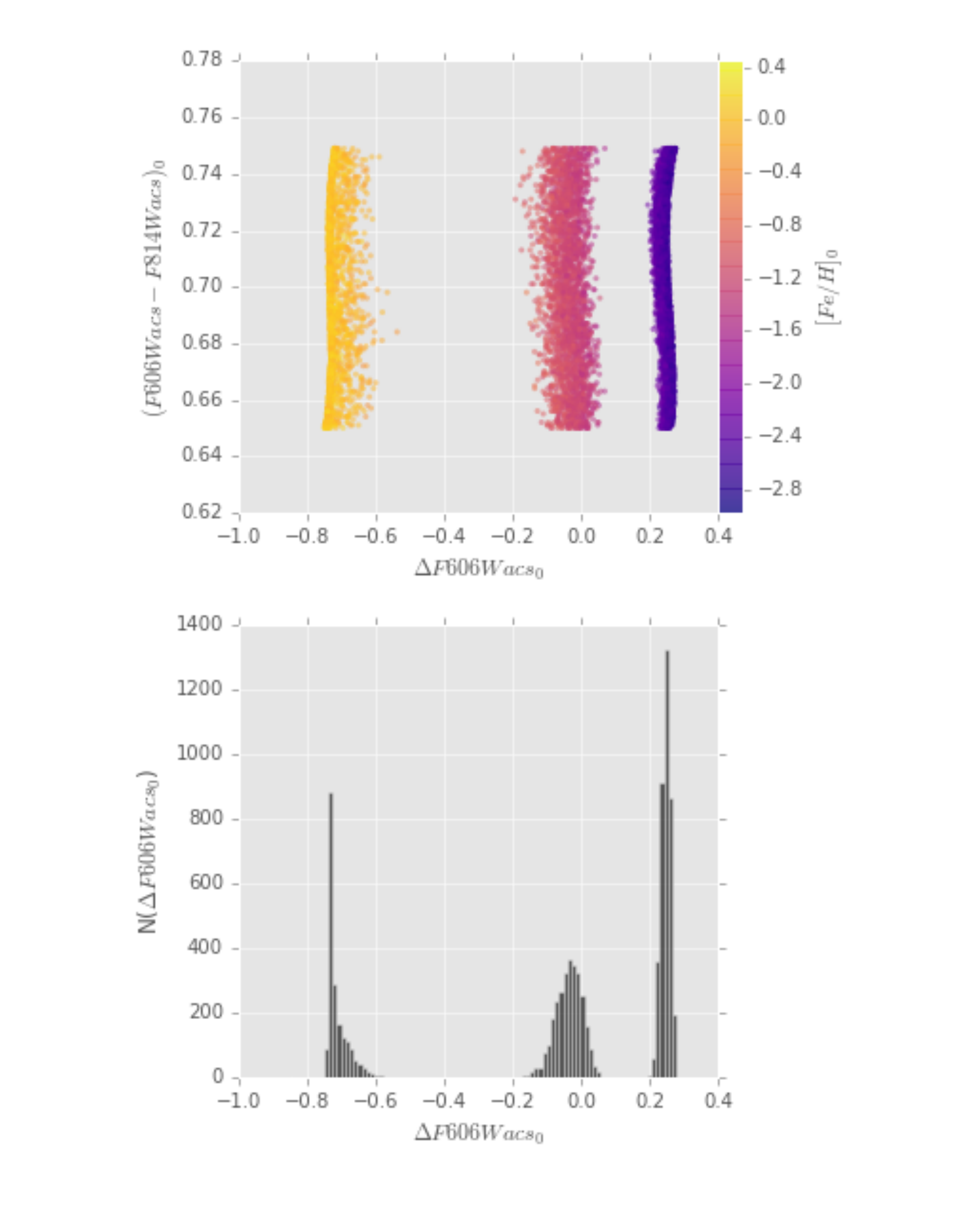}
}
\caption{Testing the behavior of \basti~simulations for populations with \feh~close to the internal boundaries. {\it Left column:} panels and symbols as \autoref{f:testBasti:median}, with specified bursts ordered bottom-top and their simulated populations ordered left-right. Here a three-component $\alpha$-enhanced population is simulated, with $\feh = \left\{ -2.5, -1.2, +0.05 \right\}$, all with specified spread $\sigma_{\feh} = 0.1$~dex. The metal-poor and metal-rich populations show sharp cut-offs in both the CMD and the $\Delta M$~distribution, which are not present in either the central population (well away from the \feh~limits), nor are the cutoffs present in the reported \feh~distributions of the metal-poor and metal-rich populations. (Curvature in the metal-poor hard-edge is likely due to differences in the shape of the median-population for $\feh = -2.5$~and that for $\feh = -1.2$.) The simulated magnitudes of the resulting populations show hard edges at the metal-rich and metal-poor ends, suggesting truncation in the delivered populations. Curiously, however, there is no such truncation in the corresponding {\it reported} \feh~values. This suggests that a truncation is being applied {\it after} the assignment of \feh~values to simulated objects. See Appendix \ref{ss:testBasti:spread}.}
\label{f:testBasti:3comp}
\end{figure*}

To investigate whether \basti~applies a truncation to \feh~values
  that are carried outside internal \feh~limits due to the specified
  population spread, test-populations were simulated including a
  single population well away from the limits, and one component each
  just inside the two limits. Components were specified with $\feh =
  \left\{ -2.5, -1.2, +0.05 \right\}$, all with specified spread
  $\sigma_{\feh} = 0.1$~dex, to ensure that the two components near the
  \feh~limits each include substantial numbers of objects outside
  these limits, while the middle population has very few such
  objects.

\autoref{f:testBasti:3comp} shows the resulting
  simulation. Curiously, although the \feh~values reported in the
  simulated populations show no truncation, the CMD and the simulated
  absolute magnitudes quite clearly do show truncation, with a hard
  edge at both the upper and lower \feh~extrema.

We therefore find that \basti~does not truncate \feh~values
  at the stage of assignment to simulated objects, and these
  non-truncated \feh~values are carried through to the output
  simulated population. However, a truncation {\it is} applied at some
  stage before the absolute magnitudes are included in the simulated
  population. This results both in a hard edge to the distribution of
  simulated absolute magnitudes, and also a discrepancy between the
  reported \feh~values and the absolute magnitudes, in the simulation
  output.

\subsection{\basti~truncation near the metal-rich limit}
\label{ss:testBasti:metalRich}

\begin{figure*}
\centerline{
  \includegraphics[height=12cm]{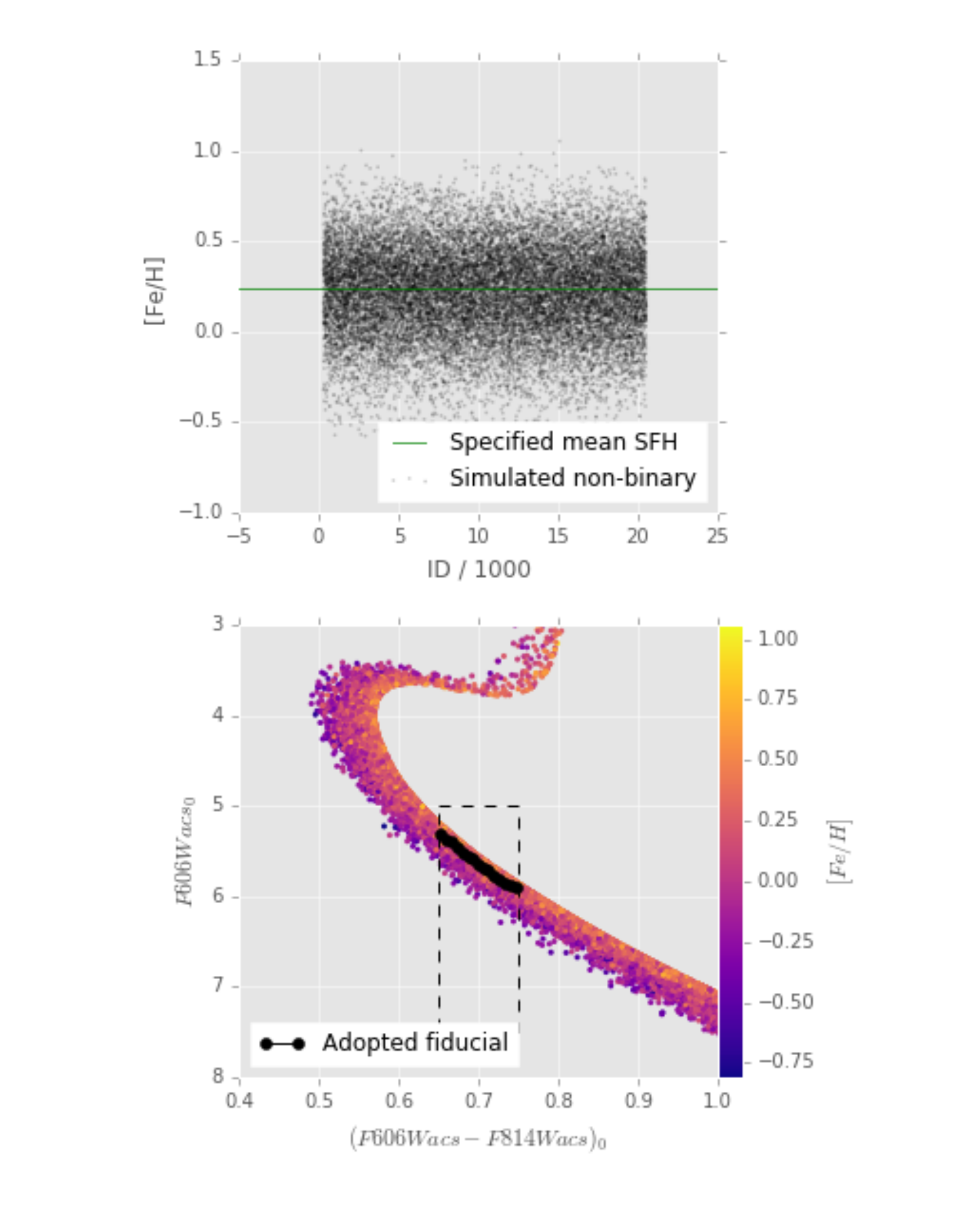}
  \includegraphics[height=12cm]{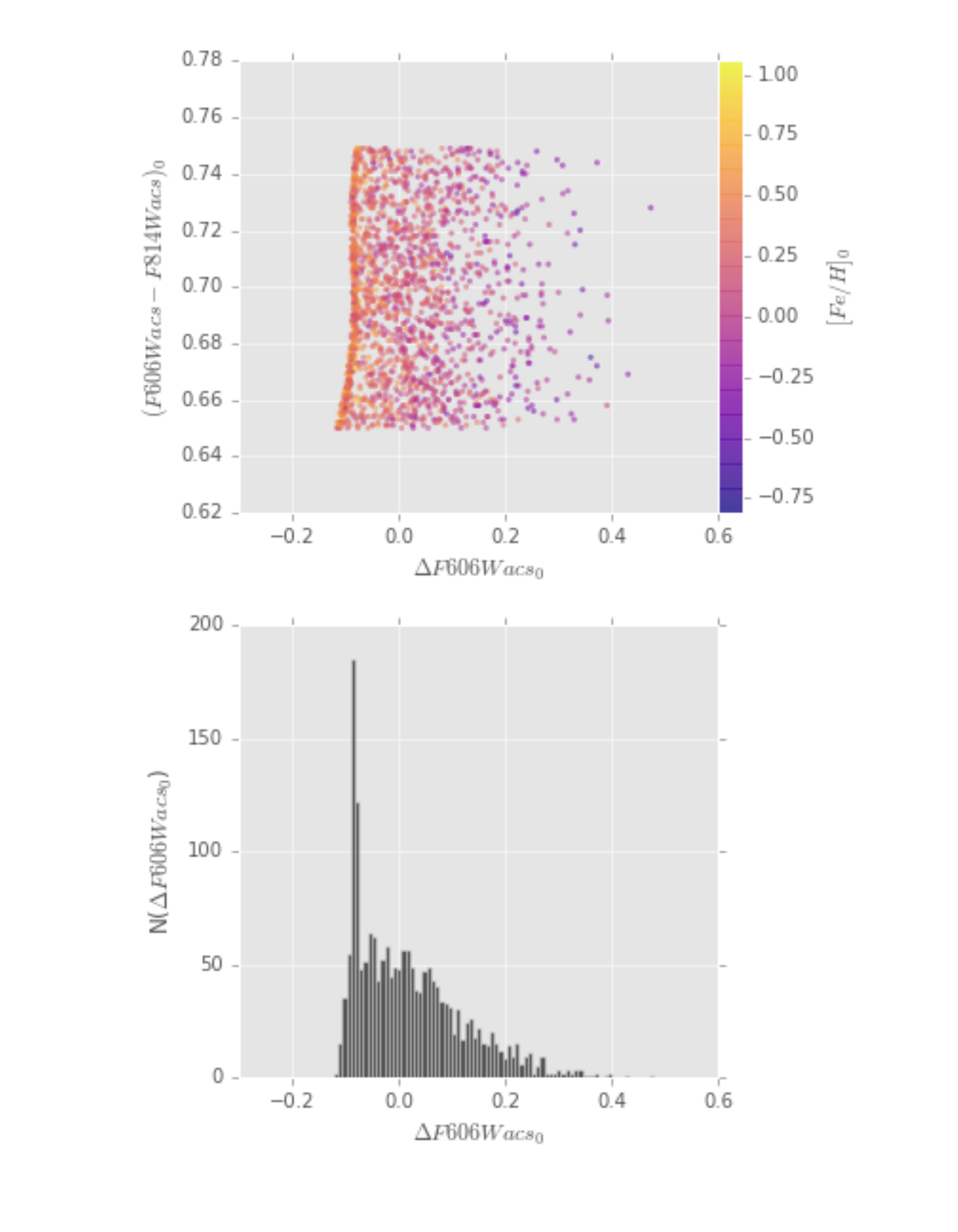}
}
\caption{Charting detailed behavior of \basti~truncation near the metal-rich limit. Panels and symbols are as in \autoref{f:testBasti:3comp}; here a single scaled-to-Solar component is simulated with $\feh=+0.24$~and scatter $\sigma_{\feh} = 0.19.$ A strong pile-up is observed at the bright end of the $\Delta M$~distribution (curvature in this component is likely due to systematics in the determination of the fiducial ridgeline, which was determined from the simulated CMD, as would be the case for observed populations, rather than specified using an isochrone). Again, while a strong truncation is observed in the simulated absolute magnitudes, no such hard edge is present in the reported \feh~values. See Appendix \ref{ss:testBasti:metalRich} and Figure \ref{f:appMetBlur:scatters}.}
\label{f:testBasti:metalRich}
\end{figure*}

To chart the behavior of the truncation near the \feh~limits in
  more detail, we simulated a single test population near the
  metal-rich limit. \autoref{f:testBasti:metalRich}~shows the result
  for a scaled-to-Solar component with $\feh=+0.24$~and scatter
  $\sigma_{\feh} = 0.19$. In this case, the truncation appears to be
  quite dramatic, with a narrow, highly over-represented component in
  the $\Delta M$~distribution.

However, the behaviour of the simulator near an \feh~limit is not
  as straightforward as a simple substitution of the \feh~limit for
  all objects beyond it. \autoref{f:appMetBlur:scatters} shows a
  simulated metal-rich population partitioned by \feh, which allows us
  to distinguish objects that were assigned \feh~values above the
  metal-rich limit (and thus would be assumed to be
  truncated). Objects with outlier \feh~values do {\it not}~only
  appear at the location where absolute magnitudes pile up; a
  substantial fraction show magnitudes deeper into the main population (see the
  bottom-left panel of \autoref{f:appMetBlur:scatters}).

That the pile-up implying truncation is also observed at the
  metal-rich edge of the population with simulated metallicities {\it
    within} the limits according to the \basti~documentation, suggests
  that the {\it effective} metallicity limits may differ from those
  documented; see the middle-left panel of
  \autoref{f:appMetBlur:scatters}. (We have not yet dissected this
  simulated population by binarity, which might offer another avenue
  for objects to wander into truncation territory.)

We therefore find that the internal truncation applied by
  \basti~is not limited to a simple pegging of values to an internal
  boundary. The behavior probably necessitates some sort of selection
  on $\Delta M$~to produce a cleaner unaffected sample. We adopt one
  such approach in Appendix \ref{ss:appMetBlur:varTruncated}.

\section{Cross-contamination in the \tindex, \mindex~diagram}
\label{s:appContam}

We consider here the mixing of the \MR~and \MP~samples (and thus
rotation curves) due to cross-contamination in the \tindex,
\mindex~space from which the two samples were drawn (\autoref{ss:mtMixingDiscussion}). 

While the formal membership probability threshold $\wik \ge \probThresh$~was chosen to be somewhat conservative, some amount of sample
contamination in \tindex, \mindex~is highly likely. Since the
(\mtindices) each represent flux ratios constructed from photometry in
three filters, it is likely that objects best characterized at one
side of the abundance range for the bulge, might be classified to an
object in the other due to photometric uncertainty. In principle, a
nearly-flat rotation curve for one sample could be polluted by samples
from another sample with a large-amplitude rotation curve, and vice
versa, sufficiently to weaken the trends in the high-amplitude sample
while imprinting a signal on the other that is not in fact present.

A rigorous exploration of the cross-contamination in
(\mtindices)~requires a somewhat involved set of computations. For
example, flat priors in observed flux (for each the five filters used
in \BTS) are unlikely to translate into flat priors in (\mtindices)
space, as suggested graphically by the degeneracy exhibited by very
metal-poor populations in the (\mtindices) diagram
\citep[e.g.][]{brown09}. To properly account for cross-contamination
likely requires simulations of the underlying metallicity and
temperature distributions (for which a range of shape parameters for
the distributions would also need exploration), then translating them
forward into the probability density function in (\mtindices)
including full accounting for the shape of the measurement uncertainty
distributions and covariances in each of the filters. We consider this
beyond the scope of the present work.

Instead, we have performed a simpler quantitative estimate of the
degree of cross-contamination in (\mtindices) space. We assume that
the four-component Gaussian Mixture Model (GMM) is indeed a reasonable
characterization of the {\it observed} distribution of
(\mtindices)~values, and also that the measurement uncertainties in
this space can be described as two-dimensional Gaussians for each
object. Samples in (\mtindices) are simulated by drawing from the
best-fit 4-component GMM and perturbing each object by an uncertainty
covariance matrix (\autoref{eq:mtCovar}) drawn randomly without
replacement from the observed population. Then a four-component GMM is
fit to each sample, and objects classified to belong to a model
component using the $\wik \ge \probThresh$~threshold that was used on the
observed dataset (an object cannot satisfy this condition for more
than one model component by construction). Finally, the model
component classification for each object is compared to the model
component from which it was originally drawn, to measure the
contamination for each component (i.e. the fraction of objects
classified with component $K$~but drawn from $k \ne K$).

\begin{figure*}
\centerline{
  \includegraphics[width=8cm]{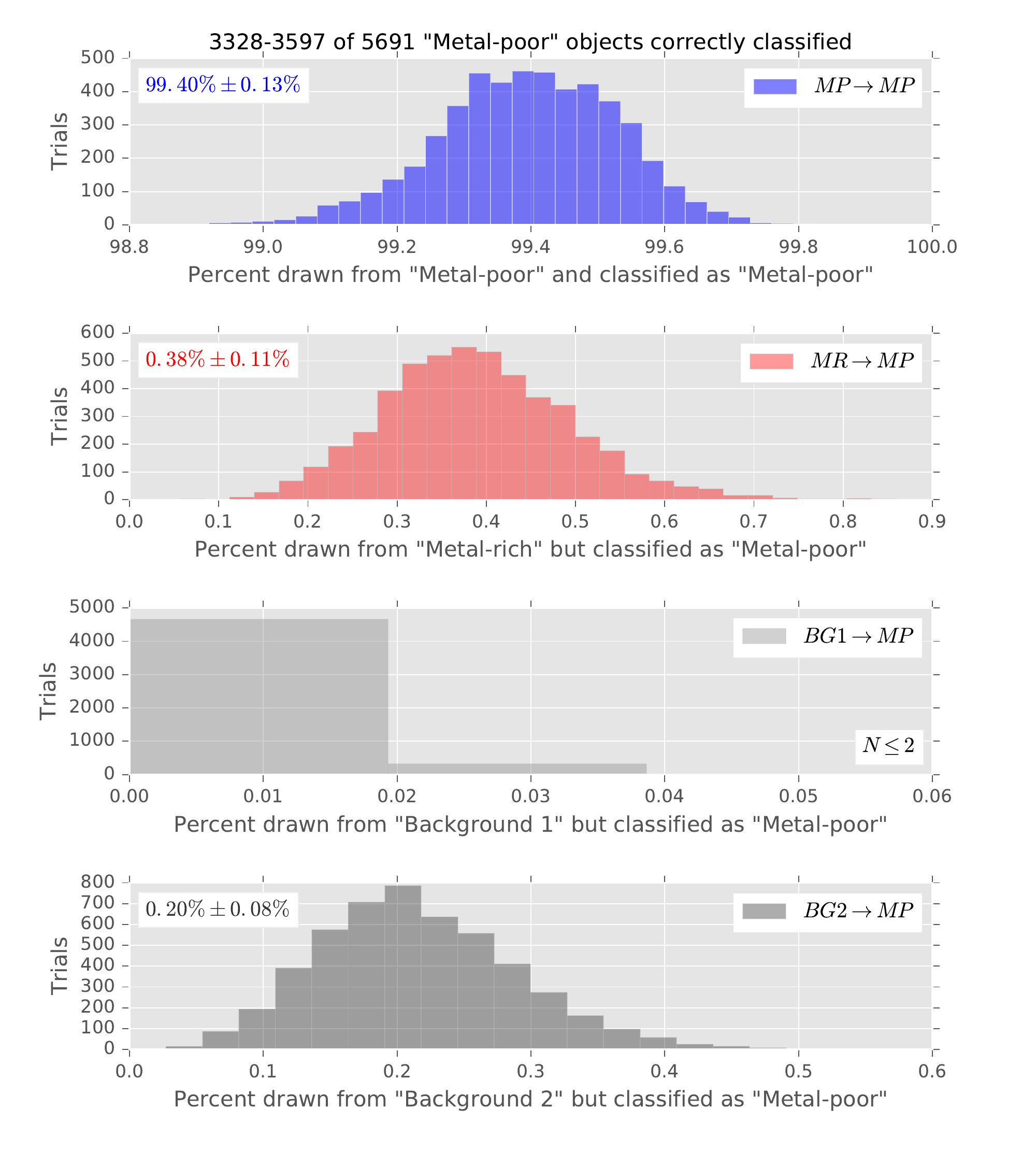}
  \includegraphics[width=8cm]{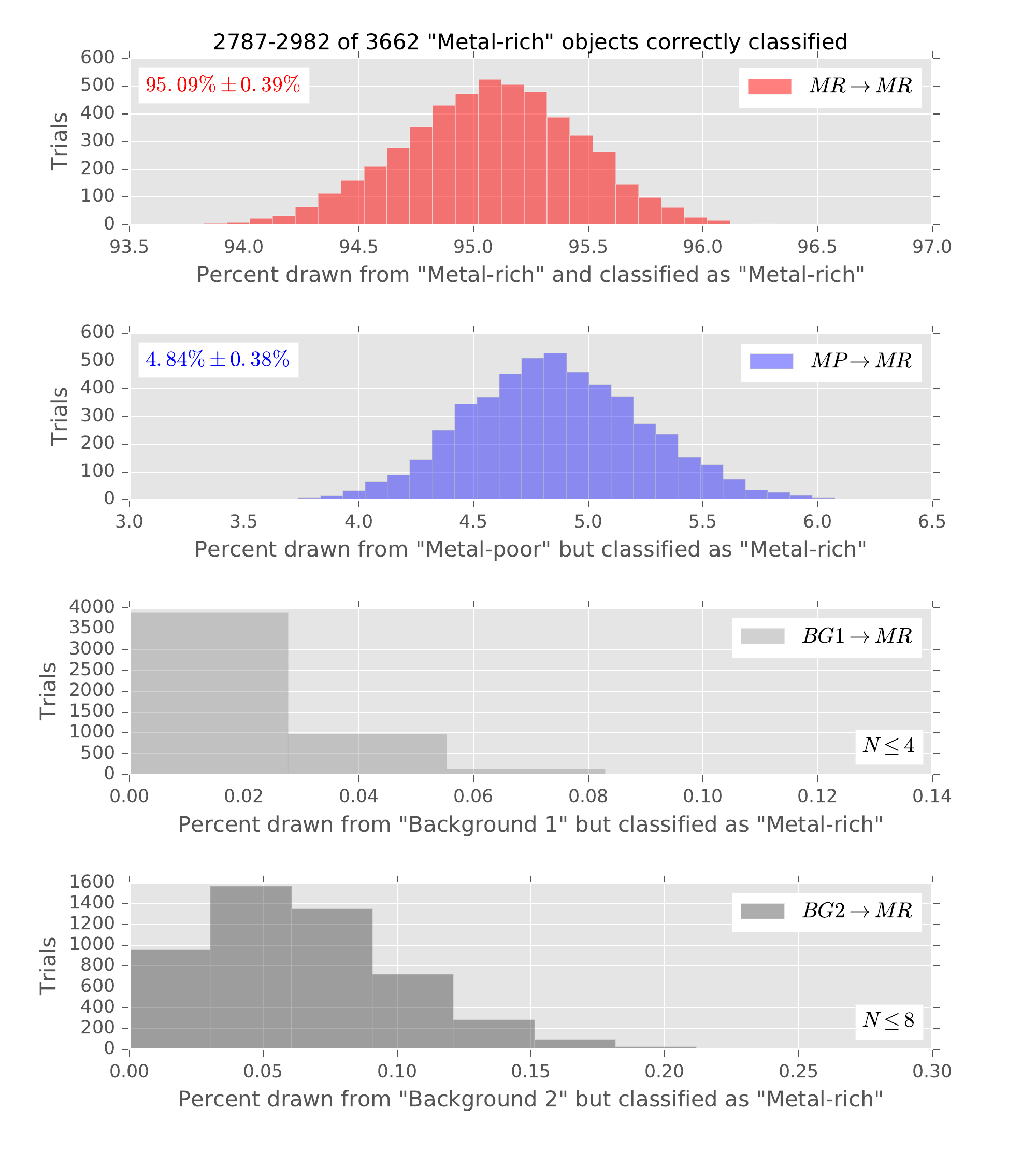}
}
\caption{Simple Monte Carlo test for cross-contamination of the
  \MR~and \MP~samples in (\mtindices) space. Objects are simulated
  from the best-fit 4-component GMM in (\mtindices) space (whose
  parameters are given in \autoref{tab:gmm}), perturbed by measurement
  uncertainty, and re-characterized using another 4-component GMM and
  classified by membership probability ($\wik \ge \probThresh$)~in the
  same way as the observed data (\autoref{ss:mtClassify}). The model
  component assigned to each object in the characterization is then
  compared to the component from which the object was drawn. Each
  panel shows the distribution of trials (out of 5,000 total) in which
  a given percentage of objects were classified with the indicated
  mixture component. The {\it left column}~shows the distribution of
  origin components for objects classified as \MP~(top left, blue),
  the {\it right column} shows the origin components for objects
  classified as \MR~(top right, red). In each column the top panel
  shows the distributions of objects classified correctly, the others
  show the distributions of objects classified with a different
  (indicated) component. Only a handful of objects from the
  ``Background-1'' component ($k=2$~in \autoref{tab:gmm}) are
  mis-identified with either the \MR~or the \MP~sample in any of the
  trials. See Appendix \ref{s:appContam}.}
\label{f:mtContam}
\end{figure*}

\autoref{f:mtContam} shows the results of 5,000 simulation
sets. Generally, the \MP~component is relatively uncontaminated by any
other population; the total contamination from these simulations is
$\le 1\%$~in all the trials, with the strongest contamination
contributed by the \MR~component (at $\sim 0.1\% - 0.8\%)$. The
\MR~component is more strongly contaminated. Roughly $5\%$~of this
sample is contaminated by the \MP~component, which is the dominant
contaminant (the two background components together providing less
than 0.5\% in all trials).

These ranges almost certainly underestimate the true contamination
between samples in (\mtindices). The observed
(\mtindices)~distribution tends to be less centrally peaked than the
model samples (\autoref{f:mtMaps}), suggesting the model likely
generates samples whose classification by \mtindices~is artificially
less vulnerable to contamination than in reality. Furthermore, even if
the distribution in flux ratio due to measurement uncertainty is
Gaussian for a given filter, for uncertainties $\sigma(\Delta F/F_0)
\gtrsim 0.1$~the apparent magnitude uncertainty distribution will
deviate substantially from a Gaussian.

Full exploration of these effects is deferred to future work. For the
present, our limited simulation suggests that the two samples are
contaminated in (\mtindices) at the $\lesssim 5\%$~level, using the
$\wik \ge \probThresh$~threshold for classification.

\section{Cross-comparison between catalog versions}
\label{app:v1v2}

While this work was at an advanced stage, a second version of the
\BTS~catalog (hereafter ``\BTStwo'') was released to the \HST~archive,
based on a re-analysis of the first-epoch \BTS~data using improved
measurement methods.\footnote{See
  \url{https://archive.stsci.edu/prepds/wfc3bulge/}. Measurement
  details are available in the {\tt README} at the same location.}
The comparison of \BTStwo~to the first catalog version (hereafter
``\BTSone'') helps prepare the ground for the ongoing investigations
discussed in \autoref{s:conclusions}, and so we present the comparison
here.

Appendix \ref{aa:bts1v2} compares the apparent magnitudes of the two
catalogs while Appendix \ref{aa:bts2propm}~compares the \SWEEPS~proper
motions with the \BTStwo~proper motions. In Appendix \ref{aa:gaia} the
two catalogs are compared against the absolute reference frame
provided by the first {\it Gaia} data release (which contains
positions but not proper motions for these objects). Finally, in
Appendix \ref{aa:resultsV2} we present a preliminary re-determination
of the proper motion rotation curves using \BTStwo~data exclusively.

\subsection{Photometry comparision between \BTStwo~\& \BTSone}
\label{aa:bts1v2}

The \BTSone~photometric catalog was produced using {\tt daophotII} on
summed images in each filter before combining into the final catalog
(\autoref{ss:obsBTS} and references therein) while the \BTStwo~catalog
uses ``effective PSF'' methods (e.g. \citealt{ak06}), the details of
which vary depending on the brightness regime of the object
used. Objects in the brightness range of our proper motion sample
(\autoref{f:CMDsel}) were measured using the {\tt kstwo} code by
J. Anderson, which fits position and flux for each star across all
exposures simultaneously (see \citealt{bellini17a} for details). For
sufficiently bright and isolated objects, the source position and flux
were fit independently, while for fainter and/or less isolated stars
the flux was measured using forced photometry (with the star position
fixed). Because both the source brightness and degree of isolation
depend on the filter and camera used, a given star might be measured
using forced photometry in some filters but not others.

\begin{figure*}
\centerline{
  \includegraphics[width=3.5in]{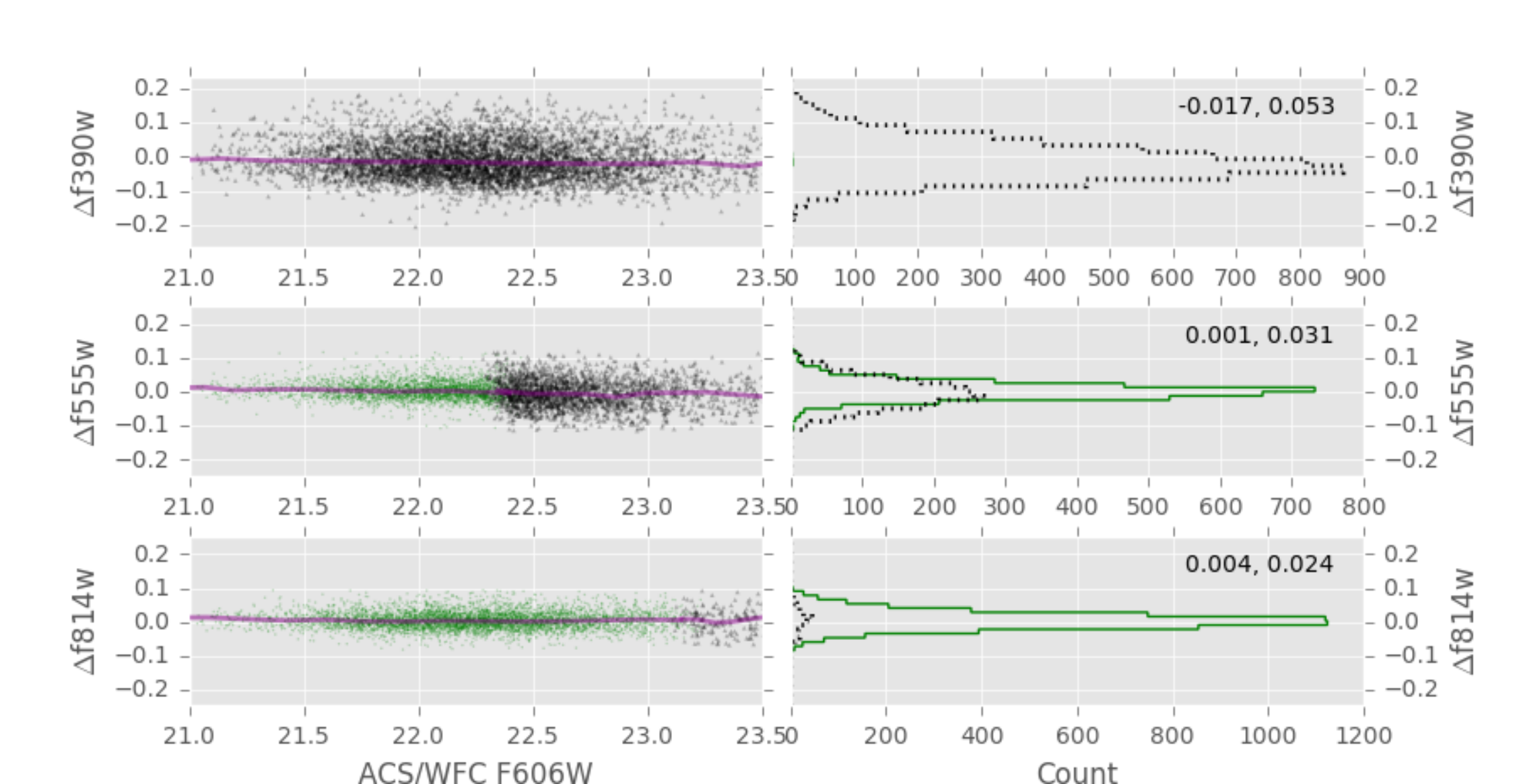}
  \includegraphics[width=3.5in]{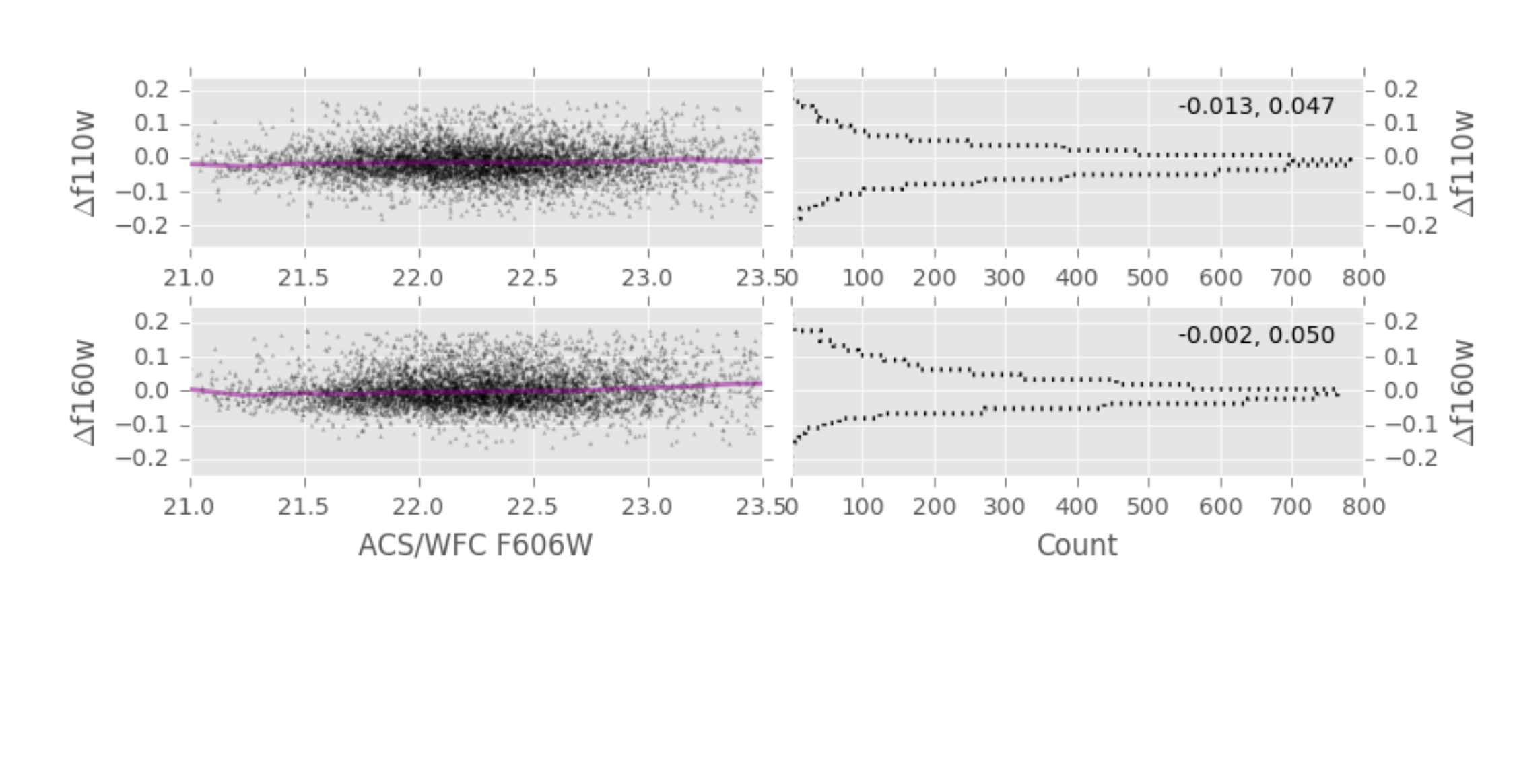}
}
\caption{Apparent magnitude comparison between the \BTS~v1 and v2~photometric catalogs for cross-matched objects within our sample of interest (see \autoref{f:CMDsel}). Each pair of panels presents the difference in apparent magnitude (in the Vegamag system), in the sense (v2-v1), showing the run against \SWEEPS~apparent magnitude (scatterplots) and the marginal distributions (histograms). Small green symbols and green solid lines represent objects whose flux and position was measured independently in \BTStwo, while larger gray symbols and the gray dashed lines represent objects whose flux was measured at fixed position. In the scatterplots, median trends are shown with a solid line. The inset annotations give the median and standard deviation of the magnitude differences. The left-hand set of panels present the comparison for WFC3/UVIS, the right-hand set represent WFC3/IR. For discussion, see Appendix \ref{aa:bts1v2}.}
\label{f:btsCompMag}
\end{figure*}

\autoref{f:btsCompMag} presents the comparison of apparent magnitude
between \BTStwo~and \BTSone, over the apparent magnitude range of
interest to the present work (stars were cross-matched between the two
catalogs using their equatorial co-ordinates). The random component of
the apparent magnitude difference is $\lesssim 0.06$~magnitudes in all
filters for most objects all objects (compare with the apparent
magnitude selection criteria in \autoref{tab:sampleSel}). Systematic
offsets between the datasets are less than $0.02$~magnitudes.

\subsection{Proper motion comparison between \BTStwo~and \SWEEPS}
\label{aa:bts2propm}

\begin{figure*}
  \begin{center}
    \includegraphics[width=18cm]{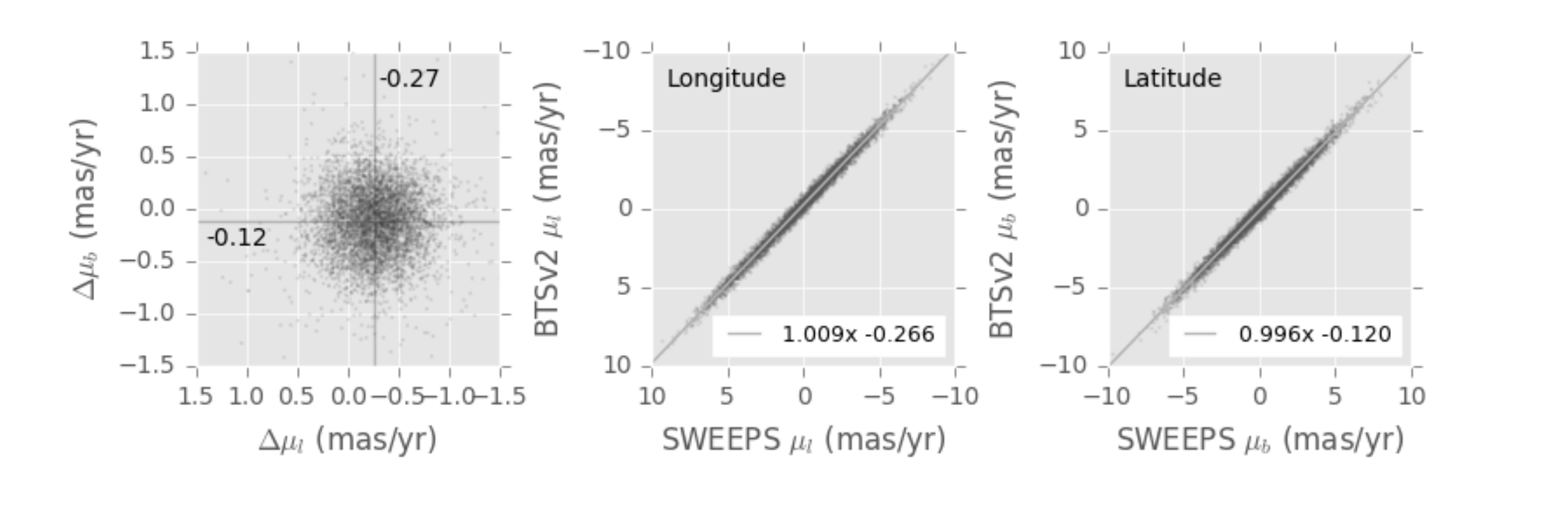}
    \end{center}
\caption{Comparison of the \BTStwo~and \SWEEPS~proper motions, for objects in our sample of interest (\autoref{f:CMDsel}). The left panel shows the proper motion differences (in the sense \BTStwo - \SWEEPS), with the mean proper motion offset between the two catalogs indicated by the cross-hair and annotation; the rms scatter in ($\Delta \mu_l, \Delta \mu_b$) is (0.28, 0.29)~\masperyear. The middle and right panels plot the proper motions  against each other, and present the best-fit straight line models, in Galactic longitude (middle) and latitude (right). See the discussion in Appendix \ref{aa:bts2propm}.}
\label{f:bts2propm}
\end{figure*}

\autoref{f:bts2propm} presents the star-by-star proper motion
comparison between the \BTStwo~and \SWEEPS~catalog for objects in our
apparent magnitude range of interest. Any difference in scale between
the proper motion determinations is below 1\%. A small offset
\muvecrefshift $\approx$~\muvecrefshiftval~between the two catalogs is
apparent, as expected if the proper motion zeropoint of the two
catalogs depends ultimately on the differing depth of the two
surveys. The proper motion differences show rms scatter $\approx
0.3$~\masperyear. Assuming the full \SWEEPS~proper motions carry
uncertainty $\epsilon_{\rm SWEEPS} \lesssim
0.12$~\masperyear~(Appendix \ref{a:unctyPM}), then the \BTStwo~proper
motions for this field would contribute approximately $\epsilon_{\rm
  BTS} \approx 0.27$~\masperyear. No trend in the proper motion
differences was found against apparent magnitude, proper motion or
position.

\subsection{Astrometric reference frame comparison with \GaiaOne}
\label{aa:gaia}

To check the orientation of the astrometric frames of the catalogs,
positions in the \SWEEPS~and \BTStwo~catalogs were matched to their
entries in the First {\it Gaia} Data Release, which should provide
absolute positions on the International Celestial Reference System
(ICRS) in the 2015.0 epoch, albeit possibly with residual distortions
in these crowded regions \citep{gaia16a, gaia16b}. Objects in the Gaia
apparent magnitude range $18.0 \le G \le 19.5$~were selected for
cross-matching, as a trade-off between quality of Gaia measurement and
the desire to avoid highly saturated objects in the \SWEEPS~and
\BTS~catalogs; this leaves a few thousand objects with which to probe
positional differences. To minimize random scatter in the comparison,
positions from the \SWEEPS~and \BTStwo~catalogs were advanced to their
positions in the 2015.0 epoch using the measured proper motions in
each catalog; the \GaiaOne~catalog does not contain proper motions for
these objects.

\begin{figure*}
  \centerline{\hbox{
      \includegraphics[width=3in]{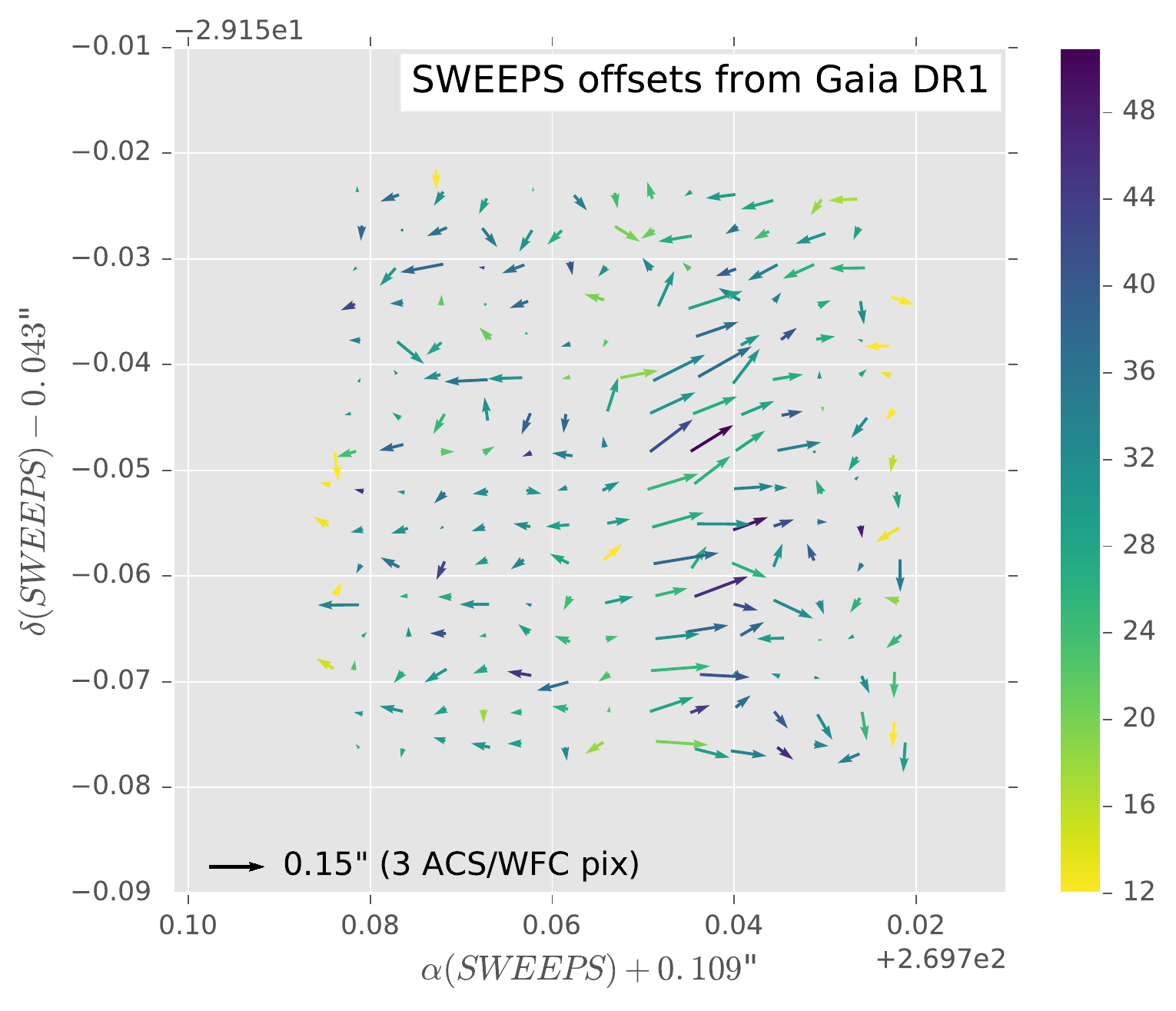}
      \includegraphics[width=3in]{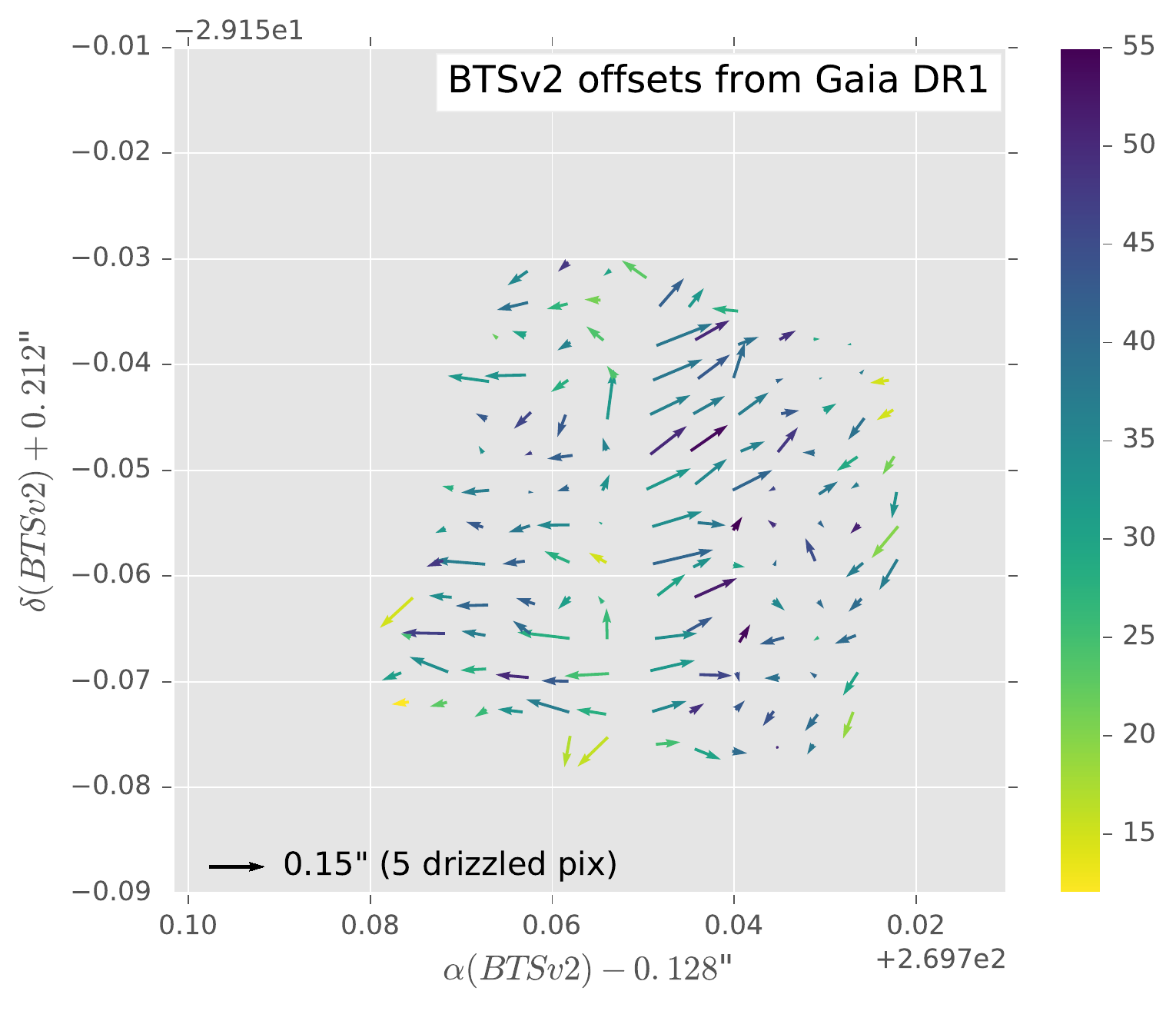}
}}
\caption{Checking the astrometric reference frame of the \SWEEPS~and
  \BTStwo~catalogs by comparing bright-star positions to the first
         {\it Gaia} Data Release \citep{gaia16a, gaia16b}. Panels show
         positional offsets between \GaiaOne~and the \SWEEPS~(left
         panel) and \BTStwo~(right panel) catalogs. Vectors show the
         median positional offsets of image regions with at least ten
         stars matched with \GaiaOne~(the shading indicates the number
         of cross-matched stars per region). Median offsets in
         arcseconds are indicated in the axis labels. While
         substantial residual structure is present in both sets of
         offsets, no frame rotation is detected at the $\lesssim
         0.05^{\circ}$~level, for either catalog. See discussion in
         Appendix \ref{aa:gaia}.}
\label{f:gaia}
\end{figure*}

\autoref{f:gaia} maps the astrometric offsets from the \GaiaOne~frame
for both the \SWEEPS~and \BTS~catalogs. While the two catalogs are
slightly offset with respect to the \GaiaOne~frame (by $\lesssim
0.3''$~in each co-ordinate), no rotational flow pattern is detected
that would suggest misalignment of either of the reference frames. The
scale of the positional residuals is surprisingly large, with flow
pattern common to both catalog-comparisons that reaches up to $\sim
0.15''$~in some regions. The largest residual structure is found at a
similar location in the comparisons to both the \SWEEPS~and
\BTStwo~catalogs, despite the two HST~catalogs being taken at
different camera orientations and with different field centers, so we
suspect the flow pattern is dominated by distortion in the
\GaiaOne~frame in these crowded regions. (Comparison of Subaru
measurements with \GaiaOne~positions near the core of the Sextans
dwarf galaxy also shows a flow pattern of offsets on a scale of $\sim
50''$~with a gap in \GaiaOne~coverage; Casetti-Dinescu 2018, private
communication.) We expect this flow pattern will vanish in comparisons
to future Gaia data releases that have included the more sophisticated
treatment for crowding outlined in \citet{pancino17}. Based on the
scale of the flow patterns near the corners of the difference-maps
($\lesssim 0.1''$), we conclude that the astrometric reference frames
of the \SWEEPS~and \BTStwo~catalogs are aligned with the ICRS~frame to
better than $\sim 0.05^{\circ}$.

\subsection{Preliminary Results using \BTStwo~only}
\label{aa:resultsV2}

Finally, to investigate whether our results qualitatively change when
moving from \BTSone~to \BTStwo, we have performed a preliminary
re-analysis using the \BTStwo~measurements only, following the
procedures of \autoref{s:analysis} as far as the production of the
proper motion rotation curves.

Not all the selection steps are common to both catalogs; for example,
\BTStwo~does not contain F606W~measurements (as were used when we
melded the \SWEEPS~and \BTStwo~catalogs in \autoref{s:analysis}),
which thus alters the initial selection of objects, and
\BTStwo~contains additional information that can be used to select
objects by measurement quality (details can be found in the
\BTStwo~README file). Additionally, \BTStwo~proper motion
uncertainties have not yet been characterized as fully as the
\ACSWFC~uncertainties in the \SWEEPS~field
(e.g. \citealt{calamida15}). Finally, the proper motion zeropoints of
the two catalogs differ, and the appropriate value of \dzer~to use for
\BTStwo~has not yet been established.

\begin{figure*}
  \centerline{\hbox{
  \includegraphics[height=7cm]{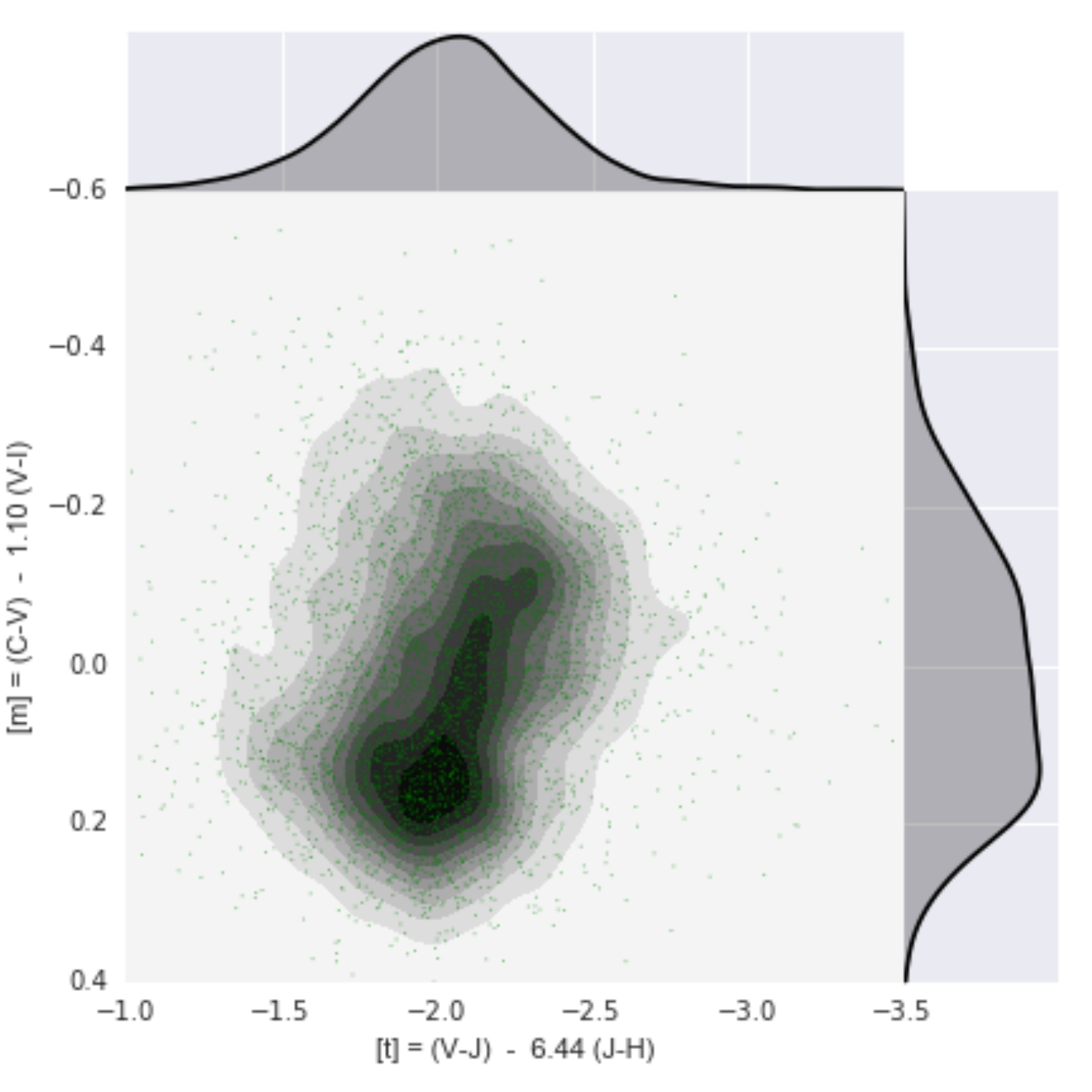}
  \includegraphics[height=7cm]{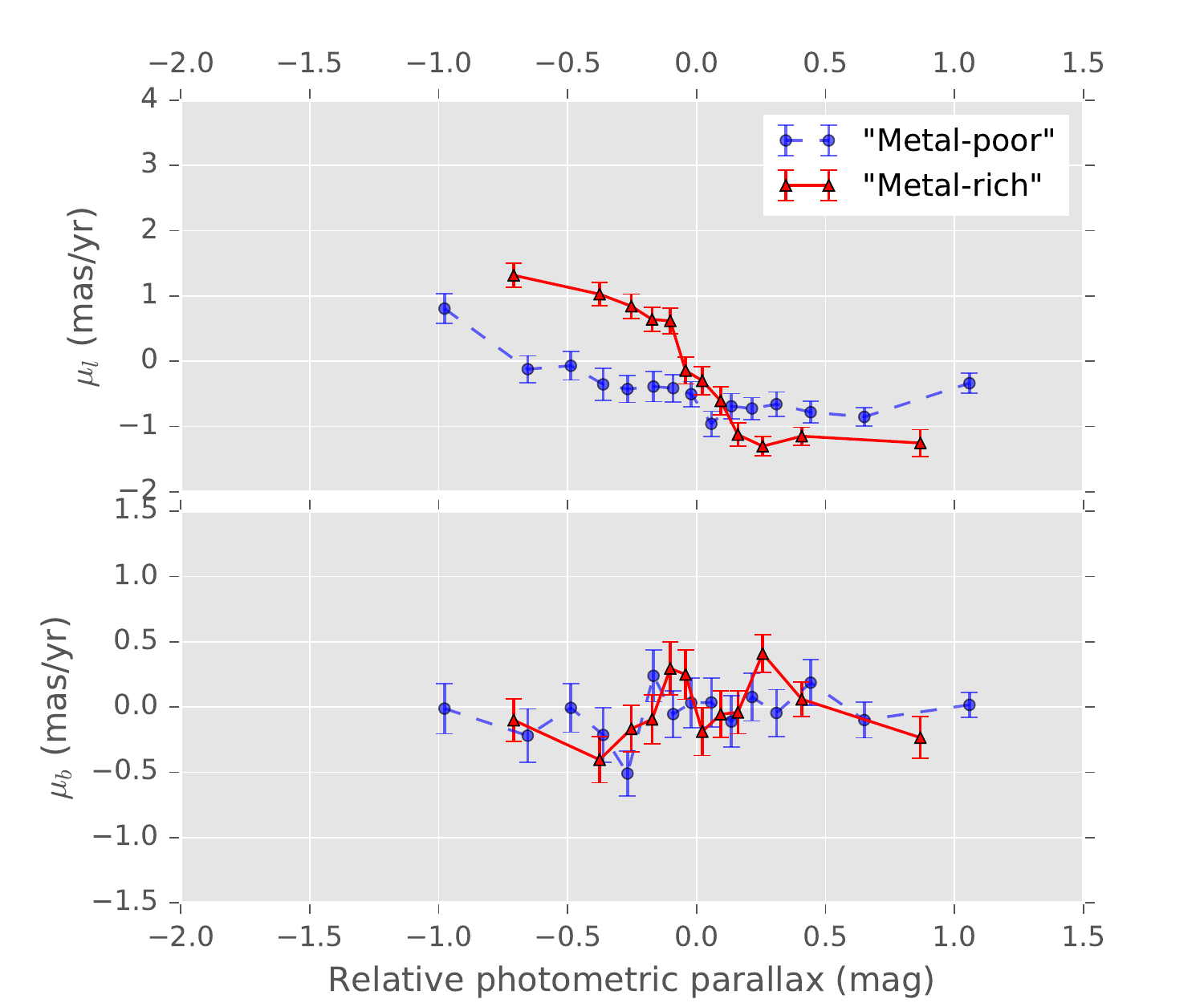}
  }}
\caption{Preliminary re-analysis using \BTStwo~data only, for both the \mtindices~and the proper motions. {\it Left
    panel:} (\mtindices)~distribution from the revised photometry
    (compare with \autoref{f:mtKDE}). {\it Right panel:} proper motion
    rotation curve for \MR~and \MP~samples using \BTStwo~proper
    motions (compare with \autoref{f:meanMotionMu}). See Appendix
    \ref{aa:resultsV2}~for discussion.}
\label{f:reanalysis}
\end{figure*}

\autoref{f:reanalysis} shows the results. While some of the fine
structure in the rotation curves appears to differ when compared to
the \BTSone-based analysis, the behavior we observe is not
substantially changed by use of the \BTStwo~catalog; the \MR~and
\MP~rotation curves still differ, with the \MR~curve showing a steeper
gradient. Full development of the chemically-dissected bulge rotation
curves will be reported in a future communication after the work has
been extended to all four \BTS~fields.

\section{Rotation curves and bin statistics in tabular form}
\label{s:rotCurveTables}

Full characterization of the variation of the proper motion ellipse
with photometric parallax for the two samples can be found in Tables
\ref{tab:trendsPho:Metal-rich} and \ref{tab:trendsPho:Metal-poor},
while Tables \ref{tab:trendsVel:Metal-rich} and
\ref{tab:trendsVel:Metal-poor}~present the same results after
converting from \rdmLong~and proper motion to distance and transverse
velocity. The bin statistics for the fine-grained binning scheme are presented in \autoref{tab:stats:Metal-rich}~and \autoref{tab:stats:Metal-poor}.

For ease of interpretation and to aid direct comparison with other work, we also tabulate the rotation curves and bin statistics for a binning scheme with constant-width bins (in photometric parallax) for each sample; see Tables \ref{tab:trendsPho:Metal-rich:unif}-\ref{tab:stats:Metal-poor:unif}.

% rotation trend, metal-rich, proper motions
\movetabledown=1.25in
\begin{longrotatetable}
\begin{deluxetable}{cccccccccccc}
\tabletypesize{\scriptsize}
\tablewidth{700pt}
\tablecaption{Observed rotation trend for the \MR~sample, with uncertainty ranges given as $1\sigma$~limits based on 1000 parametric bootstrap trials. {\bf $\langle \mu_l \rangle$, $\langle \mu_b \rangle$~represent the mean proper motions in each bin, with $a_{\mu}$, $b_{\mu}$~the major and minor axes of the velocity ellipse, respectively. $\phi_{lb}$~gives the position angle (in $l,b$) of the proper motion ellipse and $a/b$~its major:minor axis ratio. Finally, the quantities $\sigma^2_{ll}, \sigma^2_{bb}$~and $C_{lb}$~represent the three unique components of the proper motion dispersion.} See discussion in \autoref{ss:resultsCurves}. \label{tab:trendsPho:Metal-rich}}
\tablehead{\colhead{$\pi'$} & \colhead{Bin edges} & \colhead{$N$} & \colhead{$\langle \mu_l \rangle$} & \colhead{$\langle \mu_b \rangle$} & \colhead{$a_{\mu}$} & \colhead{$b_{\mu}$} & \colhead{$\phi_{lb}$} & \colhead{a/b} & \colhead{$\sigma^2_{ll}$} & \colhead{$\sigma^2_{bb}$} & \colhead{$C_{lb}$}\\ \colhead{$\mathrm{mag}$} & \colhead{$\mathrm{mag}$} & \colhead{ } & \colhead{(mas yr$^{-1}$)} & \colhead{(mas yr$^{-1}$)} & \colhead{(mas yr$^{-1}$)} & \colhead{(mas yr$^{-1}$)} & \colhead{($^{\circ}$)} & \colhead{$\mathrm{}$} & \colhead{(mas$^2$ yr$^{-2}$)} & \colhead{(mas$^2$ yr$^{-2}$)} & \colhead{(mas$^2$ yr$^{-2}$)}}
\startdata
-1.30 & -1.51,-1.18 & 20 & $ 2.96 \pm 0.591$ & $-0.01 \pm 0.287$ & $2.6 \pm 0.42$ & $1.3 \pm 0.21$ & $0.8 \pm 10.33$ & $2.00 \pm 0.112$ & $6.88 \pm 2.203$ & $1.71 \pm 0.535$ & $-0.07 \pm 0.760$ \\
-0.67 & -1.17,-0.48 & 198 & $ 1.58 \pm 0.193$ & $-0.18 \pm 0.160$ & $2.8 \pm 0.14$ & $2.2 \pm 0.11$ & $21.3 \pm 7.69$ & $1.30 \pm 0.053$ & $7.52 \pm 0.758$ & $5.10 \pm 0.512$ & $-1.11 \pm 0.434$ \\
-0.37 & -0.48,-0.31 & 199 & $ 1.14 \pm 0.188$ & $-0.06 \pm 0.161$ & $2.7 \pm 0.14$ & $2.2 \pm 0.11$ & $25.5 \pm 12.34$ & $1.20 \pm 0.058$ & $6.81 \pm 0.695$ & $5.45 \pm 0.543$ & $-0.84 \pm 0.445$ \\
-0.26 & -0.31,-0.22 & 199 & $ 1.13 \pm 0.195$ & $ 0.11 \pm 0.206$ & $3.0 \pm 0.15$ & $2.5 \pm 0.12$ & $52.7 \pm 15.78$ & $1.19 \pm 0.056$ & $8.03 \pm 0.752$ & $7.32 \pm 0.813$ & $-1.28 \pm 0.539$ \\
-0.18 & -0.22,-0.15 & 196 & $ 1.17 \pm 0.182$ & $-0.15 \pm 0.197$ & $2.8 \pm 0.14$ & $2.4 \pm 0.12$ & $59.1 \pm 25.80$ & $1.16 \pm 0.055$ & $7.45 \pm 0.664$ & $6.50 \pm 0.758$ & $-0.88 \pm 0.480$ \\
-0.12 & -0.15,-0.10 & 199 & $ 1.14 \pm 0.203$ & $-0.08 \pm 0.182$ & $3.0 \pm 0.14$ & $2.4 \pm 0.12$ & $30.5 \pm 9.48$ & $1.25 \pm 0.055$ & $8.34 \pm 0.785$ & $6.73 \pm 0.653$ & $-1.45 \pm 0.544$ \\
-0.07 & -0.10,-0.04 & 199 & $ 0.80 \pm 0.228$ & $ 0.21 \pm 0.205$ & $3.2 \pm 0.16$ & $2.7 \pm 0.13$ & $34.7 \pm 13.36$ & $1.18 \pm 0.055$ & $9.48 \pm 0.977$ & $8.43 \pm 0.837$ & $-1.40 \pm 0.627$ \\
-0.01 & -0.04, 0.02 & 199 & $-0.00 \pm 0.222$ & $ 0.26 \pm 0.189$ & $3.3 \pm 0.16$ & $2.5 \pm 0.12$ & $32.3 \pm 8.48$ & $1.29 \pm 0.053$ & $9.43 \pm 0.977$ & $7.64 \pm 0.754$ & $-1.90 \pm 0.611$ \\
0.04 & 0.02, 0.06 & 197 & $ 0.00 \pm 0.208$ & $ 0.20 \pm 0.193$ & $3.1 \pm 0.16$ & $2.6 \pm 0.13$ & $36.4 \pm 15.91$ & $1.18 \pm 0.057$ & $8.37 \pm 0.879$ & $7.59 \pm 0.773$ & $-1.27 \pm 0.572$ \\
0.09 & 0.06, 0.12 & 199 & $-0.68 \pm 0.197$ & $ 0.46 \pm 0.203$ & $3.1 \pm 0.15$ & $2.5 \pm 0.12$ & $41.0 \pm 8.84$ & $1.27 \pm 0.054$ & $8.12 \pm 0.816$ & $7.61 \pm 0.766$ & $-1.84 \pm 0.568$ \\
0.15 & 0.12, 0.18 & 198 & $-0.93 \pm 0.177$ & $ 0.42 \pm 0.190$ & $2.9 \pm 0.14$ & $2.2 \pm 0.11$ & $50.1 \pm 8.03$ & $1.30 \pm 0.051$ & $6.87 \pm 0.627$ & $6.26 \pm 0.671$ & $-1.67 \pm 0.466$ \\
0.22 & 0.18, 0.26 & 198 & $-1.03 \pm 0.164$ & $ 0.29 \pm 0.165$ & $2.4 \pm 0.10$ & $2.3 \pm 0.10$ & $63.3 \pm 54.11$ & $1.03 \pm 0.045$ & $5.47 \pm 0.515$ & $5.28 \pm 0.550$ & $-0.12 \pm 0.373$ \\
0.33 & 0.26, 0.41 & 198 & $-1.10 \pm 0.152$ & $ 0.00 \pm 0.138$ & $2.2 \pm 0.11$ & $1.9 \pm 0.09$ & $31.8 \pm 21.84$ & $1.13 \pm 0.053$ & $4.50 \pm 0.470$ & $4.04 \pm 0.413$ & $-0.46 \pm 0.291$ \\
0.53 & 0.41, 1.25 & 134 & $-0.98 \pm 0.140$ & $ 0.11 \pm 0.141$ & $1.6 \pm 0.08$ & $1.6 \pm 0.08$ & $-65.1 \pm 55.84$ & $1.03 \pm 0.049$ & $2.63 \pm 0.299$ & $2.54 \pm 0.322$ & $0.05 \pm 0.214$ \\
\enddata
\end{deluxetable}

\end{longrotatetable}

% rotation trend, metal-rich, velocities
\begin{longrotatetable}
\begin{deluxetable}{cccccccccccc}
\tabletypesize{\scriptsize}
\tablewidth{700pt}
\tablecaption{As \autoref{tab:trendsPho:Metal-rich} but with photometric parallaxes and proper motions converted into distances and velocities using a reference distance of 7.76 kpc. See discussion in \autoref{ss:resultsCurves}. \label{tab:trendsVel:Metal-rich}}
\tablehead{\colhead{$d$} & \colhead{$d_{\rm{lo}}, d_{\rm{hi}}$} & \colhead{$N$} & \colhead{$\langle v_l \rangle$} & \colhead{$\langle v_b \rangle$} & \colhead{$a_v$} & \colhead{$b_v$} & \colhead{$\phi_{lb}$} & \colhead{a/b} & \colhead{$\sigma^2_{ll,v}$} & \colhead{$\sigma^2_{bb,v}$} & \colhead{$\sigma^2_{lb,v}$}\\ \colhead{$\mathrm{kpc}$} & \colhead{$\mathrm{kpc}$} & \colhead{ } & \colhead{(km s$^{-1}$)} & \colhead{(km s$^{-1}$)} & \colhead{(km s$^{-1}$)} & \colhead{(km s$^{-1}$)} & \colhead{($^{\circ}$)} & \colhead{$\mathrm{}$} & \colhead{(km$^2$ s$^{-2}$)} & \colhead{(km$^2$ s$^{-2}$)} & \colhead{(km$^2$ s$^{-2}$)}}
\startdata
4.26 & 3.87,4.51 & 20 & $59.73 \pm 11.939$ & $-0.28 \pm 5.786$ & $53 \pm 8.4$ & $26 \pm 4.2$ & $0.8 \pm 10.33$ & $2.00 \pm 0.112$ & $138.9 \pm 44.48$ & $34.6 \pm 10.80$ & $-1.5 \pm 15.34$ \\
5.69 & 4.53,6.21 & 198 & $42.65 \pm 5.198$ & $-4.94 \pm 4.326$ & $76 \pm 3.8$ & $58 \pm 2.9$ & $21.3 \pm 7.69$ & $1.30 \pm 0.053$ & $202.6 \pm 20.42$ & $137.6 \pm 13.81$ & $-29.8 \pm 11.70$ \\
6.55 & 6.22,6.72 & 199 & $35.56 \pm 5.850$ & $-1.84 \pm 5.003$ & $83 \pm 4.2$ & $70 \pm 3.5$ & $25.5 \pm 12.34$ & $1.20 \pm 0.058$ & $211.6 \pm 21.59$ & $169.2 \pm 16.88$ & $-26.1 \pm 13.82$ \\
6.90 & 6.72,7.02 & 199 & $36.87 \pm 6.390$ & $3.48 \pm 6.723$ & $98 \pm 4.9$ & $82 \pm 4.0$ & $52.7 \pm 15.78$ & $1.19 \pm 0.056$ & $262.6 \pm 24.59$ & $239.4 \pm 26.59$ & $-42.0 \pm 17.62$ \\
7.14 & 7.02,7.23 & 196 & $39.71 \pm 6.163$ & $-5.06 \pm 6.656$ & $96 \pm 4.6$ & $83 \pm 4.0$ & $59.1 \pm 25.80$ & $1.16 \pm 0.055$ & $252.0 \pm 22.47$ & $219.9 \pm 25.64$ & $-29.9 \pm 16.25$ \\
7.33 & 7.23,7.42 & 199 & $39.69 \pm 7.058$ & $-2.84 \pm 6.330$ & $105 \pm 5.0$ & $84 \pm 4.1$ & $30.5 \pm 9.48$ & $1.25 \pm 0.055$ & $289.8 \pm 27.29$ & $234.1 \pm 22.70$ & $-50.4 \pm 18.90$ \\
7.51 & 7.42,7.62 & 199 & $28.35 \pm 8.121$ & $7.48 \pm 7.290$ & $115 \pm 5.9$ & $97 \pm 4.5$ & $34.7 \pm 13.36$ & $1.18 \pm 0.055$ & $337.3 \pm 34.78$ & $299.9 \pm 29.78$ & $-49.8 \pm 22.31$ \\
7.71 & 7.62,7.82 & 199 & $-0.17 \pm 8.097$ & $9.63 \pm 6.920$ & $119 \pm 6.0$ & $93 \pm 4.5$ & $32.3 \pm 8.48$ & $1.29 \pm 0.053$ & $344.6 \pm 35.72$ & $279.1 \pm 27.54$ & $-69.3 \pm 22.32$ \\
7.90 & 7.82,7.99 & 197 & $0.07 \pm 7.775$ & $7.36 \pm 7.235$ & $114 \pm 5.9$ & $97 \pm 4.7$ & $36.4 \pm 15.91$ & $1.18 \pm 0.057$ & $313.5 \pm 32.91$ & $284.1 \pm 28.95$ & $-47.4 \pm 21.43$ \\
8.10 & 7.99,8.20 & 199 & $-26.04 \pm 7.561$ & $17.79 \pm 7.789$ & $120 \pm 5.9$ & $94 \pm 4.7$ & $41.0 \pm 8.84$ & $1.27 \pm 0.054$ & $311.7 \pm 31.33$ & $292.0 \pm 29.41$ & $-70.5 \pm 21.80$ \\
8.30 & 8.20,8.44 & 198 & $-36.52 \pm 6.951$ & $16.47 \pm 7.486$ & $113 \pm 5.5$ & $87 \pm 4.2$ & $50.1 \pm 8.03$ & $1.30 \pm 0.051$ & $270.3 \pm 24.67$ & $246.5 \pm 26.42$ & $-65.8 \pm 18.32$ \\
8.58 & 8.44,8.75 & 198 & $-41.91 \pm 6.672$ & $11.88 \pm 6.714$ & $96 \pm 4.1$ & $93 \pm 4.0$ & $63.3 \pm 54.11$ & $1.03 \pm 0.045$ & $222.4 \pm 20.95$ & $214.9 \pm 22.37$ & $-5.0 \pm 15.17$ \\
9.02 & 8.75,9.35 & 198 & $-47.09 \pm 6.515$ & $0.04 \pm 5.912$ & $93 \pm 4.5$ & $83 \pm 3.9$ & $31.8 \pm 21.84$ & $1.13 \pm 0.053$ & $192.2 \pm 20.10$ & $172.7 \pm 17.64$ & $-19.6 \pm 12.46$ \\
9.89 & 9.36,13.82 & 134 & $-45.84 \pm 6.561$ & $5.37 \pm 6.624$ & $76 \pm 3.8$ & $74 \pm 3.7$ & $-65.1 \pm 55.84$ & $1.03 \pm 0.049$ & $123.3 \pm 14.03$ & $119.1 \pm 15.10$ & $2.5 \pm 10.05$ \\
\enddata
\end{deluxetable}

\end{longrotatetable}

% rotation trend, metal-poor, proper motions
\begin{longrotatetable}
\begin{deluxetable}{cccccccccccc}
\tabletypesize{\scriptsize}
\tablewidth{700pt}
\tablecaption{Observed rotation trend for the \MP~sample, with uncertainty ranges given as $1\sigma$~limits based on 1000 parametric bootstrap trials. See discussion in \autoref{ss:resultsCurves}. \label{tab:trendsPho:Metal-poor}}
\tablehead{\colhead{$\pi'$} & \colhead{Bin edges} & \colhead{$N$} & \colhead{$\langle \mu_l \rangle$} & \colhead{$\langle \mu_b \rangle$} & \colhead{$a_{\mu}$} & \colhead{$b_{\mu}$} & \colhead{$\phi_{lb}$} & \colhead{a/b} & \colhead{$\sigma^2_{ll}$} & \colhead{$\sigma^2_{bb}$} & \colhead{$C_{lb}$}\\ \colhead{$\mathrm{mag}$} & \colhead{$\mathrm{mag}$} & \colhead{ } & \colhead{(mas yr$^{-1}$)} & \colhead{(mas yr$^{-1}$)} & \colhead{(mas yr$^{-1}$)} & \colhead{(mas yr$^{-1}$)} & \colhead{($^{\circ}$)} & \colhead{$\mathrm{}$} & \colhead{(mas$^2$ yr$^{-2}$)} & \colhead{(mas$^2$ yr$^{-2}$)} & \colhead{(mas$^2$ yr$^{-2}$)}}
\startdata
-1.65 & -1.83,-1.42 & 20 & $ 2.96 \pm 0.544$ & $ 0.69 \pm 0.641$ & $3.2 \pm 0.49$ & $2.0 \pm 0.30$ & $-55.2 \pm 20.77$ & $1.61 \pm 0.125$ & $8.41 \pm 1.895$ & $6.15 \pm 2.686$ & $3.03 \pm 1.676$ \\
-1.01 & -1.39,-0.79 & 199 & $ 0.57 \pm 0.238$ & $-0.04 \pm 0.210$ & $3.3 \pm 0.16$ & $2.8 \pm 0.13$ & $22.9 \pm 12.17$ & $1.20 \pm 0.054$ & $10.52 \pm 1.039$ & $8.15 \pm 0.834$ & $-1.22 \pm 0.649$ \\
-0.65 & -0.78,-0.56 & 198 & $ 0.61 \pm 0.214$ & $-0.34 \pm 0.211$ & $3.4 \pm 0.17$ & $2.6 \pm 0.13$ & $40.6 \pm 9.08$ & $1.29 \pm 0.053$ & $9.47 \pm 0.984$ & $8.76 \pm 0.868$ & $-2.27 \pm 0.659$ \\
-0.47 & -0.56,-0.39 & 198 & $ 0.38 \pm 0.223$ & $-0.05 \pm 0.223$ & $3.5 \pm 0.17$ & $2.5 \pm 0.12$ & $43.8 \pm 6.82$ & $1.37 \pm 0.052$ & $9.40 \pm 0.934$ & $9.16 \pm 0.892$ & $-2.81 \pm 0.686$ \\
-0.33 & -0.39,-0.29 & 199 & $ 0.30 \pm 0.224$ & $-0.26 \pm 0.208$ & $3.2 \pm 0.15$ & $2.8 \pm 0.13$ & $14.3 \pm 18.70$ & $1.14 \pm 0.053$ & $10.32 \pm 1.012$ & $8.16 \pm 0.809$ & $-0.59 \pm 0.645$ \\
-0.24 & -0.29,-0.20 & 198 & $ 0.14 \pm 0.223$ & $ 0.42 \pm 0.216$ & $3.2 \pm 0.15$ & $3.0 \pm 0.13$ & $29.9 \pm 33.12$ & $1.08 \pm 0.049$ & $9.75 \pm 1.014$ & $9.05 \pm 0.904$ & $-0.59 \pm 0.632$ \\
-0.16 & -0.20,-0.13 & 198 & $ 0.18 \pm 0.223$ & $ 0.14 \pm 0.217$ & $3.4 \pm 0.16$ & $2.9 \pm 0.14$ & $50.4 \pm 24.56$ & $1.15 \pm 0.054$ & $10.29 \pm 0.980$ & $9.75 \pm 1.063$ & $-1.40 \pm 0.721$ \\
-0.09 & -0.13,-0.04 & 196 & $-0.01 \pm 0.208$ & $ 0.10 \pm 0.212$ & $3.1 \pm 0.15$ & $2.8 \pm 0.13$ & $50.3 \pm 26.77$ & $1.13 \pm 0.054$ & $8.95 \pm 0.842$ & $8.55 \pm 0.878$ & $-1.06 \pm 0.622$ \\
-0.01 & -0.04, 0.02 & 199 & $-0.15 \pm 0.215$ & $ 0.48 \pm 0.188$ & $3.0 \pm 0.14$ & $2.6 \pm 0.13$ & $13.9 \pm 20.05$ & $1.14 \pm 0.055$ & $8.84 \pm 0.873$ & $6.97 \pm 0.709$ & $-0.49 \pm 0.552$ \\
0.06 & 0.03, 0.09 & 200 & $-0.35 \pm 0.210$ & $ 0.08 \pm 0.195$ & $3.2 \pm 0.15$ & $2.8 \pm 0.14$ & $22.9 \pm 17.39$ & $1.15 \pm 0.056$ & $9.75 \pm 0.962$ & $8.03 \pm 0.784$ & $-0.89 \pm 0.629$ \\
0.13 & 0.09, 0.17 & 197 & $-0.38 \pm 0.198$ & $ 0.02 \pm 0.194$ & $2.8 \pm 0.12$ & $2.8 \pm 0.11$ & $59.5 \pm 53.65$ & $1.02 \pm 0.044$ & $7.84 \pm 0.746$ & $7.68 \pm 0.771$ & $-0.15 \pm 0.551$ \\
0.21 & 0.17, 0.26 & 197 & $-0.63 \pm 0.231$ & $ 0.28 \pm 0.196$ & $3.2 \pm 0.16$ & $2.7 \pm 0.13$ & $-10.7 \pm 15.33$ & $1.17 \pm 0.055$ & $10.20 \pm 1.026$ & $7.60 \pm 0.765$ & $0.51 \pm 0.639$ \\
0.32 & 0.26, 0.37 & 199 & $-0.58 \pm 0.161$ & $ 0.28 \pm 0.190$ & $2.7 \pm 0.13$ & $2.1 \pm 0.10$ & $63.7 \pm 12.58$ & $1.28 \pm 0.053$ & $6.72 \pm 0.487$ & $5.01 \pm 0.690$ & $-1.12 \pm 0.416$ \\
0.44 & 0.37, 0.54 & 197 & $-0.65 \pm 0.188$ & $ 0.21 \pm 0.173$ & $2.5 \pm 0.11$ & $2.5 \pm 0.10$ & $11.2 \pm 43.30$ & $1.03 \pm 0.043$ & $6.46 \pm 0.668$ & $6.08 \pm 0.592$ & $-0.08 \pm 0.466$ \\
0.69 & 0.54, 1.08 & 191 & $-0.38 \pm 0.129$ & $ 0.12 \pm 0.122$ & $1.8 \pm 0.09$ & $1.6 \pm 0.08$ & $18.0 \pm 22.81$ & $1.12 \pm 0.055$ & $3.24 \pm 0.324$ & $2.69 \pm 0.272$ & $-0.20 \pm 0.210$ \\
\enddata
\end{deluxetable}

\end{longrotatetable}

% rotation trend, metal-poor, velocities
\begin{longrotatetable}
\begin{deluxetable}{cccccccccccc}
\tabletypesize{\scriptsize}
\tablewidth{700pt}
\tablecaption{As \autoref{tab:trendsPho:Metal-poor} but with photometric parallaxes and proper motions converted into distances and velocities using a reference distance of 7.76 kpc. See discussion in \autoref{ss:resultsCurves}. \label{tab:trendsVel:Metal-poor}}
\tablehead{\colhead{$d$} & \colhead{$d_{\rm{lo}}, d_{\rm{hi}}$} & \colhead{$N$} & \colhead{$\langle v_l \rangle$} & \colhead{$\langle v_b \rangle$} & \colhead{$a_v$} & \colhead{$b_v$} & \colhead{$\phi_{lb}$} & \colhead{a/b} & \colhead{$\sigma^2_{ll,v}$} & \colhead{$\sigma^2_{bb,v}$} & \colhead{$\sigma^2_{lb,v}$}\\ \colhead{$\mathrm{kpc}$} & \colhead{$\mathrm{kpc}$} & \colhead{ } & \colhead{(km s$^{-1}$)} & \colhead{(km s$^{-1}$)} & \colhead{(km s$^{-1}$)} & \colhead{(km s$^{-1}$)} & \colhead{($^{\circ}$)} & \colhead{$\mathrm{}$} & \colhead{(km$^2$ s$^{-2}$)} & \colhead{(km$^2$ s$^{-2}$)} & \colhead{(km$^2$ s$^{-2}$)}}
\startdata
3.64 & 3.34,4.03 & 20 & $50.99 \pm 9.375$ & $11.92 \pm 11.049$ & $56 \pm 8.5$ & $35 \pm 5.2$ & $-55.2 \pm 20.77$ & $1.61 \pm 0.125$ & $144.9 \pm 32.67$ & $106.0 \pm 46.30$ & $52.2 \pm 28.89$ \\
4.88 & 4.10,5.40 & 199 & $13.11 \pm 5.512$ & $-0.96 \pm 4.846$ & $77 \pm 3.7$ & $64 \pm 3.1$ & $22.9 \pm 12.17$ & $1.20 \pm 0.054$ & $243.1 \pm 24.01$ & $188.4 \pm 19.29$ & $-28.2 \pm 15.00$ \\
5.74 & 5.41,5.99 & 198 & $16.61 \pm 5.827$ & $-9.19 \pm 5.748$ & $92 \pm 4.6$ & $71 \pm 3.6$ & $40.6 \pm 9.08$ & $1.29 \pm 0.053$ & $257.7 \pm 26.78$ & $238.4 \pm 23.61$ & $-61.9 \pm 17.93$ \\
6.24 & 6.00,6.48 & 198 & $11.35 \pm 6.605$ & $-1.46 \pm 6.591$ & $103 \pm 5.1$ & $75 \pm 3.7$ & $43.8 \pm 6.82$ & $1.37 \pm 0.052$ & $278.1 \pm 27.64$ & $271.1 \pm 26.38$ & $-83.0 \pm 20.30$ \\
6.66 & 6.48,6.80 & 199 & $9.57 \pm 7.064$ & $-8.07 \pm 6.555$ & $102 \pm 4.8$ & $89 \pm 4.2$ & $14.3 \pm 18.70$ & $1.14 \pm 0.053$ & $325.7 \pm 31.92$ & $257.5 \pm 25.52$ & $-18.6 \pm 20.34$ \\
6.94 & 6.80,7.07 & 198 & $4.68 \pm 7.332$ & $13.71 \pm 7.106$ & $104 \pm 4.8$ & $97 \pm 4.3$ & $29.9 \pm 33.12$ & $1.08 \pm 0.049$ & $320.4 \pm 33.33$ & $297.7 \pm 29.70$ & $-19.5 \pm 20.77$ \\
7.19 & 7.07,7.32 & 198 & $6.14 \pm 7.621$ & $4.89 \pm 7.400$ & $115 \pm 5.6$ & $100 \pm 4.8$ & $50.4 \pm 24.56$ & $1.15 \pm 0.054$ & $350.8 \pm 33.43$ & $332.6 \pm 36.26$ & $-47.9 \pm 24.59$ \\
7.45 & 7.32,7.60 & 196 & $-0.19 \pm 7.344$ & $3.42 \pm 7.488$ & $111 \pm 5.2$ & $98 \pm 4.5$ & $50.3 \pm 26.77$ & $1.13 \pm 0.054$ & $315.8 \pm 29.73$ & $301.7 \pm 31.01$ & $-37.6 \pm 21.94$ \\
7.72 & 7.61,7.85 & 199 & $-5.52 \pm 7.878$ & $17.53 \pm 6.867$ & $110 \pm 5.2$ & $96 \pm 4.6$ & $13.9 \pm 20.05$ & $1.14 \pm 0.055$ & $323.5 \pm 31.94$ & $255.3 \pm 25.95$ & $-18.0 \pm 20.20$ \\
7.99 & 7.85,8.10 & 200 & $-13.12 \pm 7.942$ & $2.91 \pm 7.377$ & $120 \pm 5.7$ & $105 \pm 5.1$ & $22.9 \pm 17.39$ & $1.15 \pm 0.056$ & $369.0 \pm 36.42$ & $303.8 \pm 29.67$ & $-33.6 \pm 23.83$ \\
8.23 & 8.10,8.38 & 197 & $-14.86 \pm 7.709$ & $0.73 \pm 7.574$ & $110 \pm 4.6$ & $107 \pm 4.4$ & $59.5 \pm 53.65$ & $1.02 \pm 0.044$ & $305.9 \pm 29.12$ & $299.5 \pm 30.08$ & $-5.8 \pm 21.49$ \\
8.54 & 8.38,8.74 & 197 & $-25.42 \pm 9.355$ & $11.16 \pm 7.915$ & $130 \pm 6.4$ & $111 \pm 5.4$ & $-10.7 \pm 15.33$ & $1.17 \pm 0.055$ & $412.6 \pm 41.53$ & $307.5 \pm 30.96$ & $20.6 \pm 25.85$ \\
8.98 & 8.74,9.20 & 199 & $-24.84 \pm 6.840$ & $11.82 \pm 8.070$ & $115 \pm 5.7$ & $90 \pm 4.4$ & $63.7 \pm 12.58$ & $1.28 \pm 0.053$ & $286.1 \pm 20.71$ & $213.4 \pm 29.36$ & $-47.7 \pm 17.69$ \\
9.49 & 9.20,9.96 & 197 & $-29.17 \pm 8.464$ & $9.30 \pm 7.800$ & $115 \pm 4.9$ & $111 \pm 4.7$ & $11.2 \pm 43.30$ & $1.03 \pm 0.043$ & $290.8 \pm 30.04$ & $273.7 \pm 26.66$ & $-3.5 \pm 20.99$ \\
10.67 & 9.96,12.79 & 191 & $-19.02 \pm 6.542$ & $6.14 \pm 6.176$ & $92 \pm 4.3$ & $82 \pm 3.9$ & $18.0 \pm 22.81$ & $1.12 \pm 0.055$ & $164.1 \pm 16.38$ & $136.3 \pm 13.76$ & $-10.1 \pm 10.64$\\
\enddata
\end{deluxetable}

\end{longrotatetable}

% bin statistics - metal-rich
\begin{deluxetable*}{cccccccccc}
\tabletypesize{\scriptsize}
\tablewidth{700pt}
\tablecaption{Bin statistics for the rotation curves of the \MR~sample. Wedge volumes $V$~and densities $\rho$~assume the reference sample lies at distance 7.76~kpc. $N (\mu_l)$~and $\rho (\mu_l)$~denote the counts and number densities of objects that would pass a kinematic cut of $\mu_l < -2.0$~mas yr$^{-1}$. The binning scheme is the same as \autoref{tab:trendsPho:Metal-rich}. The uncertainties quoted refer to $1\sigma$~ranges from 1000~parameteric bootstrap trials. See \autoref{ss:pmSelection}. \label{tab:stats:Metal-rich}}
\tablehead{\colhead{$\pi'$} & \colhead{$\pi'_{\rm{hi}}-\pi'_{\rm{lo}}$} & \colhead{$d$} & \colhead{$d_{\rm{hi}}-d_{\rm{lo}}$} & \colhead{$N$} & \colhead{$V$} & \colhead{$\rho$} & \colhead{$N(\mu_l)$} & \colhead{$f(\mu_l)$} & \colhead{$\rho (\mu_l)$}\\ \colhead{$\mathrm{mag}$} & \colhead{$\mathrm{mag}$} & \colhead{$\mathrm{kpc}$} & \colhead{$\mathrm{kpc}$} & \colhead{ } & \colhead{(pc$^3$)} & \colhead{(pc$^{-3}$)} & \colhead{ } & \colhead{ } & \colhead{(pc$^{-3}$)}}
\startdata
-1.30 & 0.330 & 4.26 & 0.636 & 20 & 1316.1 & 0.015 & $1 \pm 0.8$ & $0.05 \pm 0.040$ & $0.001 \pm 0.0006$ \\
-0.67 & 0.687 & 5.69 & 1.684 & 198 & 5556.9 & 0.036 & $21 \pm 4.2$ & $0.11 \pm 0.021$ & $0.004 \pm 0.0007$ \\
-0.37 & 0.170 & 6.55 & 0.507 & 199 & 2413.5 & 0.082 & $23 \pm 4.6$ & $0.12 \pm 0.023$ & $0.010 \pm 0.0019$ \\
-0.26 & 0.092 & 6.90 & 0.292 & 199 & 1558.6 & 0.128 & $27 \pm 4.8$ & $0.14 \pm 0.024$ & $0.017 \pm 0.0031$ \\
-0.18 & 0.065 & 7.14 & 0.213 & 196 & 1227.1 & 0.160 & $21 \pm 4.3$ & $0.11 \pm 0.022$ & $0.017 \pm 0.0035$ \\
-0.12 & 0.055 & 7.33 & 0.186 & 199 & 1138.2 & 0.175 & $31 \pm 4.7$ & $0.16 \pm 0.023$ & $0.027 \pm 0.0041$ \\
-0.07 & 0.057 & 7.51 & 0.198 & 199 & 1290.6 & 0.154 & $35 \pm 5.5$ & $0.18 \pm 0.028$ & $0.027 \pm 0.0042$ \\
-0.01 & 0.058 & 7.71 & 0.206 & 199 & 1387.8 & 0.143 & $53 \pm 6.3$ & $0.27 \pm 0.032$ & $0.038 \pm 0.0045$ \\
0.04 & 0.046 & 7.90 & 0.168 & 197 & 1202.2 & 0.164 & $47 \pm 6.0$ & $0.24 \pm 0.030$ & $0.039 \pm 0.0050$ \\
0.09 & 0.056 & 8.10 & 0.209 & 199 & 1556.8 & 0.128 & $72 \pm 6.5$ & $0.36 \pm 0.033$ & $0.046 \pm 0.0042$ \\
0.15 & 0.061 & 8.30 & 0.235 & 198 & 1843.1 & 0.107 & $65 \pm 6.7$ & $0.33 \pm 0.034$ & $0.035 \pm 0.0036$ \\
0.22 & 0.079 & 8.58 & 0.311 & 198 & 2626.0 & 0.075 & $71 \pm 6.6$ & $0.36 \pm 0.034$ & $0.027 \pm 0.0025$ \\
0.33 & 0.144 & 9.02 & 0.598 & 198 & 5651.9 & 0.035 & $76 \pm 6.7$ & $0.38 \pm 0.034$ & $0.013 \pm 0.0012$ \\
0.53 & 0.846 & 9.89 & 4.457 & 134 & 86926.5 & 0.002 & $36 \pm 5.2$ & $0.27 \pm 0.038$ & $0.000 \pm 0.0001$\\
\enddata
\end{deluxetable*}

% bin statistics - metal-poor
\begin{deluxetable*}{cccccccccc}
\tabletypesize{\scriptsize}
\tablewidth{700pt}
\tablecaption{As \autoref{tab:stats:Metal-rich} but for the \MP~sample. See \autoref{ss:pmSelection}. \label{tab:stats:Metal-poor}}
\tablehead{\colhead{$\pi'$} & \colhead{$\pi'_{\rm{hi}}-\pi'_{\rm{lo}}$} & \colhead{$d$} & \colhead{$d_{\rm{hi}}-d_{\rm{lo}}$} & \colhead{$N$} & \colhead{$V$} & \colhead{$\rho$} & \colhead{$N(\mu_l)$} & \colhead{$f(\mu_l)$} & \colhead{$\rho (\mu_l)$}\\ \colhead{$\mathrm{mag}$} & \colhead{$\mathrm{mag}$} & \colhead{$\mathrm{kpc}$} & \colhead{$\mathrm{kpc}$} & \colhead{ } & \colhead{(pc$^3$)} & \colhead{(pc$^{-3}$)} & \colhead{ } & \colhead{ } & \colhead{(pc$^{-3}$)}}
\startdata
-1.65 & 0.410 & 3.64 & 0.693 & 20 & 1196.6 & 0.017 & $0 \pm 0.7$ & $0.00 \pm 0.034$ & $0.000 \pm 0.0006$ \\
-1.01 & 0.600 & 4.88 & 1.306 & 199 & 3365.2 & 0.059 & $37 \pm 5.9$ & $0.19 \pm 0.030$ & $0.011 \pm 0.0018$ \\
-0.65 & 0.223 & 5.74 & 0.585 & 198 & 2166.4 & 0.091 & $43 \pm 5.6$ & $0.22 \pm 0.028$ & $0.020 \pm 0.0026$ \\
-0.47 & 0.168 & 6.24 & 0.484 & 198 & 2146.4 & 0.092 & $46 \pm 6.0$ & $0.23 \pm 0.030$ & $0.021 \pm 0.0028$ \\
-0.33 & 0.104 & 6.66 & 0.318 & 199 & 1594.7 & 0.125 & $54 \pm 6.0$ & $0.27 \pm 0.030$ & $0.034 \pm 0.0037$ \\
-0.24 & 0.085 & 6.94 & 0.272 & 198 & 1482.6 & 0.134 & $52 \pm 6.3$ & $0.26 \pm 0.032$ & $0.035 \pm 0.0043$ \\
-0.16 & 0.073 & 7.19 & 0.242 & 198 & 1422.1 & 0.139 & $49 \pm 6.1$ & $0.25 \pm 0.031$ & $0.034 \pm 0.0043$ \\
-0.09 & 0.083 & 7.45 & 0.287 & 196 & 1823.4 & 0.107 & $45 \pm 6.0$ & $0.23 \pm 0.031$ & $0.025 \pm 0.0033$ \\
-0.01 & 0.068 & 7.72 & 0.244 & 199 & 1654.0 & 0.120 & $59 \pm 6.6$ & $0.30 \pm 0.033$ & $0.036 \pm 0.0040$ \\
0.06 & 0.067 & 7.99 & 0.245 & 200 & 1768.9 & 0.113 & $63 \pm 6.3$ & $0.32 \pm 0.031$ & $0.036 \pm 0.0036$ \\
0.13 & 0.075 & 8.23 & 0.283 & 197 & 2189.7 & 0.090 & $56 \pm 6.4$ & $0.28 \pm 0.032$ & $0.026 \pm 0.0029$ \\
0.21 & 0.090 & 8.54 & 0.356 & 197 & 2980.7 & 0.066 & $72 \pm 6.8$ & $0.37 \pm 0.034$ & $0.024 \pm 0.0023$ \\
0.32 & 0.111 & 8.98 & 0.458 & 199 & 4176.0 & 0.048 & $57 \pm 6.2$ & $0.29 \pm 0.031$ & $0.014 \pm 0.0015$ \\
0.44 & 0.172 & 9.49 & 0.758 & 197 & 7880.7 & 0.025 & $59 \pm 6.6$ & $0.30 \pm 0.033$ & $0.007 \pm 0.0008$ \\
0.69 & 0.543 & 10.67 & 2.829 & 191 & 42078.2 & 0.005 & $29 \pm 5.1$ & $0.15 \pm 0.027$ & $0.001 \pm 0.0001$\\
\enddata
\end{deluxetable*}

% Tables with uniform binning scheme follow

% rotation trend, uniform, metal-rich, proper motions
\begin{longrotatetable}
\begin{deluxetable}{cccccccccccc}
\tabletypesize{\scriptsize}
\tablewidth{700pt}
\tablecaption{Observed rotation trend for the \MR~sample, using a constant-width binning scheme and with uncertainty ranges given as $1\sigma$~limits based on 1000 parametric bootstrap trials. See discussion in \autoref{ss:resultsCurves}. \label{tab:trendsPho:Metal-rich:unif}}
\tablehead{\colhead{$\pi'$} & \colhead{Bin edges} & \colhead{$N$} & \colhead{$\langle \mu_l \rangle$} & \colhead{$\langle \mu_b \rangle$} & \colhead{$a_{\mu}$} & \colhead{$b_{\mu}$} & \colhead{$\phi_{lb}$} & \colhead{a/b} & \colhead{$\sigma^2_{ll}$} & \colhead{$\sigma^2_{bb}$} & \colhead{$C_{lb}$}\\ \colhead{$\mathrm{mag}$} & \colhead{$\mathrm{mag}$} & \colhead{ } & \colhead{(mas yr$^{-1}$)} & \colhead{(mas yr$^{-1}$)} & \colhead{(mas yr$^{-1}$)} & \colhead{(mas yr$^{-1}$)} & \colhead{($^{\circ}$)} & \colhead{$\mathrm{}$} & \colhead{(mas$^2$ yr$^{-2}$)} & \colhead{(mas$^2$ yr$^{-2}$)} & \colhead{(mas$^2$ yr$^{-2}$)}}
\startdata
-0.81 & -0.86,-0.78 & 25 & $ 1.51 \pm 0.572$ & $-0.91 \pm 0.430$ & $2.8 \pm 0.36$ & $2.1 \pm 0.28$ & $13.0 \pm 27.17$ & $1.33 \pm 0.120$ & $7.77 \pm 2.154$ & $4.68 \pm 1.313$ & $-0.76 \pm 1.187$ \\
-0.73 & -0.77,-0.70 & 31 & $ 1.04 \pm 0.613$ & $-0.52 \pm 0.568$ & $4.4 \pm 0.55$ & $1.9 \pm 0.24$ & $41.2 \pm 5.75$ & $2.34 \pm 0.082$ & $12.46 \pm 3.047$ & $10.36 \pm 2.611$ & $-7.82 \pm 2.457$ \\
-0.65 & -0.68,-0.61 & 32 & $ 1.75 \pm 0.478$ & $ 0.40 \pm 0.320$ & $2.6 \pm 0.32$ & $1.8 \pm 0.21$ & $-4.8 \pm 15.91$ & $1.50 \pm 0.112$ & $6.88 \pm 1.733$ & $3.10 \pm 0.761$ & $0.32 \pm 0.792$ \\
-0.56 & -0.60,-0.51 & 51 & $ 1.57 \pm 0.413$ & $-0.22 \pm 0.404$ & $2.9 \pm 0.25$ & $2.8 \pm 0.23$ & $33.9 \pm 48.98$ & $1.03 \pm 0.079$ & $8.09 \pm 1.606$ & $7.91 \pm 1.504$ & $-0.22 \pm 1.115$ \\
-0.46 & -0.51,-0.43 & 69 & $ 0.91 \pm 0.320$ & $-0.22 \pm 0.295$ & $2.8 \pm 0.22$ & $2.4 \pm 0.18$ & $23.6 \pm 27.06$ & $1.17 \pm 0.078$ & $7.32 \pm 1.266$ & $5.90 \pm 0.986$ & $-0.77 \pm 0.795$ \\
-0.36 & -0.42,-0.34 & 107 & $ 1.52 \pm 0.253$ & $ 0.18 \pm 0.216$ & $2.7 \pm 0.18$ & $2.3 \pm 0.15$ & $21.2 \pm 20.41$ & $1.19 \pm 0.072$ & $7.05 \pm 0.965$ & $5.48 \pm 0.717$ & $-0.72 \pm 0.611$ \\
-0.29 & -0.34,-0.25 & 155 & $ 1.13 \pm 0.213$ & $-0.20 \pm 0.222$ & $2.9 \pm 0.16$ & $2.4 \pm 0.14$ & $46.0 \pm 19.28$ & $1.20 \pm 0.062$ & $7.19 \pm 0.795$ & $7.10 \pm 0.824$ & $-1.27 \pm 0.582$ \\
-0.20 & -0.25,-0.16 & 267 & $ 1.15 \pm 0.181$ & $ 0.01 \pm 0.184$ & $3.2 \pm 0.12$ & $2.8 \pm 0.12$ & $60.7 \pm 37.07$ & $1.11 \pm 0.048$ & $9.54 \pm 0.722$ & $8.55 \pm 0.815$ & $-0.81 \pm 0.549$ \\
-0.12 & -0.16,-0.08 & 306 & $ 0.85 \pm 0.170$ & $-0.03 \pm 0.161$ & $3.1 \pm 0.13$ & $2.6 \pm 0.11$ & $22.7 \pm 10.42$ & $1.20 \pm 0.048$ & $9.22 \pm 0.768$ & $7.18 \pm 0.588$ & $-1.03 \pm 0.482$ \\
-0.03 & -0.07, 0.01 & 293 & $ 0.43 \pm 0.171$ & $ 0.17 \pm 0.178$ & $3.4 \pm 0.14$ & $2.7 \pm 0.11$ & $43.1 \pm 8.16$ & $1.24 \pm 0.046$ & $9.58 \pm 0.798$ & $9.31 \pm 0.743$ & $-2.01 \pm 0.579$ \\
0.05 & 0.01, 0.10 & 346 & $-0.10 \pm 0.155$ & $ 0.34 \pm 0.156$ & $3.2 \pm 0.12$ & $2.7 \pm 0.10$ & $43.2 \pm 8.11$ & $1.21 \pm 0.043$ & $8.95 \pm 0.680$ & $8.74 \pm 0.668$ & $-1.69 \pm 0.481$ \\
0.14 & 0.10, 0.19 & 289 & $-0.96 \pm 0.153$ & $ 0.28 \pm 0.166$ & $3.0 \pm 0.12$ & $2.3 \pm 0.10$ & $52.2 \pm 5.67$ & $1.34 \pm 0.044$ & $7.63 \pm 0.547$ & $6.61 \pm 0.641$ & $-1.98 \pm 0.433$ \\
0.23 & 0.19, 0.27 & 210 & $-0.99 \pm 0.162$ & $ 0.22 \pm 0.154$ & $2.4 \pm 0.10$ & $2.3 \pm 0.09$ & $0.8 \pm 38.19$ & $1.05 \pm 0.045$ & $5.63 \pm 0.540$ & $5.13 \pm 0.479$ & $-0.01 \pm 0.382$ \\
0.32 & 0.28, 0.36 & 128 & $-0.87 \pm 0.212$ & $ 0.18 \pm 0.177$ & $2.4 \pm 0.15$ & $1.8 \pm 0.11$ & $31.9 \pm 9.71$ & $1.32 \pm 0.064$ & $5.22 \pm 0.648$ & $4.11 \pm 0.493$ & $-1.13 \pm 0.417$ \\
0.39 & 0.36, 0.45 & 79 & $-1.09 \pm 0.239$ & $-0.13 \pm 0.249$ & $2.3 \pm 0.16$ & $2.0 \pm 0.14$ & $-62.9 \pm 48.55$ & $1.14 \pm 0.074$ & $5.06 \pm 0.665$ & $4.33 \pm 0.767$ & $0.50 \pm 0.518$ \\
0.50 & 0.45, 0.53 & 44 & $-1.13 \pm 0.309$ & $ 0.47 \pm 0.321$ & $2.2 \pm 0.20$ & $2.0 \pm 0.18$ & $69.2 \pm 56.46$ & $1.08 \pm 0.088$ & $4.65 \pm 0.839$ & $4.15 \pm 0.990$ & $-0.22 \pm 0.670$ \\
0.67 & 0.64, 0.70 & 20 & $-1.19 \pm 0.521$ & $ 0.18 \pm 0.244$ & $2.2 \pm 0.36$ & $1.0 \pm 0.16$ & $-10.9 \pm 8.31$ & $2.17 \pm 0.110$ & $4.81 \pm 1.555$ & $1.19 \pm 0.362$ & $0.73 \pm 0.558$ \\
\enddata
\end{deluxetable}

\end{longrotatetable}

% rotation trend, uniform, metal-rich, velocities
\begin{longrotatetable}
\begin{deluxetable}{cccccccccccc}
\tabletypesize{\scriptsize}
\tablewidth{700pt}
\tablecaption{As \autoref{tab:trendsPho:Metal-rich:unif} (i.e., using a constant-width binning scheme) but with photometric parallaxes and proper motions converted into distances and velocities using a reference distance of 7.76 kpc. See discussion in \autoref{ss:resultsCurves}. \label{tab:trendsVel:Metal-rich:unif}}
\tablehead{\colhead{$d$} & \colhead{$d_{\rm{lo}}, d_{\rm{hi}}$} & \colhead{$N$} & \colhead{$\langle v_l \rangle$} & \colhead{$\langle v_b \rangle$} & \colhead{$a_v$} & \colhead{$b_v$} & \colhead{$\phi_{lb}$} & \colhead{a/b} & \colhead{$\sigma^2_{ll,v}$} & \colhead{$\sigma^2_{bb,v}$} & \colhead{$\sigma^2_{lb,v}$}\\ \colhead{$\mathrm{kpc}$} & \colhead{$\mathrm{kpc}$} & \colhead{ } & \colhead{(km s$^{-1}$)} & \colhead{(km s$^{-1}$)} & \colhead{(km s$^{-1}$)} & \colhead{(km s$^{-1}$)} & \colhead{($^{\circ}$)} & \colhead{$\mathrm{}$} & \colhead{(km$^2$ s$^{-2}$)} & \colhead{(km$^2$ s$^{-2}$)} & \colhead{(km$^2$ s$^{-2}$)}}
\startdata
5.35 & 5.23,5.43 & 25 & $38.37 \pm 14.507$ & $-23.14 \pm 10.905$ & $71 \pm 9.0$ & $54 \pm 7.2$ & $13.0 \pm 27.17$ & $1.33 \pm 0.120$ & $196.9 \pm 54.59$ & $118.5 \pm 33.27$ & $-19.1 \pm 30.07$ \\
5.55 & 5.44,5.63 & 31 & $27.43 \pm 16.119$ & $-13.76 \pm 14.947$ & $116 \pm 14.5$ & $49 \pm 6.3$ & $41.2 \pm 5.75$ & $2.34 \pm 0.082$ & $327.6 \pm 80.14$ & $272.4 \pm 68.65$ & $-205.7 \pm 64.61$ \\
5.76 & 5.68,5.86 & 32 & $47.82 \pm 13.048$ & $10.88 \pm 8.724$ & $72 \pm 8.9$ & $48 \pm 5.8$ & $-4.8 \pm 15.91$ & $1.50 \pm 0.112$ & $187.7 \pm 47.30$ & $84.7 \pm 20.78$ & $8.7 \pm 21.62$ \\
6.01 & 5.89,6.12 & 51 & $44.67 \pm 11.751$ & $-6.37 \pm 11.500$ & $82 \pm 7.2$ & $79 \pm 6.5$ & $33.9 \pm 48.98$ & $1.03 \pm 0.079$ & $230.3 \pm 45.71$ & $225.1 \pm 42.83$ & $-6.4 \pm 31.75$ \\
6.26 & 6.13,6.38 & 69 & $27.05 \pm 9.494$ & $-6.39 \pm 8.755$ & $82 \pm 6.5$ & $70 \pm 5.4$ & $23.6 \pm 27.06$ & $1.17 \pm 0.078$ & $217.4 \pm 37.61$ & $175.1 \pm 29.27$ & $-22.9 \pm 23.59$ \\
6.56 & 6.38,6.64 & 107 & $47.34 \pm 7.874$ & $5.50 \pm 6.724$ & $84 \pm 5.6$ & $71 \pm 4.6$ & $21.2 \pm 20.41$ & $1.19 \pm 0.072$ & $219.3 \pm 30.00$ & $170.3 \pm 22.29$ & $-22.4 \pm 19.01$ \\
6.80 & 6.64,6.92 & 155 & $36.29 \pm 6.861$ & $-6.39 \pm 7.150$ & $93 \pm 5.0$ & $78 \pm 4.4$ & $46.0 \pm 19.28$ & $1.20 \pm 0.062$ & $231.7 \pm 25.62$ & $228.7 \pm 26.54$ & $-40.8 \pm 18.74$ \\
7.06 & 6.92,7.20 & 267 & $38.54 \pm 6.063$ & $0.37 \pm 6.159$ & $106 \pm 4.2$ & $95 \pm 3.9$ & $60.7 \pm 37.07$ & $1.11 \pm 0.048$ & $319.3 \pm 24.18$ & $286.1 \pm 27.29$ & $-27.1 \pm 18.37$ \\
7.35 & 7.20,7.50 & 306 & $29.54 \pm 5.915$ & $-0.89 \pm 5.612$ & $108 \pm 4.4$ & $91 \pm 3.8$ & $22.7 \pm 10.42$ & $1.20 \pm 0.048$ & $321.3 \pm 26.77$ & $250.3 \pm 20.48$ & $-36.1 \pm 16.81$ \\
7.64 & 7.50,7.80 & 293 & $15.71 \pm 6.204$ & $6.20 \pm 6.456$ & $123 \pm 5.2$ & $99 \pm 3.9$ & $43.1 \pm 8.16$ & $1.24 \pm 0.046$ & $347.1 \pm 28.91$ & $337.4 \pm 26.92$ & $-72.9 \pm 20.97$ \\
7.94 & 7.80,8.12 & 346 & $-3.90 \pm 5.835$ & $12.73 \pm 5.859$ & $122 \pm 4.6$ & $101 \pm 3.9$ & $43.2 \pm 8.11$ & $1.21 \pm 0.043$ & $337.1 \pm 25.58$ & $329.1 \pm 25.16$ & $-63.7 \pm 18.11$ \\
8.27 & 8.13,8.46 & 289 & $-37.52 \pm 6.007$ & $10.89 \pm 6.490$ & $119 \pm 4.9$ & $88 \pm 3.8$ & $52.2 \pm 5.67$ & $1.34 \pm 0.044$ & $299.1 \pm 21.45$ & $259.1 \pm 25.15$ & $-77.7 \pm 16.96$ \\
8.61 & 8.46,8.81 & 210 & $-40.35 \pm 6.605$ & $9.00 \pm 6.265$ & $97 \pm 4.2$ & $92 \pm 3.7$ & $0.8 \pm 38.19$ & $1.05 \pm 0.045$ & $229.7 \pm 22.04$ & $209.3 \pm 19.53$ & $-0.3 \pm 15.59$ \\
8.98 & 8.81,9.17 & 128 & $-37.20 \pm 9.032$ & $7.47 \pm 7.539$ & $104 \pm 6.3$ & $79 \pm 4.8$ & $31.9 \pm 9.71$ & $1.32 \pm 0.064$ & $222.3 \pm 27.60$ & $175.0 \pm 21.00$ & $-48.3 \pm 17.73$ \\
9.28 & 9.17,9.54 & 79 & $-47.75 \pm 10.531$ & $-5.60 \pm 10.950$ & $101 \pm 7.1$ & $89 \pm 6.3$ & $-62.9 \pm 48.55$ & $1.14 \pm 0.074$ & $222.6 \pm 29.28$ & $190.7 \pm 33.77$ & $22.1 \pm 22.78$ \\
9.77 & 9.55,9.92 & 44 & $-52.53 \pm 14.310$ & $21.90 \pm 14.861$ & $101 \pm 9.3$ & $93 \pm 8.2$ & $69.2 \pm 56.46$ & $1.08 \pm 0.088$ & $215.5 \pm 38.88$ & $192.1 \pm 45.83$ & $-10.4 \pm 31.03$ \\
10.57 & 10.40,10.72 & 20 & $-59.62 \pm 26.087$ & $9.01 \pm 12.234$ & $111 \pm 17.9$ & $51 \pm 8.0$ & $-10.9 \pm 8.31$ & $2.17 \pm 0.110$ & $241.0 \pm 77.94$ & $59.6 \pm 18.14$ & $36.4 \pm 27.94$ \\
\enddata
\end{deluxetable}

\end{longrotatetable}

% rotation trend, uniform, metal-poor, proper motions
\begin{longrotatetable}
\begin{deluxetable}{cccccccccccc}
\tabletypesize{\scriptsize}
\tablewidth{700pt}
\tablecaption{Observed rotation trend for the \MP~sample, using a constant-width binning scheme and with uncertainty ranges given as $1\sigma$~limits based on 1000 parametric bootstrap trials. See discussion in \autoref{ss:resultsCurves}. \label{tab:trendsPho:Metal-poor:unif}}
\tablehead{\colhead{$\pi'$} & \colhead{Bin edges} & \colhead{$N$} & \colhead{$\langle \mu_l \rangle$} & \colhead{$\langle \mu_b \rangle$} & \colhead{$a_{\mu}$} & \colhead{$b_{\mu}$} & \colhead{$\phi_{lb}$} & \colhead{a/b} & \colhead{$\sigma^2_{ll}$} & \colhead{$\sigma^2_{bb}$} & \colhead{$C_{lb}$}\\ \colhead{$\mathrm{mag}$} & \colhead{$\mathrm{mag}$} & \colhead{ } & \colhead{(mas yr$^{-1}$)} & \colhead{(mas yr$^{-1}$)} & \colhead{(mas yr$^{-1}$)} & \colhead{(mas yr$^{-1}$)} & \colhead{($^{\circ}$)} & \colhead{$\mathrm{}$} & \colhead{(mas$^2$ yr$^{-2}$)} & \colhead{(mas$^2$ yr$^{-2}$)} & \colhead{(mas$^2$ yr$^{-2}$)}}
\startdata
-1.17 & -1.21,-1.13 & 26 & $ 2.08 \pm 0.672$ & $-0.02 \pm 0.607$ & $3.6 \pm 0.46$ & $2.9 \pm 0.37$ & $25.6 \pm 30.86$ & $1.27 \pm 0.117$ & $12.25 \pm 3.393$ & $9.15 \pm 2.517$ & $-1.93 \pm 2.108$ \\
-1.06 & -1.12,-1.04 & 31 & $-0.42 \pm 0.683$ & $-0.27 \pm 0.475$ & $3.8 \pm 0.43$ & $2.7 \pm 0.32$ & $-0.2 \pm 17.83$ & $1.41 \pm 0.111$ & $14.19 \pm 3.301$ & $7.11 \pm 1.756$ & $0.02 \pm 1.756$ \\
-1.00 & -1.04,-0.95 & 34 & $ 0.64 \pm 0.537$ & $ 0.64 \pm 0.394$ & $3.3 \pm 0.40$ & $2.2 \pm 0.25$ & $17.2 \pm 14.89$ & $1.50 \pm 0.106$ & $10.05 \pm 2.525$ & $5.21 \pm 1.220$ & $-1.66 \pm 1.256$ \\
-0.91 & -0.95,-0.86 & 35 & $ 0.02 \pm 0.443$ & $-0.21 \pm 0.536$ & $3.3 \pm 0.37$ & $2.4 \pm 0.28$ & $57.9 \pm 28.81$ & $1.37 \pm 0.110$ & $9.57 \pm 1.742$ & $7.32 \pm 2.218$ & $-2.32 \pm 1.422$ \\
-0.81 & -0.86,-0.78 & 47 & $-0.17 \pm 0.432$ & $ 0.35 \pm 0.492$ & $3.4 \pm 0.31$ & $2.9 \pm 0.27$ & $86.4 \pm 68.08$ & $1.17 \pm 0.090$ & $11.85 \pm 1.786$ & $8.65 \pm 2.384$ & $-0.20 \pm 1.534$ \\
-0.73 & -0.77,-0.69 & 67 & $ 0.58 \pm 0.387$ & $ 0.37 \pm 0.363$ & $3.8 \pm 0.33$ & $2.3 \pm 0.20$ & $40.0 \pm 7.64$ & $1.61 \pm 0.077$ & $10.53 \pm 1.822$ & $9.02 \pm 1.601$ & $-4.25 \pm 1.336$ \\
-0.64 & -0.69,-0.60 & 75 & $ 0.36 \pm 0.367$ & $-0.43 \pm 0.300$ & $3.2 \pm 0.25$ & $2.6 \pm 0.21$ & $5.6 \pm 17.79$ & $1.24 \pm 0.083$ & $10.50 \pm 1.707$ & $6.85 \pm 1.127$ & $-0.36 \pm 0.957$ \\
-0.56 & -0.60,-0.51 & 109 & $ 0.88 \pm 0.311$ & $-0.51 \pm 0.323$ & $3.9 \pm 0.26$ & $2.5 \pm 0.17$ & $45.5 \pm 6.01$ & $1.58 \pm 0.060$ & $10.93 \pm 1.502$ & $10.76 \pm 1.465$ & $-4.67 \pm 1.113$ \\
-0.47 & -0.51,-0.43 & 99 & $ 0.46 \pm 0.307$ & $ 0.02 \pm 0.315$ & $3.4 \pm 0.24$ & $2.7 \pm 0.19$ & $50.7 \pm 17.36$ & $1.28 \pm 0.076$ & $9.82 \pm 1.248$ & $8.94 \pm 1.381$ & $-2.21 \pm 0.988$ \\
-0.37 & -0.42,-0.34 & 138 & $ 0.19 \pm 0.254$ & $-0.07 \pm 0.259$ & $3.3 \pm 0.18$ & $2.9 \pm 0.16$ & $47.4 \pm 27.52$ & $1.15 \pm 0.061$ & $9.85 \pm 1.080$ & $9.62 \pm 1.145$ & $-1.35 \pm 0.819$ \\
-0.29 & -0.34,-0.25 & 192 & $ 0.23 \pm 0.240$ & $-0.16 \pm 0.196$ & $3.4 \pm 0.17$ & $2.8 \pm 0.14$ & $0.4 \pm 11.94$ & $1.22 \pm 0.054$ & $11.71 \pm 1.186$ & $7.92 \pm 0.789$ & $-0.03 \pm 0.702$ \\
-0.20 & -0.25,-0.16 & 222 & $ 0.11 \pm 0.206$ & $ 0.61 \pm 0.206$ & $3.2 \pm 0.14$ & $3.0 \pm 0.13$ & $51.1 \pm 39.87$ & $1.08 \pm 0.047$ & $9.69 \pm 0.888$ & $9.37 \pm 0.904$ & $-0.74 \pm 0.642$ \\
-0.12 & -0.16,-0.08 & 226 & $-0.10 \pm 0.221$ & $-0.21 \pm 0.205$ & $3.5 \pm 0.16$ & $3.0 \pm 0.14$ & $34.9 \pm 16.87$ & $1.16 \pm 0.055$ & $10.98 \pm 1.028$ & $9.92 \pm 0.930$ & $-1.44 \pm 0.718$ \\
-0.03 & -0.07, 0.01 & 226 & $-0.05 \pm 0.211$ & $ 0.40 \pm 0.184$ & $3.2 \pm 0.15$ & $2.9 \pm 0.13$ & $-1.5 \pm 20.98$ & $1.12 \pm 0.050$ & $10.38 \pm 1.017$ & $8.34 \pm 0.773$ & $0.05 \pm 0.623$ \\
0.06 & 0.01, 0.10 & 255 & $-0.36 \pm 0.184$ & $ 0.25 \pm 0.178$ & $3.1 \pm 0.13$ & $2.7 \pm 0.11$ & $25.8 \pm 15.68$ & $1.15 \pm 0.049$ & $9.11 \pm 0.784$ & $7.70 \pm 0.693$ & $-0.89 \pm 0.539$ \\
0.14 & 0.10, 0.19 & 234 & $-0.41 \pm 0.193$ & $ 0.00 \pm 0.201$ & $3.1 \pm 0.13$ & $2.9 \pm 0.12$ & $-68.7 \pm 56.01$ & $1.07 \pm 0.045$ & $9.35 \pm 0.756$ & $8.46 \pm 0.888$ & $0.41 \pm 0.552$ \\
0.23 & 0.19, 0.27 & 181 & $-0.55 \pm 0.230$ & $ 0.09 \pm 0.225$ & $3.3 \pm 0.16$ & $2.9 \pm 0.14$ & $34.7 \pm 25.90$ & $1.12 \pm 0.056$ & $10.18 \pm 1.053$ & $9.38 \pm 0.981$ & $-1.07 \pm 0.728$ \\
0.32 & 0.28, 0.36 & 154 & $-0.63 \pm 0.185$ & $ 0.38 \pm 0.207$ & $2.7 \pm 0.15$ & $2.1 \pm 0.12$ & $58.7 \pm 11.70$ & $1.29 \pm 0.061$ & $6.59 \pm 0.597$ & $5.25 \pm 0.722$ & $-1.29 \pm 0.477$ \\
0.40 & 0.36, 0.45 & 122 & $-0.51 \pm 0.226$ & $ 0.35 \pm 0.211$ & $2.5 \pm 0.15$ & $2.4 \pm 0.14$ & $-10.2 \pm 35.70$ & $1.07 \pm 0.061$ & $6.34 \pm 0.825$ & $5.56 \pm 0.717$ & $0.15 \pm 0.557$ \\
0.49 & 0.45, 0.54 & 90 & $-0.57 \pm 0.271$ & $ 0.25 \pm 0.292$ & $2.7 \pm 0.18$ & $2.5 \pm 0.16$ & $80.5 \pm 65.21$ & $1.09 \pm 0.065$ & $7.49 \pm 0.960$ & $6.41 \pm 1.139$ & $-0.19 \pm 0.714$ \\
0.57 & 0.54, 0.62 & 65 & $-0.28 \pm 0.455$ & $-0.24 \pm 0.335$ & $3.6 \pm 0.30$ & $2.6 \pm 0.23$ & $17.5 \pm 12.41$ & $1.39 \pm 0.083$ & $12.66 \pm 2.141$ & $7.43 \pm 1.328$ & $-1.82 \pm 1.229$ \\
0.66 & 0.63, 0.71 & 50 & $-0.40 \pm 0.247$ & $ 0.29 \pm 0.305$ & $2.1 \pm 0.21$ & $1.7 \pm 0.16$ & $83.6 \pm 70.83$ & $1.22 \pm 0.095$ & $4.56 \pm 0.600$ & $3.08 \pm 0.940$ & $-0.17 \pm 0.520$ \\
0.76 & 0.71, 0.80 & 29 & $ 0.12 \pm 0.471$ & $ 0.69 \pm 0.331$ & $2.6 \pm 0.33$ & $1.7 \pm 0.21$ & $-8.5 \pm 16.79$ & $1.51 \pm 0.115$ & $6.75 \pm 1.741$ & $3.09 \pm 0.774$ & $0.56 \pm 0.846$ \\
0.83 & 0.80, 0.88 & 27 & $-0.69 \pm 0.246$ & $ 0.58 \pm 0.245$ & $1.6 \pm 0.21$ & $0.8 \pm 0.11$ & $-45.1 \pm 9.36$ & $1.90 \pm 0.103$ & $1.64 \pm 0.430$ & $1.63 \pm 0.427$ & $0.93 \pm 0.348$ \\
\enddata
\end{deluxetable}

\end{longrotatetable}

% rotation trend, uniform, metal-poor, velocities
\begin{longrotatetable}
\begin{deluxetable}{cccccccccccc}
\tabletypesize{\scriptsize}
\tablewidth{700pt}
\tablecaption{As \autoref{tab:trendsPho:Metal-poor:unif} (i.e., using a constant-width binning scheme) but with photometric parallaxes and proper motions converted into distances and velocities using a reference distance of 7.76 kpc. See discussion in \autoref{ss:resultsCurves}. \label{tab:trendsVel:Metal-poor:unif}}
\tablehead{\colhead{$d$} & \colhead{$d_{\rm{lo}}, d_{\rm{hi}}$} & \colhead{$N$} & \colhead{$\langle v_l \rangle$} & \colhead{$\langle v_b \rangle$} & \colhead{$a_v$} & \colhead{$b_v$} & \colhead{$\phi_{lb}$} & \colhead{a/b} & \colhead{$\sigma^2_{ll,v}$} & \colhead{$\sigma^2_{bb,v}$} & \colhead{$\sigma^2_{lb,v}$}\\ \colhead{$\mathrm{kpc}$} & \colhead{$\mathrm{kpc}$} & \colhead{ } & \colhead{(km s$^{-1}$)} & \colhead{(km s$^{-1}$)} & \colhead{(km s$^{-1}$)} & \colhead{(km s$^{-1}$)} & \colhead{($^{\circ}$)} & \colhead{$\mathrm{}$} & \colhead{(km$^2$ s$^{-2}$)} & \colhead{(km$^2$ s$^{-2}$)} & \colhead{(km$^2$ s$^{-2}$)}}
\startdata
4.53 & 4.44,4.62 & 26 & $44.70 \pm 14.437$ & $-0.53 \pm 13.031$ & $78 \pm 9.8$ & $62 \pm 7.9$ & $25.6 \pm 30.86$ & $1.27 \pm 0.117$ & $263.1 \pm 72.84$ & $196.4 \pm 54.04$ & $-41.4 \pm 45.25$ \\
4.76 & 4.62,4.81 & 31 & $-9.54 \pm 15.400$ & $-6.19 \pm 10.718$ & $85 \pm 9.7$ & $60 \pm 7.2$ & $-0.2 \pm 17.83$ & $1.41 \pm 0.111$ & $320.1 \pm 74.49$ & $160.5 \pm 39.63$ & $0.5 \pm 39.63$ \\
4.89 & 4.82,5.00 & 34 & $14.90 \pm 12.436$ & $14.93 \pm 9.121$ & $75 \pm 9.2$ & $50 \pm 5.8$ & $17.2 \pm 14.89$ & $1.50 \pm 0.106$ & $232.9 \pm 58.51$ & $120.8 \pm 28.26$ & $-38.5 \pm 29.11$ \\
5.11 & 5.01,5.21 & 35 & $0.39 \pm 10.748$ & $-5.08 \pm 13.002$ & $80 \pm 9.0$ & $59 \pm 6.8$ & $57.9 \pm 28.81$ & $1.37 \pm 0.110$ & $231.9 \pm 42.23$ & $177.5 \pm 53.75$ & $-56.1 \pm 34.48$ \\
5.34 & 5.22,5.42 & 47 & $-4.41 \pm 10.930$ & $8.90 \pm 12.457$ & $87 \pm 7.9$ & $74 \pm 6.8$ & $86.4 \pm 68.08$ & $1.17 \pm 0.090$ & $300.1 \pm 45.22$ & $219.0 \pm 60.35$ & $-5.1 \pm 38.82$ \\
5.56 & 5.44,5.65 & 67 & $15.27 \pm 10.189$ & $9.71 \pm 9.545$ & $99 \pm 8.7$ & $62 \pm 5.3$ & $40.0 \pm 7.64$ & $1.61 \pm 0.077$ & $277.2 \pm 47.98$ & $237.6 \pm 42.15$ & $-111.9 \pm 35.17$ \\
5.78 & 5.66,5.89 & 75 & $9.75 \pm 10.050$ & $-11.72 \pm 8.208$ & $89 \pm 7.0$ & $72 \pm 5.7$ & $5.6 \pm 17.79$ & $1.24 \pm 0.083$ & $287.8 \pm 46.76$ & $187.6 \pm 30.88$ & $-10.0 \pm 26.21$ \\
5.99 & 5.89,6.13 & 109 & $25.05 \pm 8.836$ & $-14.60 \pm 9.163$ & $112 \pm 7.5$ & $71 \pm 4.9$ & $45.5 \pm 6.01$ & $1.58 \pm 0.060$ & $310.4 \pm 42.66$ & $305.7 \pm 41.61$ & $-132.6 \pm 31.60$ \\
6.25 & 6.13,6.38 & 99 & $13.57 \pm 9.096$ & $0.71 \pm 9.318$ & $101 \pm 7.0$ & $79 \pm 5.6$ & $50.7 \pm 17.36$ & $1.28 \pm 0.076$ & $291.0 \pm 36.96$ & $264.7 \pm 40.91$ & $-65.4 \pm 29.26$ \\
6.53 & 6.38,6.64 & 138 & $5.95 \pm 7.879$ & $-2.18 \pm 8.016$ & $103 \pm 5.4$ & $90 \pm 5.1$ & $47.4 \pm 27.52$ & $1.15 \pm 0.061$ & $304.8 \pm 33.44$ & $298.0 \pm 35.45$ & $-41.7 \pm 25.37$ \\
6.79 & 6.65,6.92 & 192 & $7.50 \pm 7.726$ & $-5.15 \pm 6.299$ & $110 \pm 5.4$ & $91 \pm 4.4$ & $0.4 \pm 11.94$ & $1.22 \pm 0.054$ & $376.8 \pm 38.17$ & $255.0 \pm 25.38$ & $-0.8 \pm 22.60$ \\
7.07 & 6.92,7.20 & 222 & $3.73 \pm 6.914$ & $20.34 \pm 6.905$ & $107 \pm 4.5$ & $99 \pm 4.3$ & $51.1 \pm 39.87$ & $1.08 \pm 0.047$ & $324.7 \pm 29.75$ & $314.0 \pm 30.30$ & $-24.8 \pm 21.50$ \\
7.34 & 7.20,7.49 & 226 & $-3.42 \pm 7.697$ & $-7.45 \pm 7.125$ & $120 \pm 5.6$ & $104 \pm 4.7$ & $34.9 \pm 16.87$ & $1.16 \pm 0.055$ & $382.2 \pm 35.76$ & $345.3 \pm 32.38$ & $-50.0 \pm 24.99$ \\
7.65 & 7.50,7.80 & 226 & $-1.83 \pm 7.646$ & $14.52 \pm 6.688$ & $117 \pm 5.4$ & $105 \pm 4.6$ & $-1.5 \pm 20.98$ & $1.12 \pm 0.050$ & $376.5 \pm 36.88$ & $302.7 \pm 28.03$ & $1.9 \pm 22.61$ \\
7.97 & 7.81,8.12 & 255 & $-13.55 \pm 6.942$ & $9.55 \pm 6.727$ & $117 \pm 4.9$ & $102 \pm 4.3$ & $25.8 \pm 15.68$ & $1.15 \pm 0.049$ & $343.9 \pm 29.61$ & $290.7 \pm 26.18$ & $-33.6 \pm 20.35$ \\
8.28 & 8.13,8.46 & 234 & $-16.18 \pm 7.567$ & $0.11 \pm 7.888$ & $121 \pm 5.0$ & $113 \pm 4.7$ & $-68.7 \pm 56.01$ & $1.07 \pm 0.045$ & $367.1 \pm 29.70$ & $332.1 \pm 34.86$ & $16.1 \pm 21.69$ \\
8.61 & 8.46,8.81 & 181 & $-22.48 \pm 9.407$ & $3.63 \pm 9.170$ & $135 \pm 6.7$ & $120 \pm 5.9$ & $34.7 \pm 25.90$ & $1.12 \pm 0.056$ & $415.7 \pm 42.99$ & $382.9 \pm 40.06$ & $-43.6 \pm 29.72$ \\
9.00 & 8.81,9.17 & 154 & $-26.91 \pm 7.898$ & $16.32 \pm 8.828$ & $116 \pm 6.4$ & $90 \pm 5.1$ & $58.7 \pm 11.70$ & $1.29 \pm 0.061$ & $281.1 \pm 25.47$ & $223.8 \pm 30.80$ & $-55.1 \pm 20.34$ \\
9.34 & 9.17,9.55 & 122 & $-22.38 \pm 9.992$ & $15.41 \pm 9.362$ & $112 \pm 6.4$ & $104 \pm 6.1$ & $-10.2 \pm 35.70$ & $1.07 \pm 0.061$ & $280.8 \pm 36.52$ & $246.3 \pm 31.73$ & $6.4 \pm 24.68$ \\
9.75 & 9.55,9.93 & 90 & $-26.17 \pm 12.539$ & $11.45 \pm 13.468$ & $127 \pm 8.3$ & $117 \pm 7.5$ & $80.5 \pm 65.21$ & $1.09 \pm 0.065$ & $345.8 \pm 44.35$ & $296.1 \pm 52.61$ & $-8.6 \pm 33.00$ \\
10.10 & 9.96,10.34 & 65 & $-13.45 \pm 21.765$ & $-11.59 \pm 16.043$ & $174 \pm 14.4$ & $125 \pm 11.0$ & $17.5 \pm 12.41$ & $1.39 \pm 0.083$ & $606.0 \pm 102.47$ & $355.8 \pm 63.55$ & $-87.3 \pm 58.84$ \\
10.50 & 10.37,10.76 & 50 & $-19.99 \pm 12.274$ & $14.41 \pm 15.196$ & $107 \pm 10.3$ & $87 \pm 7.8$ & $83.6 \pm 70.83$ & $1.22 \pm 0.095$ & $227.2 \pm 29.89$ & $153.1 \pm 46.81$ & $-8.4 \pm 25.89$ \\
11.03 & 10.78,11.21 & 29 & $6.11 \pm 24.610$ & $36.28 \pm 17.299$ & $137 \pm 17.2$ & $91 \pm 11.0$ & $-8.5 \pm 16.79$ & $1.51 \pm 0.115$ & $352.7 \pm 90.97$ & $161.6 \pm 40.44$ & $29.1 \pm 44.22$ \\
11.39 & 11.22,11.66 & 27 & $-37.48 \pm 13.268$ & $31.33 \pm 13.255$ & $86 \pm 11.1$ & $46 \pm 6.1$ & $-45.1 \pm 9.36$ & $1.90 \pm 0.103$ & $88.5 \pm 23.20$ & $88.1 \pm 23.04$ & $50.0 \pm 18.77$ \\
\enddata
\end{deluxetable}

\end{longrotatetable}

% 2017-12-13 force one-column grid to remove the blank page that used
% to be inserted between this and the following tables.
\onecolumngrid

% bin statistics, uniform, metal-rich
\begin{deluxetable*}{cccccccccc}
\tabletypesize{\scriptsize}
\tablewidth{700pt}
\tablecaption{Bin statistics for the rotation curves of the \MR~sample, using the same constant-width binning scheme as \autoref{tab:trendsPho:Metal-rich:unif}. Wedge volumes $V$~and densities $\rho$~assume the reference sample lies at distance 7.76~kpc. $N (\mu_l)$~and $\rho (\mu_l)$~denote the counts and number densities of objects that would pass a kinematic cut of $\mu_l < -2.0$~mas yr$^{-1}$. The uncertainties quoted refer to $1\sigma$~ranges from 1000~parameteric bootstrap trials. See \autoref{ss:resultsCurves}. \label{tab:stats:Metal-rich:unif}}
\tablehead{\colhead{$\pi'$} & \colhead{$\pi'_{\rm{hi}}-\pi'_{\rm{lo}}$} & \colhead{$d$} & \colhead{$d_{\rm{hi}}-d_{\rm{lo}}$} & \colhead{$N$} & \colhead{$V$} & \colhead{$\rho$} & \colhead{$N(\mu_l)$} & \colhead{$f(\mu_l)$} & \colhead{$\rho (\mu_l)$}\\ \colhead{$\mathrm{mag}$} & \colhead{$\mathrm{mag}$} & \colhead{$\mathrm{kpc}$} & \colhead{$\mathrm{kpc}$} & \colhead{ } & \colhead{(pc$^3$)} & \colhead{(pc$^{-3}$)} & \colhead{ } & \colhead{ } & \colhead{(pc$^{-3}$)}}
\startdata
-0.81 & 0.080 & 5.35 & 0.196 & 25 & 688.1 & 0.036 & $3 \pm 1.5$ & $0.12 \pm 0.062$ & $0.004 \pm 0.0022$ \\
-0.73 & 0.073 & 5.55 & 0.186 & 31 & 776.5 & 0.040 & $8 \pm 2.2$ & $0.26 \pm 0.070$ & $0.010 \pm 0.0028$ \\
-0.65 & 0.071 & 5.76 & 0.189 & 32 & 876.3 & 0.037 & $2 \pm 1.5$ & $0.06 \pm 0.047$ & $0.002 \pm 0.0017$ \\
-0.56 & 0.085 & 6.01 & 0.234 & 51 & 988.9 & 0.052 & $5 \pm 2.3$ & $0.10 \pm 0.045$ & $0.005 \pm 0.0023$ \\
-0.46 & 0.087 & 6.26 & 0.250 & 69 & 1116.0 & 0.062 & $8 \pm 3.0$ & $0.12 \pm 0.043$ & $0.007 \pm 0.0027$ \\
-0.36 & 0.086 & 6.56 & 0.257 & 107 & 1259.4 & 0.085 & $10 \pm 3.0$ & $0.09 \pm 0.028$ & $0.008 \pm 0.0024$ \\
-0.29 & 0.087 & 6.80 & 0.272 & 155 & 1421.2 & 0.109 & $20 \pm 4.1$ & $0.13 \pm 0.026$ & $0.014 \pm 0.0029$ \\
-0.20 & 0.087 & 7.06 & 0.283 & 267 & 1603.8 & 0.166 & $34 \pm 5.6$ & $0.13 \pm 0.021$ & $0.021 \pm 0.0035$ \\
-0.12 & 0.087 & 7.35 & 0.294 & 306 & 1809.9 & 0.169 & $52 \pm 6.6$ & $0.17 \pm 0.022$ & $0.029 \pm 0.0037$ \\
-0.03 & 0.087 & 7.64 & 0.308 & 293 & 2042.5 & 0.143 & $66 \pm 6.8$ & $0.23 \pm 0.023$ & $0.032 \pm 0.0033$ \\
0.05 & 0.087 & 7.94 & 0.319 & 346 & 2305.0 & 0.150 & $93 \pm 8.3$ & $0.27 \pm 0.024$ & $0.040 \pm 0.0036$ \\
0.14 & 0.087 & 8.27 & 0.331 & 289 & 2601.1 & 0.111 & $101 \pm 8.1$ & $0.35 \pm 0.028$ & $0.039 \pm 0.0031$ \\
0.23 & 0.087 & 8.61 & 0.346 & 210 & 2935.4 & 0.072 & $78 \pm 6.9$ & $0.37 \pm 0.033$ & $0.027 \pm 0.0023$ \\
0.32 & 0.085 & 8.98 & 0.353 & 128 & 3312.6 & 0.039 & $44 \pm 5.5$ & $0.34 \pm 0.043$ & $0.013 \pm 0.0017$ \\
0.39 & 0.087 & 9.28 & 0.373 & 79 & 3738.2 & 0.021 & $29 \pm 4.2$ & $0.37 \pm 0.054$ & $0.008 \pm 0.0011$ \\
0.50 & 0.084 & 9.77 & 0.376 & 44 & 4218.6 & 0.010 & $13 \pm 3.1$ & $0.30 \pm 0.070$ & $0.003 \pm 0.0007$ \\
0.67 & 0.066 & 10.57 & 0.323 & 20 & 5372.3 & 0.004 & $10 \pm 2.2$ & $0.50 \pm 0.110$ & $0.002 \pm 0.0004$ \\
\enddata
\end{deluxetable*}

% bin statistics, uniform, metal-poor
\begin{deluxetable*}{cccccccccc}
\tabletypesize{\scriptsize}
\tablewidth{700pt}
\tablecaption{As \autoref{tab:stats:Metal-rich:unif} but for the \MP~sample and with the binning scheme of \autoref{tab:trendsPho:Metal-poor:unif}. See \autoref{ss:resultsCurves}. \label{tab:stats:Metal-poor:unif}}
\tablehead{\colhead{$\pi'$} & \colhead{$\pi'_{\rm{hi}}-\pi'_{\rm{lo}}$} & \colhead{$d$} & \colhead{$d_{\rm{hi}}-d_{\rm{lo}}$} & \colhead{$N$} & \colhead{$V$} & \colhead{$\rho$} & \colhead{$N(\mu_l)$} & \colhead{$f(\mu_l)$} & \colhead{$\rho (\mu_l)$}\\ \colhead{$\mathrm{mag}$} & \colhead{$\mathrm{mag}$} & \colhead{$\mathrm{kpc}$} & \colhead{$\mathrm{kpc}$} & \colhead{ } & \colhead{(pc$^3$)} & \colhead{(pc$^{-3}$)} & \colhead{ } & \colhead{ } & \colhead{(pc$^{-3}$)}}
\startdata
-1.17 & 0.085 & 4.53 & 0.177 & 26 & 424.3 & 0.061 & $3 \pm 1.6$ & $0.12 \pm 0.062$ & $0.007 \pm 0.0038$ \\
-1.06 & 0.086 & 4.76 & 0.188 & 31 & 478.8 & 0.065 & $10 \pm 2.6$ & $0.32 \pm 0.085$ & $0.021 \pm 0.0055$ \\
-1.00 & 0.081 & 4.89 & 0.183 & 34 & 540.3 & 0.063 & $6 \pm 2.3$ & $0.18 \pm 0.068$ & $0.011 \pm 0.0043$ \\
-0.91 & 0.086 & 5.11 & 0.202 & 35 & 609.8 & 0.057 & $5 \pm 2.4$ & $0.14 \pm 0.070$ & $0.008 \pm 0.0040$ \\
-0.81 & 0.082 & 5.34 & 0.200 & 47 & 688.1 & 0.068 & $13 \pm 3.0$ & $0.28 \pm 0.063$ & $0.019 \pm 0.0043$ \\
-0.73 & 0.083 & 5.56 & 0.211 & 67 & 776.5 & 0.086 & $17 \pm 3.3$ & $0.25 \pm 0.049$ & $0.022 \pm 0.0042$ \\
-0.64 & 0.086 & 5.78 & 0.229 & 75 & 876.3 & 0.086 & $18 \pm 3.7$ & $0.24 \pm 0.050$ & $0.021 \pm 0.0043$ \\
-0.56 & 0.087 & 5.99 & 0.240 & 109 & 988.9 & 0.110 & $17 \pm 4.0$ & $0.16 \pm 0.037$ & $0.017 \pm 0.0041$ \\
-0.47 & 0.085 & 6.25 & 0.246 & 99 & 1116.0 & 0.089 & $22 \pm 4.0$ & $0.22 \pm 0.041$ & $0.020 \pm 0.0036$ \\
-0.37 & 0.086 & 6.53 & 0.259 & 138 & 1259.4 & 0.110 & $37 \pm 4.7$ & $0.27 \pm 0.034$ & $0.029 \pm 0.0037$ \\
-0.29 & 0.086 & 6.79 & 0.270 & 192 & 1421.2 & 0.135 & $56 \pm 5.8$ & $0.29 \pm 0.030$ & $0.039 \pm 0.0041$ \\
-0.20 & 0.087 & 7.07 & 0.284 & 222 & 1603.8 & 0.138 & $55 \pm 6.2$ & $0.25 \pm 0.028$ & $0.034 \pm 0.0039$ \\
-0.12 & 0.087 & 7.34 & 0.294 & 226 & 1809.9 & 0.125 & $59 \pm 6.7$ & $0.26 \pm 0.030$ & $0.033 \pm 0.0037$ \\
-0.03 & 0.087 & 7.65 & 0.307 & 226 & 2042.5 & 0.111 & $58 \pm 6.5$ & $0.26 \pm 0.029$ & $0.028 \pm 0.0032$ \\
0.06 & 0.087 & 7.97 & 0.320 & 255 & 2305.0 & 0.111 & $80 \pm 7.3$ & $0.31 \pm 0.028$ & $0.035 \pm 0.0031$ \\
0.14 & 0.087 & 8.28 & 0.331 & 234 & 2601.1 & 0.090 & $73 \pm 7.1$ & $0.31 \pm 0.030$ & $0.028 \pm 0.0027$ \\
0.23 & 0.087 & 8.61 & 0.345 & 181 & 2935.4 & 0.062 & $59 \pm 6.2$ & $0.33 \pm 0.034$ & $0.020 \pm 0.0021$ \\
0.32 & 0.087 & 9.00 & 0.361 & 154 & 3312.6 & 0.046 & $46 \pm 5.6$ & $0.30 \pm 0.036$ & $0.014 \pm 0.0017$ \\
0.40 & 0.087 & 9.34 & 0.374 & 122 & 3738.2 & 0.033 & $34 \pm 5.0$ & $0.28 \pm 0.041$ & $0.009 \pm 0.0013$ \\
0.49 & 0.084 & 9.75 & 0.376 & 90 & 4218.6 & 0.021 & $27 \pm 4.4$ & $0.30 \pm 0.049$ & $0.006 \pm 0.0010$ \\
0.57 & 0.083 & 10.10 & 0.387 & 65 & 4760.6 & 0.014 & $15 \pm 3.7$ & $0.23 \pm 0.057$ & $0.003 \pm 0.0008$ \\
0.66 & 0.081 & 10.50 & 0.396 & 50 & 5372.3 & 0.009 & $10 \pm 2.7$ & $0.20 \pm 0.054$ & $0.002 \pm 0.0005$ \\
0.76 & 0.084 & 11.03 & 0.425 & 29 & 6062.7 & 0.005 & $4 \pm 2.1$ & $0.14 \pm 0.074$ & $0.001 \pm 0.0004$ \\
0.83 & 0.083 & 11.39 & 0.438 & 27 & 6841.7 & 0.004 & $3 \pm 1.8$ & $0.11 \pm 0.068$ & $0.000 \pm 0.0003$ \\
\enddata
\end{deluxetable*}

\end{document}